\documentclass[preprint,12pt]{elsarticle}

\usepackage{amsmath}
\usepackage{amssymb}
\usepackage{graphicx}
\usepackage{epstopdf, epsfig}
\usepackage{mathrsfs}
\usepackage[top=15truemm,bottom=15truemm,left=15truemm,right=15truemm]{geometry}
\usepackage{cite}
\usepackage{setspace}

\begin{document}

\begin{frontmatter}


\title{Hairpin vortex generation around a straight vortex tube in a laminar boundary-layer flow}
\author{Kazuo Matsuura\corref{cor1}}
\address{Graduate School of Science and Engineering, Ehime University,
3 Bunkyo-cho, Matsuyama, Ehime, 790-8577, Japan}
\ead{matsuura.kazuo.mm@ehime-u.ac.jp}





\begin{abstract}
Direct numerical simulation (DNS) was conducted to investigate the evolution of a straight vortex tube based on Hon \& Walker's model, which was originally proposed for a symmetric hairpin vortex tube 
(Hon, T.L. \& Walker, J.D.A, Computers \& Fluids, 20(3), pp. 343-358, 1991),  
under the shear of a laminar boundary-layer flow and its response of the near-wall flows. 
Here, the streamwise vortex tube is a lifted streamwise vortex.
The circulation and angle-to-wall of the vortex tube are varied. 
Circulation above a threshold produces multiple hairpin vortices that align in the streamwise direction over the vortex tube. 
Stability of the disturbances according to the situation of the vortex tube inside the boundary layer such as the fraction of the vortex tube inside the boundary layer in addition to the circulation and angle-to-wall is shown by a linear stability analysis. 
The contribution of linear and nonlinear terms in the dynamics of the disturbance evolution is clarified by a fully-nonlinear disturbance analysis that is conducted concurrently with the DNS.
When the circulation is sufficiently large, two unstable modes, namely the `off-wall mode' and `near-wall mode', appear. 
The presence of these two modes generates a corkscrew-like disturbance around the vortex tube, and becomes a precursor for the generation of hairpin vortices. 
The dominant structure of disturbances appearing around the vortex tube along the elapse of time is also analyzed by the proper orthogonal decomposition. 
The present DNS of the model allowing continuous variation of the characteristics of the vortex tube clarifies this new aspect of the nonlinear boundary-layer stability regarding the generation of hairpin vortices.
%

\end{abstract}


\end{frontmatter}

%
%
%
%
%
\section{Introduction}
\begin{figure}
  \centerline{\includegraphics[scale=0.6]{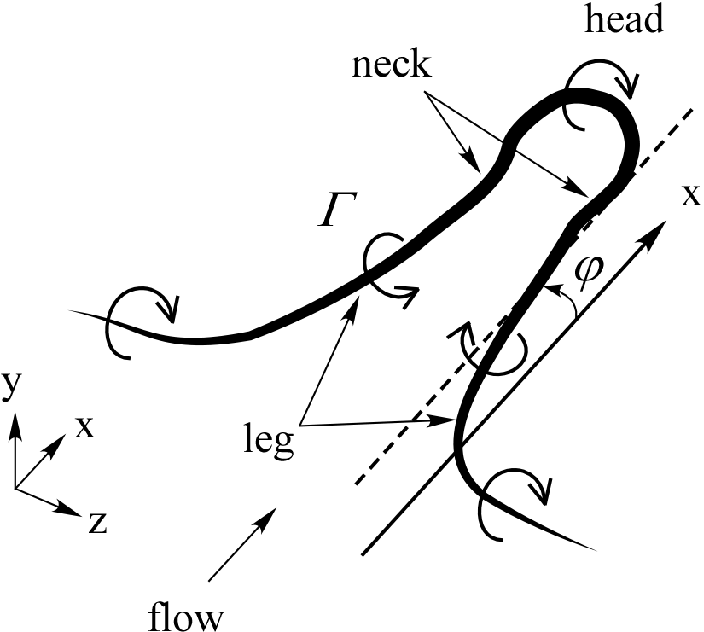}}
  \caption{Symmetric hairpin vortex, $\Gamma$ and $\phi$ are circulation and angle-to-wall, respectively.}
\label{fig:1}
\end{figure}

The hairpin vortex is considered to be the basic building block commonly observed in the dynamics of transitional and turbulent flows near a wall \citep{Smith91}. 
Interaction between vortices, the background shear flow, and the viscous flow near the surface explains the behavior of near-wall flows. 
Since the work of \citet{Theo52}, there have been many studies on hairpin/horseshoe vortices. 
Several publications provide overviews of the formation of the hairpin vortex \citep{Panton01,Adrian07,Lee08,Dennis15}. 
A typical symmetric hairpin vortex, which is shown in Fig. 1, consists of two legs, neck and head; connected legs directed away from the wall by the head constitute a warped structure of the vortex tube. 
A hairpin vortex does not necessarily have a mirror-image-symmetric shape. 
The `quasi-streamwise vortex', i.e., a vortex that consists principally of a convected section of streamwise vorticity, is treated as a `one-legged' hairpin. 
The asymmetric hairpin vortex appears more often than the symmetric vortex \citep{Robinson91a,Robinson91b}. 
Based on the literature, a hairpin vortex has many intriguing aspects in understanding the generation of turbulence. 

In computations using the Biot-Savart law, \citet{Moin86} studied the deformation of a hairpin-shaped vortex filament under self-induction and in the presence of shear, and introduced a mechanism for generating ring vortices in turbulent shear flows. 
\citet{Acarlar87} experimentally visualized the dynamics of hairpin vortices in the downstream wake of a hemisphere. 
They obtained comprehensive understanding of the behavior of hairpin vortices such as the growth of legs, head, kink on the legs, and relationship of the hairpin vortex with a low-speed streak and secondary vortices. 
\citet{Hon91} developed a stable numerical method based on the Lagrangian vortex method that can accurately compute the trajectory of a three-dimensional vortex having a small core radius. 
By this method, they showed that a two-dimensional vortex containing small three-dimensional disturbances becomes complex with subsidiary hairpin vortices forming alongside of the original hairpin vortex. 
\citet{Singer94} studied the formation and growth of a hairpin vortex in a flat-plate boundary layer and its later development into a young turbulent spot. 
Fluid injected through a slit in the wall initiated the hairpin vortex. 
They reported the formation of multiple hairpin vortex heads between stretched legs, new vortices beneath the streamwise-elongated vortex legs, and a traveling region of highly disturbed flow with an arrowhead shape similar to that of a turbulent spot. 
By direct numerical simulation (DNS) using a linear stochastic estimation procedure, 
Zhou et al. \citep{Zhou96,Zhou99} studied the evolution of a symmetric pair of quasi-streamwise vortical structures extracted from the two-point correlation tensor of the turbulent channel flow data. 
They reported that sufficiently strong hairpin vortices generate a hierarchy of secondary hairpin vortices, and the mechanism of their creation closely resembles the formation of the primary hairpin vortex. 
They addressed in detail the mechanisms for the autogeneration of hairpin vortices, and 
also proposed a parameter region for the generation of secondary hairpin vortices in a map between the strength and location of the initial hairpin vortex.
\citet{Liu11} conducted a compressible DNS for the non-linear stages of the laminar-turbulent transition. 
They discussed the coherent vortex structure appearing in the late stages of the transition and the mechanism of formation of a single vortex ring and multiple vortex rings, as well as the small length scale. 
\citet{Liu14} also explain the formation of legs and relationship with the ring-like vortices for two-legged symmetric hairpin vortices. 
At the inflow, they assumed two-dimensional waves and a pair of oblique waves in addition to the laminar boundary layer profile to reproduce transition associated with spatial array of $\Lambda$-vortices. 
\citet{Duguet12} investigated the dynamics in the region of phase space separating transitional from relaminarizing trajectories concerning to the Blasius boundary layer, and also a quasicyclic mechanism for the generation of hairpin vortex offspring. 
\citet{Cohen14} proposed a model consisting of minimal flow elements that can produce packets of hairpins. 
They showed that the three components of the model are simple shear, a counter-rotating vortex pair having a finite magnitude for the streamwise vorticity, and a two-dimensional wavy (in the streamwise direction) spanwise vortex sheet. 
\citet{Eitel15} studied the characteristics of the hairpin vortices in the turbulent boundary layers by parallel and spatially developing simulations. 
They found that shortly after the initialization only secondary hairpins are created with all rotational structures decaying later. 
They also reported that the regeneration process is rather short-lived and is not sustained once a turbulent background develops. 
With an experimental setup where fluid was injected through a narrow slot, \citet{Sabatino15} studied hairpin vortex formation in a laminar boundary layer. 
They analyzed the hairpin vortex head and legs as well as secondary hairpin vortex focusing on its circulation strength. 
Their experiment reports that the legs of the primary vortex continue to strengthen up to regeneration although the circulation strength of the head is comparable to the strength of the primary head at the time of regeneration. 
Based on the importance of circulation, they propose a threshold condition for generating a secondary vortex using an injection momentum ratio and a nondimensionalized injection time.

Although there are many studies on hairpin vortices, effects produced by varying the physical parameters associated with the hairpin vortices on its stability and near-wall dynamics have not necessarily been investigated separately and systematically. 
In this regard, the hairpin model proposed by \citet{Hon91} is particularly interesting.
Firstly, it is because the degrees of freedom of the system such as vortex curve, circulation, size, angle-to-wall, core radius of a vortex cross-section, 
and background velocity field appears to enable a continuous and systematic investigation directly. 
Secondly, it is because tertiary instability \citep{Morkovin91} is directly induced. 
However, the model has not been studied in detail within the framework of DNS. 
Recently, the present author conducted a DNS on the dynamics of single hairpin vortex and a straight vortex tube using the model, and reported the capability of generating numerous secondary hairpin vortices much more than previously reported along with some results of a linear stability analysis (LSA) of the vortex tube \citep{Matsuura16}. 
As confirmed in \citet{Matsuura16}, instability of the leg is crucial for the autogeneration of the secondary hairpin vortices. 
However, such instability was not observed in Hon and Walker's computation based on the Lagrangian vortex method \citep{Hon91}. 
On the other hand, a straight vortex tube appears as a leg of a symmetric hairpin vortices or an asymmetric one-legged hairpin vortex. 
In this study, the evolution of the straight vortex tube under the shear of background laminar boundary-layer flows and the response of near-wall flows are investigated in detail by DNS.  
The focus is on the effects of circulation and the angle-to-wall. 
In addition to the DNS, LSA, fully-nonlinear disturbance analysis(FNDA) and proper orthogonal decomposition (POD) analysis are conducted to understand disturbances triggering the hairpin vortex generated around the vortex tube, and its detailed structure, respectively. 
In Section 2, computational details such as numerical methods, computational cases, initial conditions, the procedures of the present LSA, FNDA and POD are described. 
In Section 3, evolution of a vortex tube and its initial stability, and also structure of disturbance generated around the vortex tube are investigated in detail.  
In Section 4, conclusions from this study are drawn. 
In Appendix A, the details of the present LSA are explained.
In Appendix B, the system of equations used for the FNDA is derived.

\section{Computational details}
\subsection{Numerical methods and mesh}
The governing equations are the unsteady three-dimensional fully compressible Navier-Stokes equations in general coordinates $(\xi, \eta,\zeta)$, written in conservative variables. 
The perfect gas law closes the system of equations. 
Viscosity is evaluated by Sutherland's formula and a constant Prandtl number of $Pr=0.72$ is assumed. 
The equations are solved using the finite-difference method. 
Spatial derivatives that appear in the metrics, convective and viscous terms are evaluated using the sixth-order tridiagonal compact scheme \citep{Lele92}. 
The fourth-order one-sided and classical Pad$\mathrm{\acute{e}}$ schemes are used on boundaries and at one point internal to them, respectively. 
Time-dependent solutions to the governing equations are obtained using the third-order explicit Runge-Kutta (RK) scheme. 
The time increment is constant and set to $\Delta t=42.4\times10^{-4}\delta_{in}/u_{\infty}$ in all flow fields. 
The time increment is selected considering mainly the stability of the RK scheme. 
In this study, the Courant-Friedrichs-Lewy (CFL) number, which is defined by the maximum sums of a contravariant velocity and the speed of sound scaled by the metrics as
\begin{equation}
CFL \equiv \Delta t \mathrm{max}(|U_1|+c_{loc} \sqrt{ \xi_{x_i} \xi_{x_i}}, |U_2|+c_{loc} \sqrt{ \eta_{x_i} \eta_{x_i}}, |U_3|+c_{loc} \sqrt{ \zeta_{x_i} \zeta_{x_i}})
\end{equation}
is around 0.62.  
Here, $(x, y, z)=(x_1, x_2, x_3)$ are Cartesian coordinates, $U_i$ (i=1,2,3) are the contravariant velocities, 
$\xi_i, \eta_i, \zeta_i (i=x, y, z)$ are the metrics, and $c_{loc}$ is the local speed of sound. 
In addition to the above-mentioned spatial discretization and time integration, a tenth-order implicit filtering \citep{Gaitonde00} is introduced to suppress numerical instabilities that arise from central differencing in the compact scheme. 
The filter parameters that appear in the left-hand side are set to 0.33 for $i=2$ and $i_{max}$-1 and 0.492 for $2<i<i_{max}-1$. 
Near boundaries, implicit filters of orders $p=(4,4,6,8,10)$ for $i=(2,\cdots,6)$ and $i=(i_{max}-1,\cdots,i_{max}-5)$, are used. 
Details of the present numerical method are explained in \citet{Matsuura07} and \citet{Matsuura12}, and the present method has been well validated for the prediction of transitional and turbulent subsonic flows there. 
Table I summarizes the computational settings for circulation $\Gamma$, angle-to-wall $\phi$ and domain size along with the mesh type used. 
$\Gamma$ and $\phi$, which are explained graphically in Fig. 1, associated with the straight vortex tube are defined in the next subsection. 
For $\Gamma=1.25-12.5$, the computational domain is a rectangular domain with dimensions $124.9\delta_{in}, 16.4\delta_{in}$ and $10.9\delta_{in}$ in the $x$, $y$ and $z$ directions, respectively. 
Length $\delta_{in}$ will be mentioned later.
For larger circulation $\Gamma=$18.7 and 25.0, a rectangular domain with twice the spanwise dimension is used. 
The reason for employing the extended domain is to reduce the influence of discontinuities in the generated velocity fields resulting from the large circulation of the initial vortex tubes. 
Simulations with $\Gamma$ of 1.25-25.0, and $\phi$ of $4^\circ$ and $10^\circ$ were computed except for $\Gamma$=18.7 where only $\phi=4^\circ$ is considered. 
The case of $\Gamma$=18.7 is conducted only for showing the effect of continuous variation in circulation on the eigenvalues of unstable modes.
Regarding factors determining the angle of the hairpin vortex to the wall, it is known that a hairpin vortex is under the influence of both (i) the shear flow, which tries to rotate the vortex towards the wall; and (ii) mutually induced velocities, 
which try to lift the vortex away from the wall \citep{Acarlar87}. 
At the point of formation, the angle-to-wall of the legs is less than $10^\circ$ \citep{Acarlar87}. 
\citet{Haidari94} report that the downstream portion of the hairpin vortex legs form an angle between $14^\circ$ and $32^\circ$ with the wall. 
So, the angle-to-walls used for the present study agree with the previous reports approximately. 
Three distinct meshes, i.e., Grids A, B and C are used in this study. 
Grid C is a fine mesh in which the mesh widths in all directions are half those of Grid A. 
Regarding $\Gamma$=12.5, the adequacy of Grid A has been confirmed by comparing the results for Grids A and C in terms of the generation of the train of hairpin vortices and boundary layer profiles, although not shown here. 
For larger circulation, i.e., $\Gamma$=18.7 and 25.0, Grid B is used. 
Hence, Grids A and B were mainly used in this study. 
The spanwise-averaged grid resolutions in wall units $\Delta x^+, \Delta y^+_{min}$ and $\Delta z^+$ are about 10.8, 0.362, and 4.49, respectively, and the locally maximum grid resolutions are 18.1, 0.61, and 7.6, respectively for $\Gamma=12.5$. 
Regarding the boundary condition, inflow velocity and temperature profiles are specified and pressure is extrapolated from the interior at the inlet. 
The freestream Mach number is $M_{\infty}=0.5$ which is typical subsonic Mach number for the aeronautical application but with small compressibility.
$\delta_{in}$ is the displacement thickness at a virtual place $x=300.79\delta_{in}$. 
The Reynolds number based on freestream density $\rho_{\infty}$, velocity $u_{\infty}$ and viscosity $\mu_{\infty}$, and $\delta_{in}$, i.e., $Re=\rho_{\infty}u_{\infty}\delta_{in}/\mu_{\infty}$, is 1000. 
The starting coordinate of all the computational domains is $(x,y,z)=(404.45\delta_{in}, 0, 0)$. 
The inflow profile is obtained by solving the steady laminar boundary layer equation without the pressure gradient term \citep{Cebeci74}. 
At the outflow and upper boundaries, nonreflecting boundary conditions around the freestream static pressure $p_{\infty}$ are used \citep{Kim00}. 
At the wall, no-slip, isothermal wall condition is imposed. 
The ratio of the wall temperature to the freestream static temperature is 1.0. 
Periodicity is imposed in the spanwise direction.

\begin{figure}
\begin{minipage}{1.\hsize}
\begin{center}
\includegraphics[width=100mm]{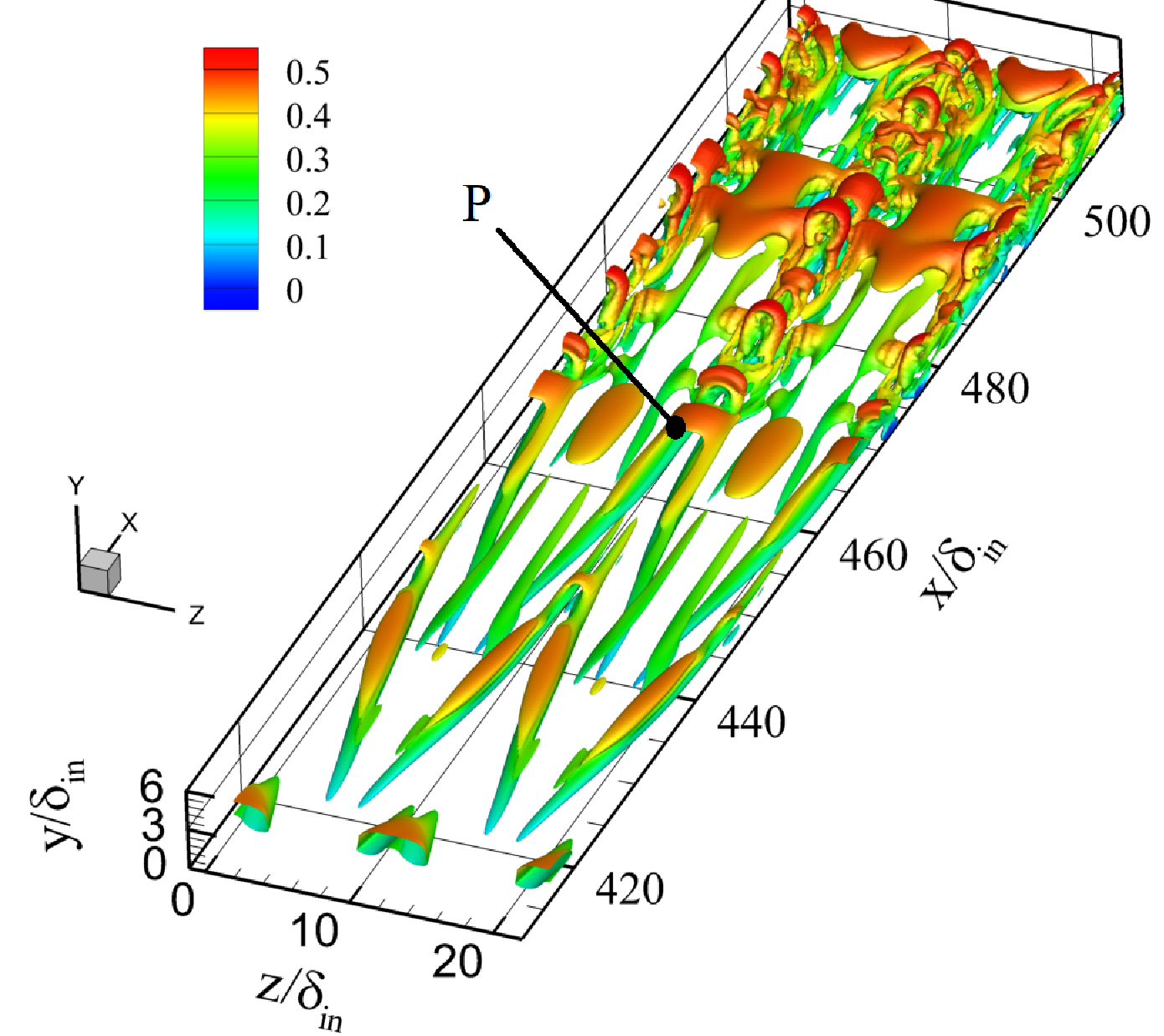}
\end{center}
(a)~Instantaneous hairpin vorticies visualized by the iso-surfaces of the second invariance of the velocity gradient tensor, See Sec. 3.1.1 for its visualization method.
The colour signifies streamwise Mach number.
The $x$ and $z$ coordinates of Point P are $x=465.6\delta_{in}$ and $z=9.7\delta_{in}$, respectively.
\label{fig:14}
\end{minipage}\\
\begin{minipage}{0.5\hsize}
\begin{center}
\includegraphics[width=65mm]{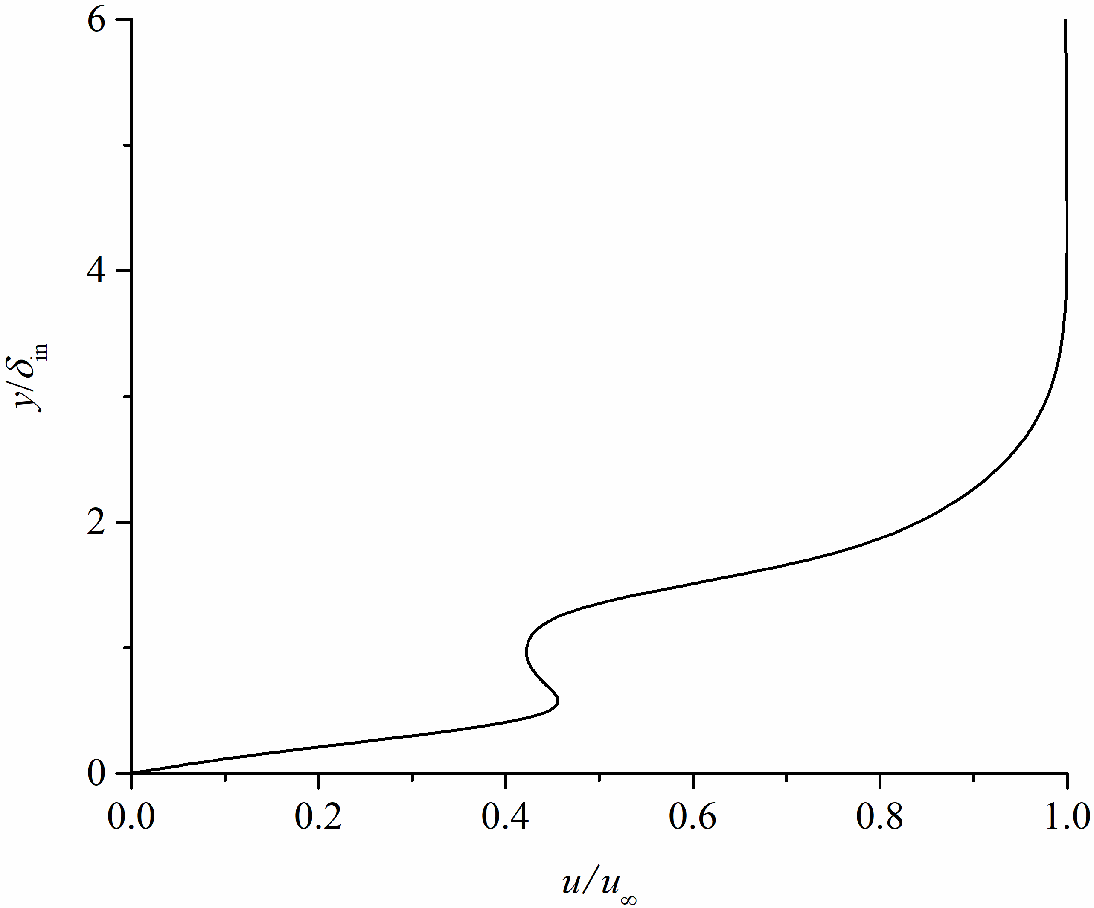}\\
(b)~The wall-normal profile of instantaneous streamwise velocity $u$ sampled at Point P.
\end{center}
\label{fig:16}
\end{minipage}
\begin{minipage}{0.5\hsize}
\begin{center}
\includegraphics[width=65mm]{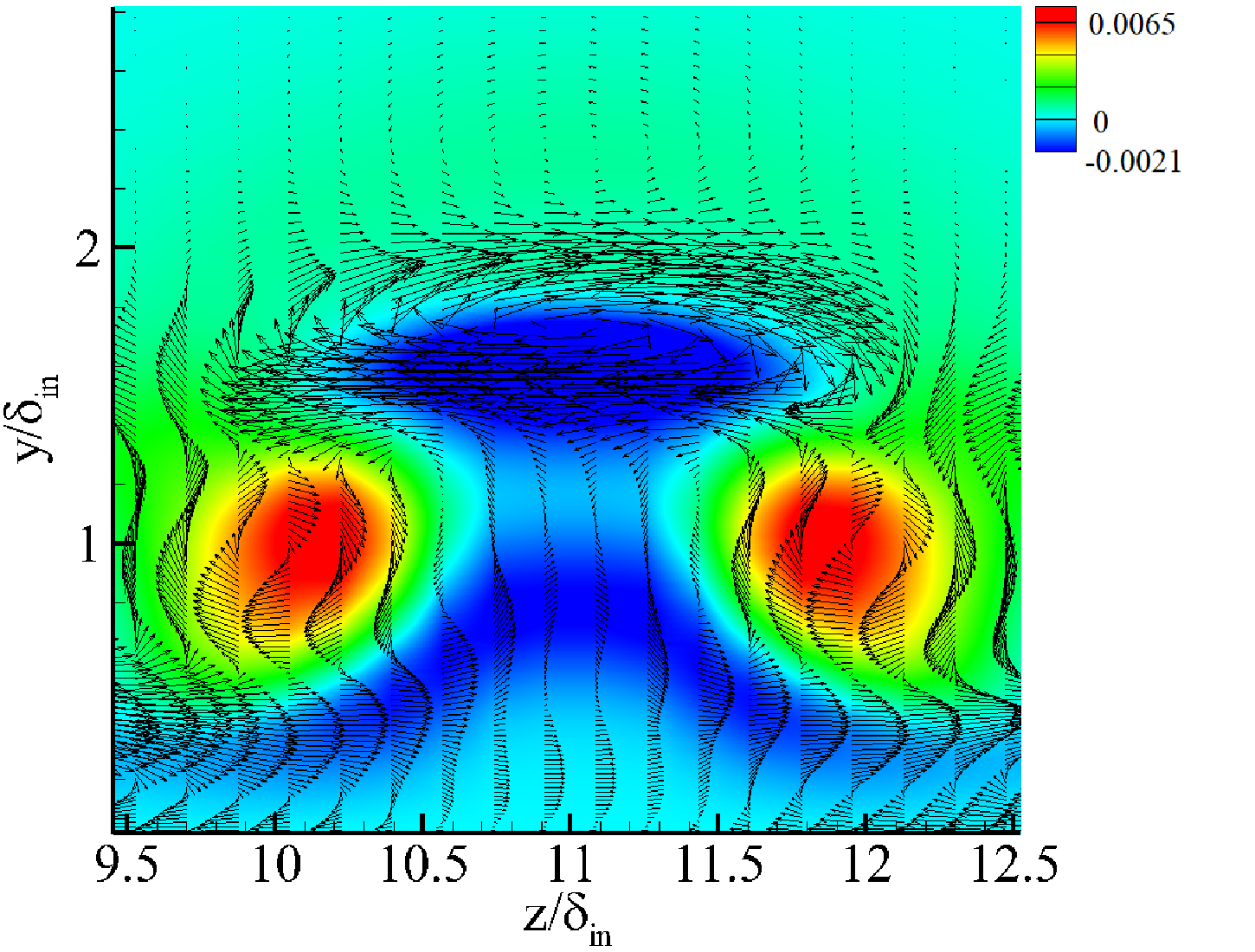}
\end{center}
(c)~Cross section of the hairpin vortex around $x=465.6\delta_{in}$. The colour signifies the second invariance of the velocity gradient tensor, and
an arrow shows a vector of the left hand side of the vorticity transport equation, projected on the plane. 
See Sec. 3.1.1 for its equation.
\label{fig:15}
\end{minipage}\\
\begin{center}
\caption{Hairpin vortices computed in the DNS of laminar-turbulent transition on a flat plate without free-stream turbulence at $M_{\infty}=0.5$ \citep{Mat16b}.}
\end{center}
\end{figure}

\subsection{Generation of initial fields and computational cases}
\begin{table}
\caption{Computational settings: the initial length of the vortex tube is $24\delta_{in}$, and the core radius is $0.25\delta_{in}$,
$\delta_{in}$: the displacement thickness of the velocity profile, $x$:~streamwise, $y$:~wall-normal, $z$:~spanwise}
\label{tab:1}
\begin{center}
\begin{tabular}{|l|l|l|l|l|}
\hline
Circulation $\Gamma$ & Angle-to-wall $\phi$ & Domain size $L_x \times L_y \times L_z$ & Grid & Grid points in $x,y,z$ dir.\\ \hline
1.25,3.13,6.25 & $4^\circ,10^\circ$ & $124.9\delta_{in}\times16.4\delta_{in}\times10.9\delta_{in}$ & A & A:$301\times 210\times64$ \\
\cline{1-1}\cline{4-4}
12.5 & & & A,C & C:$601\times 417\times128$ \\
\hline
18.7 & $4^\circ$ & $124.9\delta_{in}\times16.4\delta_{in}\times21.8\delta_{in}$ & B & $301\times 210\times128$ \\
\cline{1-2}
25.0 & $4^\circ,10^\circ$ & & &  \\
\hline
\end{tabular}
\end{center}
\end{table}

As an initial condition at $t=0$, the velocity vectors of a straight vortex tube of finite length are simultaneously superposed on the steady laminar boundary layer under a zero pressure gradient, i.e., 
\begin{equation}
(\rho,u,v,w,T)^T|_{t=0}=(\rho,u,v,w,T)^T|_{\mathrm{steady}}+(0,u,v,w,0)^T|_{\mathrm{vortex~tube}}
\end{equation}
Superscript `T' means transpose of a vector. 
Subscript `steady' means the steady and spatially growing laminar boundary layer profiles \citep{Cebeci74}, and `vortex tube' means the velocity vector fields induced by the straight vortex tube. 
Conserved variables at $t=0$ are constructed from the variables. 
This operation can also be considered as an addition of an impulsive force in the compressible Navier-Stokes equations. 
Explicit background turbulence is not considered. 

The velocity fields induced by the vortex tube are generated based on the algorithm of \citet{Hon91} represented by Eq. (2.4)-(2.6) mentioned later. 
The algorithm is a modification of Moore's algorithm, which was reported to exhibit strong numerical instability for small values of the vortex core radius \citep{Moore72}. 
Basically, the velocity vector field induced by the vortex $\pmb{u}=(u,v,w)^T|_{\mathrm{vortex~tube}}$ at any arbitrary location in the space $\pmb{X}_0$ is given by the Biot-Savart Law:
\begin{equation}
\pmb{u}(\pmb{X}_0)=-\displaystyle \frac{\Gamma_d}{4 \pi} \int_C \displaystyle \frac{(\pmb{X}_0-\pmb{X})}{|\pmb{X}_0-\pmb{X}|^3} \times d\pmb{X},
\end{equation} 
where contour $C$ is a curve defining a vortex tube, $\Gamma_d$ is circulation about the vortex core. 
However, when $\pmb{X}_0$ is close to the curve $C$, the above equation becomes singular because it is interpreted as an outer inviscid solution, which is not uniformly valid near the vortex core. 
To circumvent this singularity, the above equation is replaced with
\begin{equation}
\pmb{u}(s_0,t)=\displaystyle \frac{\Gamma_d}{4 \pi} \int_C R(s,s_0) ds+
\displaystyle \frac{\Gamma_d}{4 \pi} \displaystyle (\frac{\partial \pmb{X}}{\partial s})_0 \times (\frac{\partial^2 \pmb{X}}{\partial s^2})_0 \int_C P(s) ds,
\end{equation}
\begin{equation}
P(s)=\displaystyle \frac{\frac{1}{2}(s-s_0)^2}{[(s-s_0)^2(\displaystyle \frac{\partial \pmb{X}}{\partial s})^2_0+\mu^2]^{3/2}}.
\end{equation}
The parameter $s$ is a Lagrangian coordinate of the curve $C$. 
The integrand of $R(s,s_0)$ is given in \citet{Hon91}. 
The artificially induced small parameter $\mu$ in the denominator of the integrand is set to $\mu=\mathrm{exp}(-3/4)a$, with core radius $a$ set to $a=0.25\delta_{in}$. 

In this study, the initial coordinates of the vortex tube are defined as
\begin{equation}
\pmb{X}(s)=A s(i) \left(
    \begin{array}{c}
      \mathrm{cos} \phi \\
      \mathrm{sin} \phi \\
      0
    \end{array}
  \right)+
  \left(
    \begin{array}{c}
      x_c \\
      y_c \\
      z_c
    \end{array}
  \right),~i \in (0,\cdots,N).
\end{equation}
Here, $x_c=434\delta_{in}$, $y_c=0.4\delta_{in}$, and $z_c=5.5\delta_{in}$, and $A$ is the length of the vortex tube, i.e., $A=24\delta_{in}$. 
$\{s(i)=i/N; i\in(0,\cdots,N)\}$ measures normalized distance along the curve $C$. 
To generate velocity fields resulting from the vortex tube, the straight vortex tube is approximated by 1,001 representative points, i.e., $N=1000$. 
Although not shown here, the sufficiency of the number of the representative points is confirmed by comparing the results of $N=500$ and 1000 
regarding the evolution of the vortex tube for the case of $\Gamma$=12.5 and $\phi=4^\circ$.     
Grid points are taken as $\pmb{X}_0$. 
The vortex tube is initially placed parallel to the $x$ axis in all the instances. 
The non-dimensional circulation $\Gamma$ is defined as $\Gamma=\Gamma_d/(\delta_{in}u_{\infty})$.

The length, angle and location of the vortex tube are determined based on a previous DNS by the author of laminar-turbulent transition on a flat plate without free-stream turbulence at $M_{\infty}=0.5$ \citep{Mat16b}.
Its computational condition is same as \citet{Liu11}.
The inlet location of the computation is $x=300.79\delta_{in}$ where the displacement thickness of the velocity profile is $\delta_{in}$.
In the DNS, the inflow profile is given as
\begin{equation}
q=q_{lam}+A_{2d}q'_{2d}\mathrm{exp}{i(\alpha_{2d} x-\omega t)}+A_{3d}q'_{3d}\mathrm{exp}{i(\alpha_{3d} x \pm \beta z-\omega t)}.
\end{equation}
Here, $q$ denotes $u,v,w,p$ and $T$, and $q_{lam}$ is a self-similar solution for a two-dimensional compressible laminar flat plate boundary layer under zero pressure gradient \citep{Cebeci74}.
The streamwise wavenumber, spanwise wavenumber, frequency, and amplitude are
\begin{eqnarray}
\alpha_{2d}=0.29919-i5.09586 \times 10^{-3},~\beta=0.5712,\nonumber\\
\omega=0.114027,~A_{2d}=0.03,~A_{3d}=0.01,
\end{eqnarray}
respectively \citep{Liu11}. 
The length of the vortex tube $A=24\delta_{in}$ is close to the wavelength of the two-dimensional TS wave, i.e., $2\pi/\alpha_{2d}$.   

Figure 2 shows hairpin vortices reproduced in the DNS.
Part (a) shows the instantaneous hairpin vortices, part (b) the wall-normal profile of instantaneous streamwise velocity $u$ at Point P, i.e., $x=465.6\delta_{in}$ and $z=9.7\delta_{in}$, and 
part (c) a cross section of the hairpin vortex around $x=465.6\delta_{in}$.
The arrow shows a vector of the left hand side of the vorticity transport equation, projected on the plane. 
The method of the vortex visualization and the vorticity transport equation are explained in Sec. 3.1.1. 
The point P corresponds to a near-head region on the hairpin leg where secondary hairpin vortices are generated.
As found in part (b), the $u$ profile has an S-shaped deformation due to the swirl of the leg.
In part (c), the vectors change signatures alternately in the $y$ direction near the cross section of the hairpin leg, e.g., around $z=10\delta_{in}$.
These properties are reproduced when the straight vortex tube is placed in a laminar boundary layer as shown in Sections 3.1.2 and 3.2.1.

The presence of the vortex tube in the laminar boundary layer is intrinsically non-stationary. 
Also, from a physical perspective, the vortex curve does not vanish inside a flow. 
In this regard, using Navier-Stokes equations, the present initial condition is only a computational artifact. 
However, generating a single inclined vortex tube itself in a shear flow is realistic
as shown in \citet{Jukes13,Zhang98,Eibeck87,Shizawa92}.
They used a plasma actuator, an air jet and a half-delta-wing vortex generator for the generation. 
The initial inconsistency is complemented in the process of computation. 
During the process, two problems are of concern. 
One is satisfying the no-slip condition, which is achieved with the passage of time as also found from the velocity profiles shown later in Section 3.1.1, and the other is the effect of the artificial snipping-off condition near the two ends of the finitely long vortex tube. 
To investigate its effects on the present analysis, separate computations were performed with vortex tubes having intentionally attenuated $\Gamma$ near the ends of the vortex tubes. 
However, except for the ends, this attenuation did not change the velocity structure of the boundary layer near the interior region of the vortex tube. 
Hence, the present investigation into the mechanism generating the hairpin vortex is, within the time span considered, robust against this snipping-off condition and the finite-time evolution of the vortex tube.

One may think of differences in unstable behaviors between a streamwise vortex and a streamwise streak, which are often discussed in the same context of a turbulence self-sustaining cycle \citep{Swear87,Hamilton95,Waleffe97}. 
A pure streamwise streak has a velocity profile symmetric in the spanwise direction. 
Because unstable shear layers are created both in its wall-normal and spanwise directions \citep{Asai02,Brandt07} based on the shape, anti-symmetric sinuous and symmetric varicose modes appear. 
In contrast, the present vortex tube has swirl of only one direction, and therefore is not symmetric in the spanwise direction. 
While sinuous variation both in the streamwise and spanwise directions, and also corkscrew disturbances are observed in the present DNS as mentioned later in Section 3, the varicose mode is not confirmed. 
Thus, the presence of swirl along an elongated structure makes the present disturbance development different from that of a streamwise streak. 

\subsection{Linear stability analysis}
From the DNS results, to be presented in Section 3, the initial disturbances triggering the hairpin vortices appear near the upper and lower edges of the vortex tube, i.e., 
a shear layer forms above and beneath the vortex tube because of vortex advection in the boundary layer. 
To investigate the initial disturbances, a LSA is conducted. 
The formulation is derived from the original compressible Navier-Stokes equation. 
Instantaneous values are decomposed into a mean and a fluctuation quantity. 
Substituting these into the non-dimensional form of the governing equations and neglecting the quadratic terms of fluctuations yields the linearized perturbation equations. 
From the DNS, instantaneous boundary layer profiles of $(\rho, u, w, T)$ are extracted at a streamwise position around the center of the vortex tube in the spanwise direction. 
The profiles are used for mean base profiles in the LSA. 
Employing the boundary layer assumption, mean pressure $p$ is constant across the layer and is equal to $1/(\gamma M_{\infty}^2)$. 
Here, $\gamma$ is the ratio of specific heats.
As explained later, flow is assumed to be locally parallel; also the mean velocity fields are uniform in the spanwise direction neglecting the circumferential mode of the vortex tube. 
In addition, the disturbances are assumed to be harmonic, i.e.,
$\tilde{\phi}=\tilde{\phi(y)}\mathrm{exp}\{i(\alpha x+\beta z-\omega t)\}$
where $\alpha$ and $\beta$ are the streamwise and spanwise wavenumbers, respectively, and $\omega$ is the frequency. 
%
In this study, a temporal LSA is conducted. 
Here, the variation of disturbances as time is the concern. The complex frequency ω is evaluated when α and β are given real values. 
The linearized equations along with the boundary conditions are formulated as a matrix eigenvalue problem, and ω becomes the eigenvalue. 
The disturbances are unstable and can grow exponentially in time if $\omega_i>0$. 
Its procedure is explained in detail in \citet{Malik90}. 
Regarding the non-dimensionalization of the compressible linear stability equation, $\delta_{in}$ is used for the reference length. 
In this study, the accuracy of the finite difference scheme used to evaluate the derivative is different from \citet{Malik90}, and the difference is explained in Appendix A. 
A global method is used to select candidate eigenvalues for the unstable modes, i.e, eigenvalues $\omega$ such that its imaginary part $\omega_i>0$, and then a local method is used with Newton iteration to obtain accurate eigenvalues. 
While the second-order finite-difference method is used in the global method as in Appendix A, 
the fourth-order scheme is used in the local method.
Although there is no inherent problem in the global method in obtaining the eigenvalues, 
the local refinement is expected to improve the accuracy of the eigenvalues.
However, because computations diverged for some eigenvalues during Newton iteration,
the candidates are used to understand the trend of the eigenvalues.     
The accuracy of the program developed is successfully validated against the first five modes (phase velocity) of the compressible stability equations in the incompressible limit ($M_{\infty}=10^{-6}$), and
also the temporal stability problem at Mach number of 0.5 given in \citet{Malik90}.
The stability analysis attempted here serves only as a tool to obtain a qualitative understanding of the mechanism leading to the formation of secondary hairpin vortices because 
(a) the vortex tube is inclined, there is background shear, and the mean profiles are non-parallel, \\
(b) the flow has circulation, there is a wall beneath the vortex tube, the background flow field is not axisymmetric, not even at a cross section, and the vortex tube is not uniform in the spanwise direction,\\ 
(c) the boundary layer profiles used for the mean quantities were extracted from instantaneous fields of the DNS, and\\ 
(d) the vortex tube has finite length. \\
%
%
Based on the DNS results shown in Section 3.1.1, the dominant initial disturbances triggering the hairpin vortices around the vortex tube has the clear characteristic of a streamwise variation. 
Thus, local analysis of wall-normal profiles of streamwise velocity was considered, neglecting the spanwise component of the base flow and assuming that $\beta=0$.
%
Considering a non-zero mean spanwise velocity $W$ in the present stability analysis for the case of $\Gamma=12.5$ confirms that
the largest growth rates of disturbance for the variation of $\alpha$ monotonously decay as $\beta$ is increased, which means that disturbances with $\beta=0$ are most likely to appear.    
Moreover, the boundary layer profiles were extracted at multiple positions under each setting. 
In this regard, the present LSA corresponds to quasi-parallel analysis with $W=0$ and $\beta=0$. 

\subsection{Fully-nonlinear disturbance analysis}
The LSA mentioned above is attractive in that it is simple, and also that eigenvalues are obtained, and amplification/attenuation can be judged by them.
However, because the analysis is based on many assumptions, its validity is lost when their effects are large.
In order to overcome the inherent limitations of the LSA, and quantify the contribution of linear and nonlinear terms in the dynamics of disturbance evolution, the FNDA is conducted. 
In this analysis, the time evolution of flow deviation from a base flow profile $(\bar{\rho(y)}, \bar{u(y)}, \bar{v(y)}, \bar{w(y)}, \bar{T(y)})$ is computed.
The profile is a set of instantaneous wall-normal profiles of $(\rho, u, v, w, T)$ taken at a $(x,z)$ location.    
The deviation is represented by the independent variables of fluctuation $(\tilde{\rho},\tilde{u},\tilde{v},\tilde{w},\tilde{T})$, and 
the right hand sides (RHSs) of the partial derivatives of them with respect to time, i.e.,
$\frac{\partial}{\partial t}(\tilde{\rho},\tilde{u},\tilde{v},\tilde{w},\tilde{T})^T$ are evaluated.
The RHS terms determine the time increments of the independent variables.
A system of the time evolution equations is derived in Appendix B.
Dominant terms governing the vortex dynamics at each space-time location are evaluated concurrently with the DNS.
In order to single out dominant terms in the $y$ range [1.0,1.5] where the largest growth of streamwise velocity fluctuation is observed in Section 3.2, all averaged values of the terms within the range are compared.
In the RHSs of Eq. (B9)-(B12), time derivative terms such as 
$\frac{\partial \tilde{u}}{\partial t}, \frac{\partial \tilde{v}}{\partial t}, \frac{\partial \tilde{w}}{\partial t}, \frac{\partial \tilde{T}}{\partial t}, \frac{\partial \tilde{p}}{\partial t}$
appear.
These terms are approximated as
\begin{equation}
\frac{\partial f}{\partial t}=\frac{3f^{n}-4f^{n-1}+f^{n-2}}{2 \Delta t},~~f \in \{\tilde{u},\tilde{v},\tilde{w},\tilde{T},\tilde{p}\}.
\end{equation}
Here, superscripts $n, n-1$ and $n-2$ denote the present timestep, one timestep before the present and two timesteps before the present, respectively.
         
\subsection{Proper orthogonal decomposition}
To understand the dominant spatial structure of disturbance generated around the vortex tube in an instantaneous flow field, a snapshot POD analysis proposed by \citet{Siro87} is conducted in Section 3. 
Suppose the series of data
$\{S_{x_j}(k,l); k=k_s, \cdots, k_e, l=l_s, \cdots, l_e\}$ for $j=j_s,\cdots,j_e$ are given.
Here, $j,~k$ and $~l$ show node indices for $\xi,~\eta$ and $\zeta$ coordinates, respectively.
The method of collecting the series of data in an instantaneous flow field is explained in Section 3.2.2.
The variation along the vortex tube is orthogonally decomposed based on the $M$ snapshots of cross-sections
$\{S_{x_j}(k,l); k=k_s, \cdots, k_e, l=l_s, \cdots, l_e\}$ $(j=j_s,\cdots,j_e)$ at each time.
Here, $M=j_e-j_s+1$. 
In the method, fluctuation $\pmb{\mathscr{S}}''_{x_j}(\cdot)$($j=j_s,\cdots,j_e$) is expanded by a set of 
eigenfunctions $\{\pmb{\phi}_i(\cdot)\}^M_{i=1}$ and the corresponding coefficients $\{a_i(j)\}^M_{j=1}$ as
\begin{equation}
\pmb{\mathscr{S}}''_{x_j}(\cdot)=\sum_{i=1}^M a_i(j)\pmb{\phi}_i(\cdot),~j=j_s,\cdots,j_e.
\end{equation}
Here,
\begin{equation}
\pmb{\mathscr{S}}''_{x_j}(\cdot)=S_{x_j}(k,l)-(\sum_{j=j_s}^{j_e} S_{x_j}(k,l)/M),~k=k_s, \cdots, k_e, l=l_s, \cdots, l_e\
\end{equation}
and, $\pmb{\mathscr{S}}''_{x_j}(\cdot)$ and $\pmb{\phi}_i(\cdot)$ are 
$\pmb{\mathscr{S}}''_{x_j}(k,l)$ and $\pmb{\phi}_i(k,l)$ respectively in which the combination of indices $(k,l)$ are converted to a single index.
$\pmb{\mathscr{S}}''_{x_j}(\cdot)$ and $\pmb{\phi}_i(\cdot)$ are $N$-dimensional vectors and, $N$ is the total number of 
grid points that consist of a cross section, i.e.,
\begin{equation}
N=(k_e-k_s+1)\times(l_e-l_s+1).
\end{equation}
Further details of the procedure to obtain the eigenfunctions $\pmb{\phi}_i(\cdot)$ 
and the coefficients $a_i(j)$ are explained in \citet{Matsuura12}.

\section{Results and discussion}

\begin{figure}
\begin{minipage}{0.32\hsize}
\centering
\includegraphics[width=38mm]{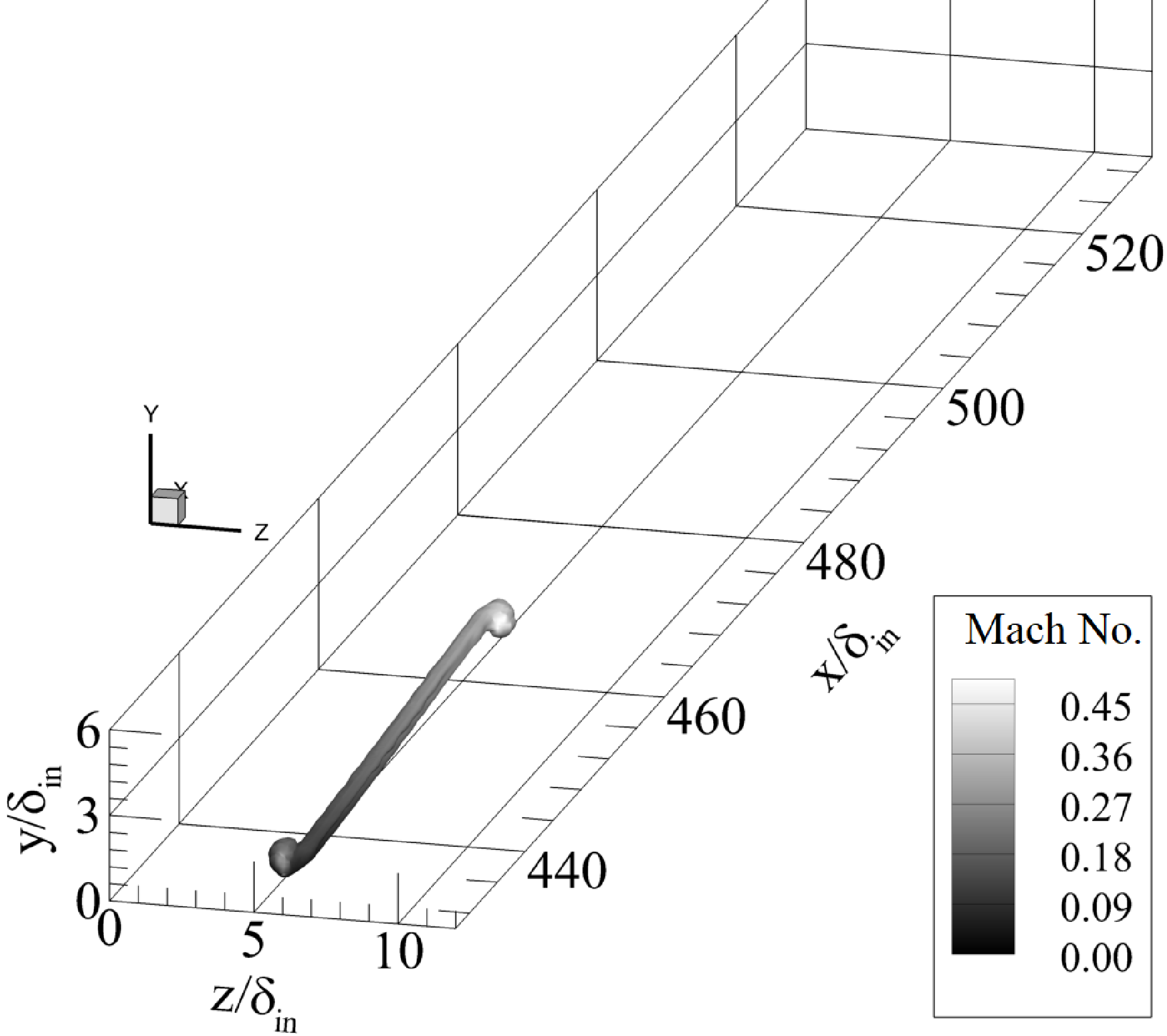}
\label{fig:2}
\end{minipage}
\begin{minipage}{0.32\hsize}
\centering
\includegraphics[width=38mm]{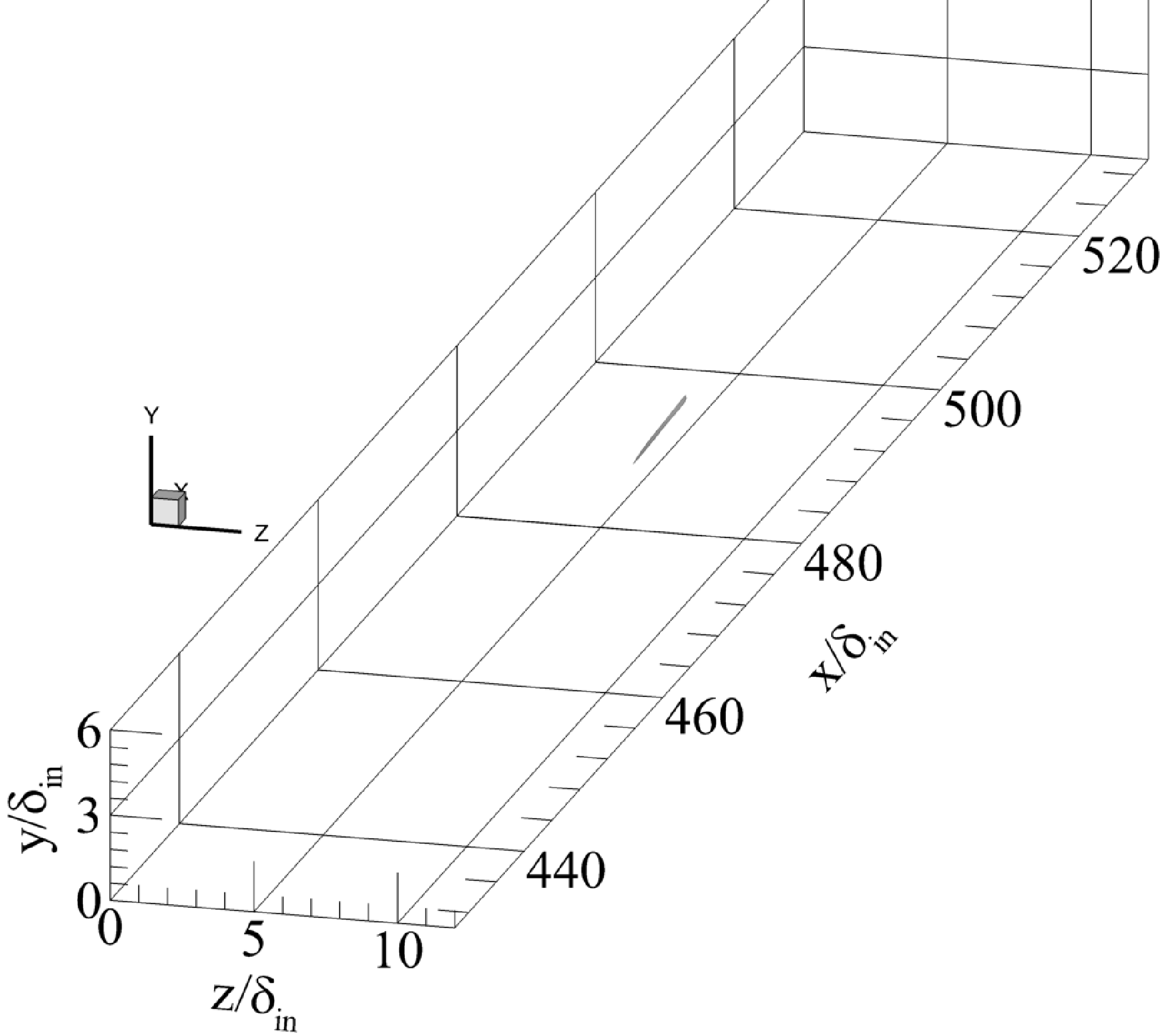}
\label{fig:3}
\end{minipage}
\begin{minipage}{0.32\hsize}
\centering
\includegraphics[width=38mm]{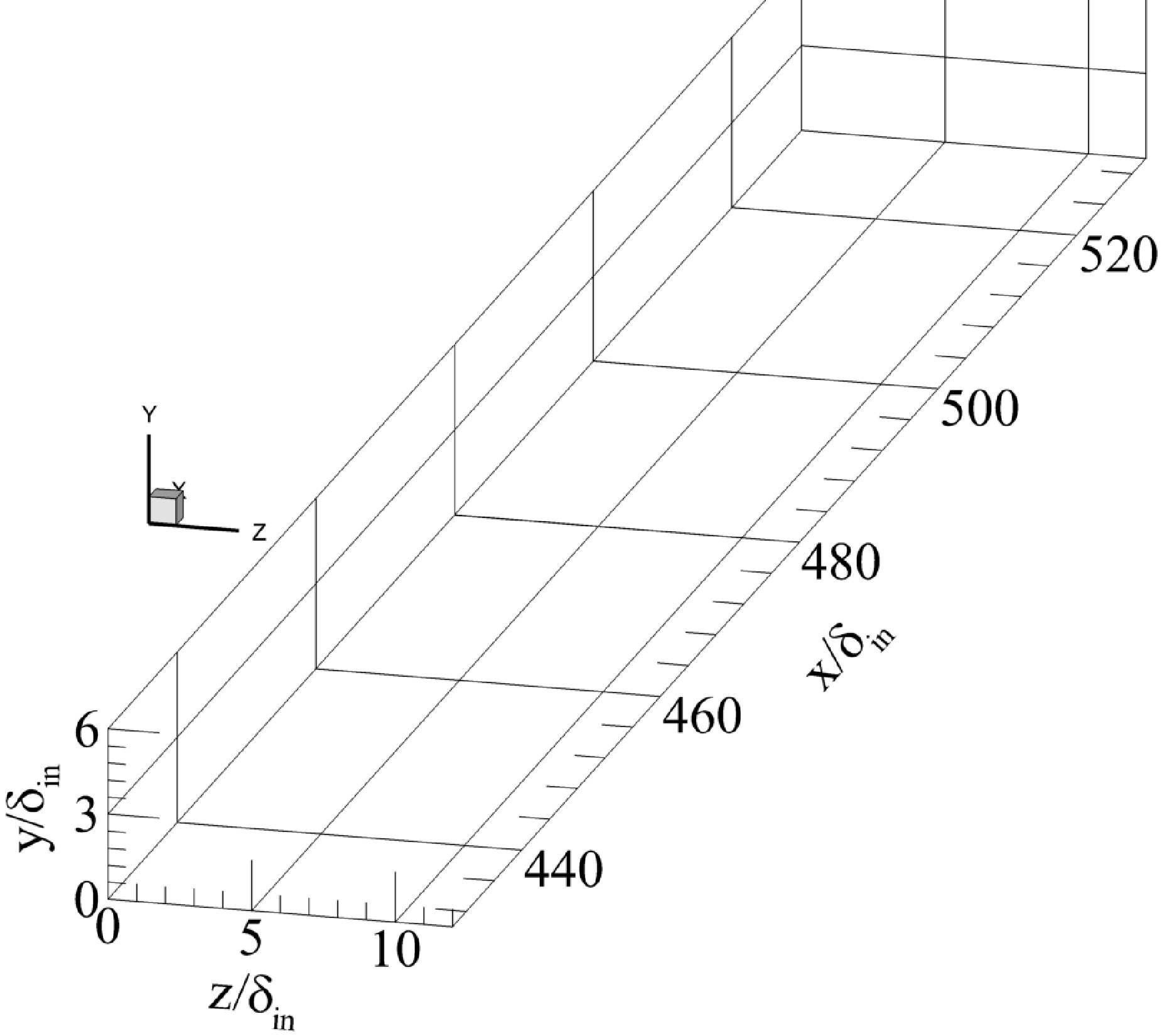}
\label{fig:4}
\end{minipage}
\centering
(i)~$\Gamma=1.25~(Q_{max}^\ast=2.32\times10^{-3})$\\
%
%
\begin{minipage}{0.32\hsize}
\centering
\includegraphics[width=38mm]{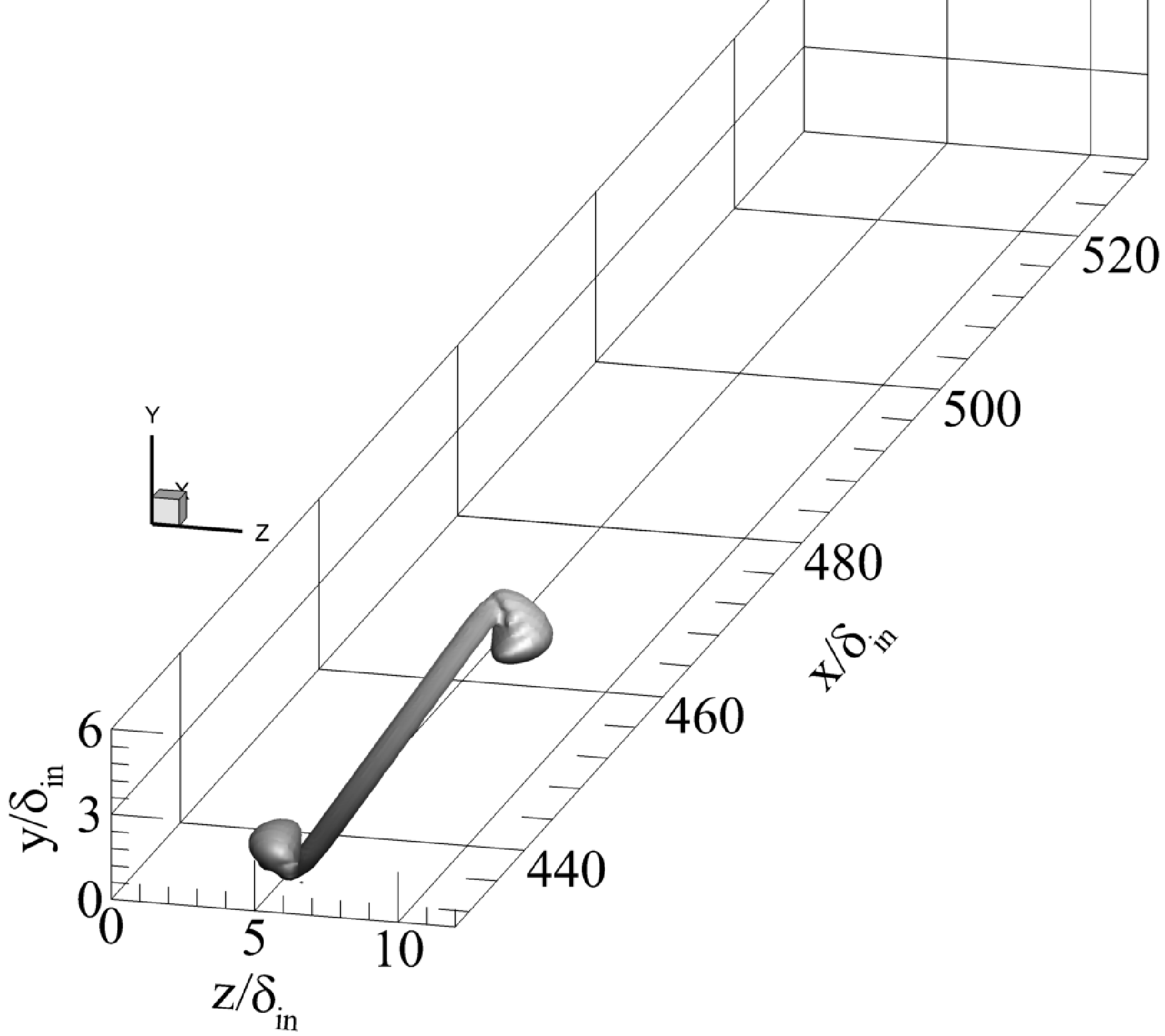}
\label{fig:5}
\end{minipage}
\begin{minipage}{0.32\hsize}
\centering
\includegraphics[width=38mm]{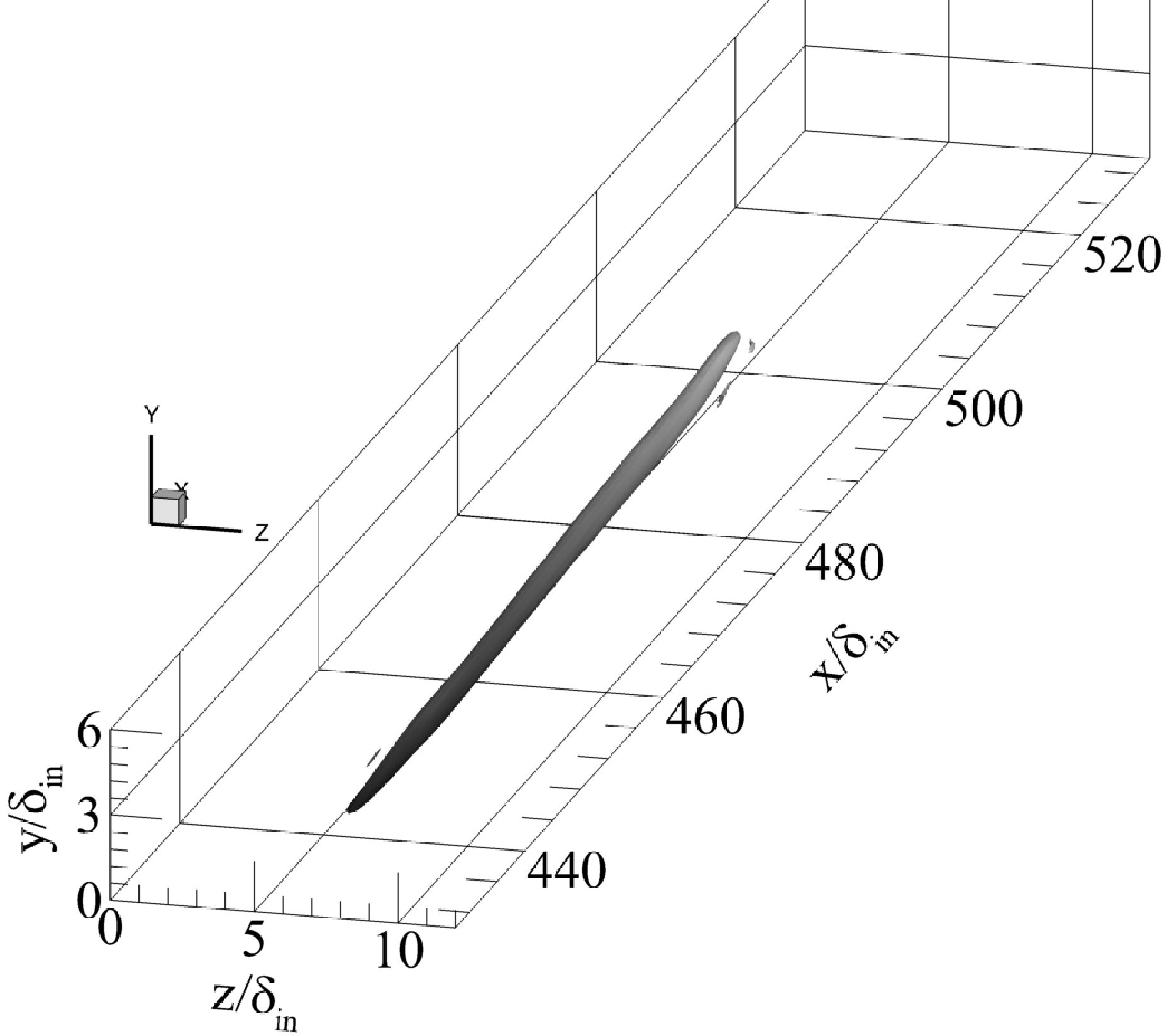}
\label{fig:6}
\end{minipage}
\begin{minipage}{0.32\hsize}
\centering
\includegraphics[width=38mm]{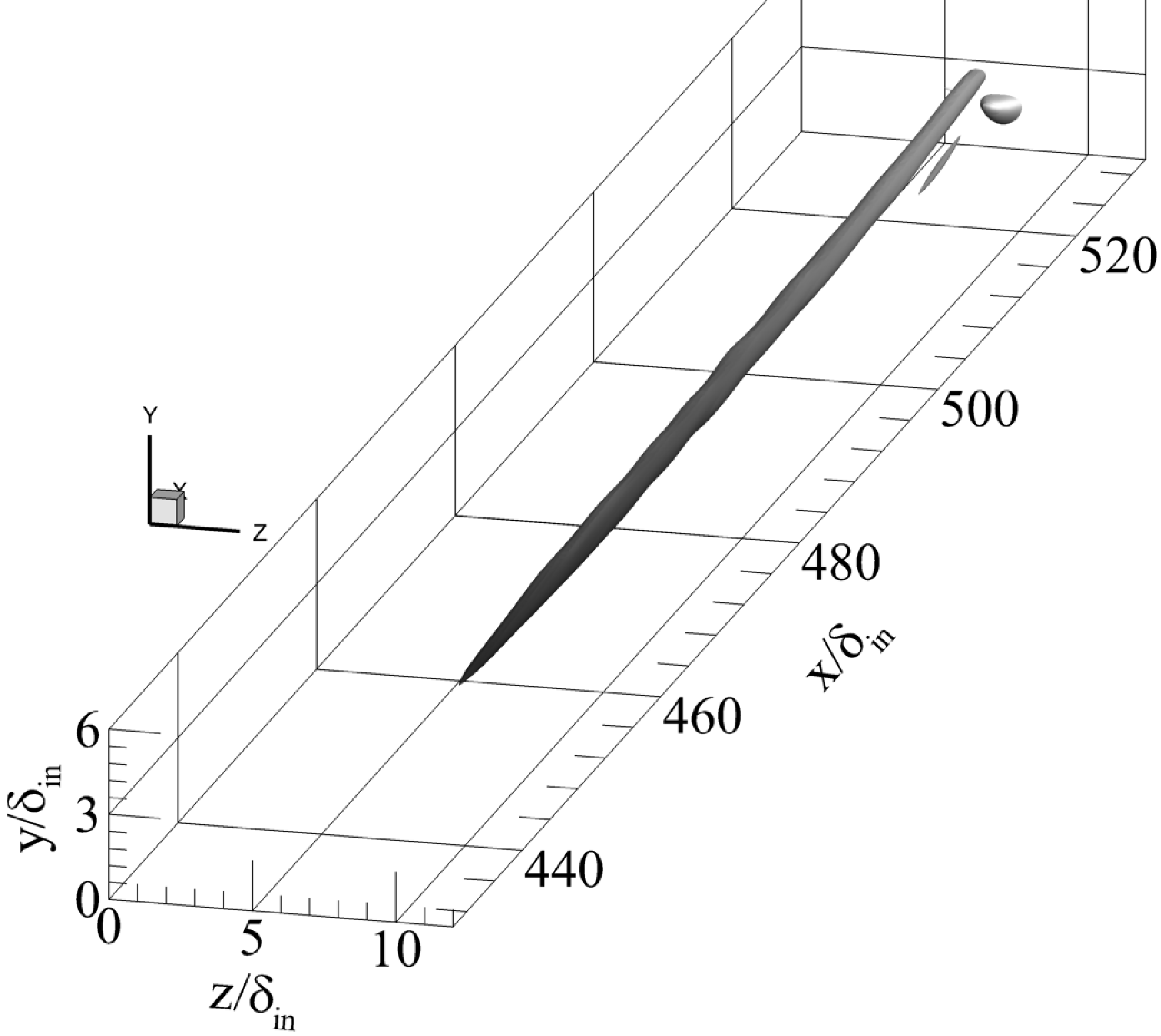}
\label{fig:7}
\end{minipage}
\centering
(ii)~$\Gamma=6.25~(Q_{max}^\ast=3.26\times10^{-2})$\\
%
\begin{minipage}{0.32\hsize}
\centering
\includegraphics[width=38mm]{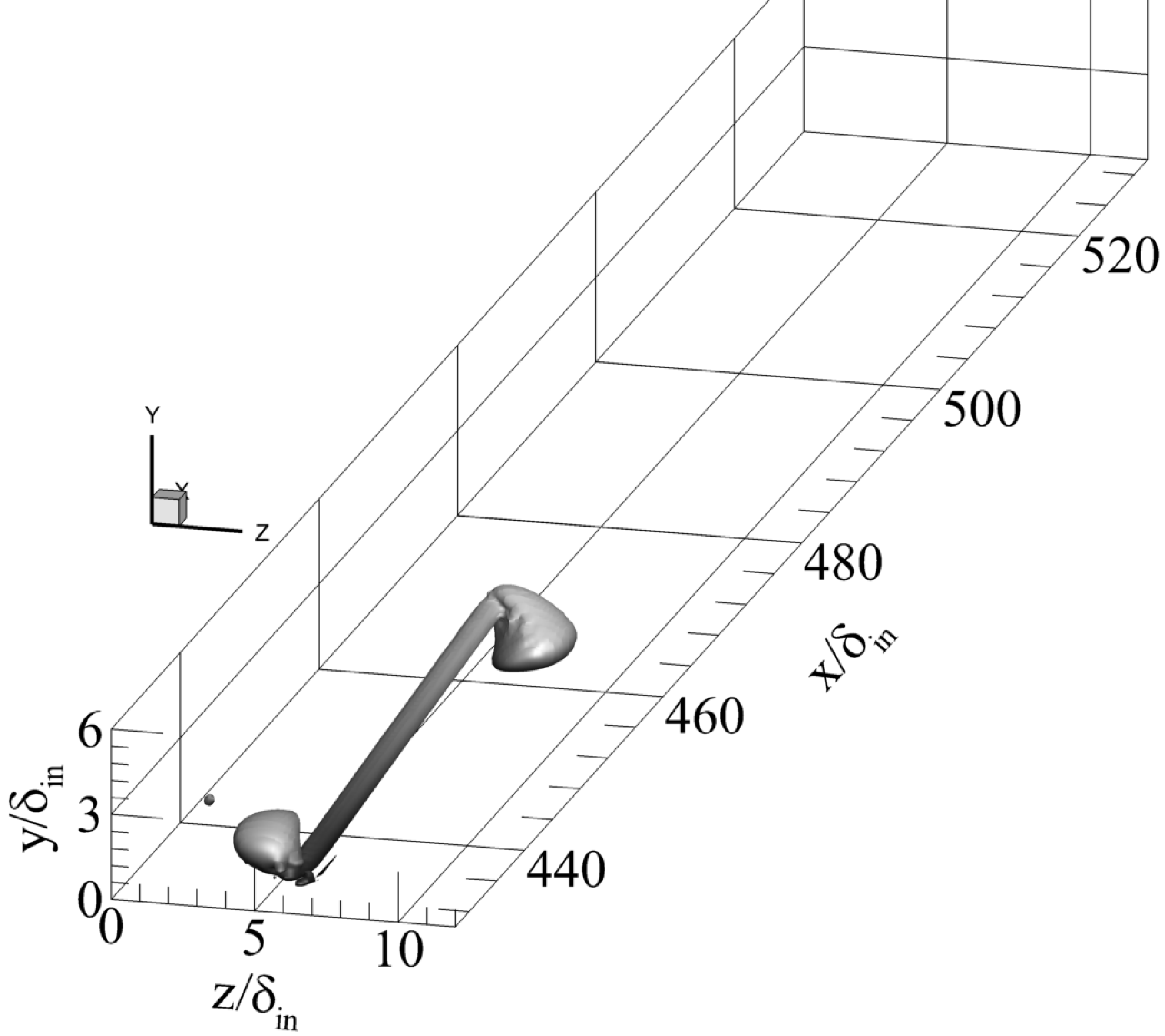}
\label{fig:5}
\end{minipage}
\begin{minipage}{0.32\hsize}
\centering
\includegraphics[width=38mm]{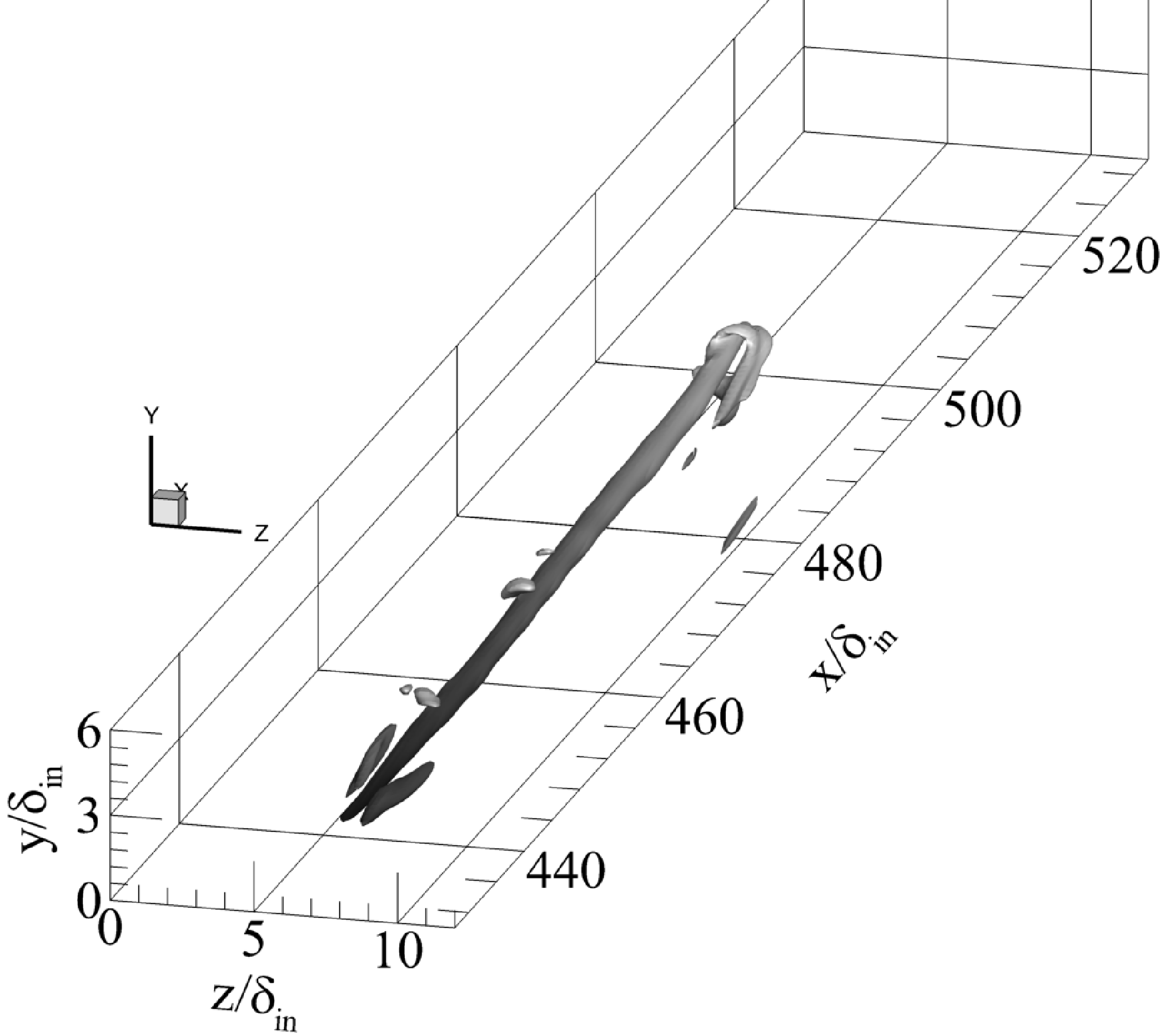}
\label{fig:6}
\end{minipage}
\begin{minipage}{0.32\hsize}
\centering
\includegraphics[width=38mm]{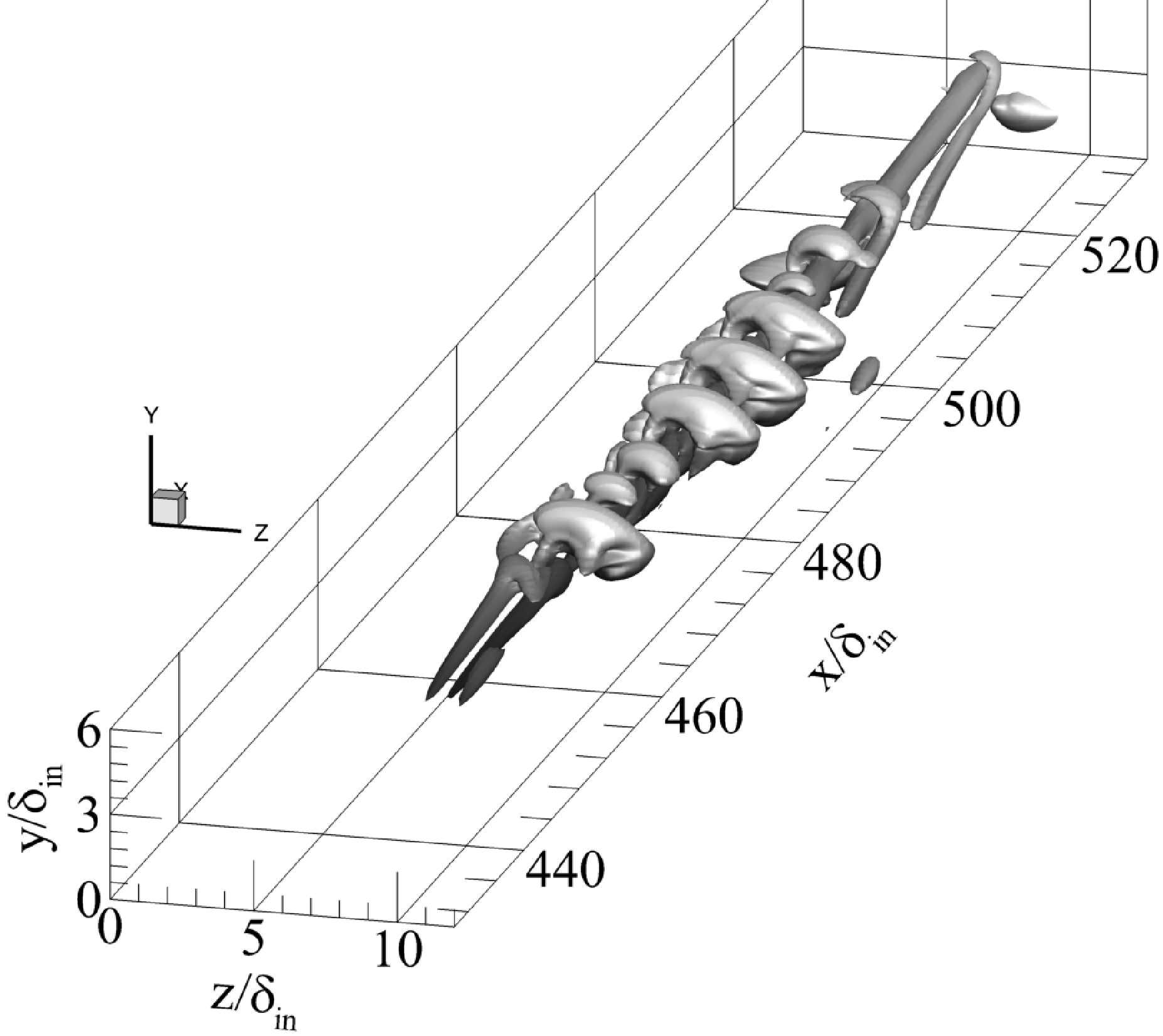}
\label{fig:7}
\end{minipage}
\centering
(iii)~$\Gamma=12.5~(Q_{max}^\ast=1.90\times10^{-1})$\\
%
%
\begin{minipage}{0.32\hsize}
\centering
\includegraphics[width=38mm]{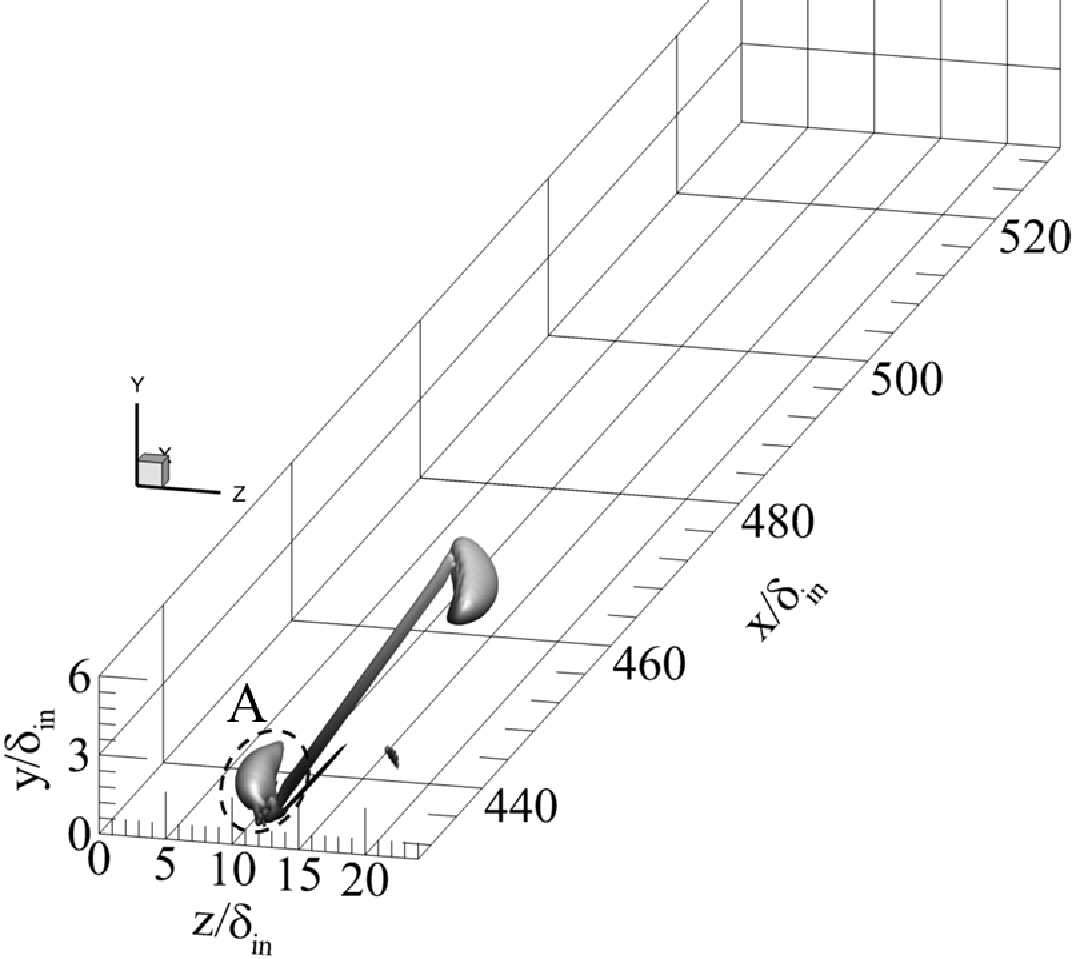}
\label{fig:8}
\end{minipage}
\begin{minipage}{0.32\hsize}
\centering
\includegraphics[width=38mm]{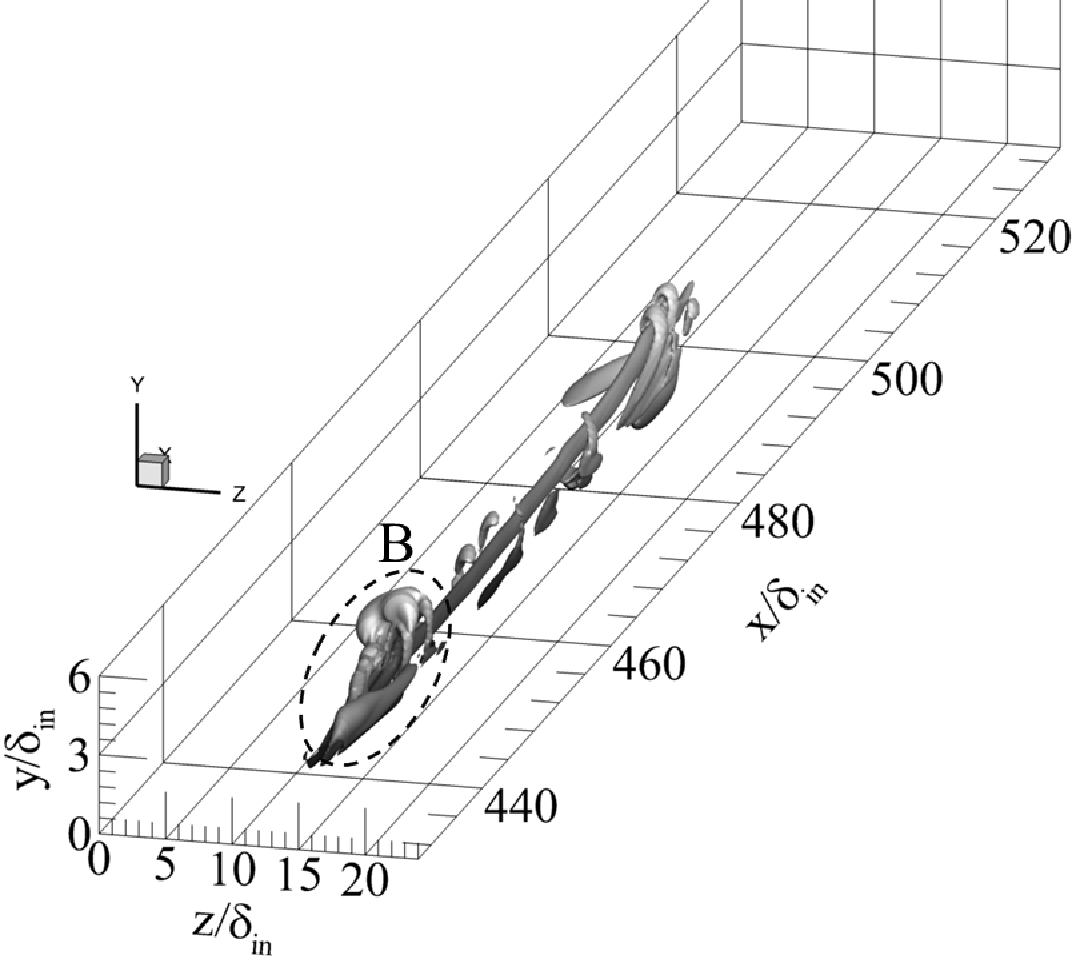}
\label{fig:9}
\end{minipage}
\begin{minipage}{0.32\hsize}
\centering
\includegraphics[width=38mm]{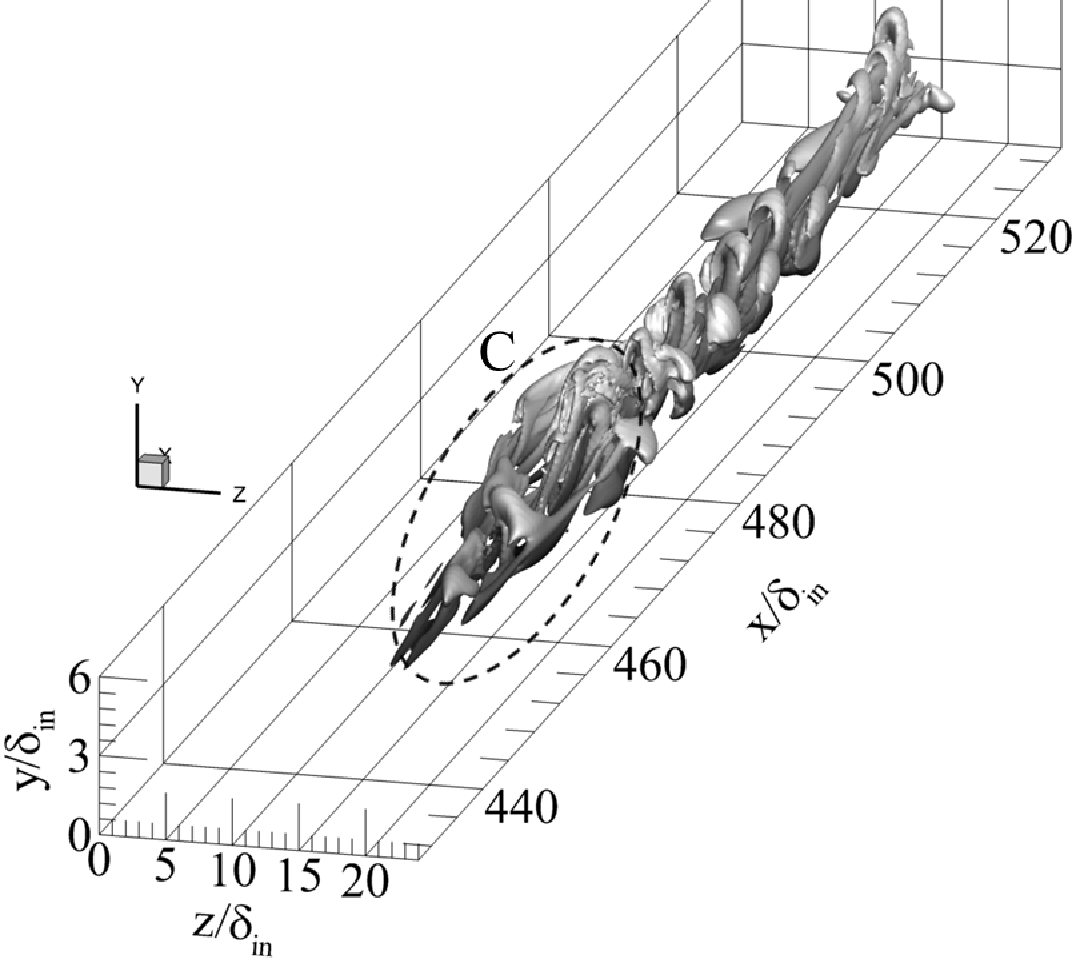}
\label{fig:10}
\end{minipage}
\begin{minipage}{0.32\hsize}
\centering
\includegraphics[width=42mm]{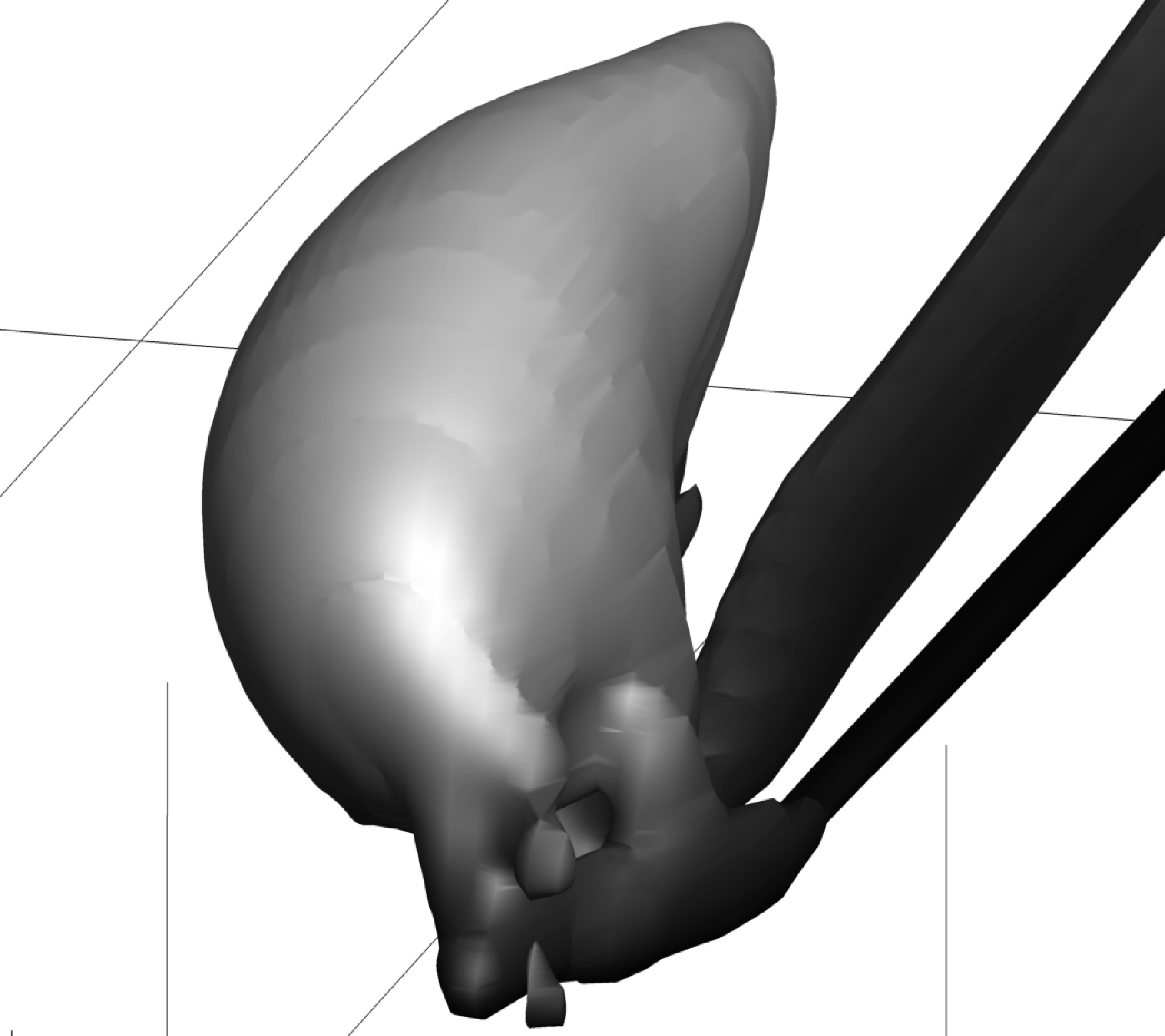}
\label{fig:11}
\end{minipage}
\begin{minipage}{0.32\hsize}
\centering
\includegraphics[width=42mm]{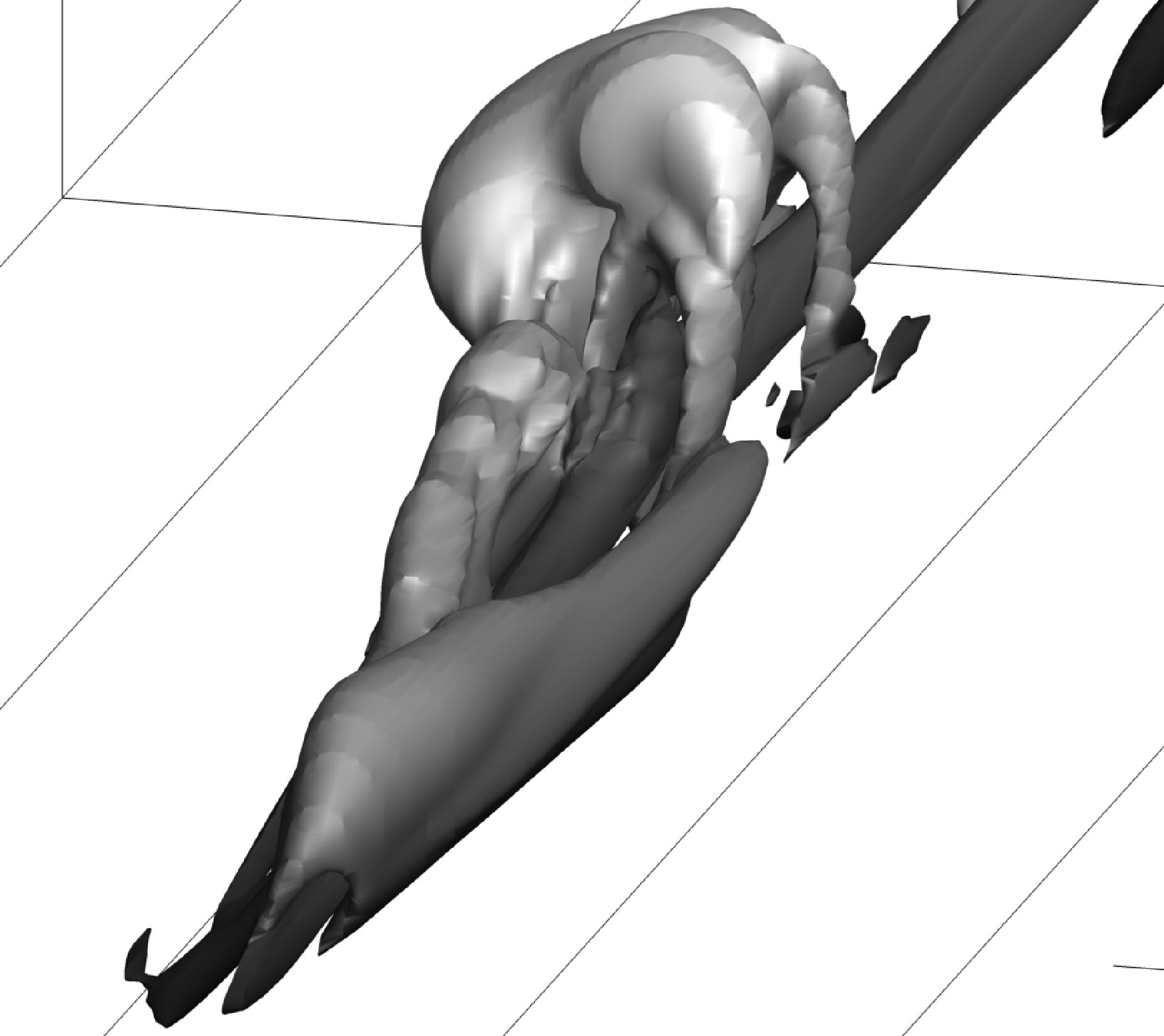}
\label{fig:12}
\end{minipage}
\begin{minipage}{0.32\hsize}
\centering
\includegraphics[width=42mm]{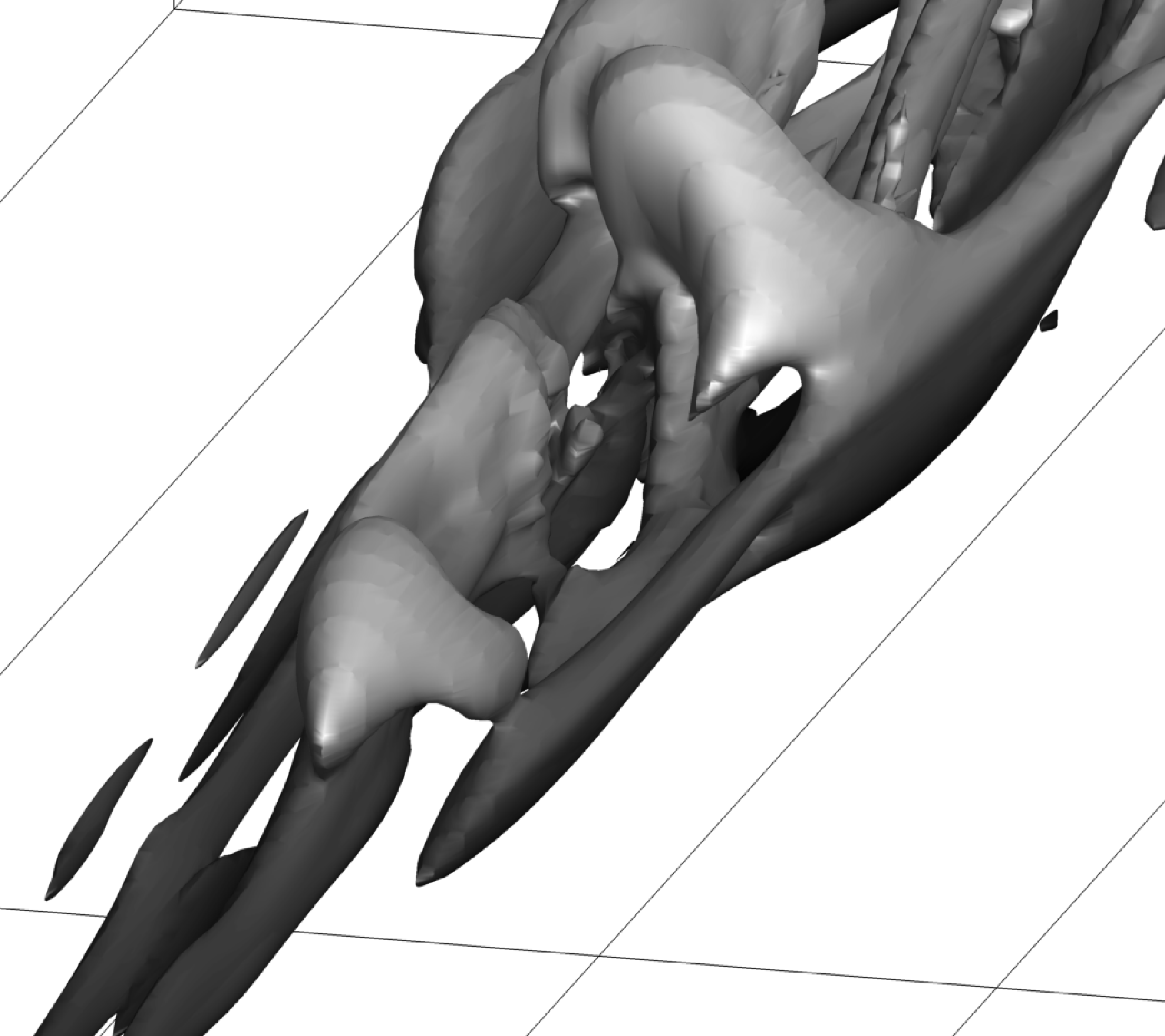}
\label{fig:13}
\end{minipage}
%
\centering
(iv)~$\Gamma=25.0~(Q_{max}^\ast=6.45\times10^{-1})$ \\(lower figures are enlargements of regions A, B, C, respectively)\\
%
\centering
(a)~$\phi=4^\circ$
\caption{Time sequence of vortex tube deformation and the dynamic response of boundary layers for 
$\Gamma=$1.25, 6.25, 12.5, and 25.0 and $\phi=4^\circ$ and $10^\circ$.  
Moving vortex tubes are visualized using the iso-surfaces of the second invariance of the velocity gradient tensor $Q^\ast=4.2\times10^{-4}$.
$Q_{max}^\ast$ is the maximum value of $Q^\ast$ in each case.
Left, center and right columns show snapshots at $t^\ast$=4.24, 42.4 and 84.8, respectively. 
The colour signifies streamwise Mach number, and regions A-F show vortices that originate from the upstream end of the vortex tube.}
\end{figure}
%
%
\begin{figure}
\begin{minipage}{0.32\hsize}
\centering
\includegraphics[width=42mm]{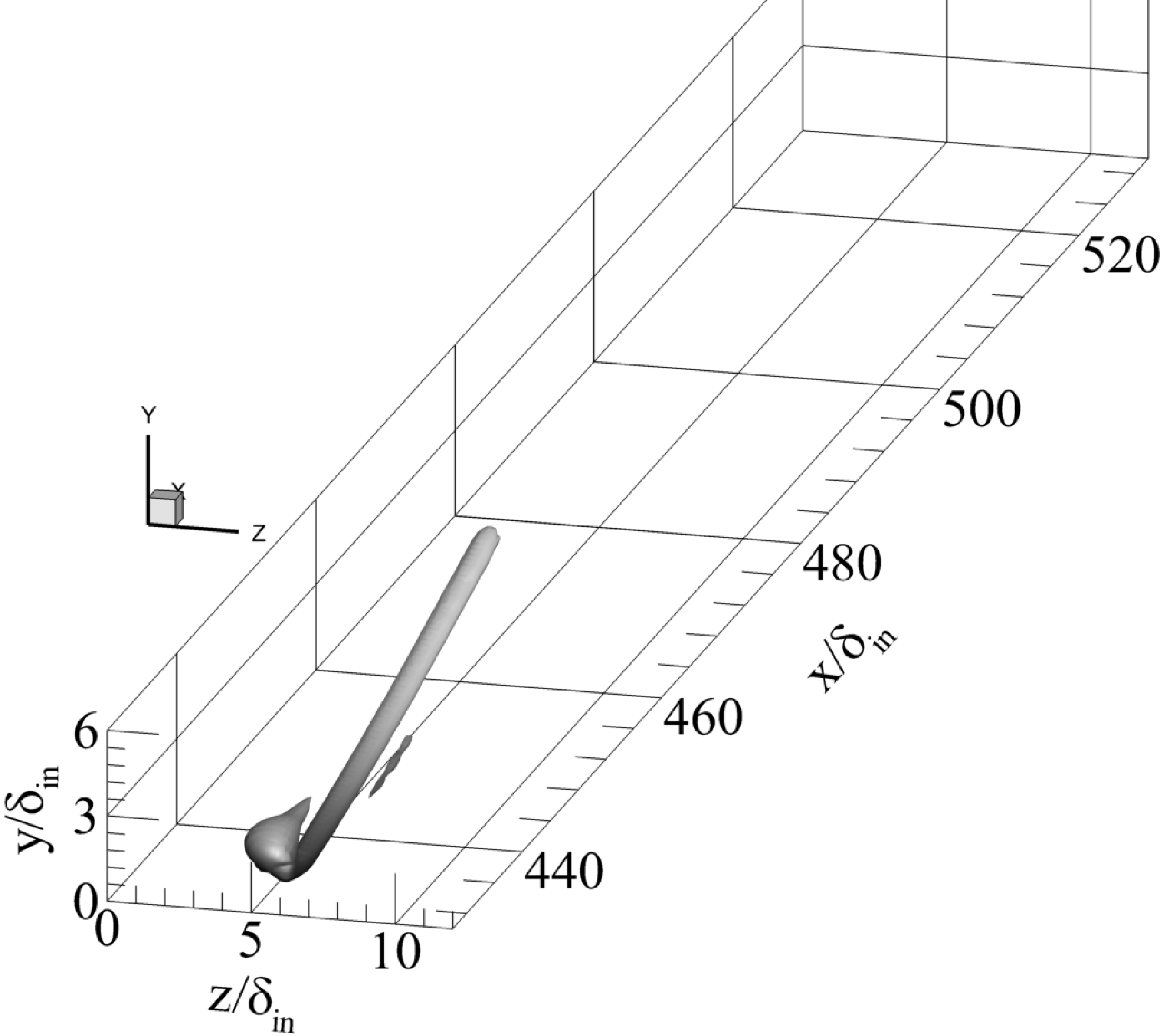}
\label{fig:14}
\end{minipage}
\begin{minipage}{0.32\hsize}
\centering
\includegraphics[width=42mm]{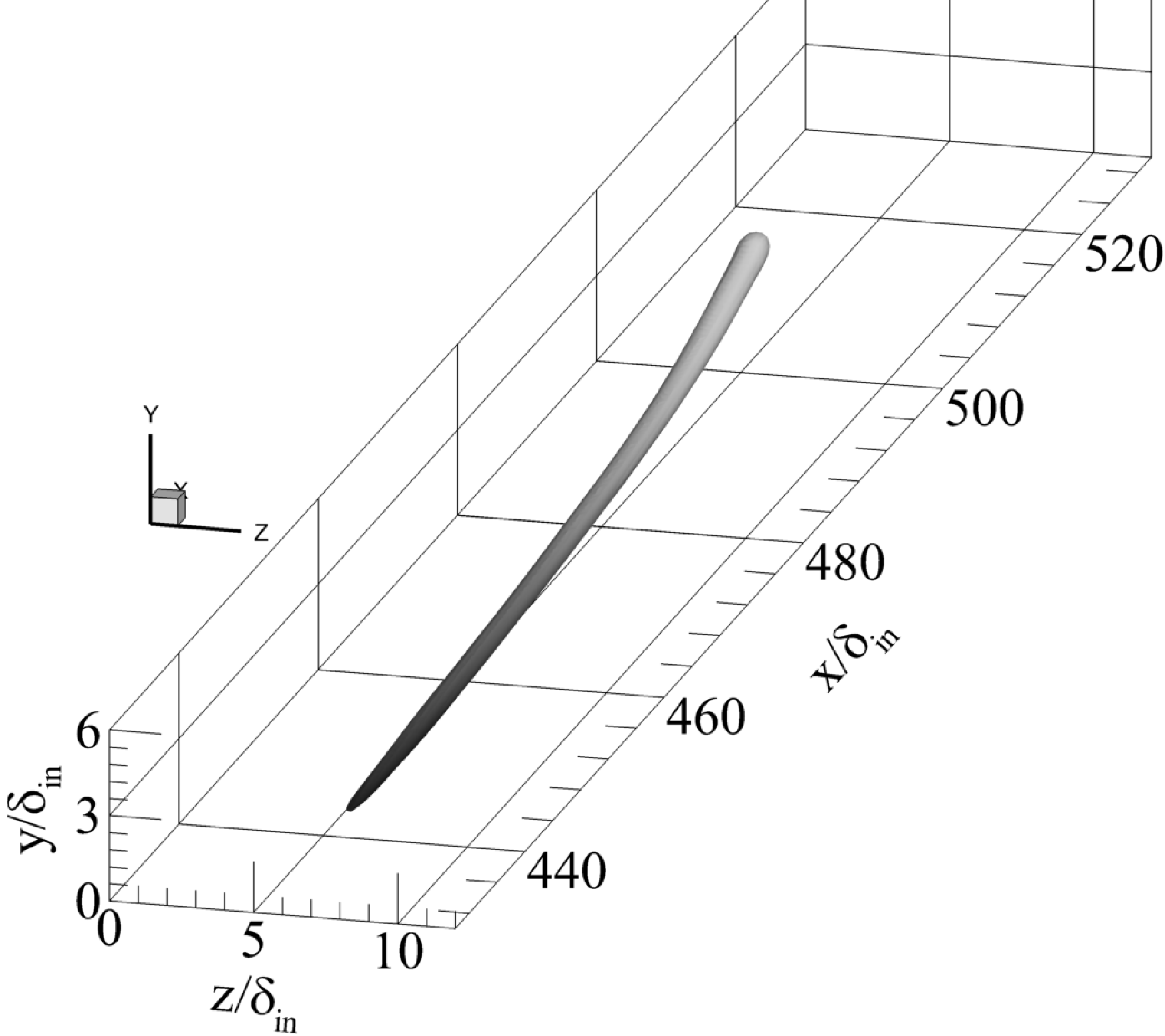}
\label{fig:15}
\end{minipage}
\begin{minipage}{0.32\hsize}
\centering
\includegraphics[width=42mm]{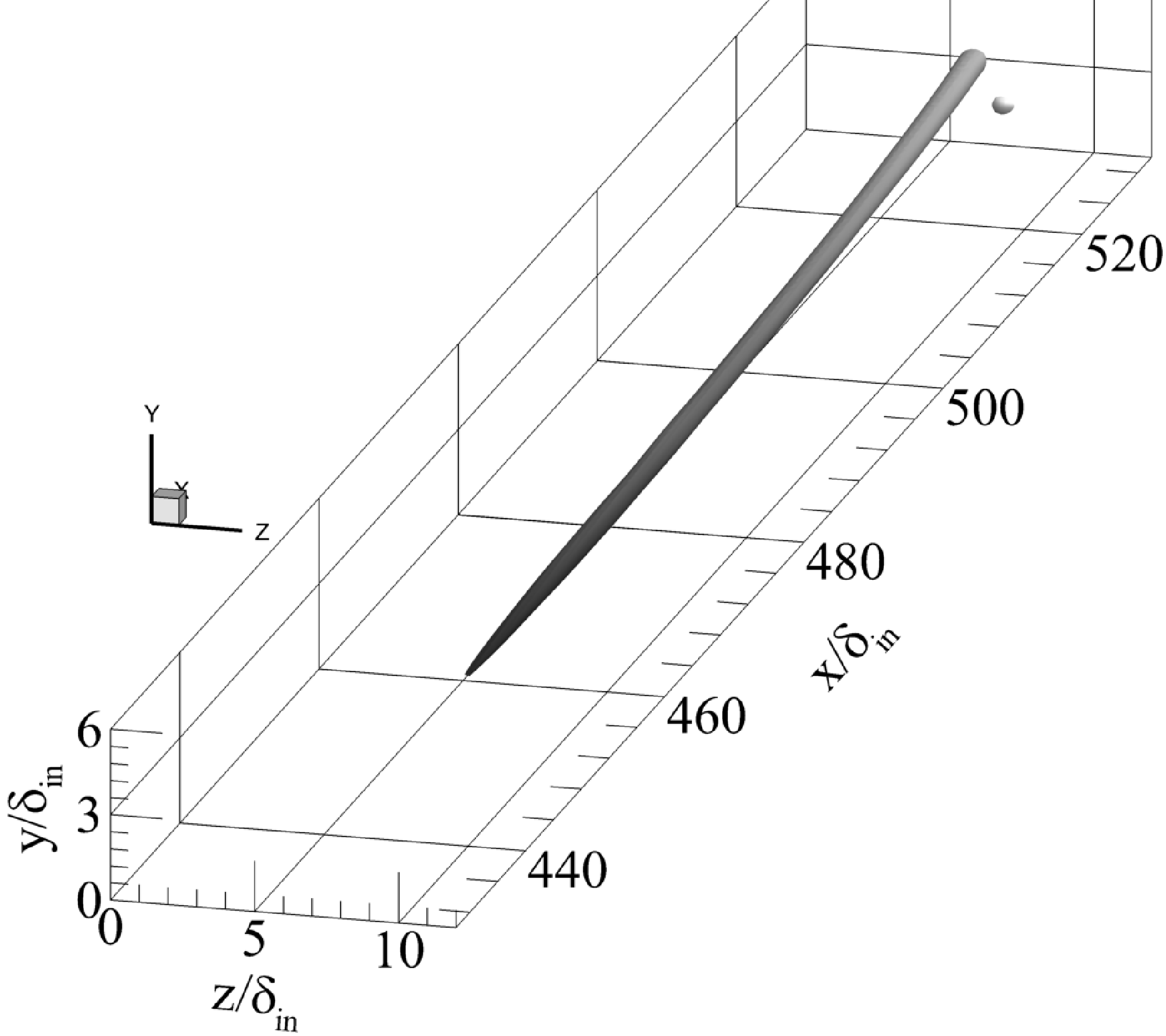}
\label{fig:16}
\end{minipage}\\
\centering
(i)~$\Gamma=6.25~(Q_{max}^\ast=3.64\times10^{-2})$\\
%
%
\begin{minipage}{0.32\hsize}
\centering
\includegraphics[width=42mm]{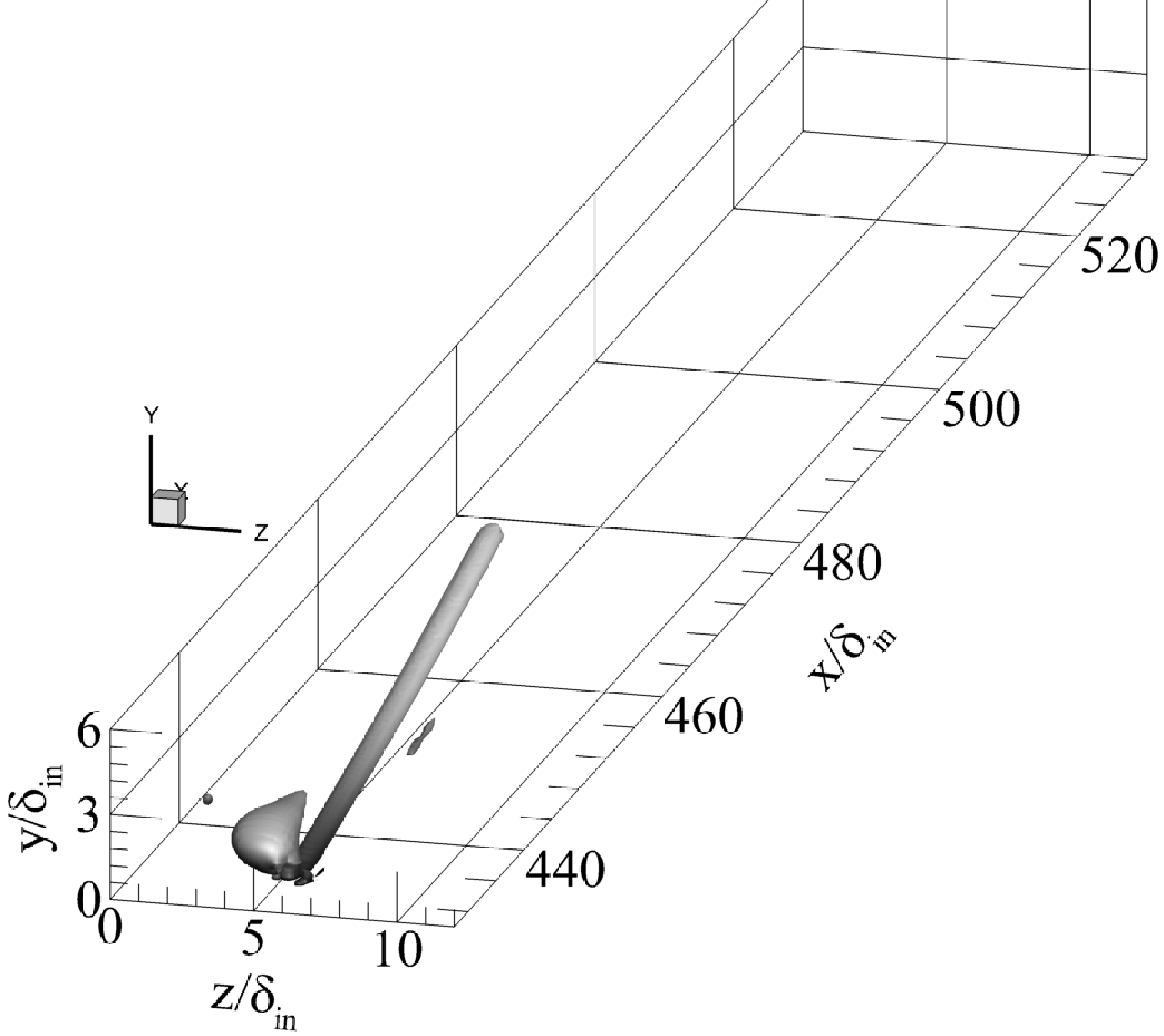}
\label{fig:17}
\end{minipage}
\begin{minipage}{0.32\hsize}
\centering
\includegraphics[width=42mm]{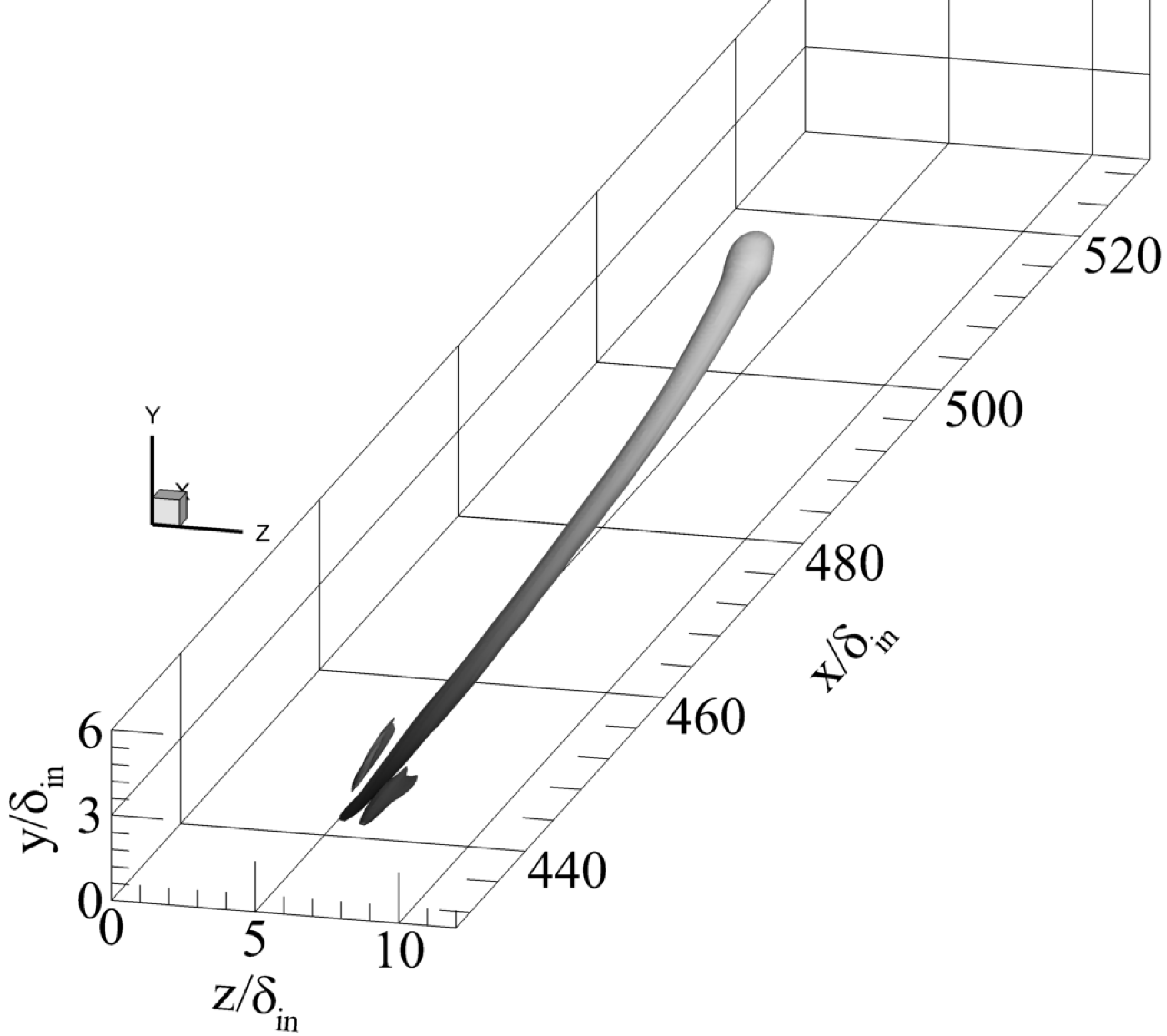}
\label{fig:18}
\end{minipage}
\begin{minipage}{0.32\hsize}
\centering
\includegraphics[width=42mm]{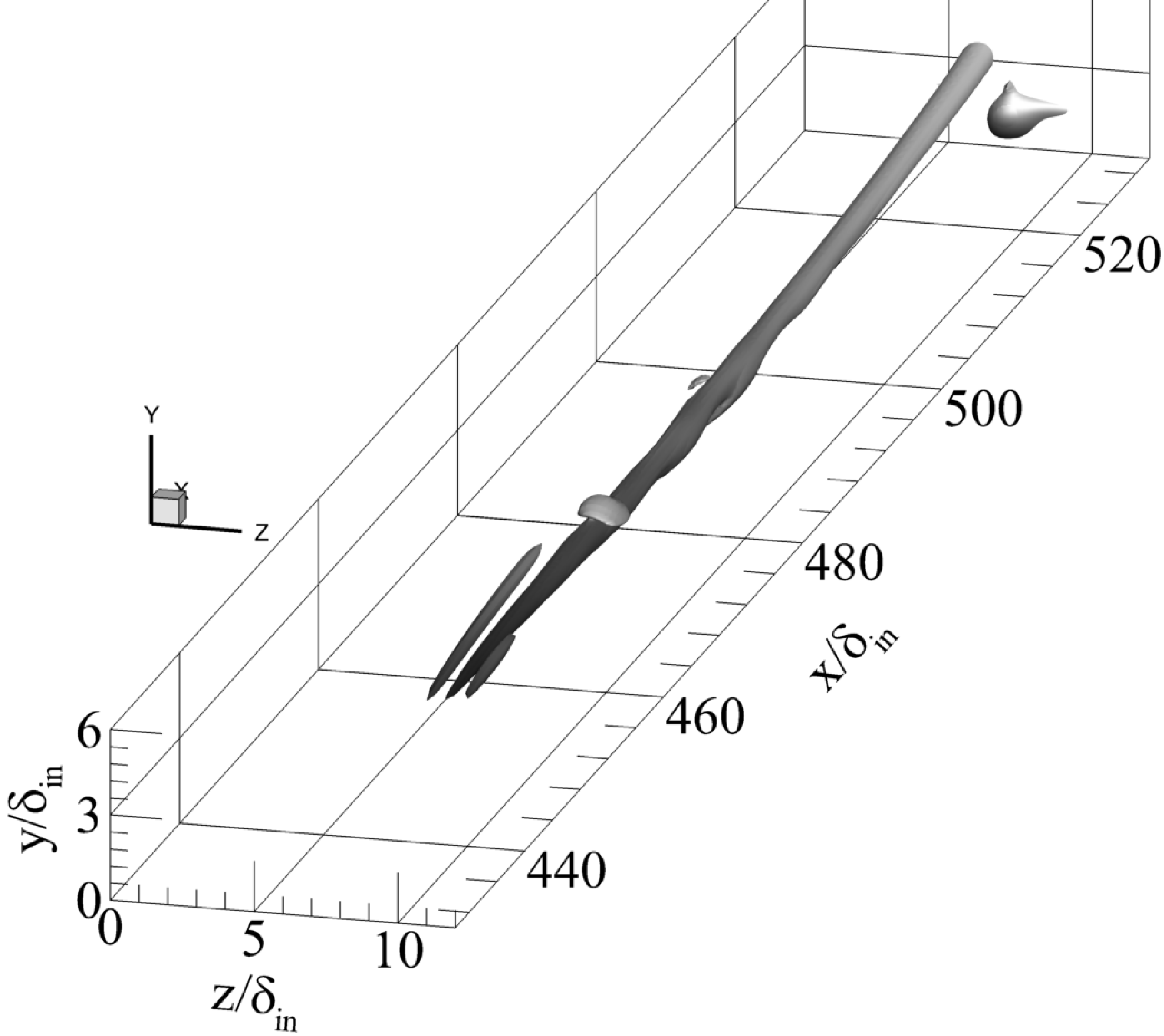}
\label{fig:19}
\end{minipage}\\
\centering
(ii)~$\Gamma=12.5~(Q_{max}^\ast=1.98\times10^{-1})$\\
%
%
\begin{minipage}{0.32\hsize}
\centering
\includegraphics[width=42mm]{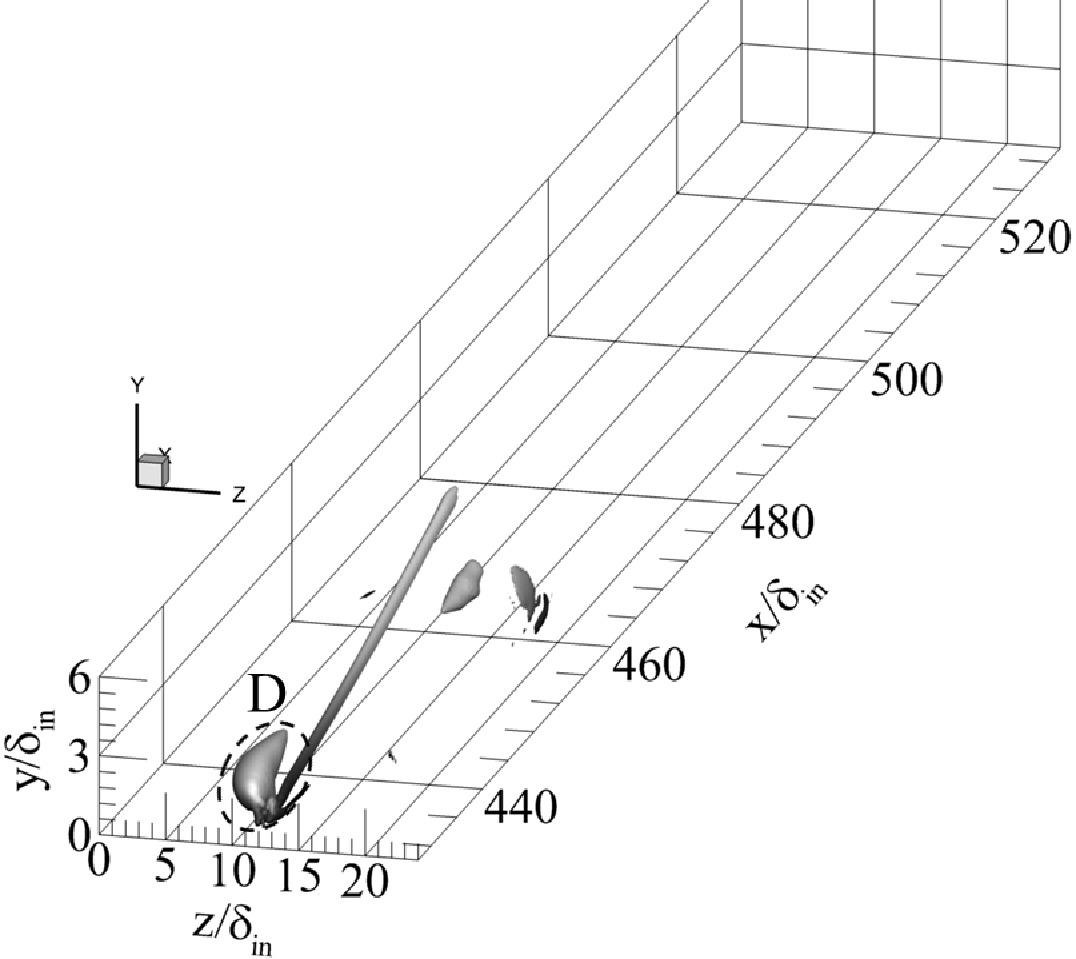}
\label{fig:17}
\end{minipage}
\begin{minipage}{0.32\hsize}
\centering
\includegraphics[width=42mm]{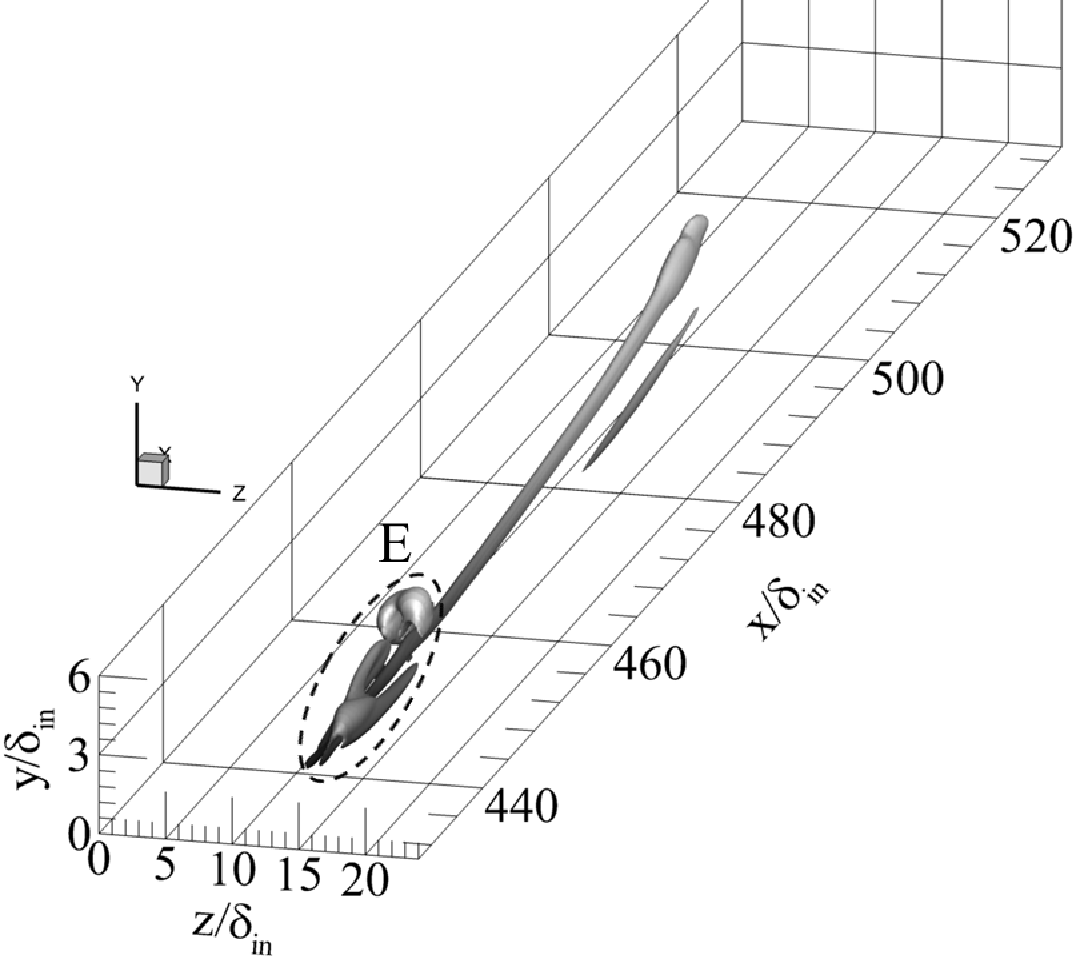}
\label{fig:18}
\end{minipage}
\begin{minipage}{0.32\hsize}
\centering
\includegraphics[width=42mm]{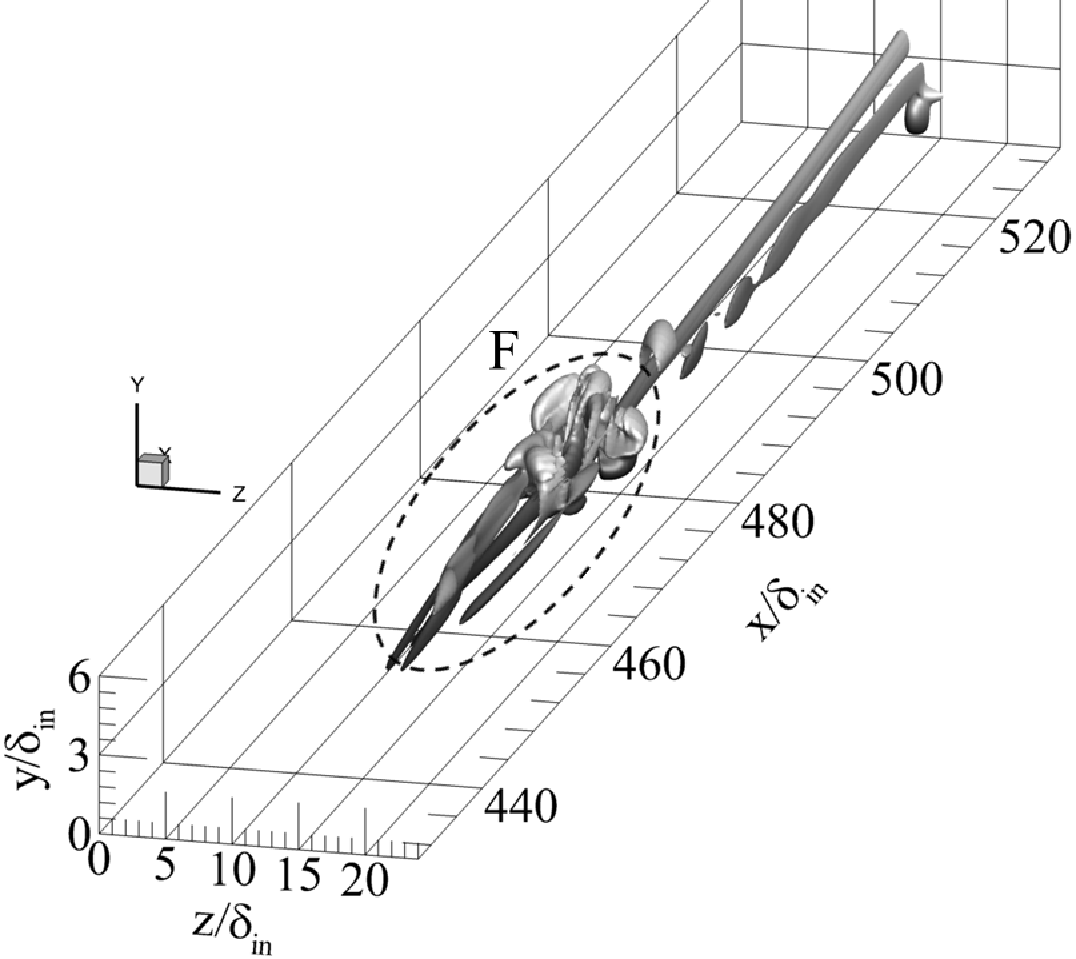}
\label{fig:19}
\end{minipage}\\
\centering
%
%
\begin{minipage}{0.32\hsize}
\centering
\includegraphics[width=42mm]{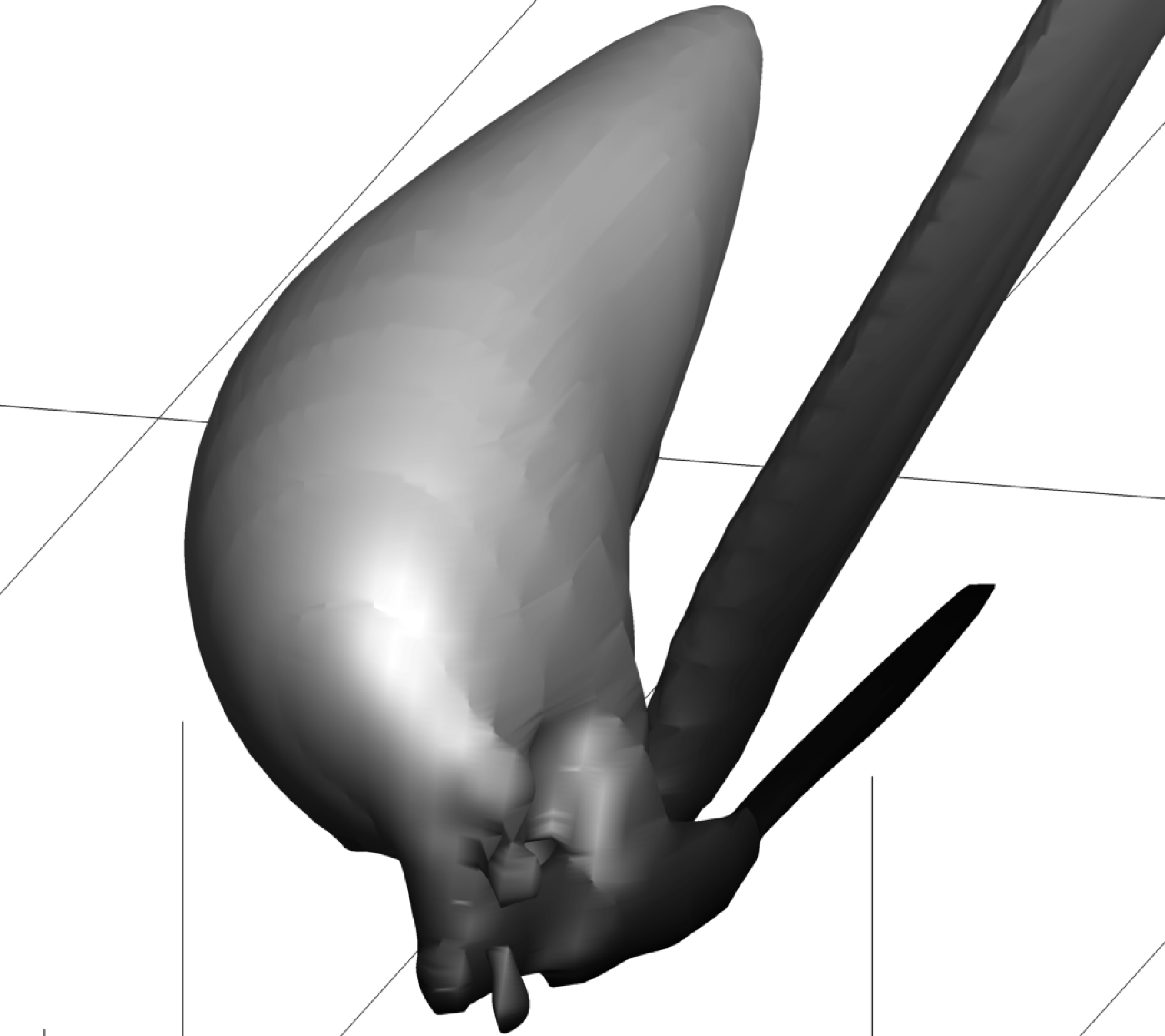}
\label{fig:17}
\end{minipage}
\begin{minipage}{0.32\hsize}
\centering
\includegraphics[width=42mm]{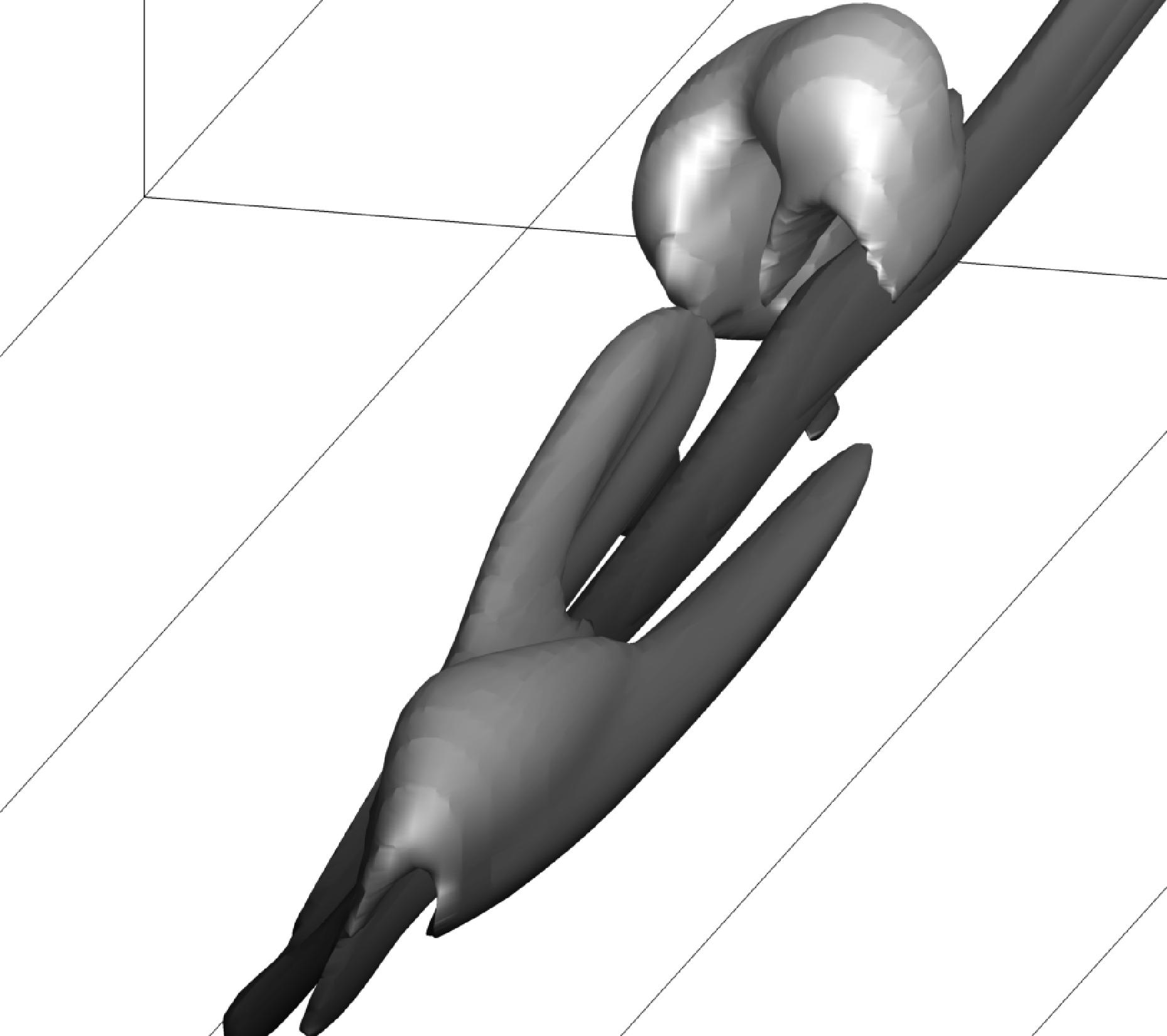}
\label{fig:18}
\end{minipage}
\begin{minipage}{0.32\hsize}
\centering
\includegraphics[width=42mm]{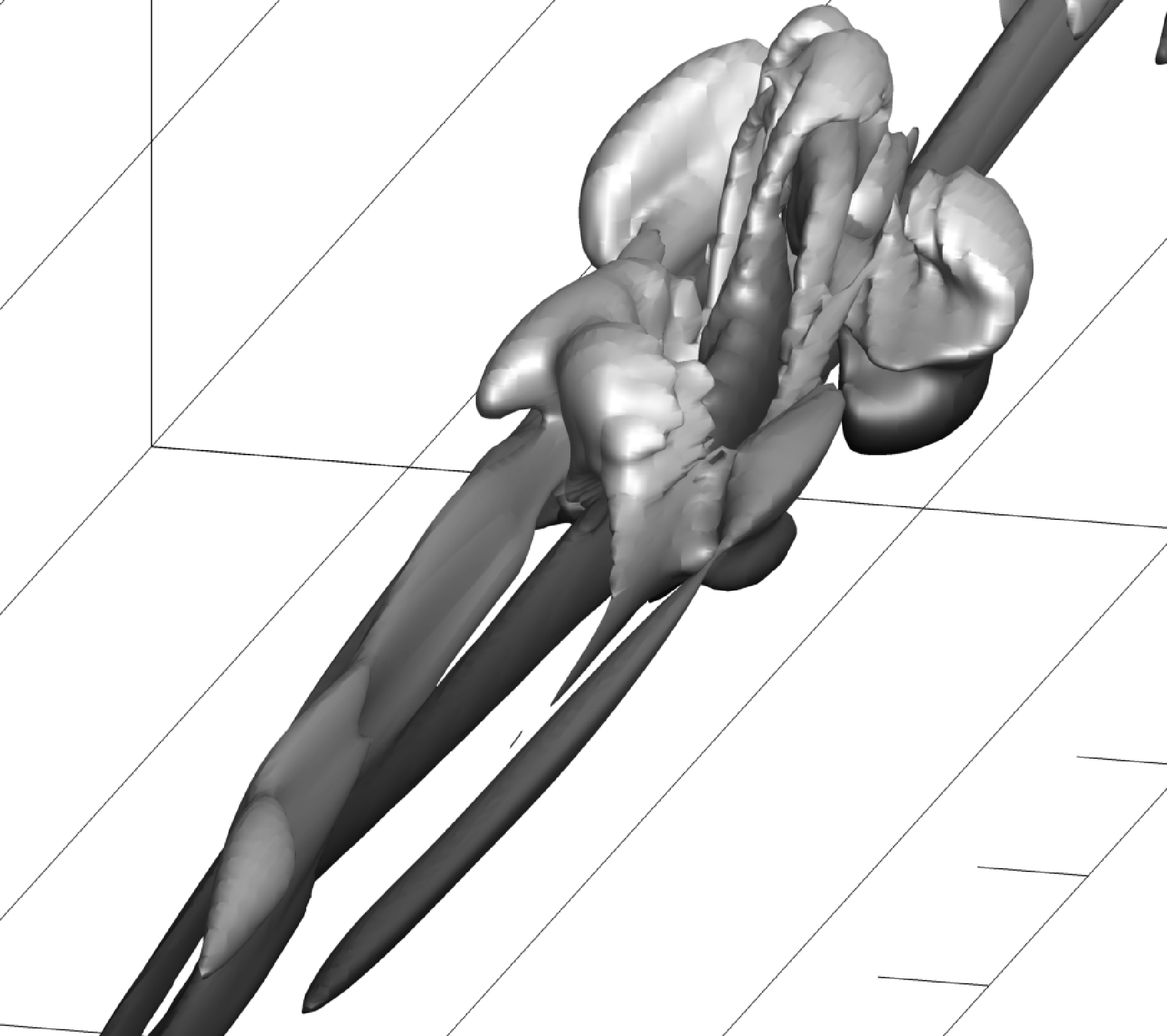}
\label{fig:19}
\end{minipage}\\
\centering
(iii)~$\Gamma=25.0~(Q_{max}^\ast=6.67\times10^{-1})$\\ (lower figures are enlargements of regions D,~E,~F, respectively)\\
\centering
(b)~$\phi=10^\circ$\\
%
\centering
FIGURE 3. cont'd
\end{figure}

In Section 3.1, an overview of the evolution of the vortex tube and its initial stability are investigated by the DNS and LSA.
In Section 3.2, detailed evolution of disturbance around the vortex tube is investigated by the FNDA and POD.
\subsection{Overview of the evolution of the vortex tube and its initial stability}
DNS results are first discussed, and the results of the LSA are then discussed.
\subsubsection{DNS}
Figure 3 shows a time sequence of the vortex tube deformation and the dynamic response of the boundary layer for $\Gamma=1.25-25.0$ when $\phi=4^\circ$ and $\Gamma=6.25-25.0$ when $\phi=10^\circ$. 
The moving vortex tube is visualized using the iso-surfaces of the second invariance of the velocity gradient tensor $Q^\ast=4.2\times10^{-4}$. Here, $Q^*$ is defined by
\begin{equation}
Q^\ast=\displaystyle \frac{1}{2}(-S_{ij}^\ast S_{ji}^\ast+\Omega_{ij}^\ast \Omega_{ij}^\ast)=\displaystyle \frac{1}{2}(||\Omega^\ast||^2-||S^\ast||^2)
\end{equation}
where $S_{ij}^\ast$ is the rate of strain tensor and $\Omega_{ij}^\ast$ is the vorticity tensor.
Quantities with `$\ast$' are non-dimensionalized by $u_{\infty}$ and $\delta_{in}$.
$||\Omega^\ast||^2 \equiv \Omega_{ij}^\ast \Omega_{ij}^\ast, ||S^\ast||^2 \equiv S_{ij}^\ast S_{ji}^\ast$.
$Q^\ast$ represents the balance between shear strain rate and vorticity magnitude, and a region where
$||\Omega^\ast||^2 > ||S^\ast||^2$ is defined as a vortex \citep{Hunt88}.
Snapshots are selected at $t^\ast=$4.24, 42.4, and 84.8. 
The color on the iso-surfaces distinguishes values of streamwise velocity divided by the local speed of sound, i.e., the streamwise Mach number. 
Fig. 3(a,b) shows time sequences when $\phi=4^\circ$ and $10^\circ$, respectively. 

Because the vortex tube studied here is of finite length with both ends snipped off as mentioned in Section 2.2, different vortical fields are generated between the interior region far from the ends and the region close to the ends. 
Because the downstream end is located near/outside the boundary-layer edge, it appears to have negligible influence on the vortex generated inside the boundary layer. 
However, the upstream end might perturb the vortical field of the interior region because vortices generated near the upstream end are transported downstream. 
Here the focus is only on the interior region unperturbed by the vortices generated near the ends. 
The presence of such a region was confirmed from the computation of attenuating circulation near the ends of the vortex tube. 
As seen from in Fig. 3(a, i-iii) and Fig. 3(b, i-ii), the upstream end attenuates when $\Gamma \le 12.5$, whereas vortices generated near the upstream end grow and are transported downstream as indicated by $A \rightarrow B \rightarrow C$ in Fig. 3(a, iv) and 
$D \rightarrow E \rightarrow F$ in Fig. 3(b, iii) when $\Gamma=25.0$. 
These regions are excluded from the present discussion.

Here, the case of $\phi=4^\circ$ is focused on. 
With $\Gamma=1.25$, the vortex is attenuated but at $\Gamma=6.25$, the vortex structure is elongated. 
With $\Gamma=12.5$ and 25.0, consecutive hairpin vortices aligned in the streamwise position evolve along the vortex tube; the details of the vortex structure are evident in Figs. 4 and 5. 

\begin{figure}
\centering
\centerline{\includegraphics[scale=0.4]{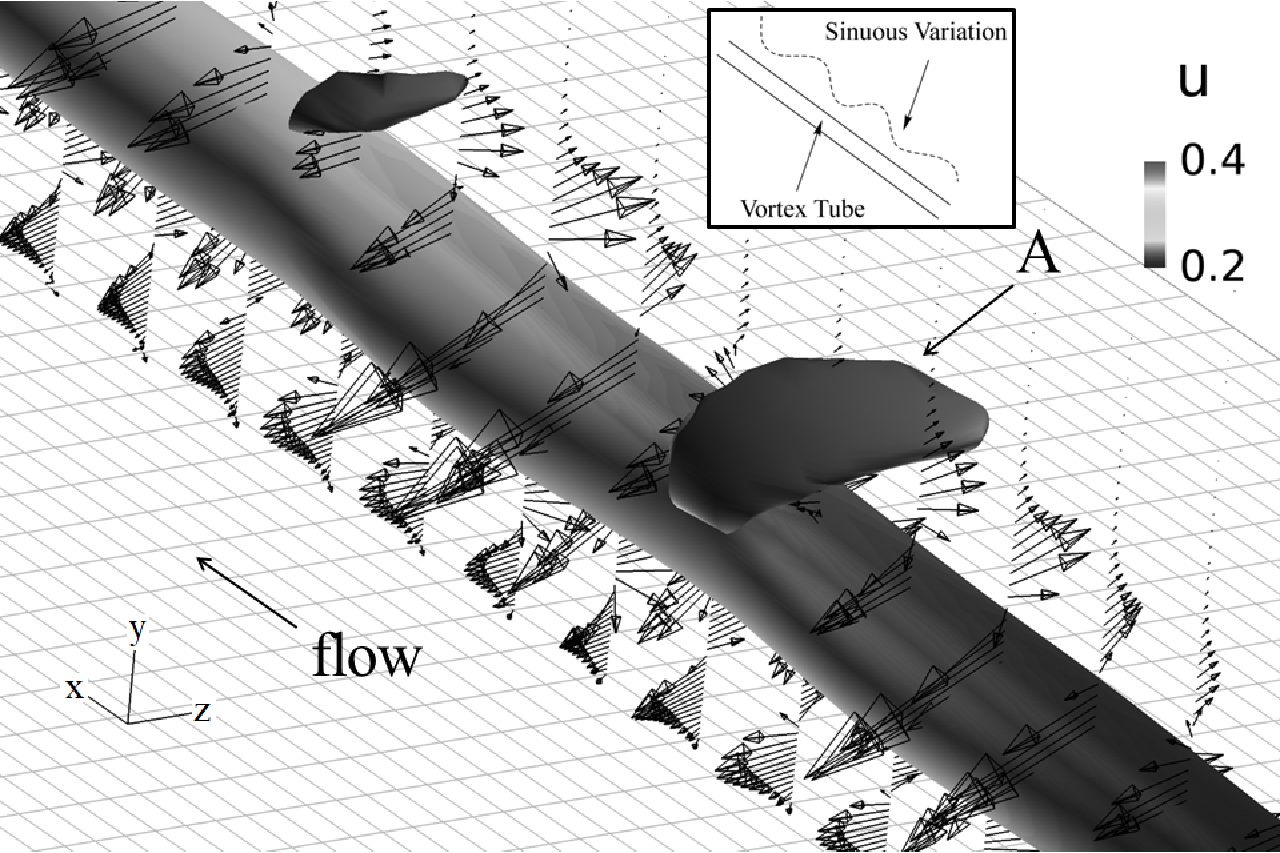}}
(a)~$t^\ast=42.4$
\centering
\centerline{\includegraphics[scale=0.4]{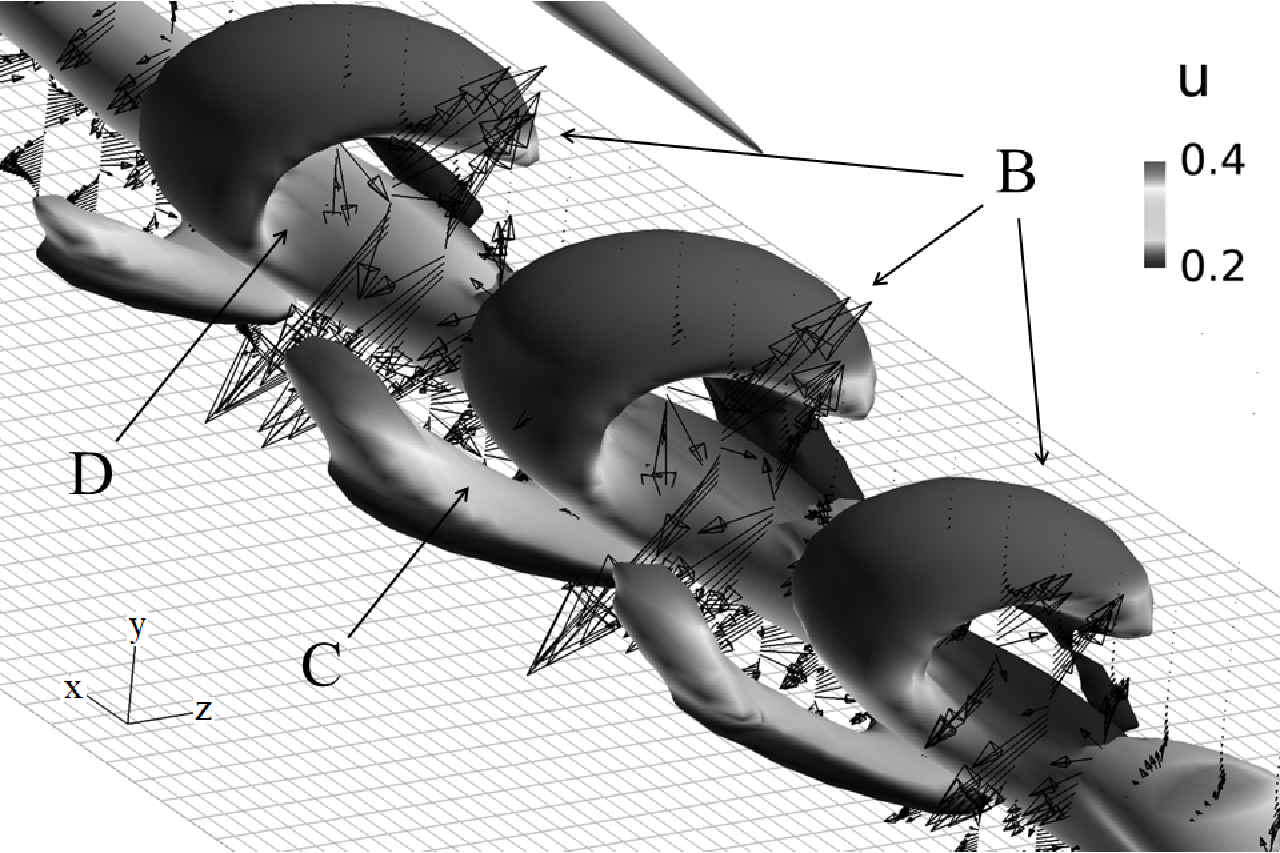}}
(b)~$t^\ast=63.6$
\caption{Incremental vectors of vorticity and iso-surfaces of the second invariant of the velocity gradient tensor for $\Gamma$=12.5 when $\phi=4^\circ$. 
The colour signifies streamwise Mach number. 
Schematic of a sinuous disturbance appearing along the vortex tube is shown in the small frame on panel (a).}
\label{fig:20}
\end{figure}
\begin{figure}
\centering
\centerline{\includegraphics[scale=0.4]{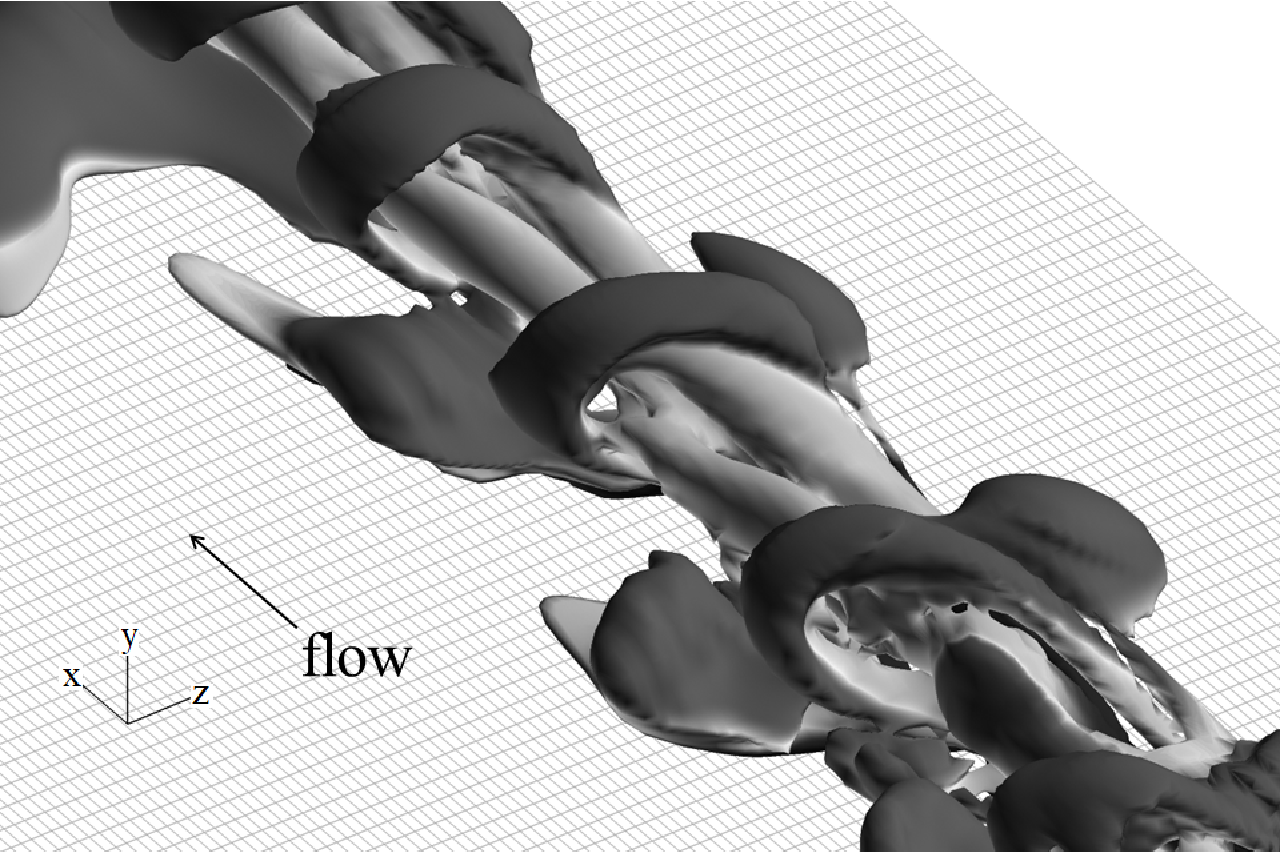}}
(a)~Left side view\\
\centering
\centerline{\includegraphics[scale=0.4]{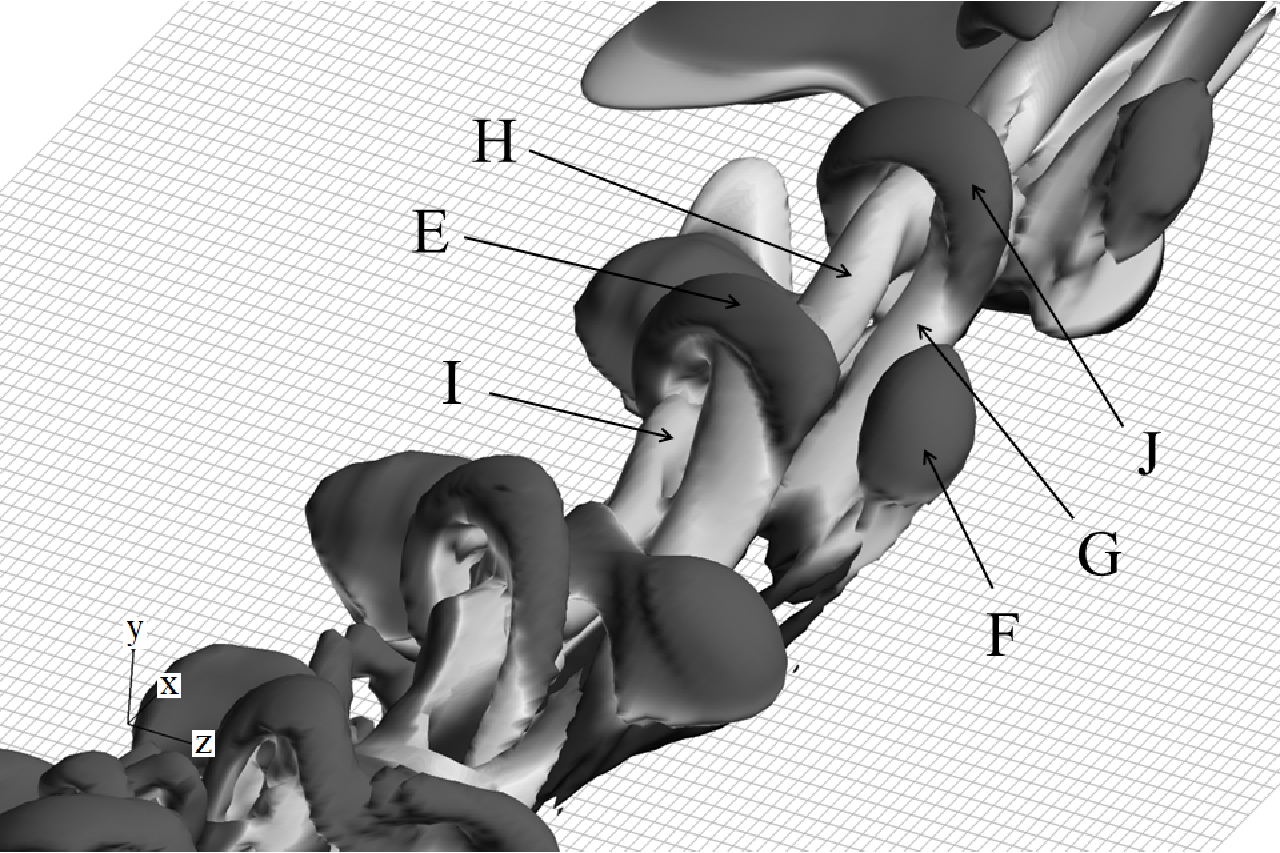}}
(b)~Right side view\\
\caption{Iso-surfaces of the second invariant of the velocity gradient tensor for $\Gamma$=25.0 at $t^\ast$=84.8 when $\phi=4^\circ$.
The colour signifies streamwise Mach number. }
\label{fig:20}
\end{figure}

Fig. 4 gives vector plots of the right-hand side of the transport equation for vorticity \citep[p. 146]{Chassaing10}, i.e.,
\begin{eqnarray}
\label{eq:eq1}
\displaystyle \frac{\partial \mbox{\boldmath $\omega$}}{\partial t}&=&-(\pmb{V} \cdot \nabla){\mbox{\boldmath $\omega$}}+(\mbox{\boldmath $\omega$} \cdot \nabla)\pmb{V}+
\nu \nabla \mbox{\boldmath $\omega$}-\mbox{\boldmath $\omega$}(\nabla \cdot \pmb{V})+\frac{\nabla \rho}{\rho} \wedge \frac{d\pmb{V}}{dt}+E_1,\\
E_1&=&\displaystyle \frac{1}{\rho} (\nabla \times E_2)-\nabla \mu \wedge (\nabla \times \mbox{\boldmath $\omega$}),\\
E_2&=&2 {\bf S} \odot \nabla \mu-\displaystyle \frac{2}{3} (\nabla \cdot \pmb{V}) \nabla \mu
\end{eqnarray}
and the instantaneous vortical structures at $t^\ast$=42.4 and 63.6 for $\Gamma=12.5$. 
In Eq. (3.2), $\pmb{V}$ is the velocity vector defined as $\pmb{V}=(u,v,w)^T=(u_1,u_2,u_3)^T$, $\mbox{\boldmath $\omega$}$ is the instantaneous vorticity vector defined as $\mbox{\boldmath $\omega$}=\nabla \times \pmb{V}$, 
$\mu$ is the dynamic viscosity, $\nu=\mu/\rho$ is the kinematic viscosity and $\pmb{S}$ is the strain rate tensor defined as
$\pmb{S}=\displaystyle \frac{1}{2}(\frac{\partial u_i}{\partial x_j}+\frac{\partial u_j}{\partial x_i})$.
The symbol `$\odot$' means the product obtained by contraction upon the derivative coordinate \citep[p. 146]{Chassaing10}. 
At $t^\ast$=42.4, i.e., in panel (a), a sinuous variation in the streamwise direction appears in the shear layers, and in particular, above the vortex tube concurrently with the appearance of a separated volume enveloped by iso-surfaces A above the vortex tube. 
The schematic of a sinuous disturbance appearing along the vortex tube is shown in the small frame on panel (a). 
At $t^\ast$=63.6, i.e., in panel (b), the streamwise variation becomes more evident along with the development of organized hairpin vortices B. 
The hairpin vortices are generated over the vortex tube D. 
The branch-like secondary structure C extends from the root of the main vortex tube. 
The generation of the hairpin vortices B accompanies the satellite structure C connecting the near-wall region beneath the vortex tube and the edge of the boundary layer. 
Fig. 5 shows the instantaneous vortical structures for $\Gamma$=25.0 at $t^\ast$=84.8; panels (a) and (b) show left and right side views, respectively. 
Compared with setting $\Gamma$=12.5, streamwise vortices G, H and I with smaller cross-section, i.e., vortices resulting from smaller-scale variation in the cross-section of the initial vortex tube, appear in this instance. 
The width of the hairpin vortex becomes smaller than that of B in Fig. 4, structures E and F are separated in the spanwise direction, and leg G of the hairpin vortex J is located between E and F. 
From Fig. 5(a), the hairpin vortices are skewed and inclined in the spanwise direction because the circulation is strong. 
The branch structures are inclined more toward the wall. 
Thus, a structural change is observed depending on the circulation.

When $\phi=10^\circ$, i.e., in Fig. 3(b), the heads of the vortex tubes are located in the freestream, as is also found from the velocity profiles discussed later in Section 3.1.2. 
The features preserve shape while being convected downstream. 
Inside the boundary layer, appearance of hairpin vortices is confirmed at $t^\ast$=84.8. 
In this regard, the deformation of the vortex tubes is similar to but much less amplified compared with that observed for $\phi=4^\circ$ shown in Fig. 3(a). 

In this study, only the results of the vortex tube length $A=24\delta_{in}$ are shown.
When a shorter vortex tube such as $A=6\delta_{in}$ or $12\delta_{in}$ is introduced initially,
it is confirmed that it tilts and elongates, reaching an angle less than $4^\circ$, 
and then secondary hairpin vortices are generated for the case of $\Gamma$=12.5 with $\phi=4^\circ$ and $10^\circ$.    
 
\subsubsection{Linear stability analysis}
\begin{figure}
\begin{minipage}{0.5\hsize}
\centering
\includegraphics[width=67mm]{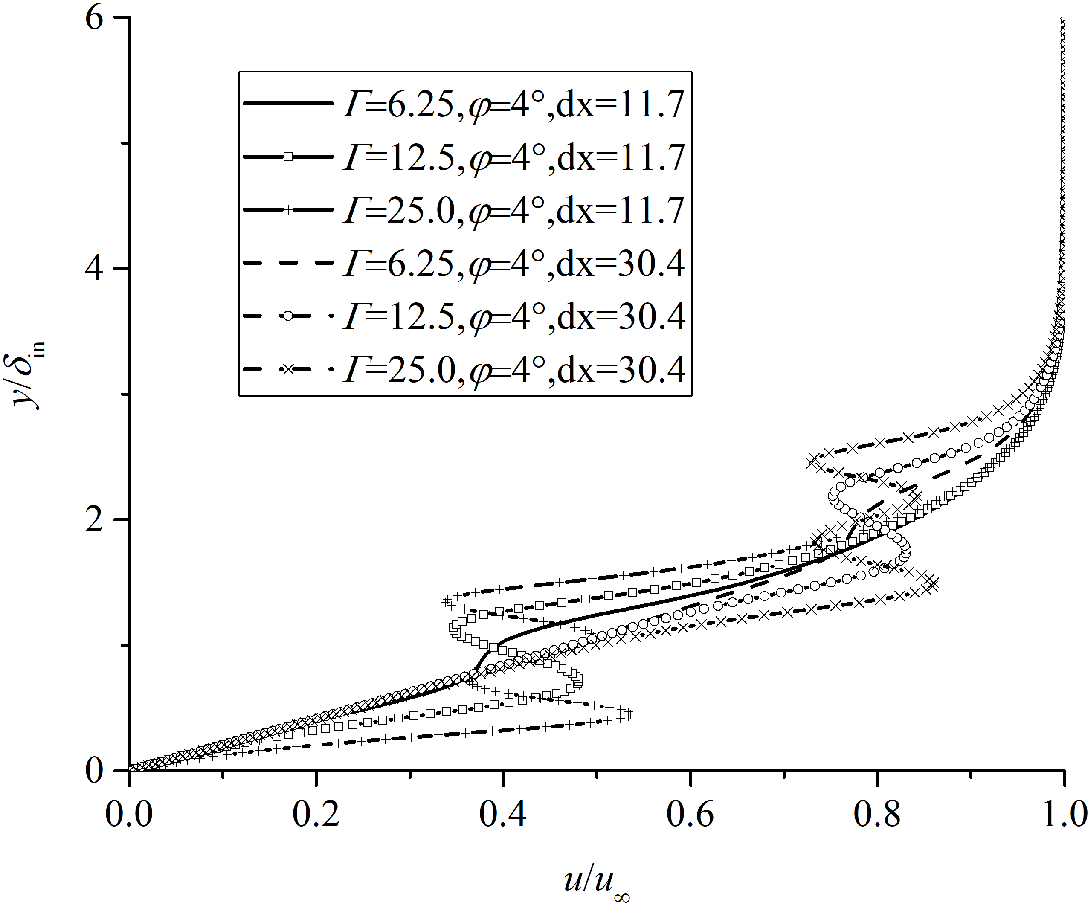}\\
(a)~$\phi=4^\circ$
\label{fig:21}
\end{minipage}
\begin{minipage}{0.5\hsize}
\centering
\includegraphics[width=67mm]{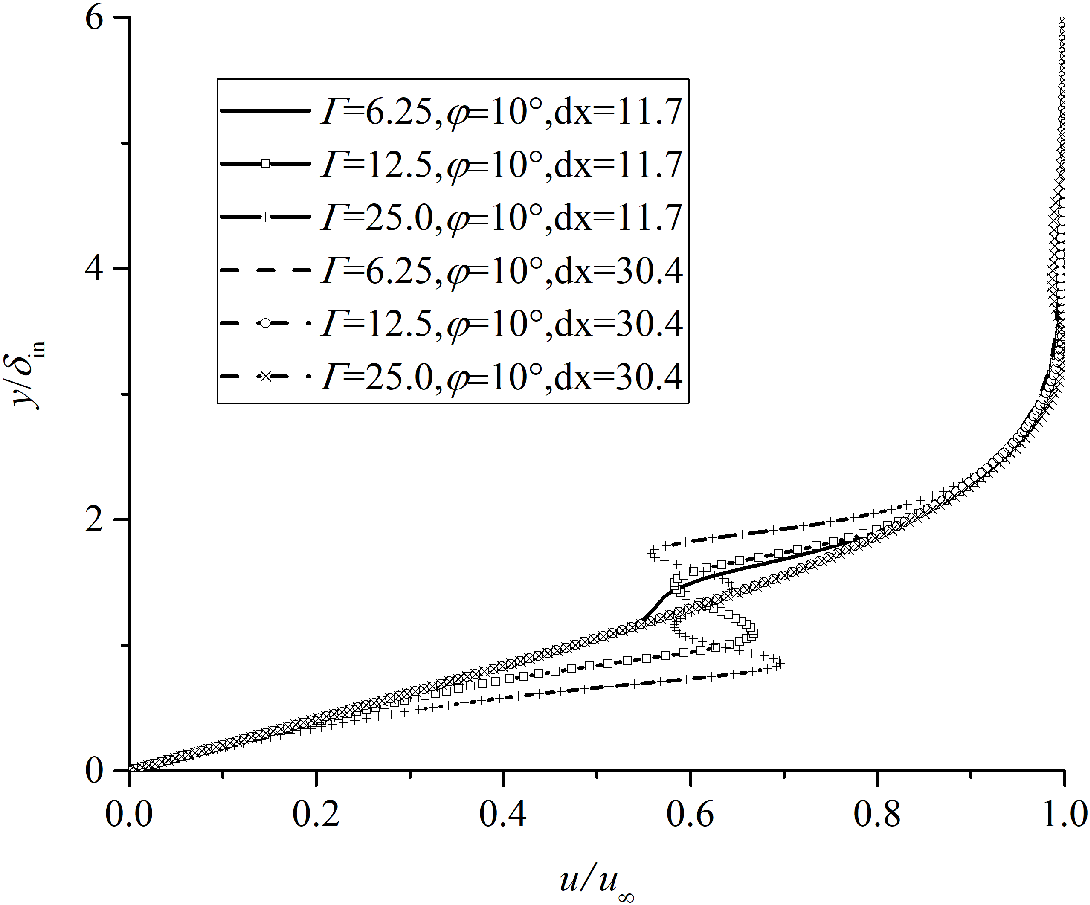}\\
(b)~$\phi=10^\circ$
\label{fig:22}
\end{minipage}
\caption{Mean velocity profiles used for the linear stability analysis for $\Gamma$=6.25-25.0 and $\phi=4^\circ$ and $10^\circ$. 
The velocity profiles are extracted at $Dx=11.6\delta_{in}$ and $30.4\delta_{in}$ when $t^\ast$=10.6.,; Part(a):
Courtesy of WIT press from international journal of computational methods
and experimental measurements, Volume (4), No. (4), November 2016, page 474} 
\end{figure}
\begin{figure}
\centerline{\includegraphics[width=100mm]{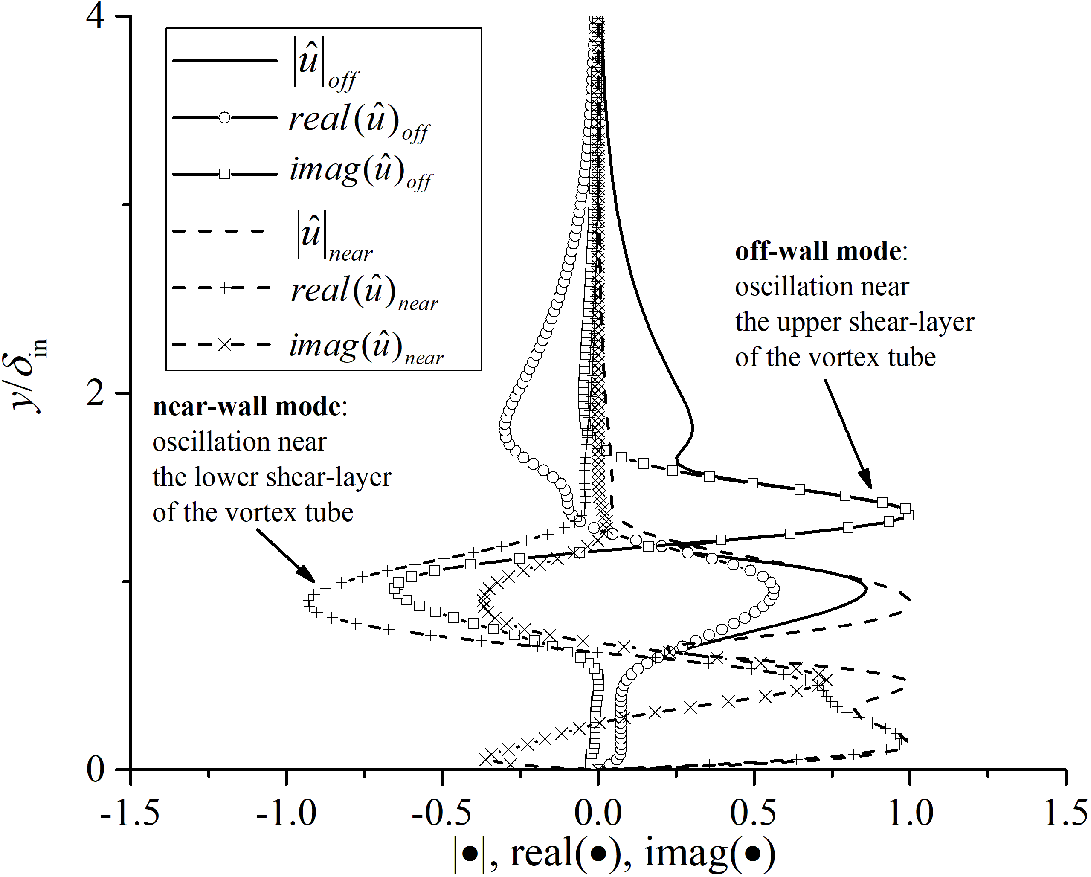}}
\caption{Absolute value, and the real and imaginary parts of the eigenfunctions of the streamwise velocity of the off-wall and near-wall modes when 
$\Gamma$=12.5,  $\phi=4^\circ$ and $t^\ast$=10.6. The off-wall mode for $\alpha=1.8$ and near-wall mode for $\alpha=2.2$ are shown.}
\label{fig:23}
\end{figure}
%
To understand initial disturbances that trigger the consecutive hairpin vortices and also estimate the wavelengths of unstable waves allowed based on the wall-normal profile, a LSA was conducted. 
Mean flows used in the stability analysis were extracted from the instantaneous flow fields of the DNS. 
Figure 6 shows the collection of boundary layer profiles obtained at two streamwise locations $Dx=11.6\delta_{in}$ and $30.4\delta_{in}$ around the center of the vortex tube in the spanwise direction for each vortex tube when $t^\ast=10.6$ given settings $\Gamma=6.25-25.0$ and $\phi=4^\circ$ and $10^\circ$. 
That is, two profiles with same $\Gamma$ and $\phi$ are extracted from a same snapshot. 
Here, $Dx=x-x_c$. 
These two locations are close to the root and tip, respectively, of the vortex tube from its center but avoid the finite-length anomalies arising from the ends of the vortex tube, i.e., they are in the interior region mentioned in Section 3.1.1. 
When $\phi=4^\circ$, S-shaped profile modification by the vortex tube, which is observed in the velocity profile through the leg of the hairpin vortex in Fig. 2, 
is generated near the middle of boundary layer at $Dx=11.6\delta_{in}$ and near the interior edge of the boundary layer at $Dx=30.4\delta_{in}$. 
While a small step is introduced in the profile when $\Gamma=6.25$, a large deformation of magnitude about 0.17$u_{\infty}$ is introduced when $\Gamma=25.0$. 
The shape of the deformation varies depending on $\Gamma$. When $\phi=10^\circ$, the modification is generated near the middle of the boundary layer at $Dx=11.6\delta_{in}$ and also near the exterior edge of the boundary layer at $Dx=30.4\delta_{in}$. 

In the present LSA, two types of unstable modes, i.e., `off-wall' and `near-wall' modes, are observed. 
Figure 7 shows the absolute value, the real and imaginary parts of the eigenfunctions of the streamwise velocity of the modes when $\Gamma=12.5$ and $\phi=4^\circ$. 
The eigenfunctions are evaluated when the growth rates of the modes becomes maximum, respectively. 
The off-wall mode for $\alpha=1.8$ and near-wall mode for $\alpha=2.2$ are shown. 
The two modes have differences in the height of oscillation; the eigenfunction associated with the oscillation further from and nearer to the wall are referred to as the `off-wall mode' and `near-wall mode', respectively \citep{Matsuura16}. 
From Figs. 4 and 6, the differences in the height of the oscillation arise from disturbances in the shear layers above and beneath the vortex tube. 

\begin{figure}
\begin{minipage}{0.32\hsize}
\centering
\includegraphics[width=52mm]{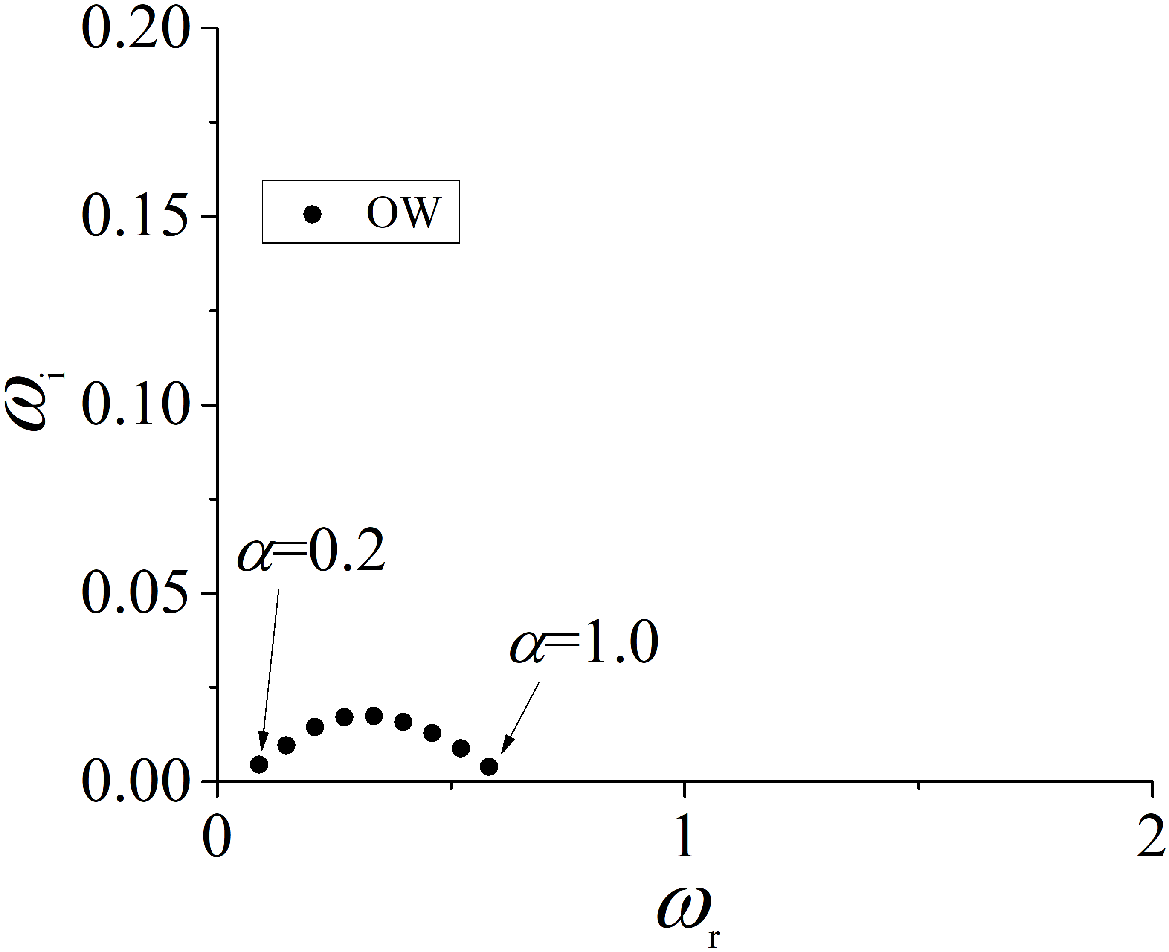}
(i)~$\Gamma=3.13$
\label{fig:24} 
\end{minipage}
\begin{minipage}{0.32\hsize}
\centering
\includegraphics[width=52mm]{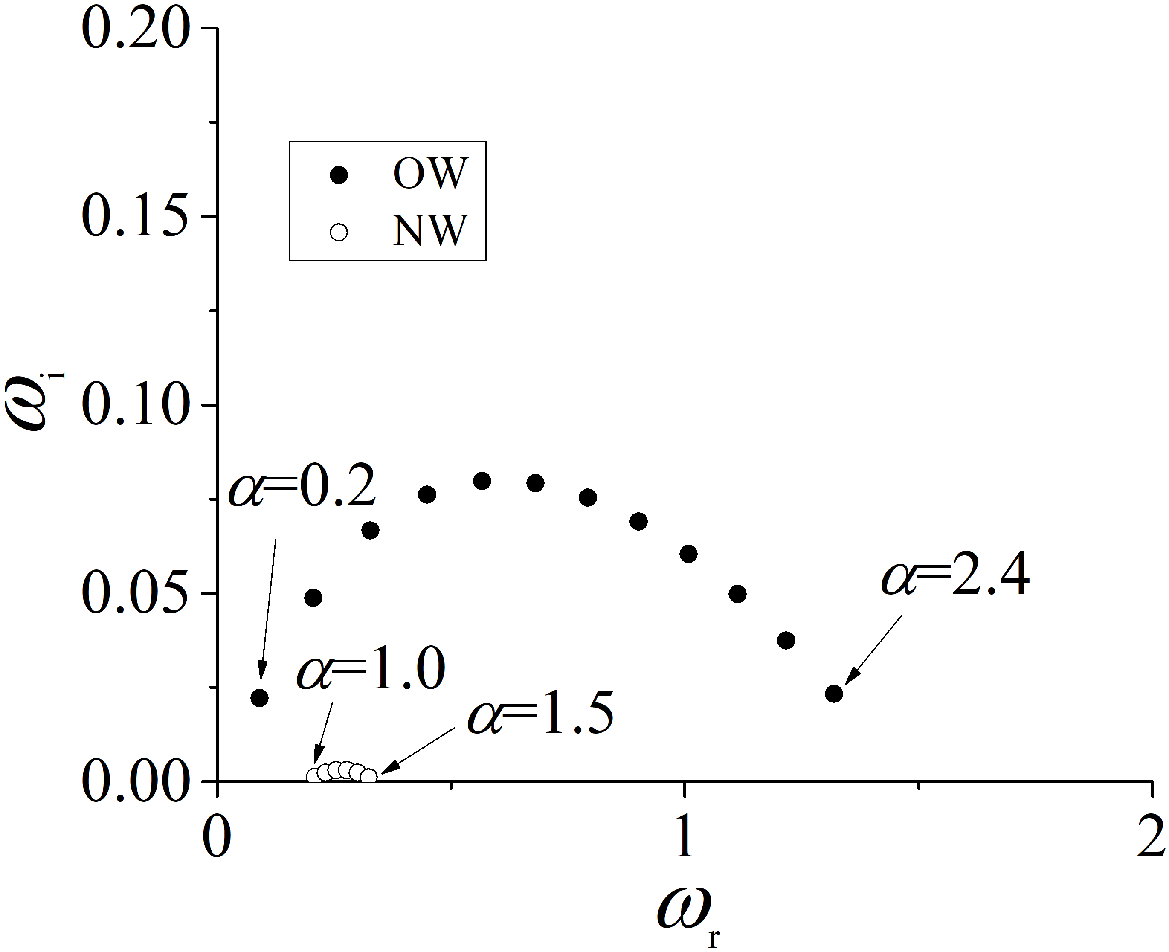}
(ii)~$\Gamma=6.25$
\label{fig:25}
\end{minipage}
\begin{minipage}{0.32\hsize}
\centering
\includegraphics[width=52mm]{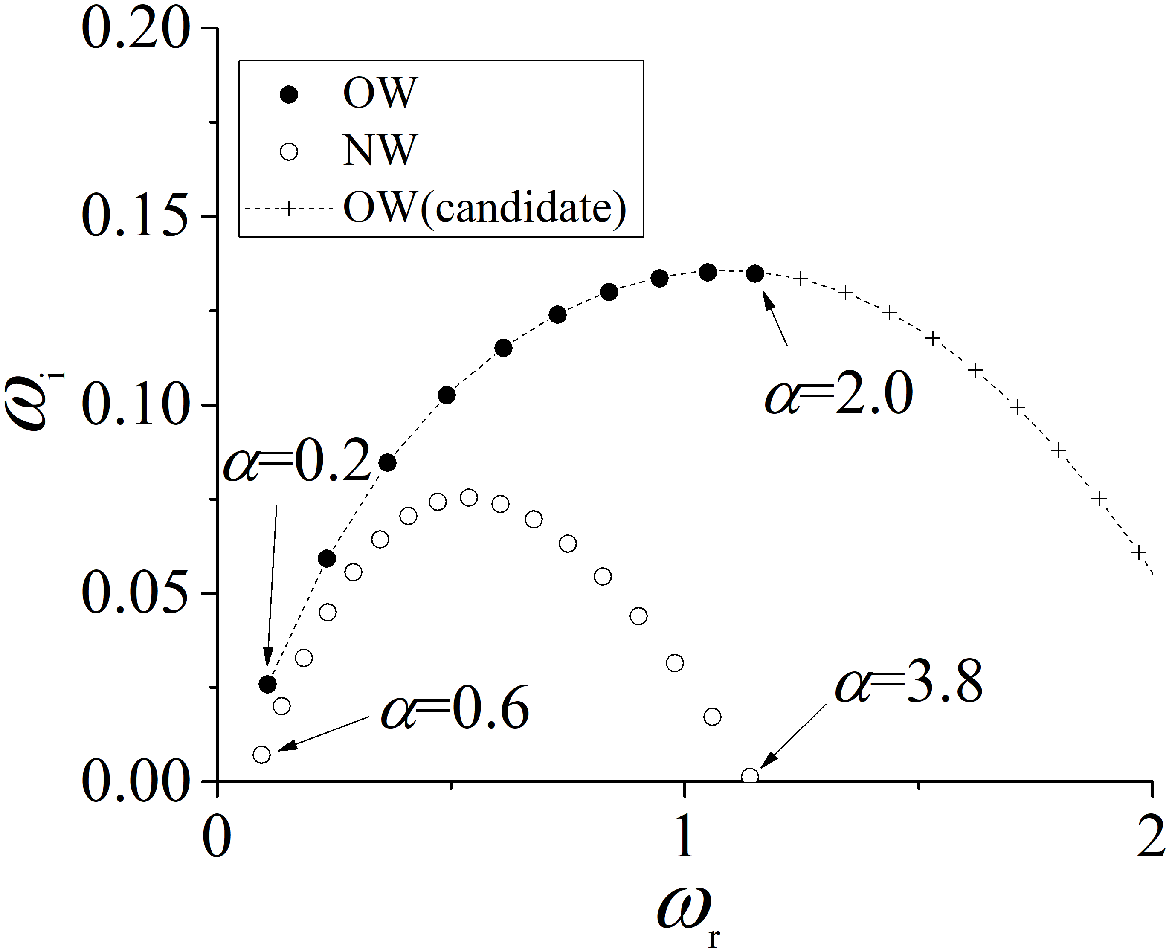}
(iii)~$\Gamma=12.5$
\label{fig:25}
\end{minipage}\\
\begin{minipage}{0.32\hsize}
\centering
\includegraphics[width=52mm]{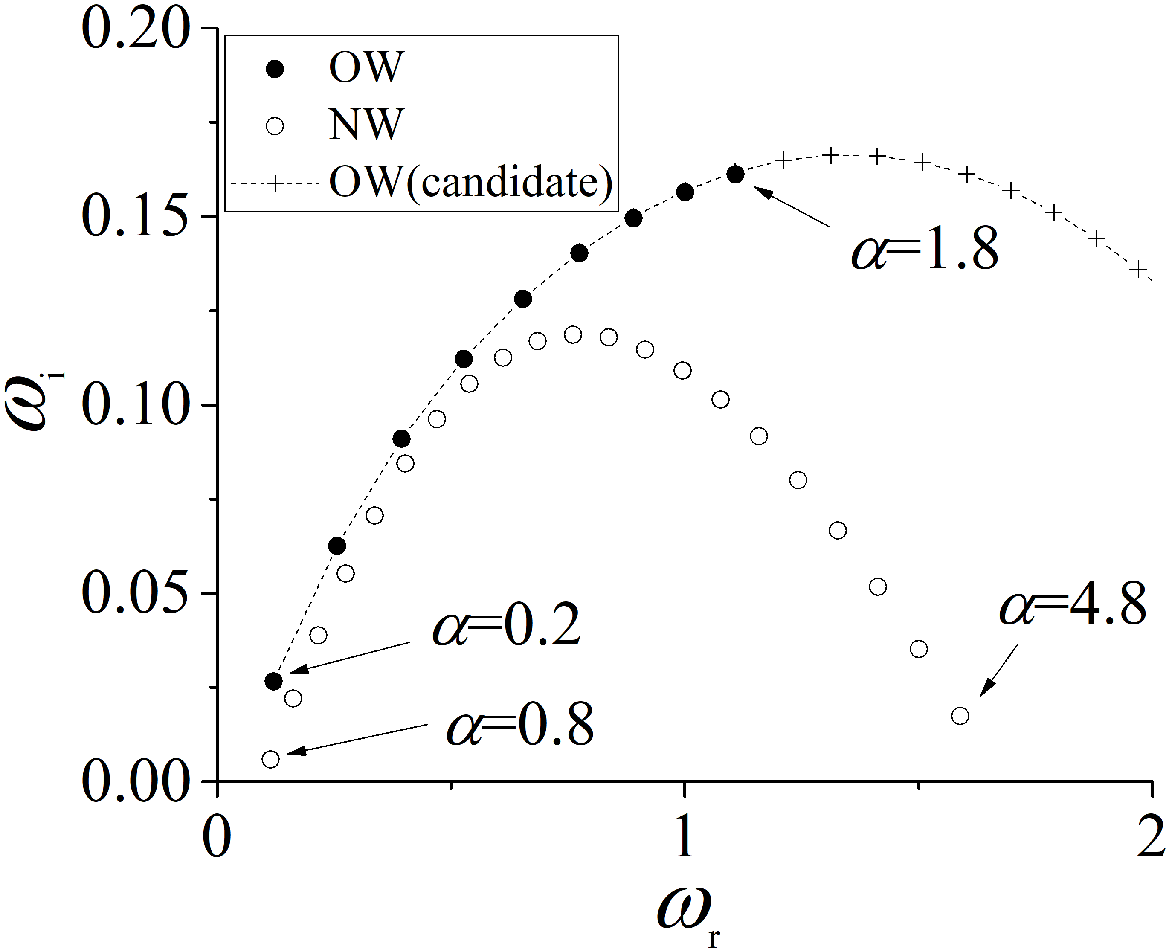}
(iv)~$\Gamma=18.7$
\label{fig:25}
\end{minipage}
\begin{minipage}{0.32\hsize}
\centering
\includegraphics[width=52mm]{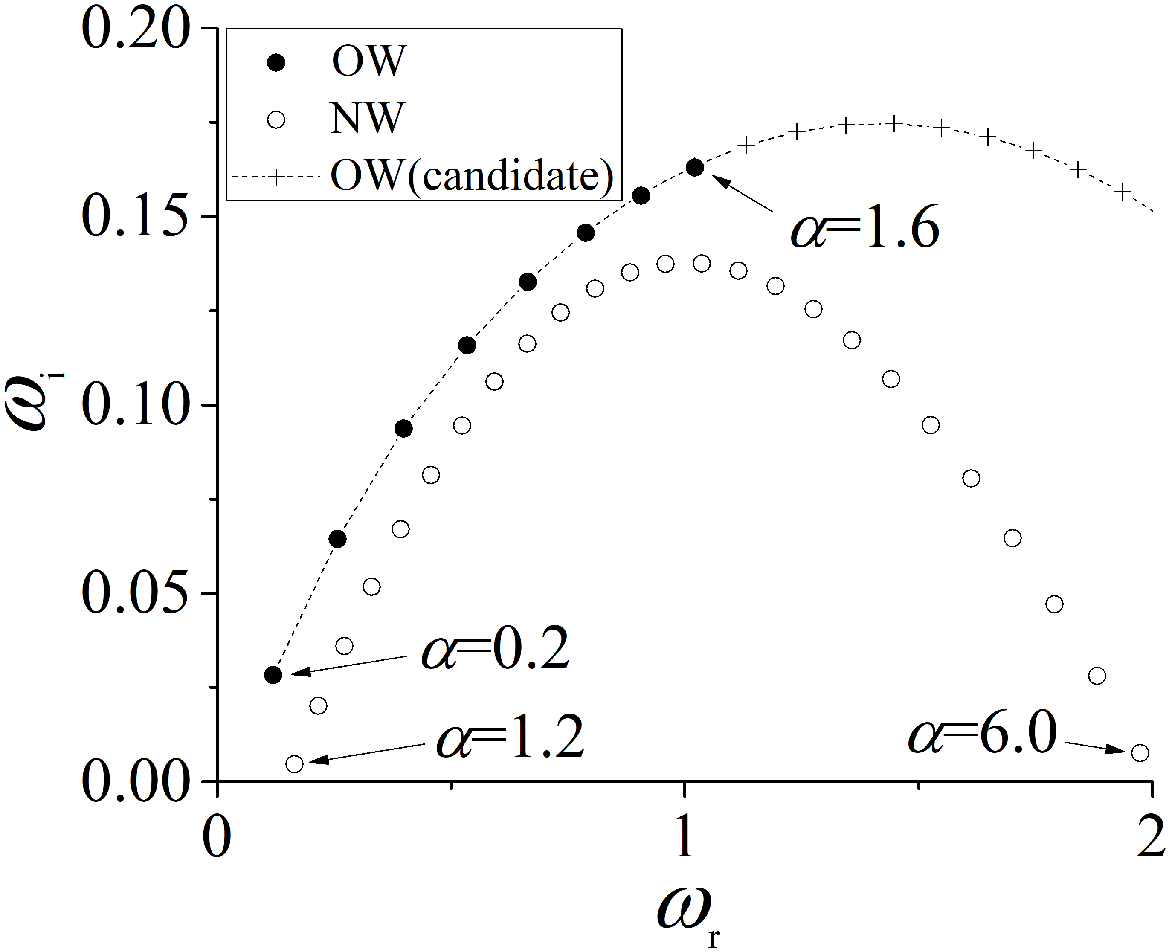}
(v)~$\Gamma=25.0$
\label{fig:25}
\end{minipage}\\
\centering
(a)~$Dx=11.6\delta_{in}$\\
%
%
\begin{minipage}{0.32\hsize}
\centering
\includegraphics[width=52mm]{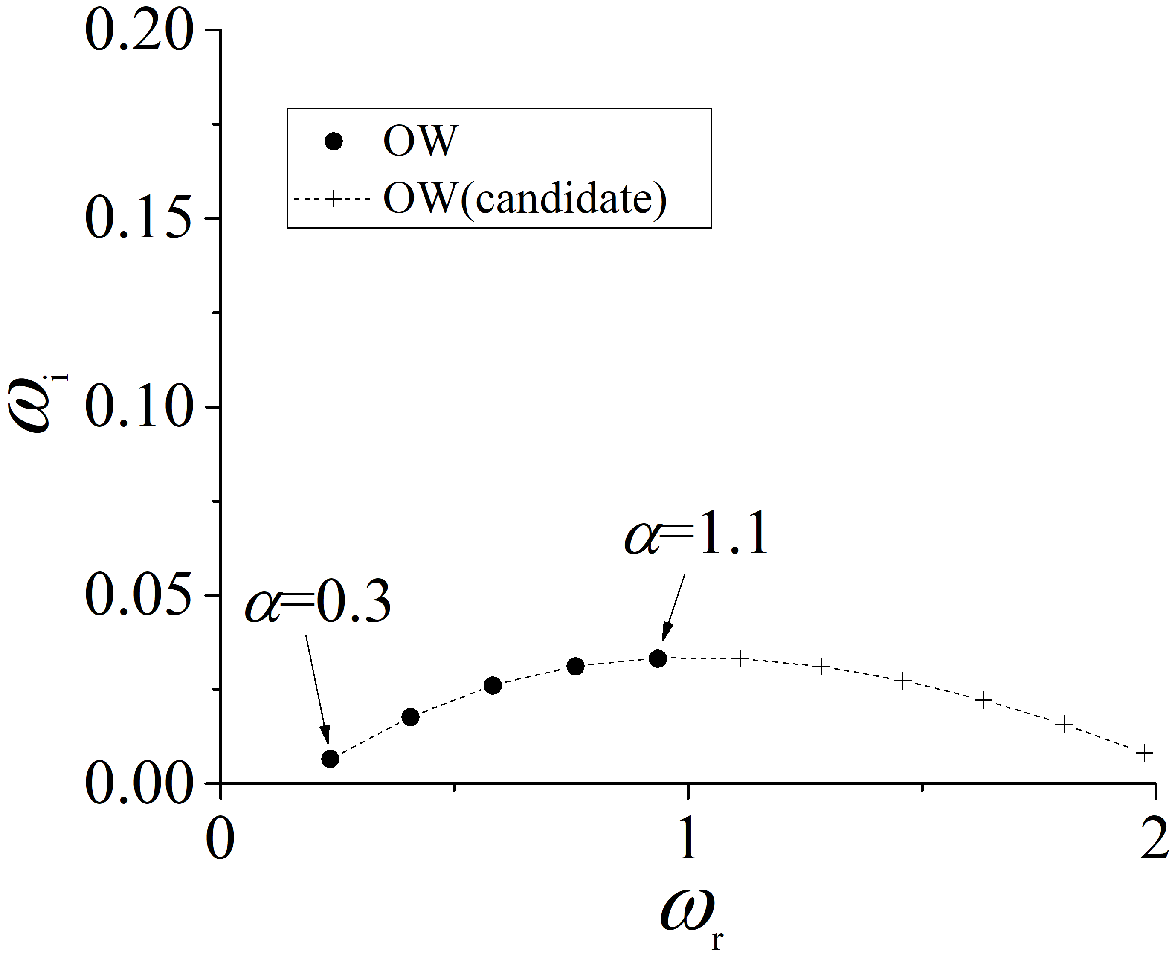}
(i)~$\Gamma=6.25$
\label{fig:24} 
\end{minipage}
\begin{minipage}{0.32\hsize}
\centering
\includegraphics[width=52mm]{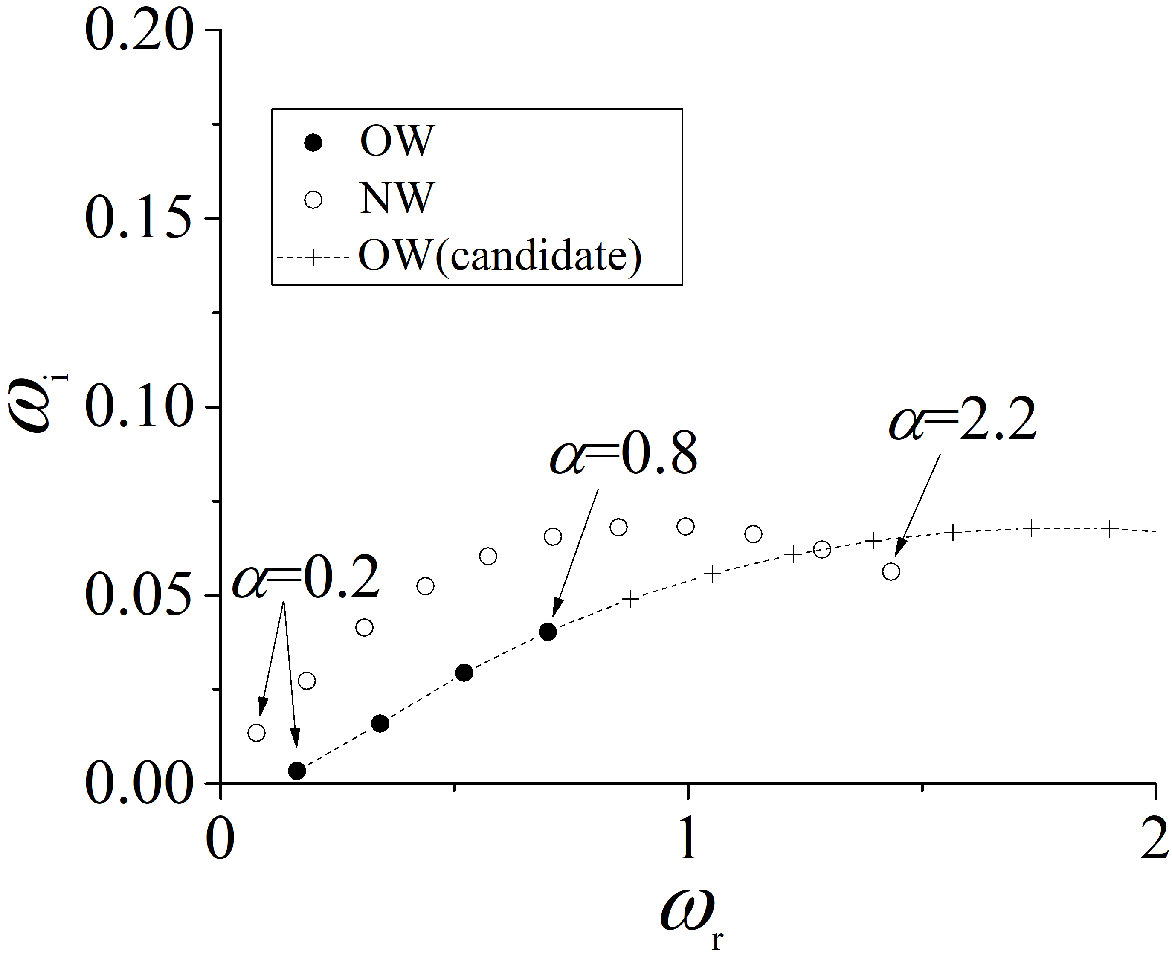}
(ii)~$\Gamma=12.5$
\label{fig:25}
\end{minipage}
\begin{minipage}{0.32\hsize}
\centering
\includegraphics[width=52mm]{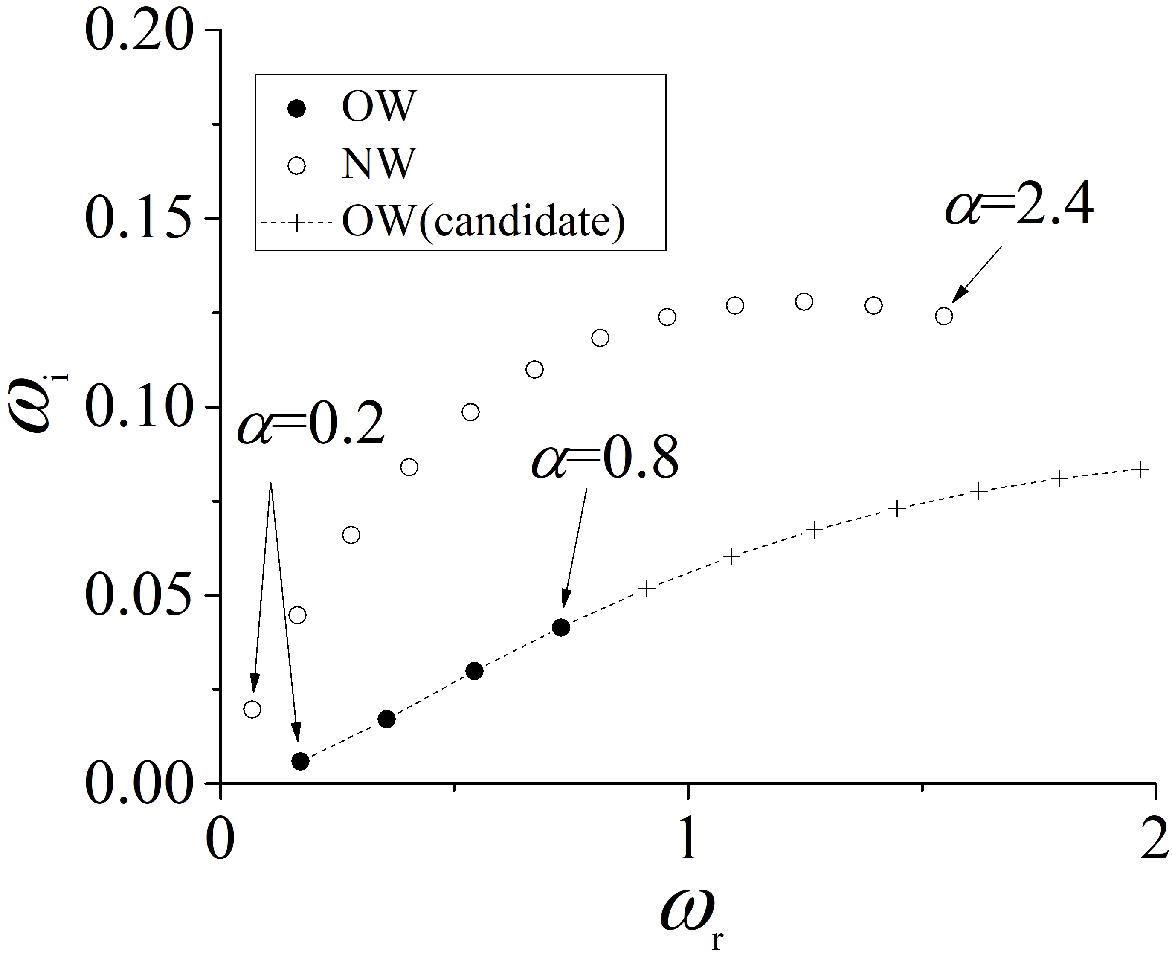}
(iii)~$\Gamma=25.0$
\label{fig:25}
\end{minipage}
%
\centering
(b)~$Dx=30.4\delta_{in}$

\caption{Variation of the eigenvalues of unstable modes when streamwise wavenumber $\alpha$ is varied when
$\phi=4^\circ$ and $t^\ast=10.6$ 
($\alpha$ is varied over ranges $\Gamma=$3.13-25.0 and $\Gamma=$6.25-25.0 at (a) $Dx=11.6\delta_{in}$ and (b) $Dx=30.4\delta_{in}$, respectively), OW: off-wall mode, NW: near-wall mode}
\end{figure}
%
\begin{figure}
\begin{minipage}{0.32\hsize}
\centering
\includegraphics[width=52mm]{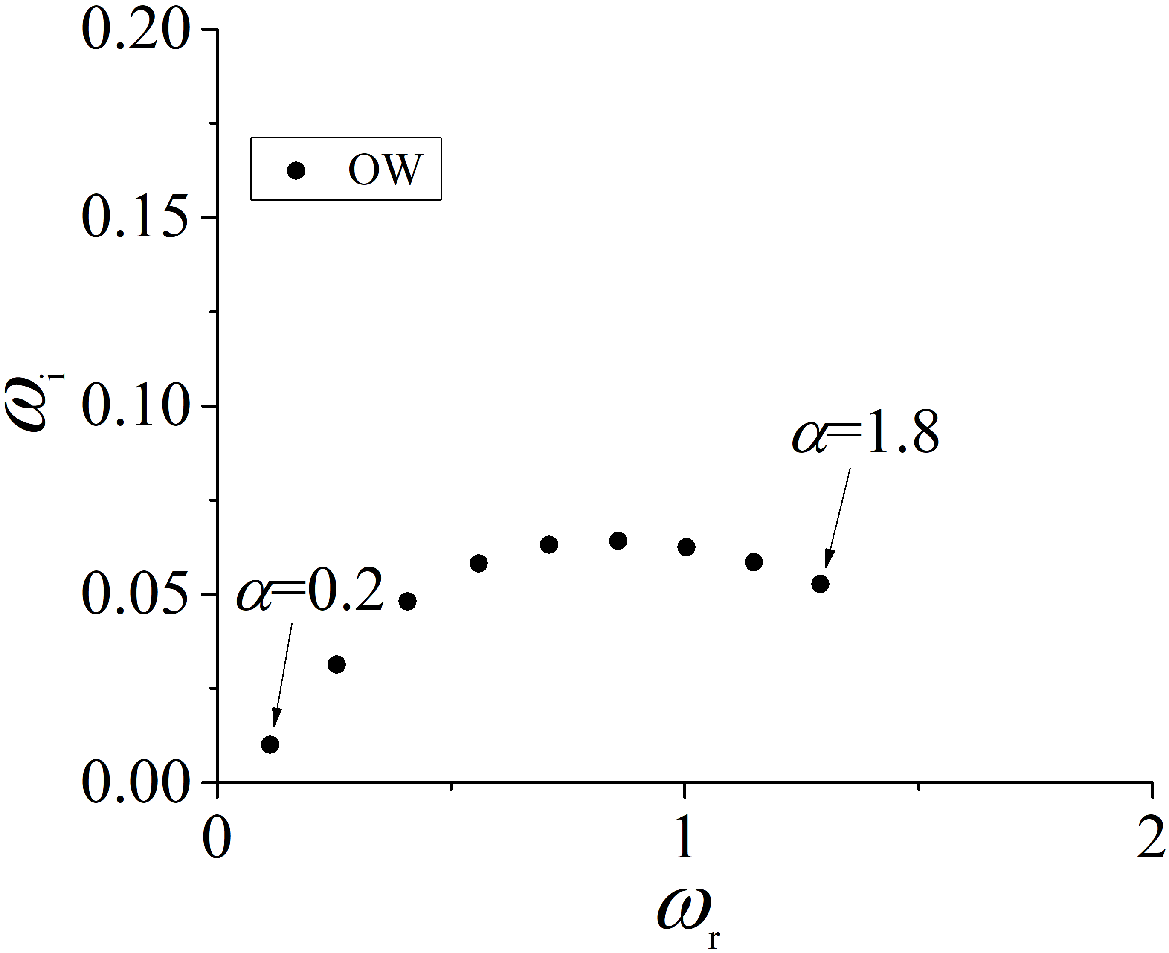}
(i)~$\Gamma=6.25$
\label{fig:24} 
\end{minipage}
\begin{minipage}{0.32\hsize}
\centering
\includegraphics[width=52mm]{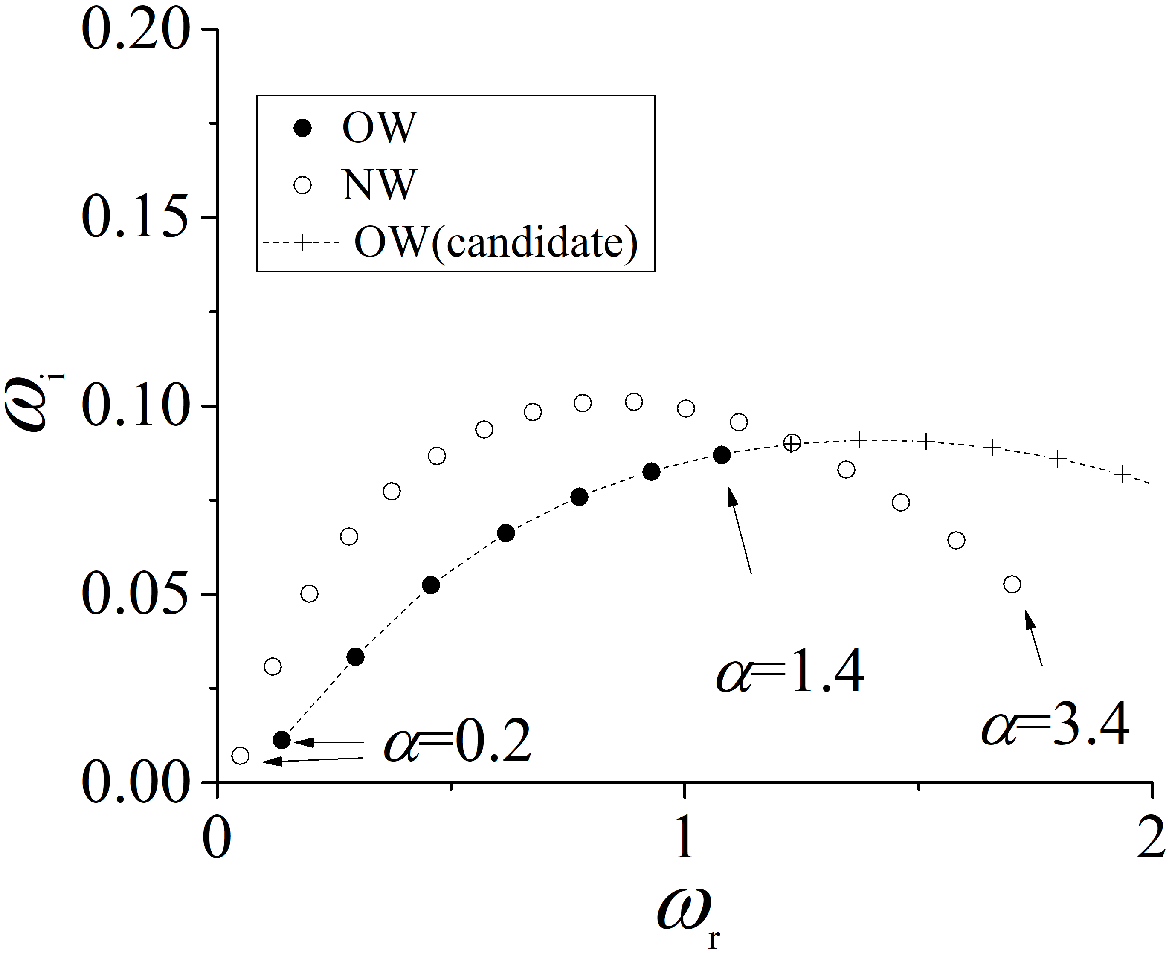}
(ii)~$\Gamma=12.5$
\label{fig:25}
\end{minipage}
\begin{minipage}{0.32\hsize}
\centering
\includegraphics[width=52mm]{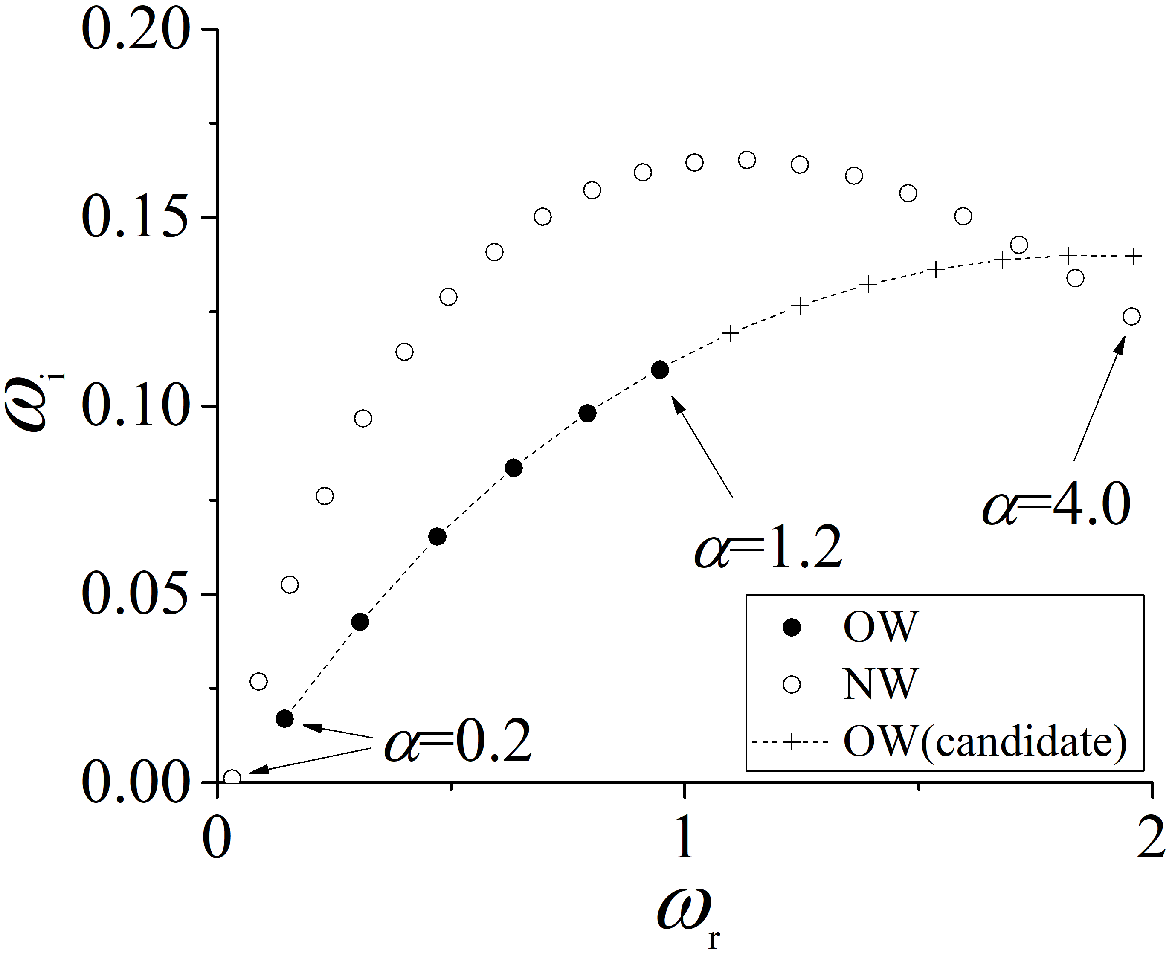}
(iii)~$\Gamma=25.0$
\label{fig:25}
\end{minipage}
%
\centering
\caption{Variation of the eigenvalues of unstable modes when the streamwise wavenumber $\alpha$ is varied over range $\Gamma=$6.25-25.0 at $Dx=11.6\delta_{in}$ when $\phi=10^\circ$ and $t^\ast=10.6$,
OW: off-wall mode, NW: near-wall mode}
\end{figure}

Figure 8 shows the distribution of the eigenvalues of unstable modes along the variation of $\alpha$ with a constant increment when $\phi=4^\circ$ and $t^\ast=10.6$; 
panels (a) and (b) show the distributions for $\Gamma=3.13-25.0$ at $Dx=11.6\delta_{in}$ and $\Gamma=6.25-25.0$ at $Dx=30.4\delta_{in}$, respectively.
The ordinate and abscissa show the imaginary and real parts, i.e., $\omega_i$ and $\omega_r$, of the eigenvalue, respectively. 
Regarding the off-wall modes, `black' points show the converged eigenvalues as a result of Newton iteration.
On the other hand, Newton iteration does not necessarily converge for a wide range of eigenvalues.
To understand the distribution trend of eigenvalues, candidate eigenvalues obtained by the global method, which are not necessarily converged by Newton iteration, are also shown.     
In the figure, values of $\alpha$ for terminal points are also shown.  
The increments of $\alpha$ are 0.1 or 0.2 for all cases.
When $\phi=4^\circ$, there is one unstable mode for $\Gamma=3.13$, and two unstable modes, i.e., `off-wall' and `near-wall' modes, for $\Gamma \ge 6.25$ at $Dx=11.6\delta_{in}$. 
At $Dx=30.4\delta_{in}$, there is one mode for $\Gamma=6.25$ at $Dx=11.6\delta_{in}$, and two modes for $\Gamma \ge 12.5$. 
When $\Gamma$ is increased, the peak of a curve of eigenvalues with constant $\Gamma$ basically shifts in the direction of larger $\omega_r$ and larger $\omega_i$ as found from Fig. 8(a). 

Figure 9 shows the distribution of the eigenvalues of unstable modes at $Dx=11.6\delta_{in}$ for $\Gamma=6.25-25.0$ when $\phi=10^\circ$ at $t^\ast=10.6$. 
In this figure, there is one mode for $\Gamma=6.25$, and two modes for $\Gamma \ge 12.5$. 
In contrast, at $Dx=30.4\delta_{in}$, there are no unstable modes for $\Gamma=6.25-25.0$ although they are not shown here. 
When $\phi=10^\circ$, because the vortex tube introduces only a slight modification to the velocity profile in the region where the freestream is at $Dx=30.4\delta_{in}$, the boundary layer remains stable at the position. 

As mentioned previously, consecutive hairpin vortices are generated for $\Gamma \ge 12.5$ in Fig. 3, and there are always two modes for such circulation as found in Fig. 8. 
When the fact is taken into consideration, one sees that presence of the two modes coincides with the generation of the hairpin vortices. 
Conversely, the presence of the two modes can be considered as a precursor for the generation of hairpin vortices. 
Indeed, for $\Gamma=6.25$ and $\phi=4^\circ$, the two modes are seen in Fig. 8. 
However, the growth rate of the near-wall mode is very small and mimics a situation where only one mode is present. 
Therefore, clear hairpin vortices do not develop as evident in Fig. 3.

Although the formation of hairpin vortices appears to fall under nonlinear stability theory, which is distinct from the present LSA, the trend in their appearance around the vortex tube agrees with the results of the present analysis. 
As found in Fig. 8(a), the maximum amplification takes place when $\alpha=1.8$ for the off-wall mode and $\alpha=2.2$ for the near-wall mode at $Dx=11.6\delta_{in}$ corresponding to wavelengths $\lambda=3.5\delta_{in}$ and $2.9\delta_{in}$, respectively, when $\phi=4^\circ$, $\Gamma=12.5$ and $t^\ast=10.6$. 
The distance between conspicuous hairpin vortices in the DNS is around $4.16\delta_{in}$ when $t^\ast=53.1$, and $2.92\delta_{in}$ when $t^\ast=74.3$. 
Approximate distance between the hairpin vortices can also be inferred from Fig. 3(a). 
Associated with Fig. 8(a), the peaks of eigenvalue curves shift as $\Gamma$ is increased as mentioned previously. 
This trend has a physical implication that the distance between hairpin vortices decreased as $\Gamma$ is increased.
This is consistent to and also justified by the fact that the distance between hairpin vortices decrease as $\Gamma$ is increased in the case of single hairpin vortex \citep{Matsuura16}.
Hence, the present analysis unexpectedly predicts the properties of the initial disturbances triggering the consecutive hairpin vortices reasonably well. 
The number of hairpin vortices generated around the vortex tube depends on the circumstances in which the vortex tube is situated inside the boundary layer such as the circulation, angle-to-wall and the partial length of the vortex tube inside the boundary layer. 

Thus, the abrupt introduction of a vortex tube of various magnitudes and inclinations to the boundary layer reveals a new aspect of the nonlinear boundary layer stability as related to whether hairpin vortices are generated or not. 

\subsection{Detailed evolution of disturbance around the vortex tube}
\subsubsection{Emergence of disturbance}
\begin{figure}
\begin{minipage}{0.32\hsize}
\centering
\includegraphics[width=42mm]{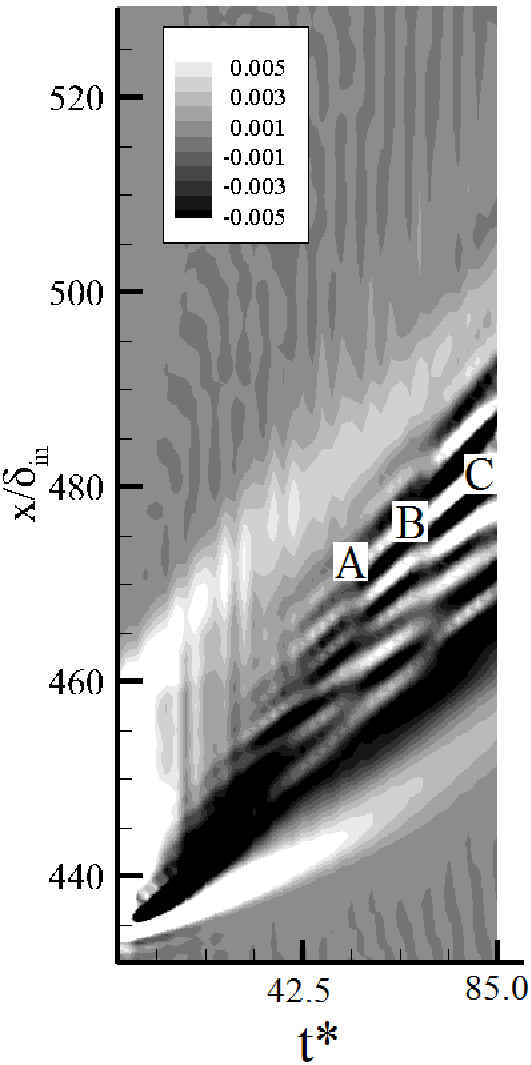}
(a)~$y=5\%\delta_{in}$
\label{fig:26} 
\end{minipage}
\begin{minipage}{0.32\hsize}
\centering
\includegraphics[width=42mm]{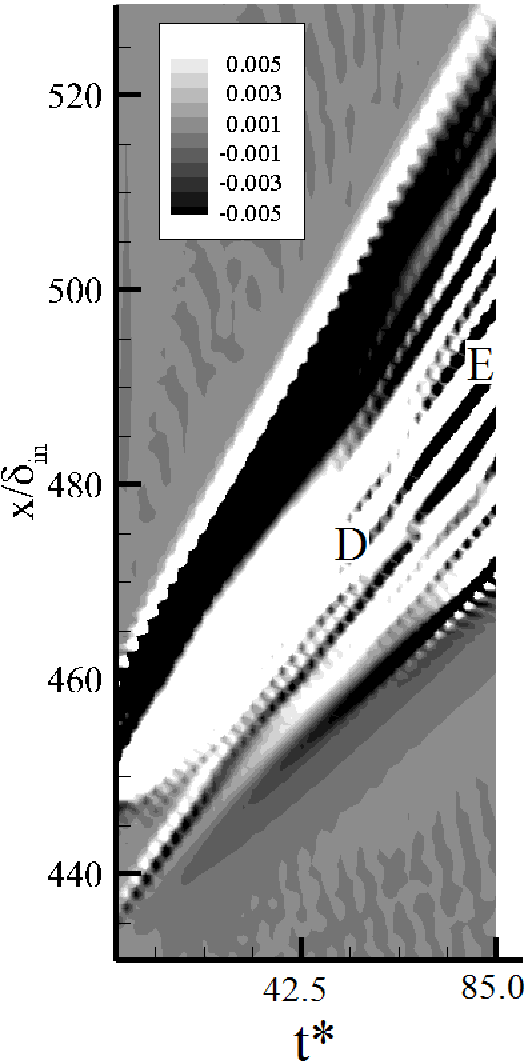}
(b)~$y=50\%\delta_{in}$
\label{fig:27}
\end{minipage}
\begin{minipage}{0.32\hsize}
\centering
\includegraphics[width=42mm]{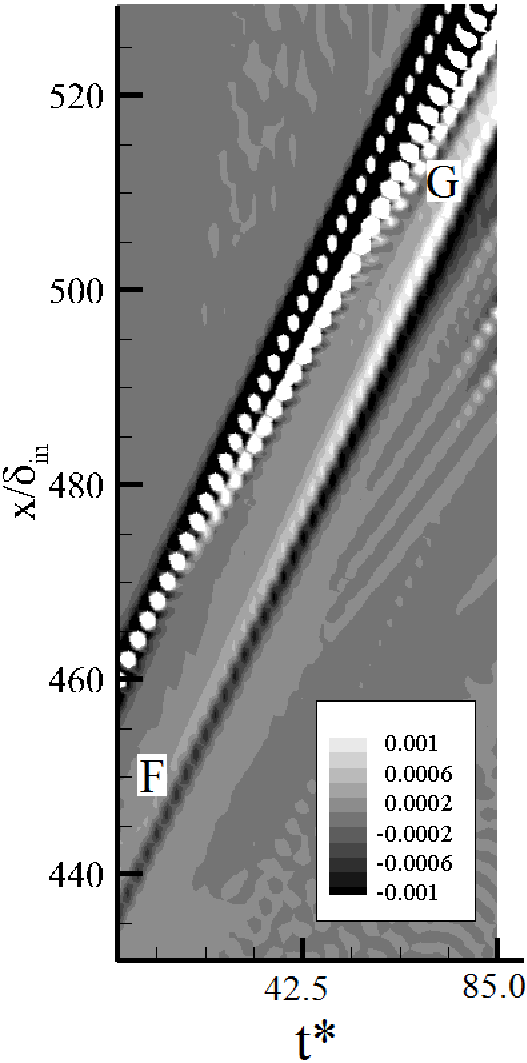}
(c)~$y=100\%\delta_{in}$
\label{fig:28}
\end{minipage}\\
\caption{Space-time plots of $S_x^\ast$ at $y=5\%\delta_{in}$, $50\%\delta_{in}$ and $100\%\delta_{in}$ when $\Gamma=12.5$ and $\phi=4^\circ$. 
The grayscale range of part (c) is made different from those in (a) and (b) for a visualization purpose.}
\end{figure}
%
%
\begin{figure}
\begin{minipage}{0.32\hsize}
\centering
\includegraphics[width=50mm]{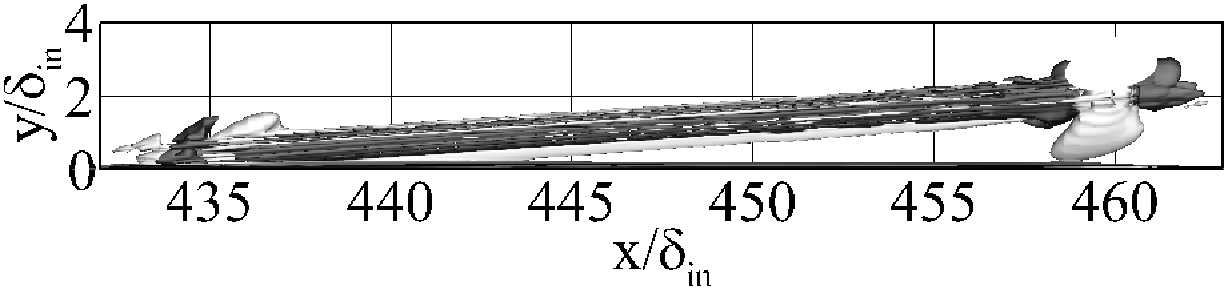}
\label{fig:29} 
\end{minipage}
\begin{minipage}{0.32\hsize}
\centering
\includegraphics[width=50mm]{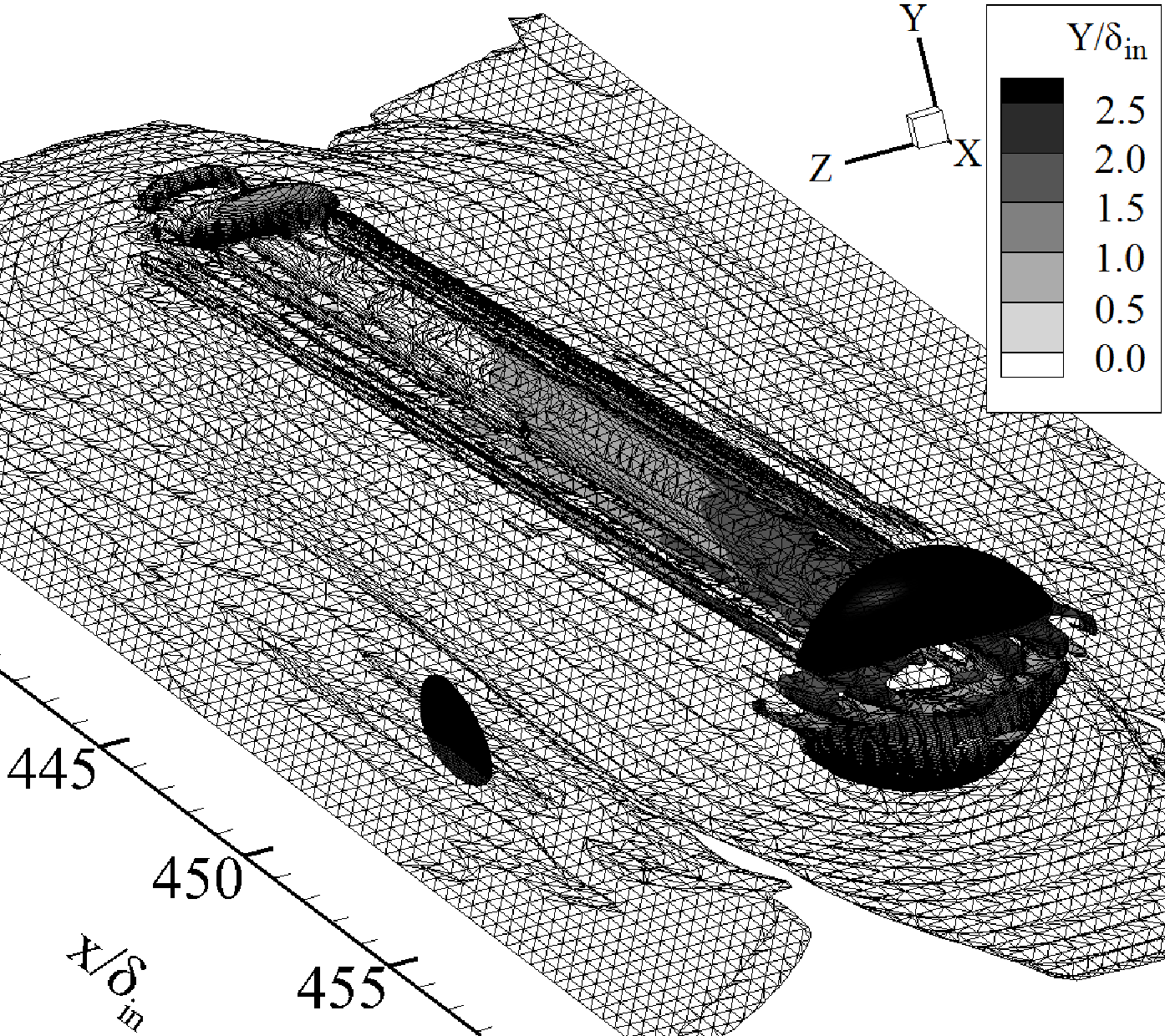}
\label{fig:29}
\end{minipage}
\begin{minipage}{0.32\hsize}
\centering
\includegraphics[width=50mm]{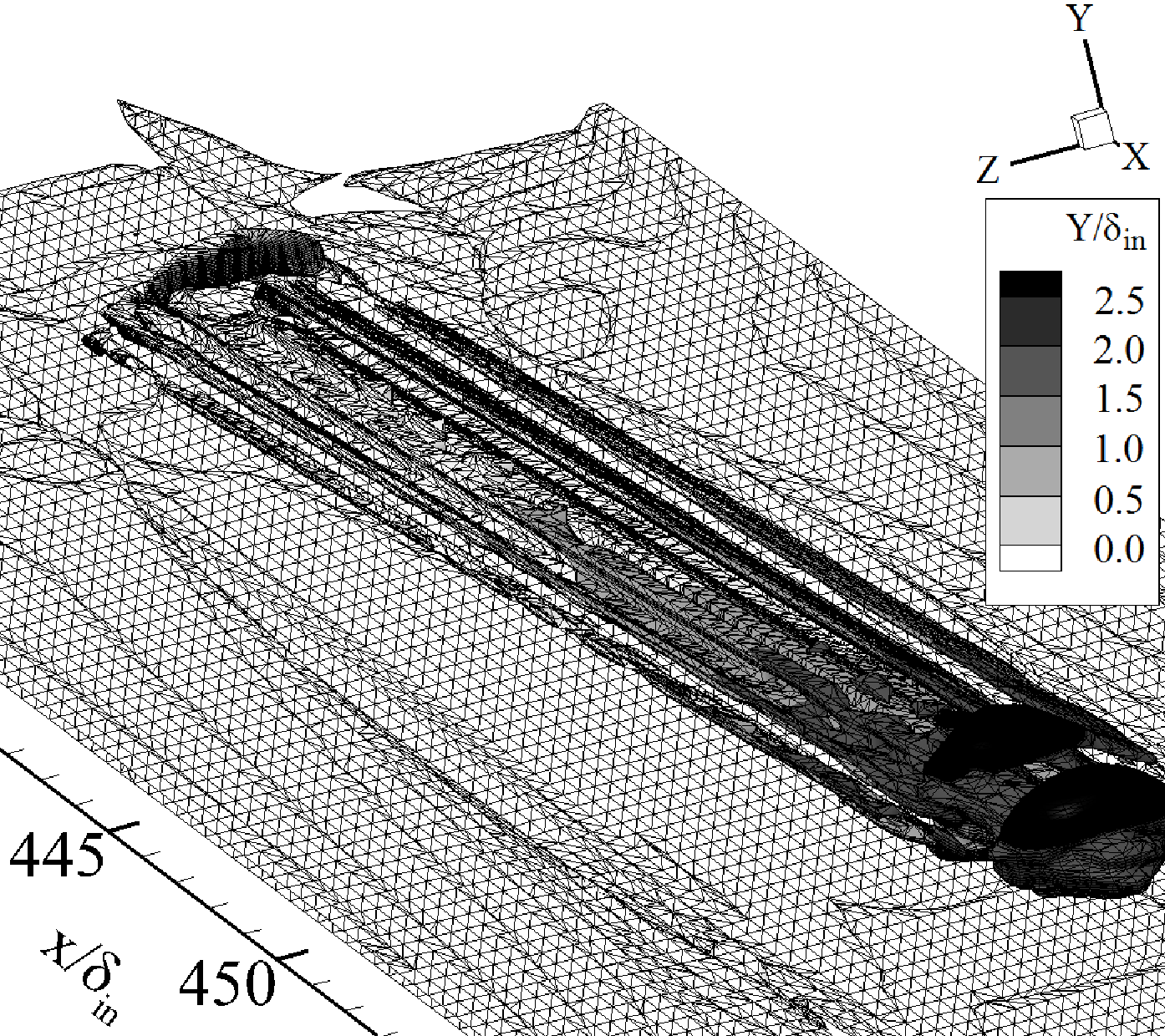}
\label{fig:29}
\end{minipage}
\centering
(a) $t^\ast=2.13$\\
%
%
\begin{minipage}{0.32\hsize}
\centering
\includegraphics[width=50mm]{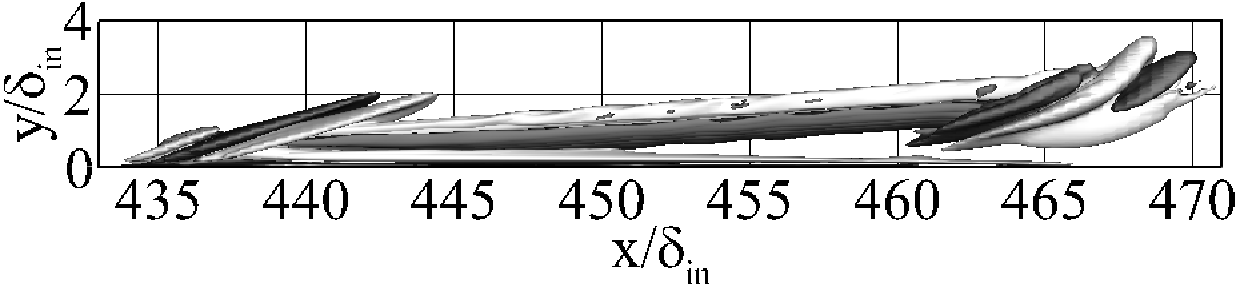}
\label{fig:30} 
\end{minipage}
\begin{minipage}{0.32\hsize}
\centering
\includegraphics[width=50mm]{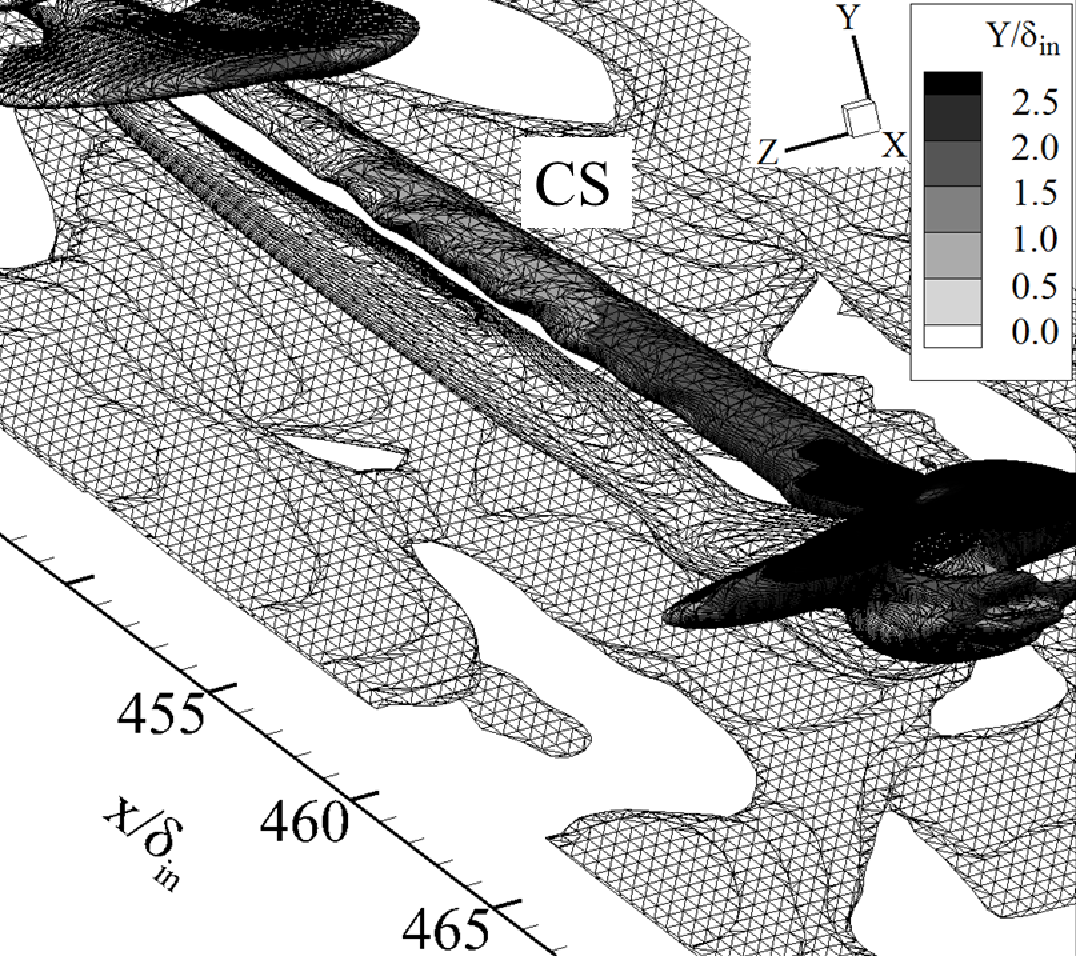}
\label{fig:31}
\end{minipage}
\begin{minipage}{0.32\hsize}
\centering
\includegraphics[width=50mm]{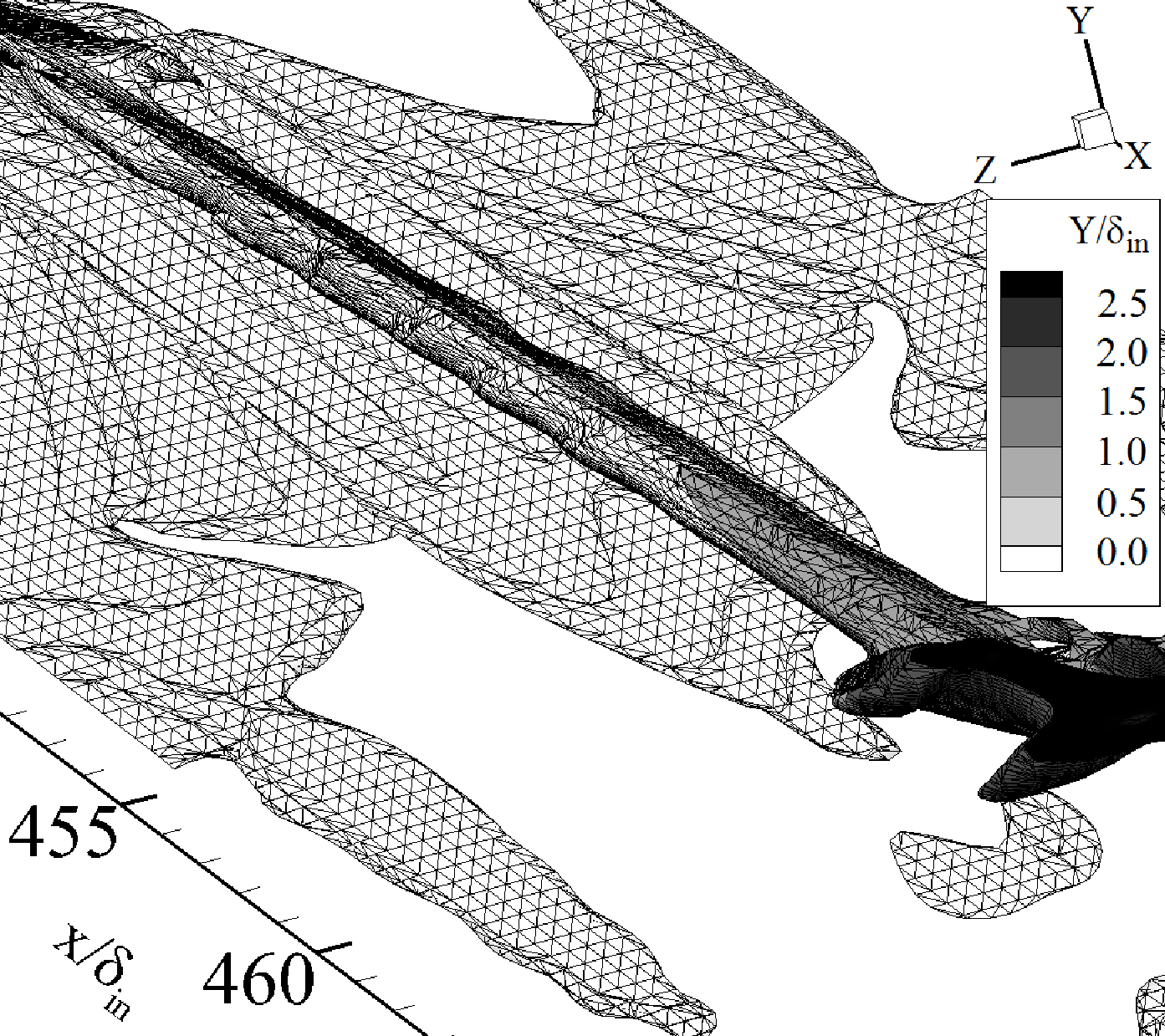}
\label{fig:32}
\end{minipage}
\centering
(b) $t^\ast=10.6$\\
%
%
\begin{minipage}{0.32\hsize}
\centering
\includegraphics[width=50mm]{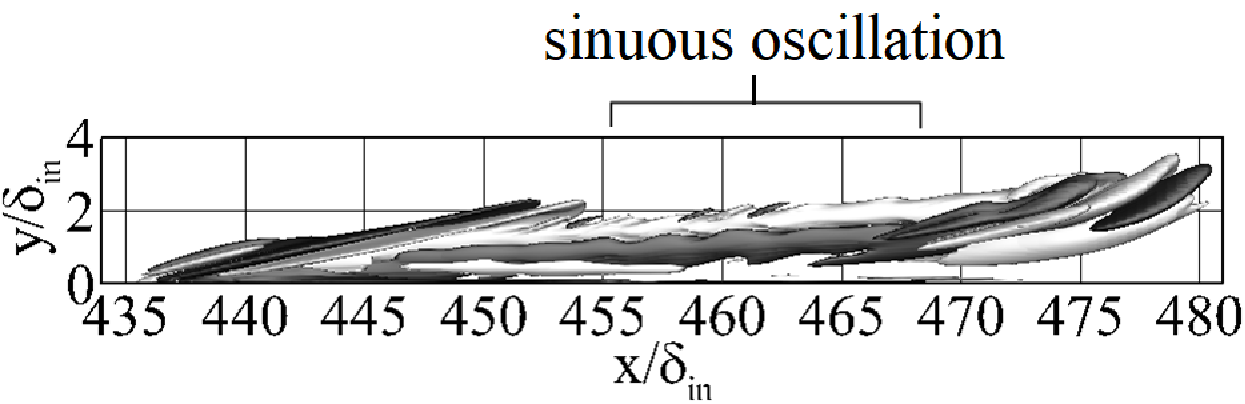}
\label{fig:33} 
\end{minipage}
\begin{minipage}{0.32\hsize}
\centering
\includegraphics[width=50mm]{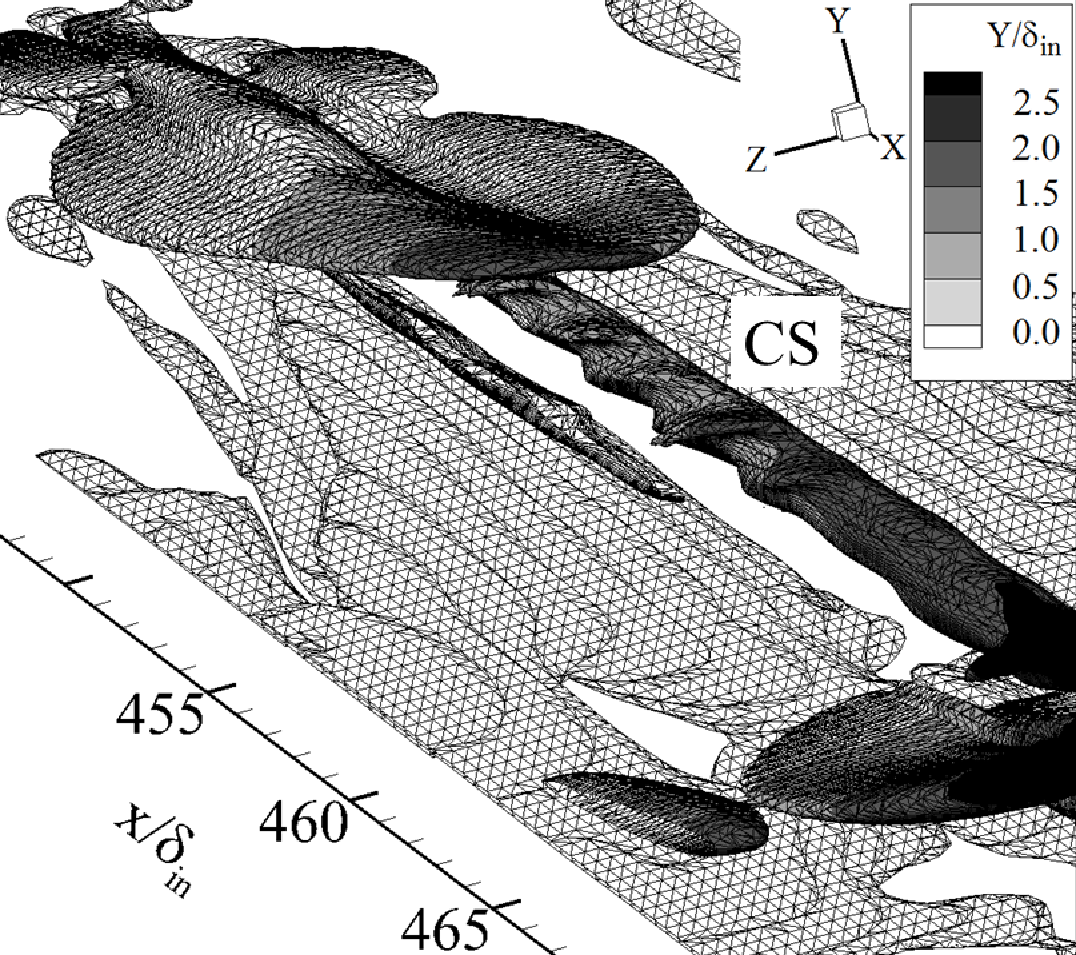}
\label{fig:34}
\end{minipage}
\begin{minipage}{0.32\hsize}
\centering
\includegraphics[width=50mm]{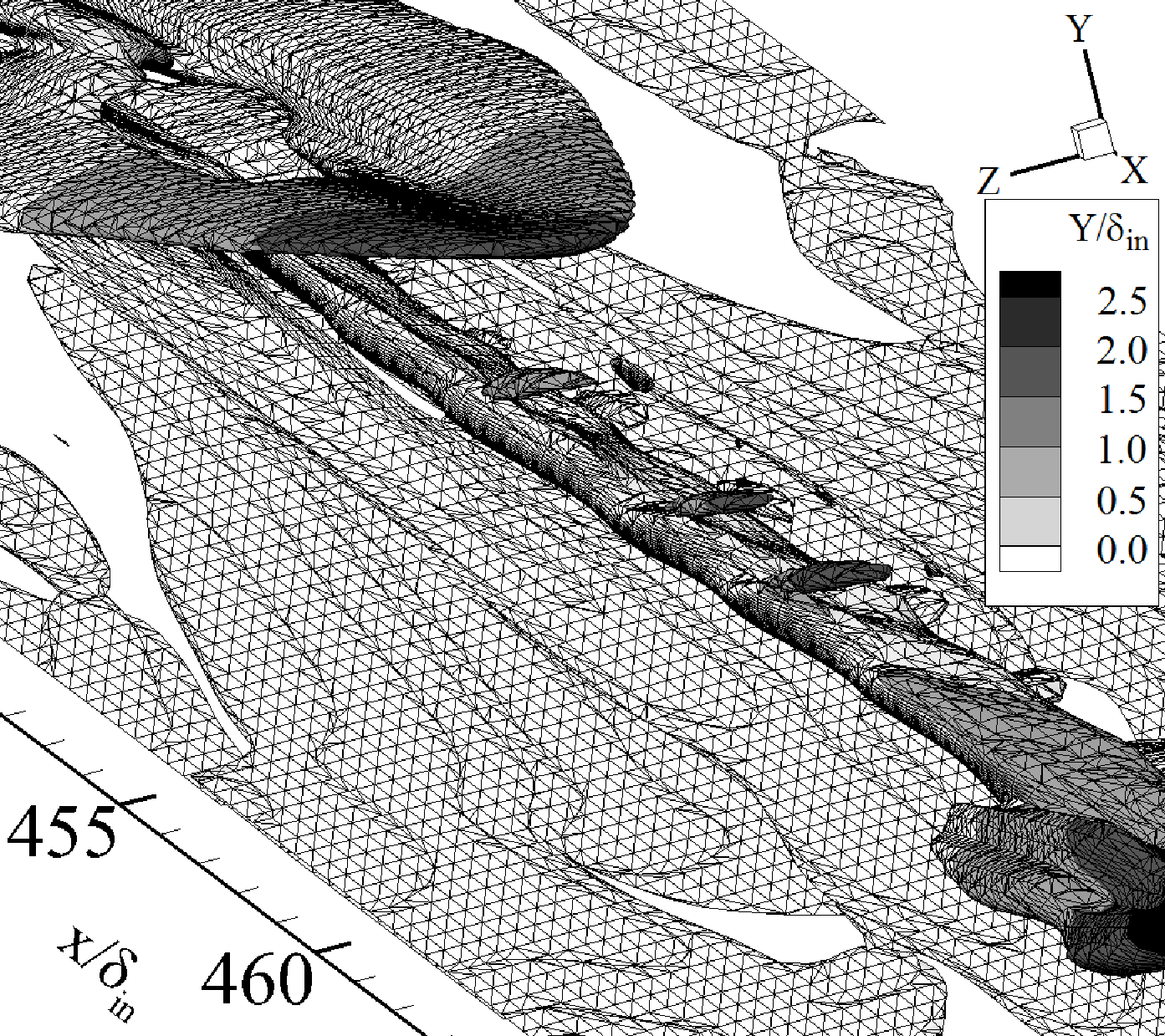}
\label{fig:35}
\end{minipage}
\centering
(c) $t^\ast=21.3$\\
%
%
\begin{minipage}{0.32\hsize}
\centering
\includegraphics[width=50mm]{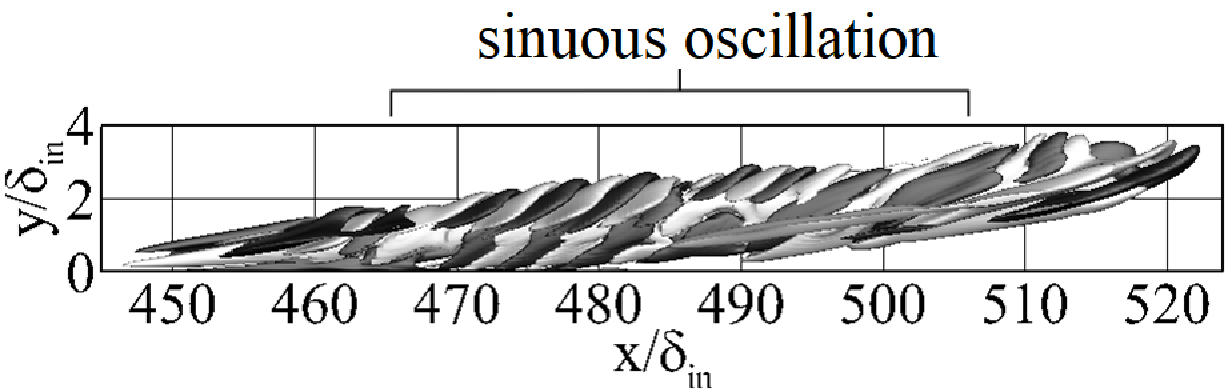}
\label{fig:36} 
\end{minipage}
\begin{minipage}{0.32\hsize}
\centering
\includegraphics[width=50mm]{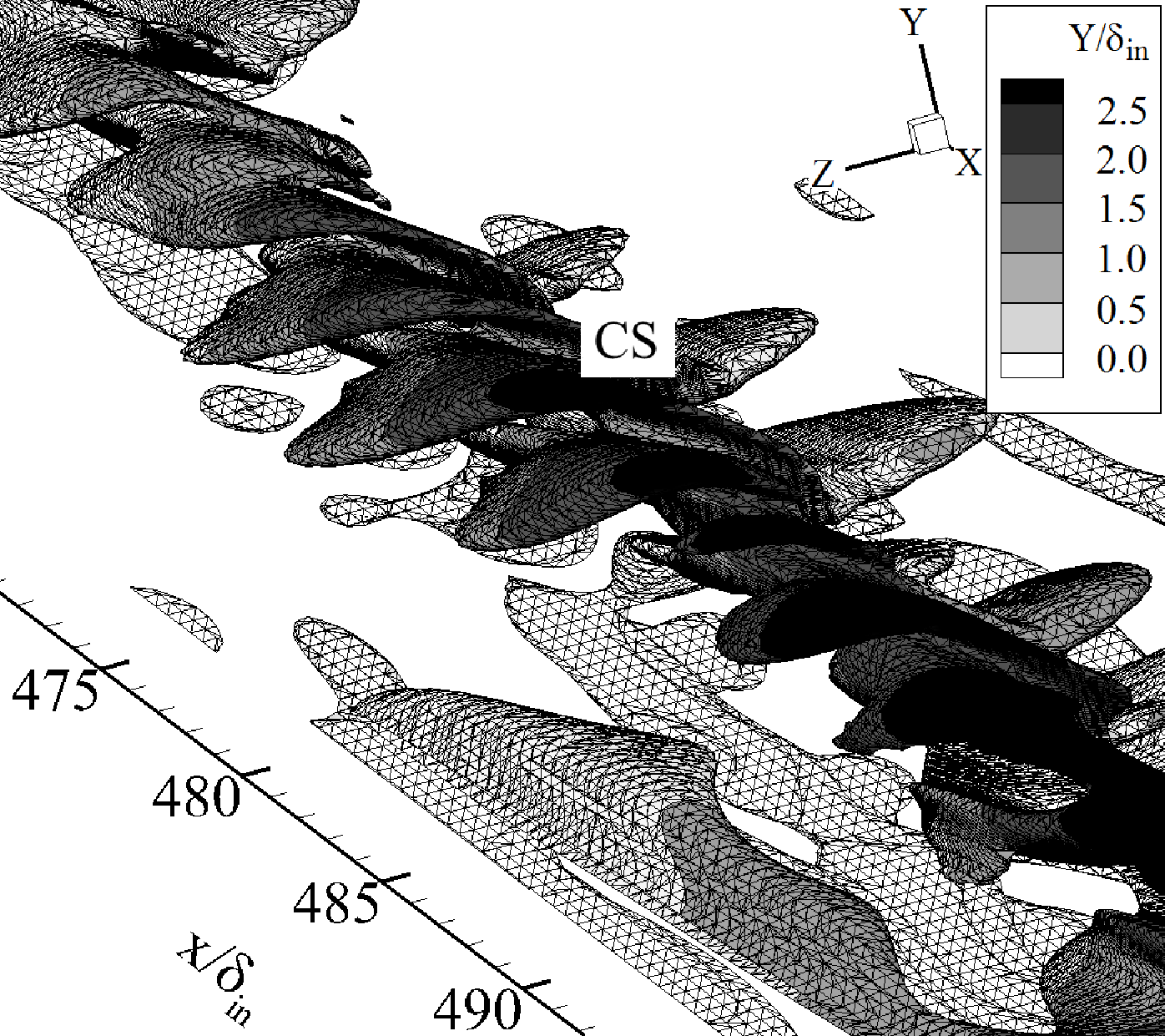}
\label{fig:37}
\end{minipage}
\begin{minipage}{0.32\hsize}
\centering
\includegraphics[width=50mm]{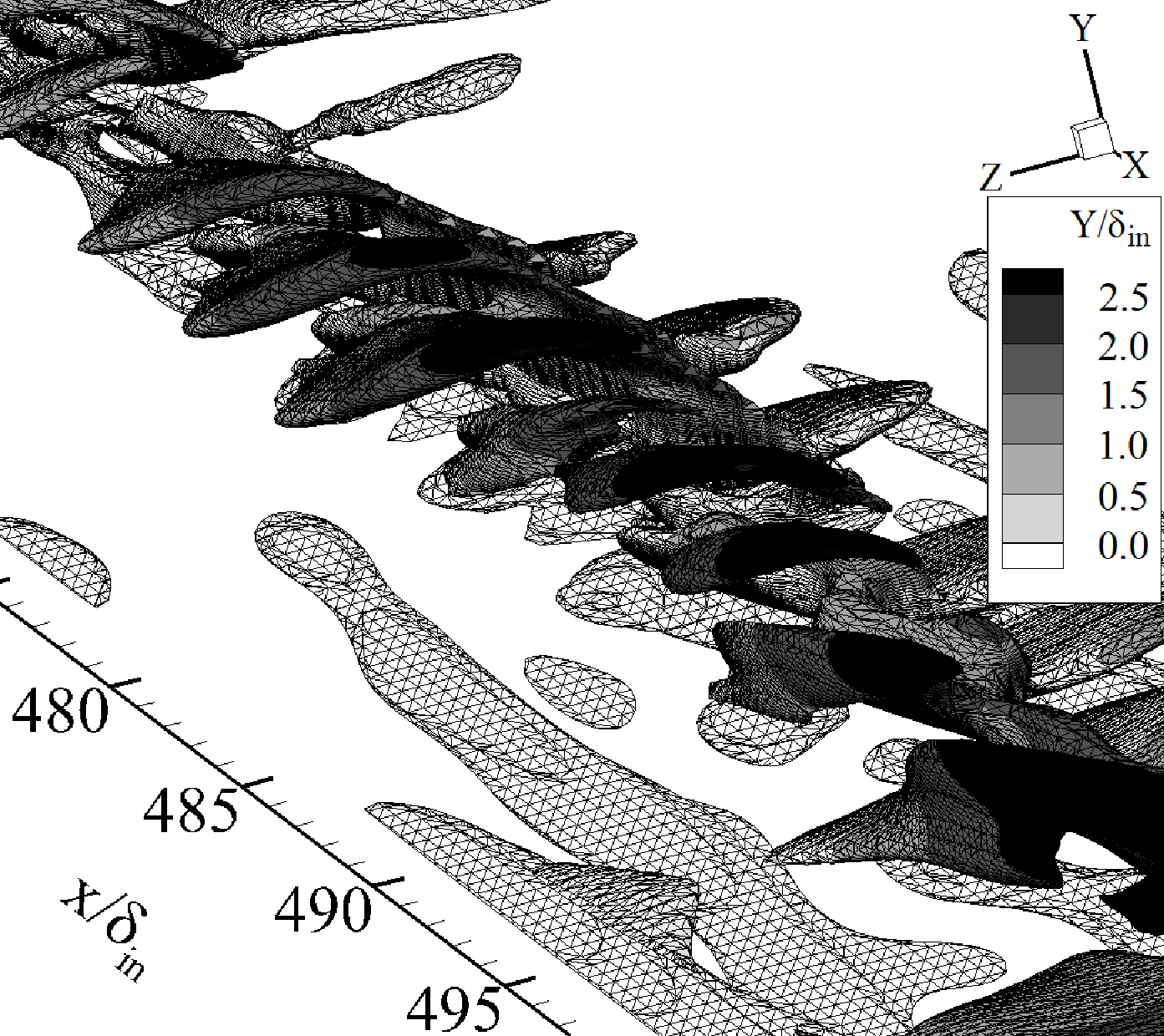}
\label{fig:38}
\end{minipage}
\centering
(d) $t^\ast=63.9$\\
%
%
\caption{Structural evolution of the vortex tube for $t^\ast$=2.13, 10.65, 21.3 and 63.9 when $\Gamma=12.5$ and $\phi=4^\circ$, 
Left: the iso-surfaces of the positive \& negative $S_x^\ast$, light gray: $S_x^\ast$=0.001, dark gray: $S_x^\ast$=-0.001, Center: the iso-surface of the positive $S_x^\ast$, 
Right: the iso-surface of the negative $S_x^\ast$}
\end{figure}
%
\begin{figure}
\centerline{\includegraphics[scale=0.3]{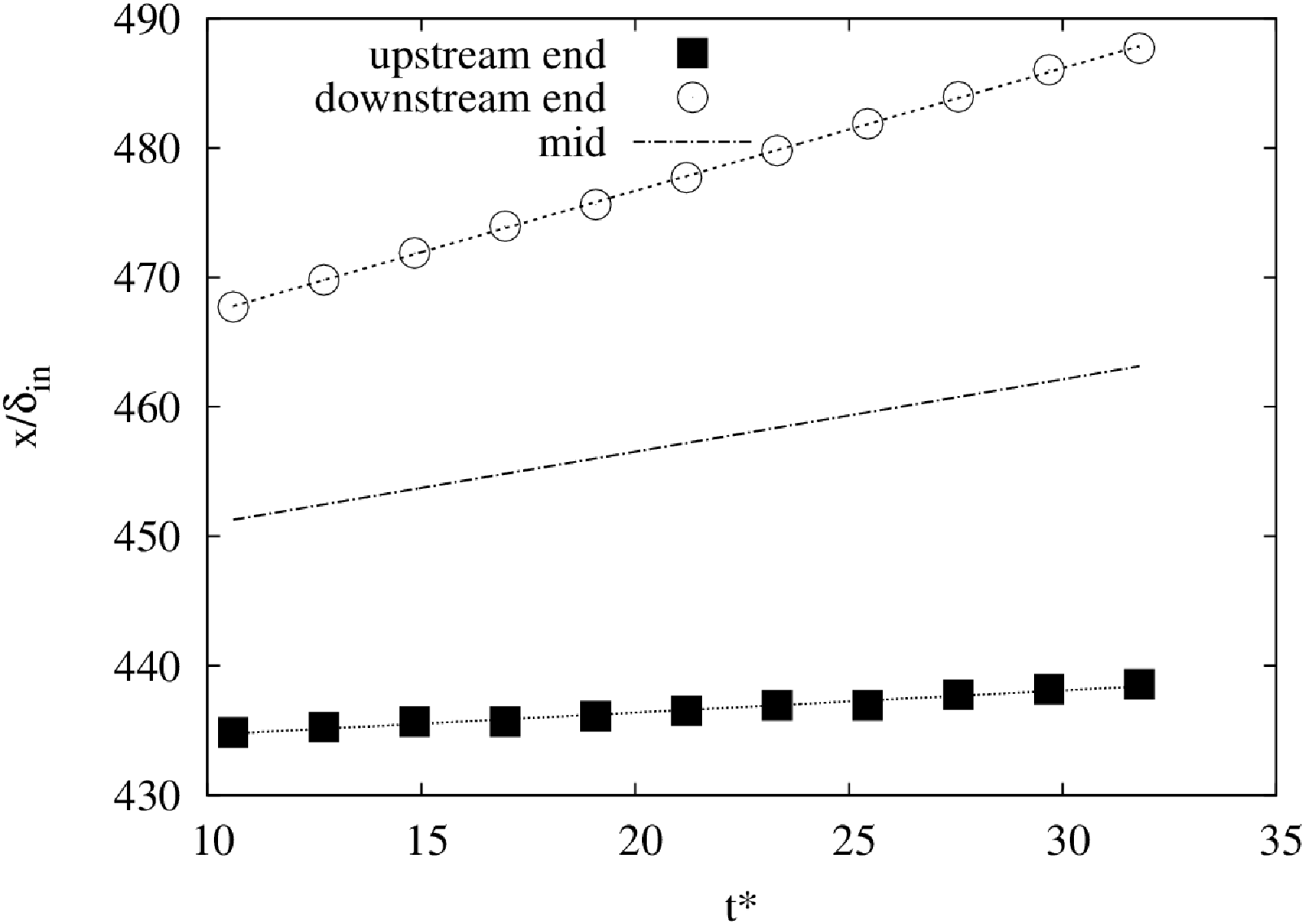}}
\caption{Locations of the upstream and downstream ends of the vortex tube along the elapse of time for $\Gamma=12.5$ and $\phi=4^\circ$. `mid' is the center locations between the ends. }
\label{fig:39}
\end{figure}
\begin{figure}
\begin{minipage}{0.5\hsize}
\begin{center}
\includegraphics[width=67mm]{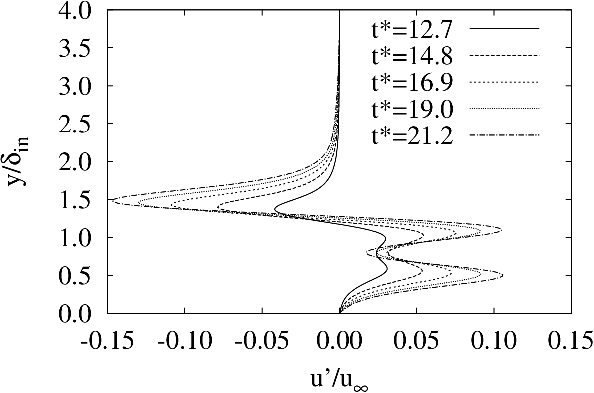}
(a)~$Dx_0=11.6\delta_{in}$
\end{center}
\label{fig:40}
\end{minipage}
\begin{minipage}{0.5\hsize}
\begin{center}
\includegraphics[width=67mm]{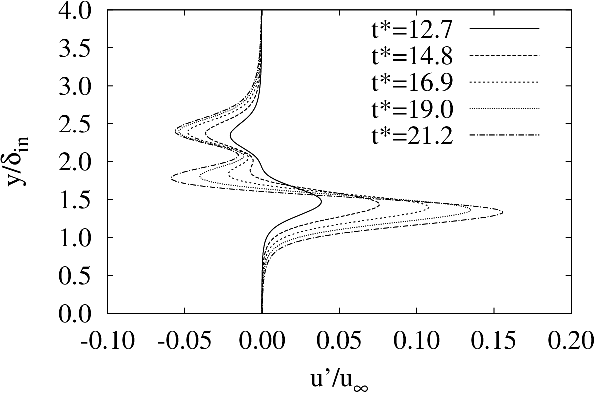}
(b)~$Dx_0=30.4\delta_{in}$
\end{center}
\label{fig:41}
\end{minipage}
\caption{Time evolution of $u'$ at $Dx_0=11.6\delta_{in}$ and $30.4\delta_{in}$. $z_0=0.524L_z$ and $t_0^\ast=10.6$.} 
\end{figure}
\begin{figure}
\centerline{\includegraphics[width=100mm]{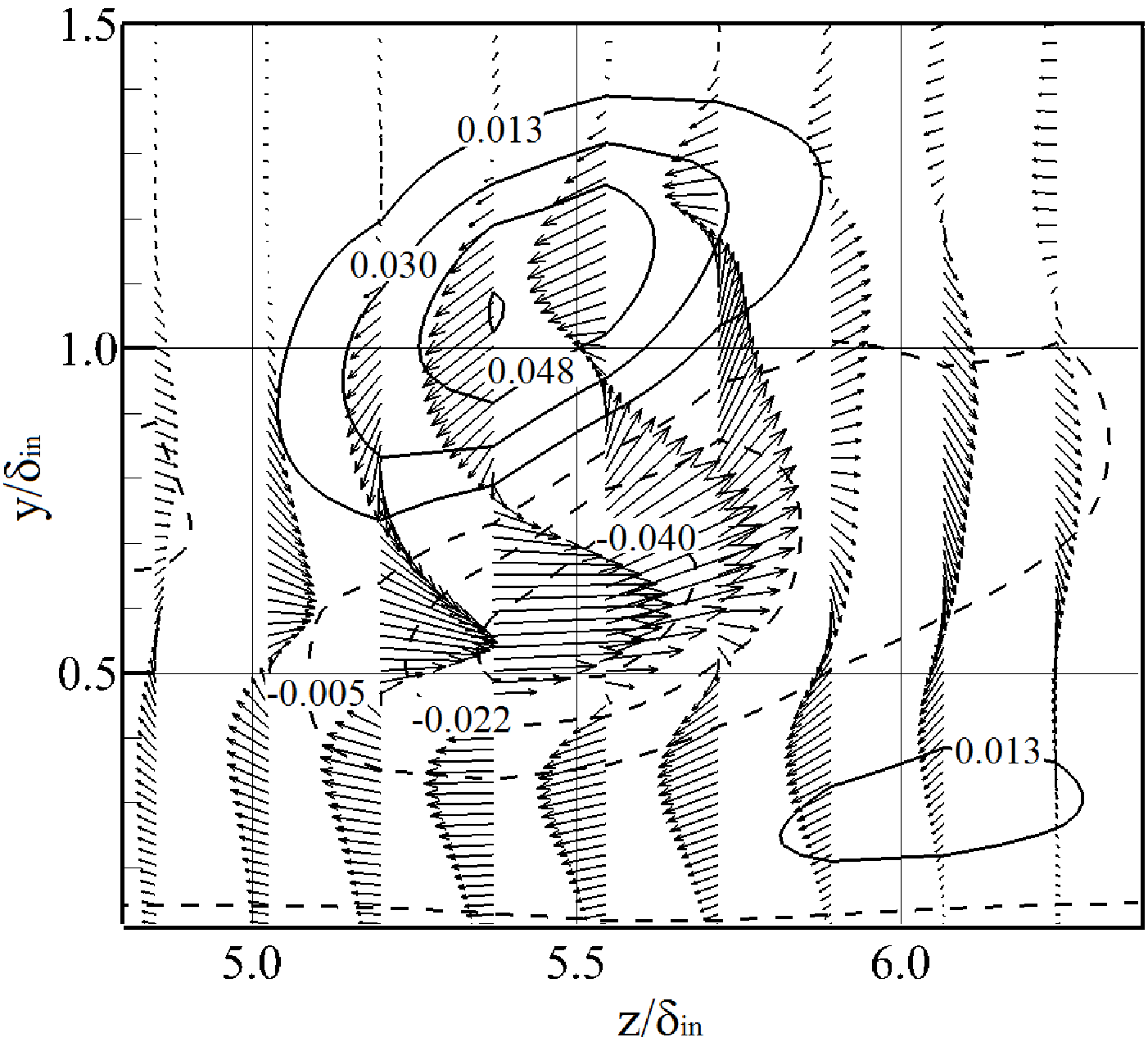}}
\caption{Distribution of $S_x^\ast$ in the $y$-$z$ plane at $Dx=11.6\delta_{in}$ and $t^\ast=10.6$. An arrow shows a vector ${\partial \mbox{\boldmath $\omega$}}/{\partial t}$ projected on the plane.}
\label{fig:42}
\end{figure}
In this section, the case of $\Gamma=12.5$ with $\phi=4^\circ$ is analyzed in further detail to investigate the disturbances evolving around the vortex tube. 
To understand when disturbances emerge, space-time plots of $S_x^\ast$ at $y=5\%\delta_{in}$, $50\%\delta_{in}$ and $100\%\delta_{in}$, i.e., 
regions near the wall, center and freestream of the boundary layer, are shown in Fig. 10(a-c). 
Here, $S_x$ is the $x$ component of the right-hand side of eq.(\ref{eq:eq1}), i.e, $(S_x,\cdot, \cdot)={\partial \mbox{\boldmath $\omega$}}/{\partial t}$;
the ordinate is the $x$ coordinate and the abscissa is time.
$S_x$ is plotted because this quantity is the time increment in streamwise rotation, and clearly shows the evolution of the disturbances around the streamwise-oriented vortex tube. 
In the figure, labels `A'-`G' mark points used in estimating the convection speeds of disturbances.

In Fig. 10(c), black and white streaks appear as a result of vortex convection. 
Taking the black streak labeled `FG' as an example, the convection speed is $u_{\infty}$. 
In particular after $t^\ast=53.1$ in Fig. 10(b), the organized periodic vortical structures propagating at the speed of $0.7u_{\infty}$
appear as found from the emergence of the periodic narrow streaks `DE' oriented in the upper right. 
In Fig. 10(a), short distinct segments (e.g., white bent streak `ABC') are seen after about $t^\ast$=42.5 arranged in a staggered pattern
of alternating positive and negative $S_x^\ast$, a feature which is different characteristically from that seen in panel (b). 
This indicates that the structural changes are occurring in the near-wall region as the vortex tube moves downstream. 
The convection speed of the regions is $0.37u_{\infty}$.

The structural evolution of the vortex tube is shown in Fig. 11. 
The figure shows the spanwise view of the iso-surfaces of positive and negative values of $S_x^\ast$, and the iso-surfaces of the positive and negative $S_x$.
The iso-surfaces in the center and right columns are shaded based on the height from the wall so as to understand the relative position of the structural change in the direction normal to the wall. 
When the disturbance around the vortex tube is small, the regions of positive and negative $S_x^\ast$ correspond to the upper and lower shear layers of the vortex tube, respectively;
this correspondence will be detailed below.
When the disturbance begins to develop around the vortex, sinuous oscillations occur in the upper and lower shear layers. 
A corkscrew-like structure develops as seen in the iso-surfaces of $t^\ast=10.6-63.9$. 
At $t^\ast=10.6$, this corkscrew-like disturbance, labeled `CS' in the figure, begins to develop around the vortex tube. 
When $t^\ast=21.3$, the disturbances are much more evident (Fig. 11(c)). 
When $t^\ast=63.9$, well-developed corkscrew-like structures are seen.

To clarify further the relationship between the results of the DNS and LSA, 
the growth of disturbances in the streamwise velocity around the base flow is initially compared between both at $Dx=11.6\delta_{in}$ and $30.4\delta_{in}$.  
The LSA is a temporal stability analysis, and its base flow is an instantaneous boundary layer profile, as mentioned in Section 2.3. 
In contrast, disturbances develop around the vortex tube both in time and space in the DNS. 
If we hold to a strict comparison, a complex treatment is needed to perform a stability analysis. 
To manage this complexity, we evaluated the amplification of the disturbance riding on the moving frame attached to the vortex tube in the DNS. 
We assumed the frame moves at constant speed $c_g$ representing the convection speed of the vortex tube. 
Because the time period for evaluating the amplification of the disturbances is short, we assumed this coordinate system has translational speed only in the $x$ direction
and thus neglected the non-parallelism associated with the movement of the vortex tube. 
The evolution of disturbance is evaluated by the following equation.
\begin{equation}
\label{eq:eq2}
u'(t,y)=u(t,x_0+c_g(t-t_0),y,z_0)-u(t_0,x_0,y,z_0).
\end{equation}
Here, subscript `0' denotes the reference position for the moving frame. 
Eq. (\ref{eq:eq2}) requires the convection speed $c_g$ of the vortex tube to be evaluated. 
Figure 12 shows the positions of both upstream and downstream ends evolving with time through advection and elongation of the vortex tube. 
Both end positions move linearly with time. 
The time dependence of the position of the central point (labeled `mid') is obtained from both of the ends; 
the speed of `mid' is taken as $c_g$. 
The equation of the mid line is
\begin{equation}
x/\delta_{in}=0.5596t^\ast+445.3,
\end{equation}
the slope giving $c_g^\ast=0.5596$.

Figure 13 shows the growth of $u'$ at $Dx_0=11.6\delta_{in}$ and $30.4\delta_{in}$ for $t^\ast$=10.6-21.2. 
Here, $Dx_0=x_0-x_c$. 
In both locations, the disturbances are evaluated at $z_0=0.524L_z$. 
When $Dx_0=11.6\delta_{in}$, the growth of the disturbance is seen at three heights, i.e., $y=0.5\delta_{in}$, $1.0\delta_{in}$ and $1.5\delta_{in}$. 
The largest growth can be seen around the highest position, i.e., $y=1.5\delta_{in}$. 
Moreover, when $Dx_0=30.4\delta_{in}$, the growth of disturbances can be seen at $y=1.3\delta_{in}$, $1.7\delta_{in}$ and $2.4\delta_{in}$. 
The largest growth can be seen around the lowest position, i.e., $y=1.3\delta_{in}$. 
As mentioned in Section 3.1.2, the LSA shows that the growth of the off-wall mode is larger than that of the near-wall mode at $Dx=11.6\delta_{in}$, and the growth of the near-wall mode is larger than that of the off-wall mode at $Dx=30.4\delta_{in}$. 
Therefore, the trend in the growth of the disturbance in the DNS agrees well with that of the LSA. 

In the DNS, at $Dx_0=11.6\delta_{in}$, the growth rate $\omega_i$ of $u'$ around $y/\delta_{in}=1.5$ computed from Fig. 13(a) varies from 0.2997 to 0.056227 for $t^\ast$=12.7 to 19.0. 
$\omega_i$ is evaluated by
\begin{equation}
\omega_i=\displaystyle \frac{1}{\Delta t^\ast} \mathrm{ln} \mid \displaystyle \frac{(A)_{t^\ast=t_2^\ast}}{(A)_{t^\ast=t_1^\ast}} \mid,~\Delta t^\ast=t^\ast_2-t^\ast_1.
\end{equation}
Here, `A' is the amplitude of disturbance.  
At $Dx_0=30.4\delta_{in}$, $\omega_i$ around $y/\delta_{in}=1.2$ computed from Fig. 13(b) varies from 0.3234 to 0.064 for $t^\ast$=12.7 to 19.0. 
In the LSA, the peak values of $\omega_i$ for the off-wall and near-wall modes are $\omega_i$=0.13 and 0.07, respectively, as found from Figs. 8(a,b).
While difference between the DNS and LSA is only within a few times of the growth rates as a whole,
the agreement between the DNS and LSA improves with increasing elapsed time. 
One possible reason is that the linear instability mechanism related to the wall-normal profile becomes influential as time proceeds.
This inference is supported by the results of the FNDA mentioned later. 

The cross-sectional structure of the vortex tube can be clearly understood from $S_x$. 
In addition, the upper and lower shear layers of the vortex tube are distinguished by the quantity even when the disturbance around the vortex tube is small. 
Figure 14 shows the distribution of $S_x$ in the $y-z$ plane at $Dx=11.6\delta_{in}$ and $t^\ast=10.6$. 
An arrow shows a vector ${\partial \mbox{\boldmath $\omega$}}/{\partial t}=(S_x,\cdot, \cdot)$ projected on the plane. 
There are two large regions of positive and negative $S_x$. 
In comparing Figs. 14 and 11(b), the upper and lower shear layers of the vortex tube are seen to correspond to the positive and negative regions, respectively. 
\begin{figure}
\begin{minipage}{0.48\hsize}
\centering
\includegraphics[width=77mm]{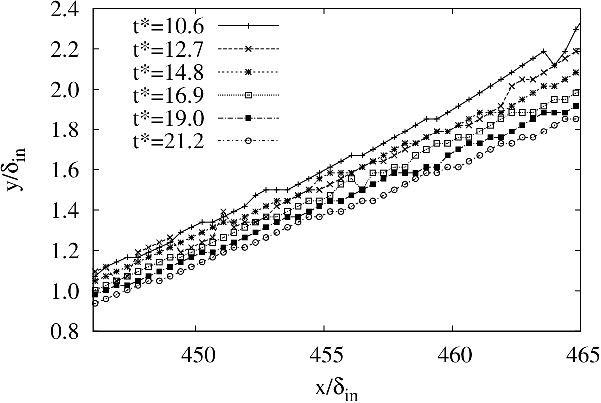}\\
(i)~$y$ coordinate of the positive peak
\label{fig:43}
\end{minipage}
\begin{minipage}{0.48\hsize}
\centering
\includegraphics[width=77mm]{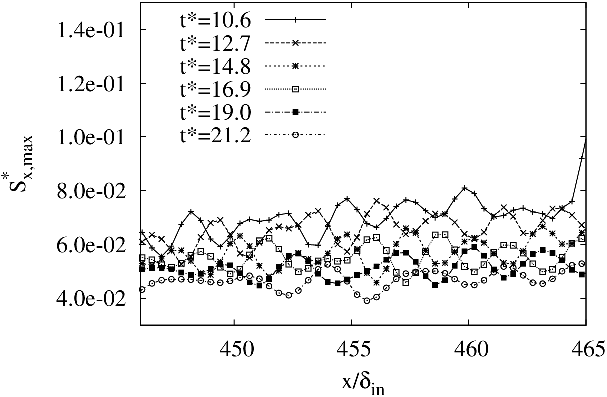}\\
(ii)~$S_{x,max}^\ast$ along $x$
\label{fig:44}
\end{minipage}\\
\centering
(a) Largest positive peak
%
%
\\
\begin{minipage}{0.48\hsize}
\centering
\includegraphics[width=77mm]{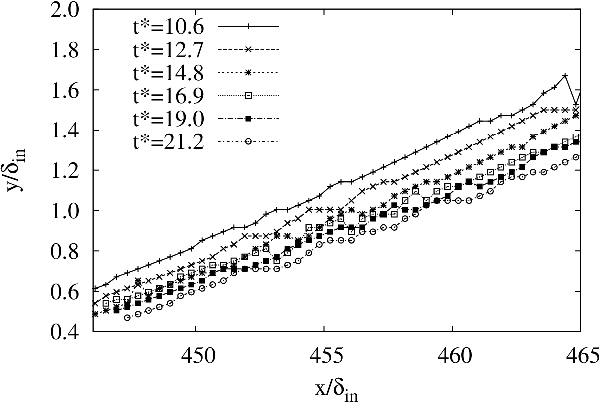}\\
(i)~$y$ coordinate of the negative peak
\label{fig:45}
\end{minipage}
\begin{minipage}{0.48\hsize}
\centering
\includegraphics[width=77mm]{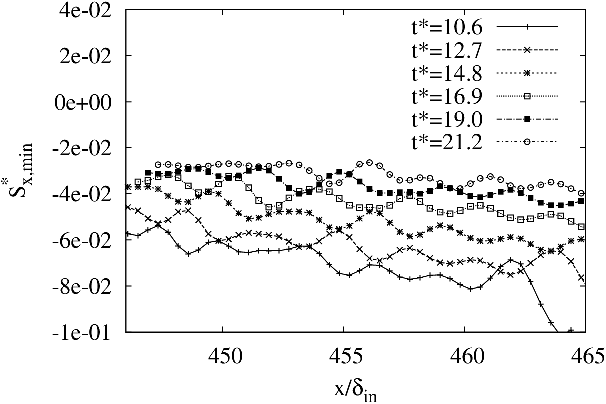}\\
(ii)~$S_{x,min}^\ast$ along $x$
\label{fig:46}
\end{minipage}\\
\centering
(b) Largest negative peak
\caption{Heights of the largest positive and negative peaks, and the variations of $S_{x,max}$ and $S_{x,min}$ along $x$}
\end{figure}

To clarify the time evolution of the disturbances in the upper and lower shear layers along the vortex tube, the maximum and minimum peaks in each cross section with $x=$const. are plotted.
Figures 15 show the variation of the $y$ coordinates of the positive and negative peaks, and $S^\ast_{x,max}(x)$ and $S^\ast_{x,min}(x)$ along the vortex tube for $t^\ast$=10.6-21.2. 
Here, $S^\ast_{x,max}(x) \equiv \displaystyle \max_{(y,z) \in \Sigma_j} S_x^\ast(x,y,z)$ and $S^\ast_{x,min}(x) \equiv \displaystyle \min_{(y,z) \in \Sigma_j} S_x^\ast(x,y,z)$.
$\Sigma_j$ shows a cross section at $x=x_j$, i.e., $\Sigma_j=\{(y,z);(y,z)\in[0,L_y]\times[0,L_z], x=x_j\}$.
The $y$ coordinates of both the positive and negative peaks increase almost linearly with $x$ due to the obliqueness of the vortex tube despite some oscillations. 
$S^\ast_{x,max}(x)$ and $S^\ast_{x,min}(x)$ along $x$ show sinuous disturbances originating from the corkscrew-like disturbance structure mentioned in Fig. 11.
These sinuous disturbances support the assumption of the sinuous disturbances in the LSA.
In the upper and lower shear layers, the respective wavelengths of the disturbances are $2.5\delta_{in}$ and $3.12\delta_{in}$, at $t^\ast=10.6$, and $3.12\delta_{in}$ and $2.5\delta_{in}$, at $t^\ast$=19.0 and 21.2. 
In the LSA, the most amplified wavelengths of the off-wall and near-wall modes are $3.5\delta_{in}$ and $2.9\delta_{in}$, respectively. 
Although the wavelengths of the most amplified disturbances predicted in the LSA for the off-wall and near-wall modes do not agree precisely with those by the DNS at $t^\ast$=10.6, 
they agree more with those by the DNS at $t^\ast$=19.0 and 21.2. 
Here again, the agreement between the LSA and DNS improves with the elapsed time similar to the growth rate mentioned above. 

In order to clarify the contribution of the linear and nonlinear terms in the dynamics of the disturbance evolution, the FNDA is conducted around 
$Dx_0=11.6\delta_{in}-49.1\delta_{in}$ of $z=0.524L_z$ for $t^\ast=10.6-44.5$.
The instantaneous wall-normal profiles of $(\rho,u,v,w,T)$ extracted at $t^\ast=10.6$, $Dx=11.6\delta_{in}$ and $z=0.524L_z$, which are the same profiles used for the LSA, are used for the base profiles.
In the case of $\tilde{u}$, locally top five ranked dominant terms are selected in the range $1.0 \le y \le 1.5$ as follows.
At each time and streamwise location of $j=j_s,\cdots,j_e$, a set of top five dominant terms $D_j(t)\equiv\{d_1,d_2,\cdots,d_5\} \subset \{u_1,u_2,\cdots,u_{37}\}$ are 
selected by comparing the absolute values of the average of each term in the period $1.0 \le y \le 1.5$.
Uniting the sets $D_j(t)$ for $j=j_s,\cdots,j_e$ at each time, the enlarged set $E(t)\equiv\bigcup_{j=j_s}^{j_e}D_j(t)$ covering all streamwise locations $j=j_s,\cdots,j_e$ is finally obtained at the time.     
$u_1, \cdots, u_{13}$ are linear terms, $u_{14}, \cdots, u_{20}$ are base flow terms and $u_{21}, \cdots, u_{37}$ are nonlinear terms.
In the present system of the equations, the base flow terms, which are removed from the system of the LSA, are remained because
these terms do not satisfy the steady Navier-Stokes equations.
In addition, terms that contain $\tilde{\mu}$ are categorized as nonlinear terms because $\mu$ is a nonlinear function of $T$.     
In the present analysis, $v$ and $w$ components are not neglected in the base profiles, and incorporated in the linear terms $u_4$ and $u_5$.

By uniting the sets $\bigcup_{t^\ast=10.6}^{44.5}E(t)$ and  
considering similarly for the other variables, a reduced system of the evolution equations that approximate the original system acurately in the range $1.0 \le y \le 1.5$
is obtained as follows for $t^\ast=10.6-44.5$ and 
$Dx_0=11.6\delta_{in}-49.1\delta_{in}$ of $z=0.524L_z$:
\begin{eqnarray}
\frac{\partial \tilde{\rho}}{\partial t}&=&r_{4}+r_{8}+r_{12}+\cdots\\
\frac{\partial \tilde{u}}{\partial t}&=&u_{1}+u_{3}+u_{4}+u_{5}+u_{12}+u_{13}+u_{15}+u_{19}+u_{22}+u_{23}+u_{24}+\cdots\\
\frac{\partial \tilde{v}}{\partial t}&=&v_{1}+v_{4}+v_{5}+v_{10}+v_{12}+v_{17}+v_{23}+v_{24}+\cdots\\
\frac{\partial \tilde{w}}{\partial t}&=&w_{1}+w_{3}+w_{4}+w_{5}+w_{10}+w_{12}+w_{13}+w_{15}+w_{19}+\nonumber\\
&&w_{22}+w_{23}+w_{24}+\cdots\\
\frac{\partial \tilde{T}}{\partial t}&=&T_{1}+T_{3}+T_{4}+T_{5}+T_{11}+T_{13}+T_{14}+T_{17}+T_{18}+T_{23}+T_{39}+\nonumber\\
&&T_{44}+T_{45}+T_{46}+T_{63}+T_{64}+T_{71}+\cdots
\end{eqnarray}

Figure 16 shows the streamwise variation of each average values of the dominant terms of the $\tilde{u}$ equation in the $y$ range $[1.0,1.5]$,
and the averaged value of the RHS in the same range for $t^\ast=$10.6, 27.6 and 44.5 at $z=0.524L_z$. 
The streamwise ranges of the locations of the vortex tube are around [434.4,468.1], [437.3,483.1] and [441.5,497.7], respectively, for the times.
$E(t)$ is plotted at each time.
At $t^\ast=10.6$, the nonlinear terms $u_{23}$ and $u_{24}$ have large positive values.
However, because they are cancelled out by the linear terms $u_{4}$ and $u_{5}$, the total value of all terms, i.e., the RHS, becomes intermediate.
Although the trend of the RHS seems to be similar to that of $u_1$, its coincidence is not so clear at this time.
At $t^\ast=27.6$, the similar trend between the RHS and $u_1$ becomes much clearer, 
and a sinuous deformation with a length of about $2.0\delta_{in}$ appear around $x=457-461$.
Similar deformation of the same size also appears on the trends of $u_{4}$ and $u_{23}$.
At $t^\ast=44.5$, the value of $u_{24}$ becomes small, the number of sinuous deformation increase, and the generation of periodically sinuous structures become much more evident.
The trend of the RHS is almost the same as that of $u_1$.
Near $x=463\delta_{in}-472\delta_{in}$, short sinuous deformation that has a length close to a distance between consecutive hairpin vortices appear.
The deformation can also be observed on the curves of $u_4$ and $u_{23}$.
From these results, it is found that linear stability mechanism of the wall-normal base profile has a great influence for the determination of the streamwise wavelengths of the sinuous waves,
and it appears to be the reason for the reasonably good agreement between the LSA and DNS.   
 
\begin{figure}
\centering
\centerline{\includegraphics[width=90mm]{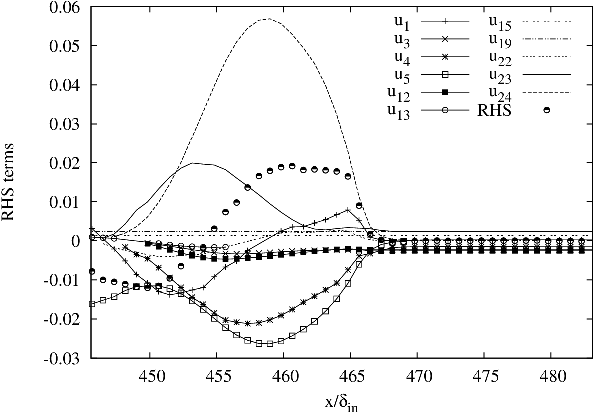}}
(a)~$t^\ast=10.6$\\
\centering
\centerline{\includegraphics[width=90mm]{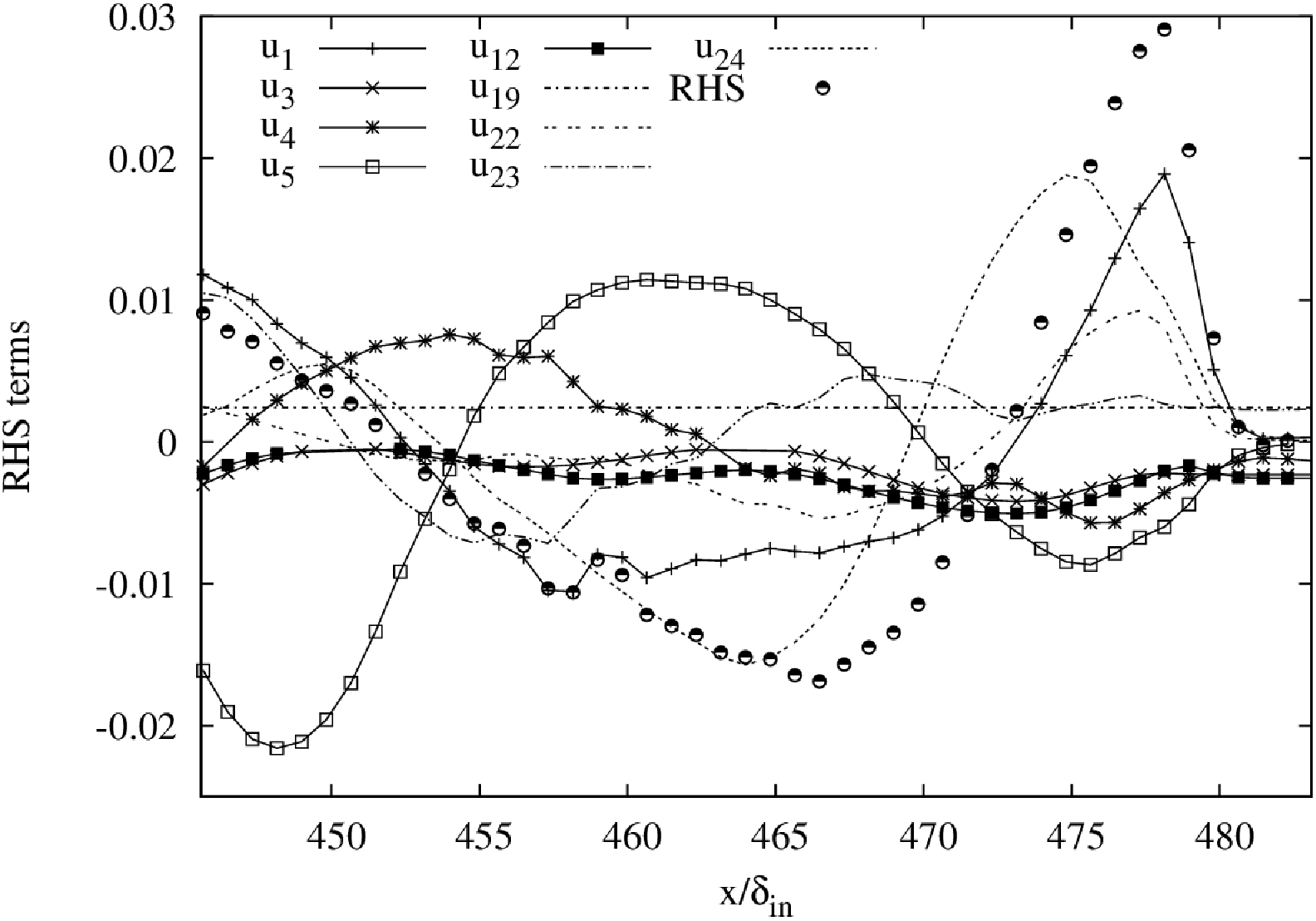}}
(b)~$t^\ast=27.6$\\
\centering
\centerline{\includegraphics[width=90mm]{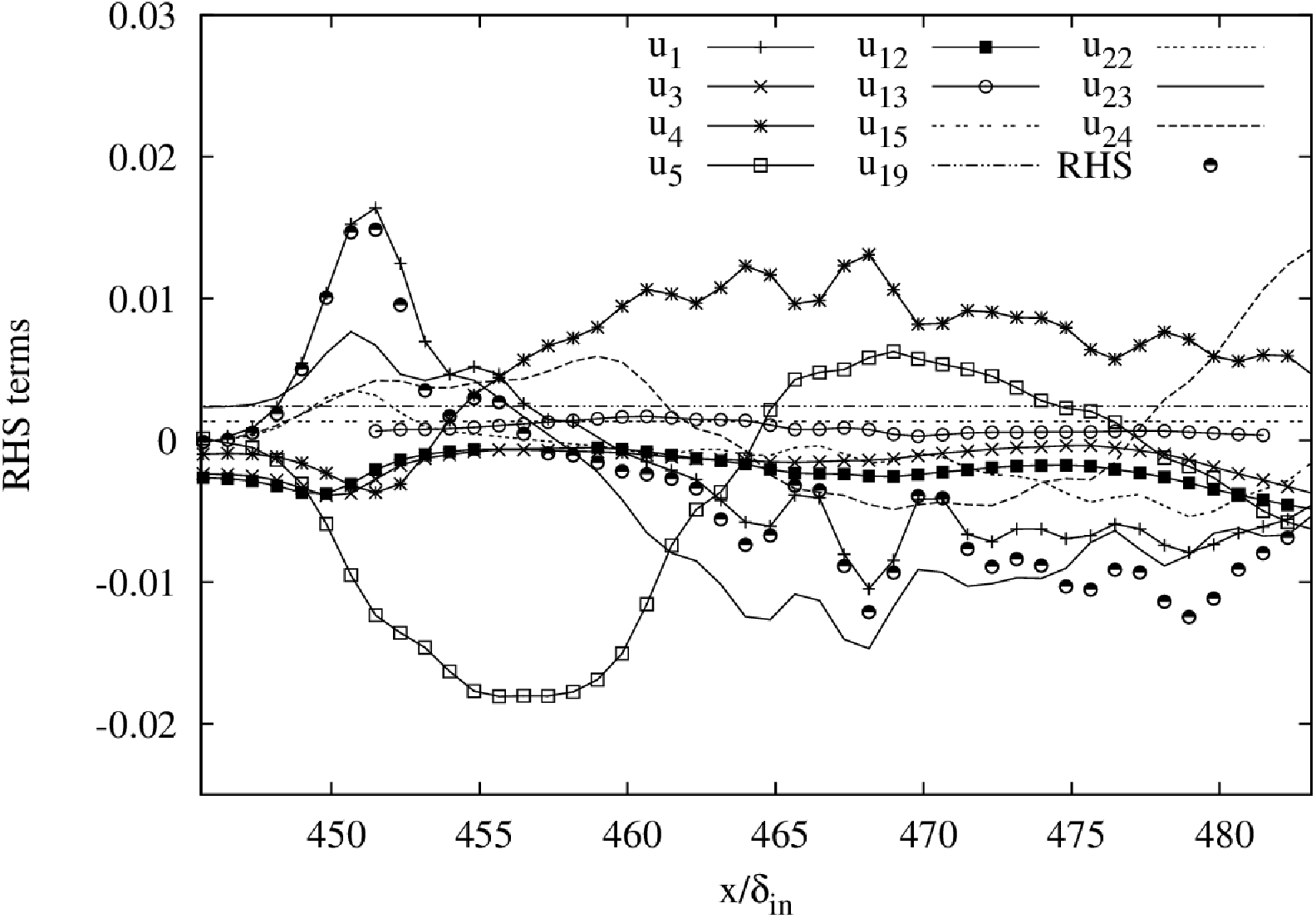}}
(c)~$t^\ast=44.5$\\
\caption{Variation of the terms ranked as the locally top five ranked dominant terms at some streamwise location along the votex tube at $t^\ast=$10.6, 27.6 and 44.5 for $\Gamma=12.5$ and $\phi=4^\circ$}
\label{fig:20}
\end{figure}
 
\subsubsection{Dominant structure extracted by the proper orthogonal decomposition}
To understand the dominant spatial structure of disturbance generated around the vortex tube at each instant of time, 
a snapshot POD analysis is performed using the series of data $S_{x_j} (j=j_s,\cdots,j_e)$ of $S_x$ obtained from the series of cross sections
\begin{equation}
\Sigma_j'=\{(y,z);(y,z)\in[y_s(x),y_e(x)]\times[z_s,z_e],~x=x_j\},~j=j_s,\cdots,j_e,
\end{equation}
each of which is extracted at the $j$th position of the CFD mesh, i.e., $x=x_j$, in an instantaneous flow field. 
The range for the $j$-index $[j_s,j_e]$ is selected for the interior region of the vortex tube away from the terminal ends of the vortex tube.

Because the vortex tube is oblique to the wall, the POD analysis of the data sets extracted simply at each $j$th position includes a trend resulting from obliqueness. 
To exclude the trend and fix the locations of the vortex tube near the centers of the cross sections, the $y$ range $[y_s, y_e]$ of each extracted plane is varied according to the $j$-index. 
The $z$-coordinate ranges $[z_s, z_e]$ of each extracted plane are kept constant because the drift of the vortex tube in the $z$ direction or the obliqueness of the vortex tube with respect to the $x$ axis in the $x-z$ plane is small. 

Here, the procedure to extract the series of $j$-planes is explained. 
The range of the $j$-index $[j_s,j_e]$ is determined near the spanwise center of the vortex tube. 
At $x=x_{j_s}$ and $x=x_{j_e}$, the heights of the centers of the vortex tube at the spanwise location, i.e., $y_c(j_s)$ and $y_c(j_e)$, are evaluated, 
and the central line passing through these two centers is determined. 
Then, $R_s \equiv [y_s(x_{j_s}),y_e(x_{j_s})]$, i.e., the $y$ range containing the upper and lower shear layers at $x=x_{j_s}$ is established. 
The $y$-coordinate ranges $[y_s(x),y_e(x)]$ for other $x_{j}(j=j_s+1,\cdots,j_e)$ are evaluated by sweeping the range $R_s$ along the central line. 
The $z$-coordinate range is always $[z_s, z_e]=[0.380L_z, 0.619L_z]$. 
The $y$ and $z$-coordinate ranges correspond to the $k$-index range $[k_s,k_e]$ and the $l$-index range $[l_s,l_e]$, respectively, at $x=x_{j_s}$. 
By sweeping the mesh points $\{(y_k,z_l); (k,l)\in[k_s,k_e]\times[l_s,l_e]\}$ along the central line, the number of mesh points $N$ is the same for all the extracted cross sections for $j\in[j_s, j_e]$. 
The data sets $\{S_{x_j}(k,l);(k,l)\in[k_s, k_e]\times[l_s, l_e]\}$ for $j_s \le j \le j_e$ are linearly interpolated from the data set $\{S_x(j,k,l); j_s \le j \le j_e, (k,l) \in [1,\mathrm{k_{max}}]\times[1,\mathrm{l_{max}}]\}$ defined on the nodes of the CFD mesh.

The POD results for times $t^\ast$=2.13, 10.6 and 63.8 are discussed as they correspond, respectively to times 
before the generation of the corkscrew disturbance, its emergence and after its clear development. 
The coordinate ranges extracted for the POD analysis and also the values of $M$ are shown in Table 2.
\begin{table}
\centering
\caption{Coordinate ranges extracted for the POD analysis, $(\cdot)^\dagger=(\cdot)/\delta_{in}$}
\begin{tabular}{|l|l|l|l|l|l|l|l|l|l|} \hline
$t^\ast$ & $x_{j_s}^\dagger$ & $x_{j_e}^\dagger$ & $M$ & $y_{c}(x_{j_s})^\dagger$ & $y_{c}(x_{j_e})^\dagger$ & $y_{s}(x_{j_s})^\dagger$ & $y_{e}(x_{j_s})^\dagger$ & $y_{s}(x_{j_e})^\dagger$ & $y_{e}(x_{j_e})^\dagger$ \\ \hline
2.13 & 438.1 & 457.3 & 47 & 0.469 & 1.760 & 0.190 & 1.583 & 1.481 & 2.874 \\ \hline
10.6 & 445.6 & 465.6 & 48 & 0.487 & 2.049 & 0.160 & 1.949 & 1.350 & 3.140 \\ \hline
63.8 & 472.3 & 501.0 & 70 & 0.0   & 0.633 & 0.0 & 2.569 & 0.624 & 3.193 \\ \hline
\end{tabular}
\label{tab:podcoord}
\end{table}
\begin{figure}
\centerline{\includegraphics[width=100mm]{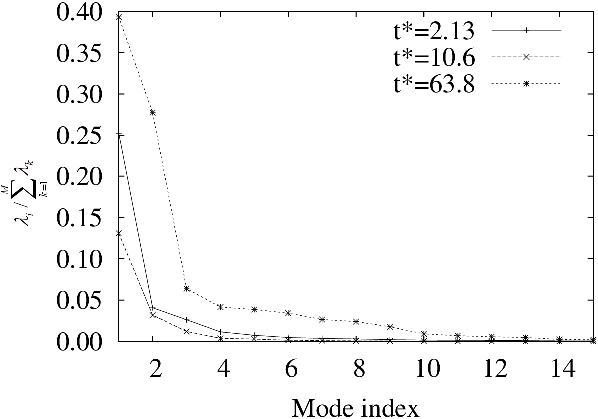}}
\caption{Fractional energies of the first 15 dominant POD modes for $t^\ast$=2.13, 10.6 and 63.8}
\label{fig:47}
\end{figure}
%
%
\begin{figure}
\begin{minipage}{0.24\hsize}
\centering
\includegraphics[width=43mm]{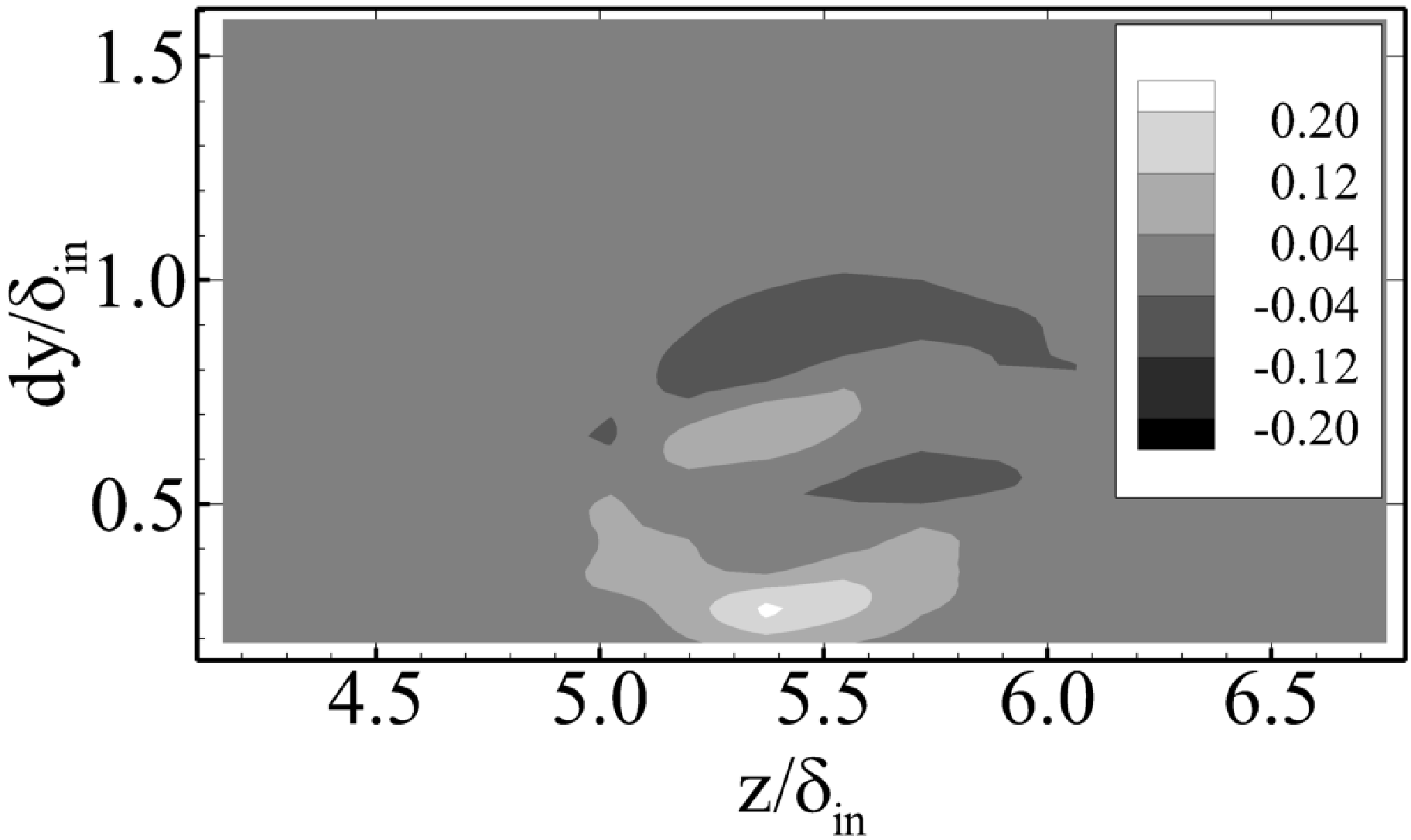}
\label{fig:48} 
\end{minipage}
\begin{minipage}{0.24\hsize}
\centering
\includegraphics[width=43mm]{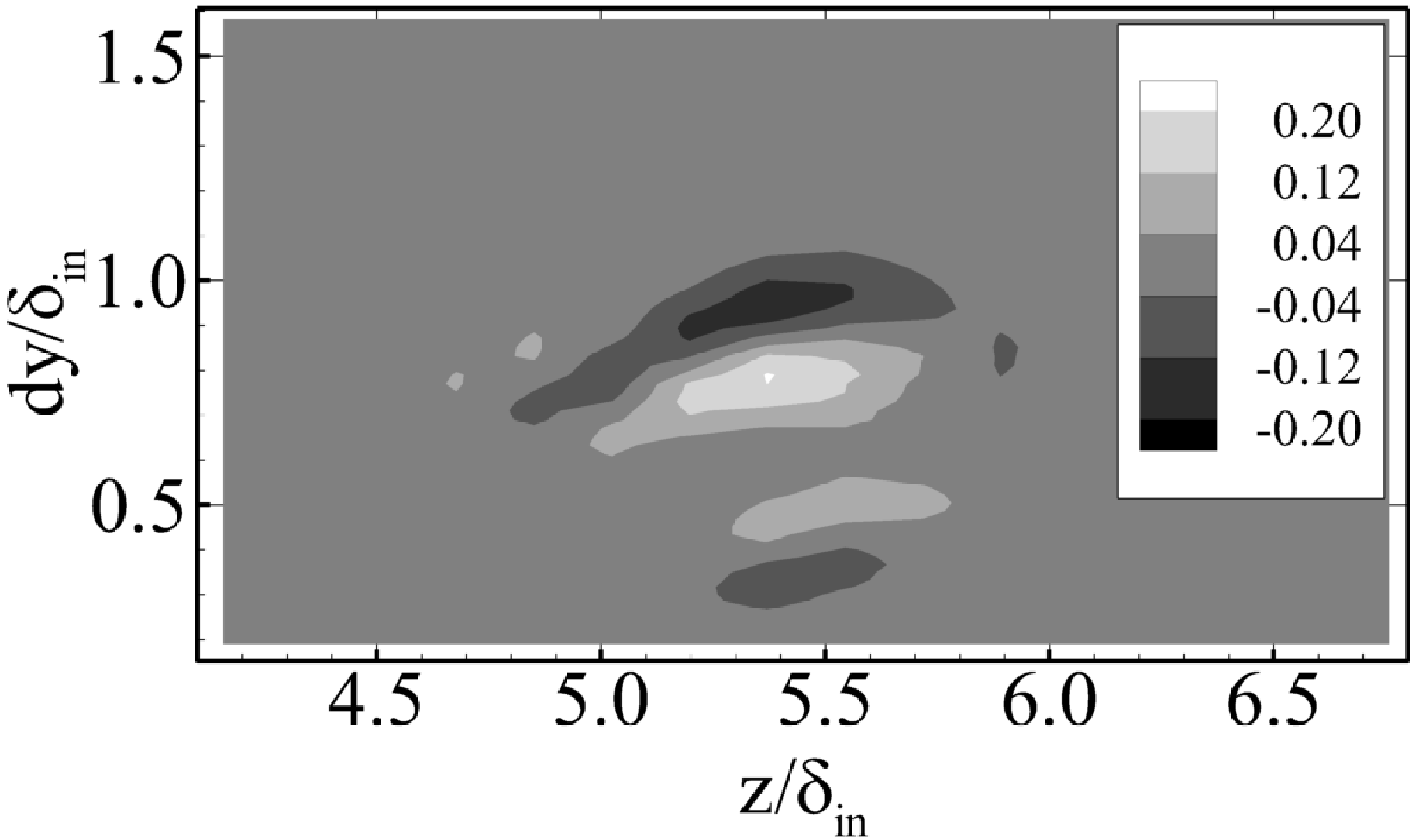}
\label{fig:49}
\end{minipage}
\begin{minipage}{0.24\hsize}
\centering
\includegraphics[width=43mm]{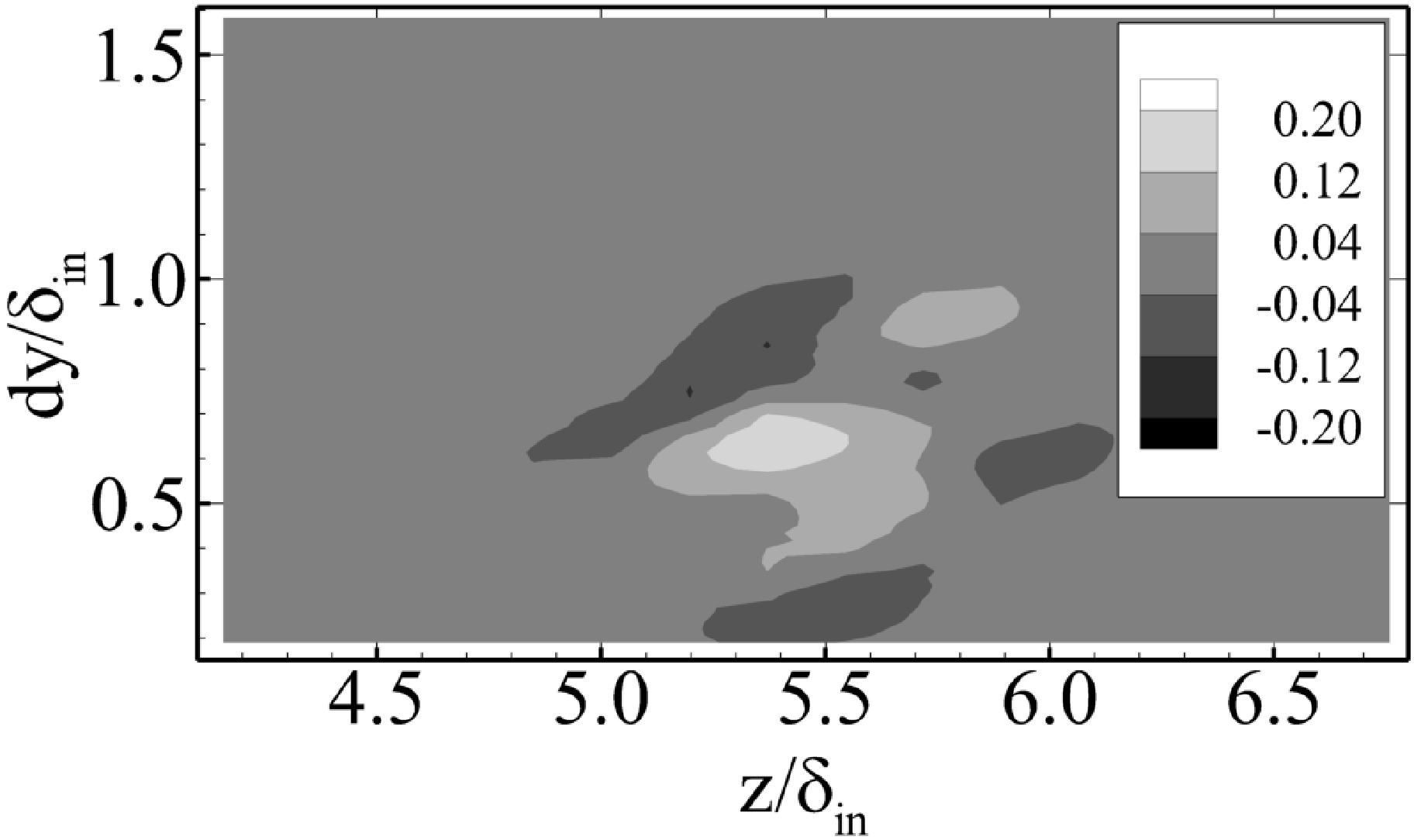}
\label{fig:50}
\end{minipage}
\begin{minipage}{0.24\hsize}
\centering
\includegraphics[width=43mm]{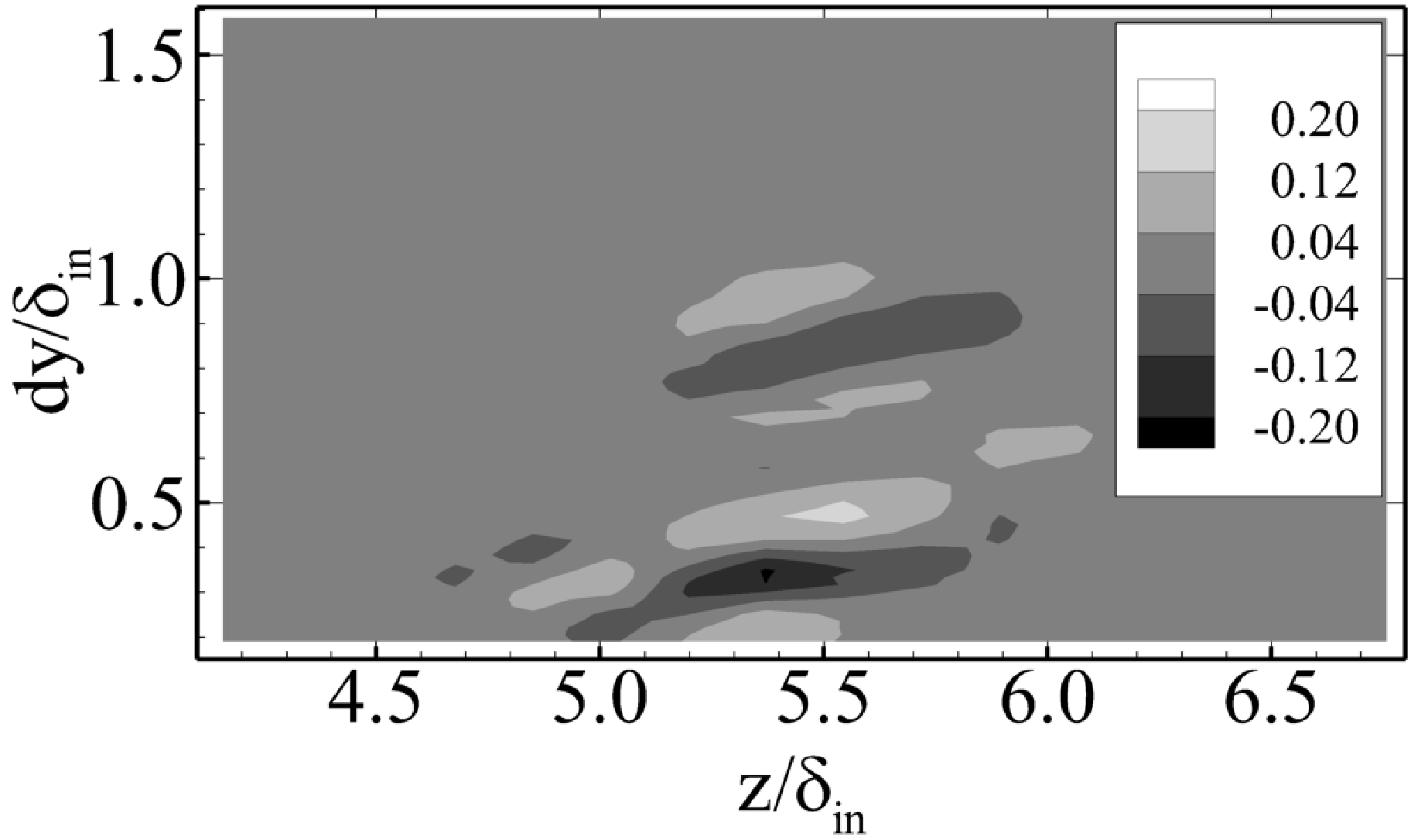}
\label{fig:51}
\end{minipage}
%
\\
\centering
(a) $t^\ast=2.13$\\
%
%
\begin{minipage}{0.24\hsize}
\centering
\includegraphics[width=43mm]{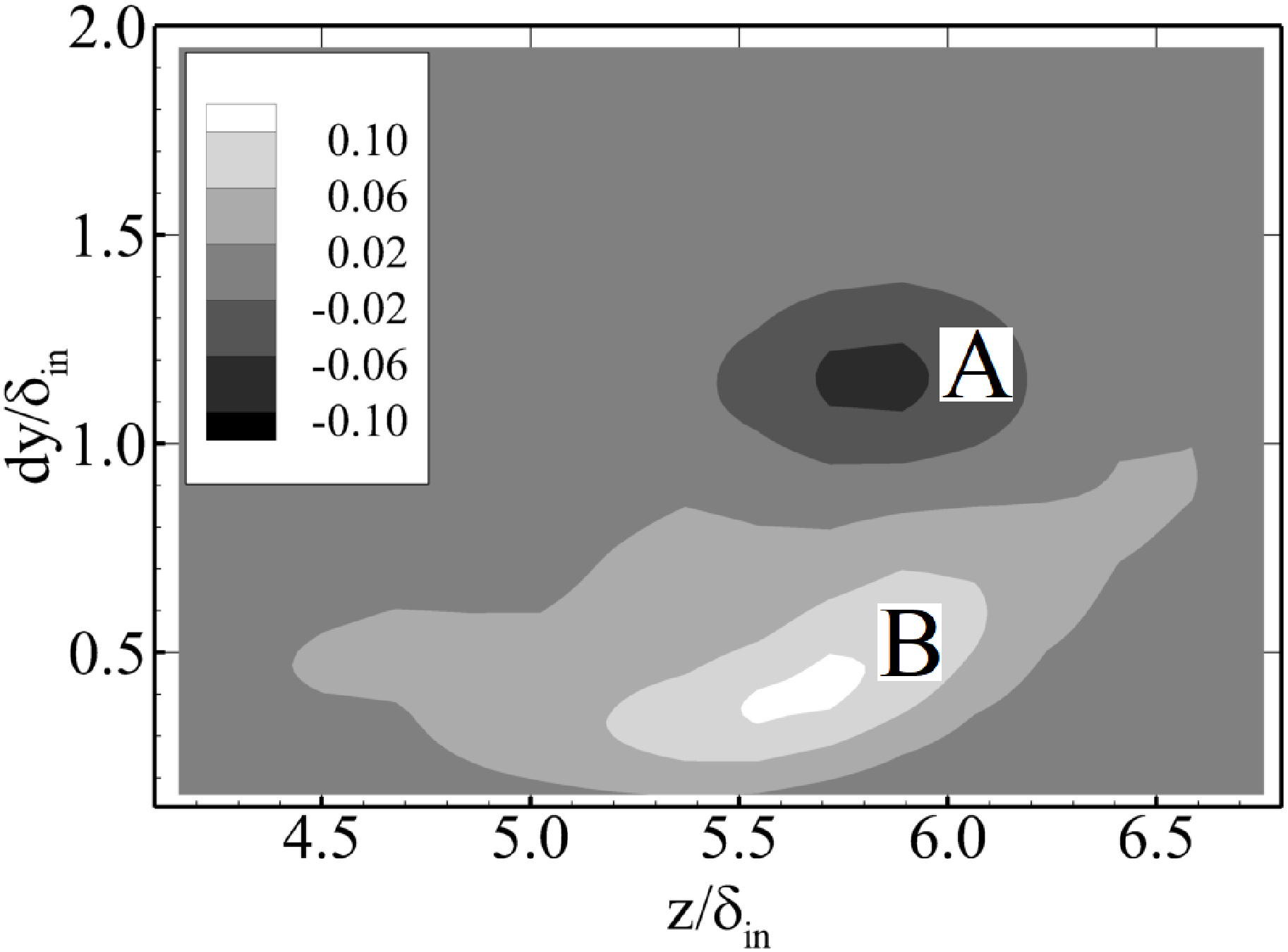}
\label{fig:52} 
\end{minipage}
\begin{minipage}{0.24\hsize}
\centering
\includegraphics[width=43mm]{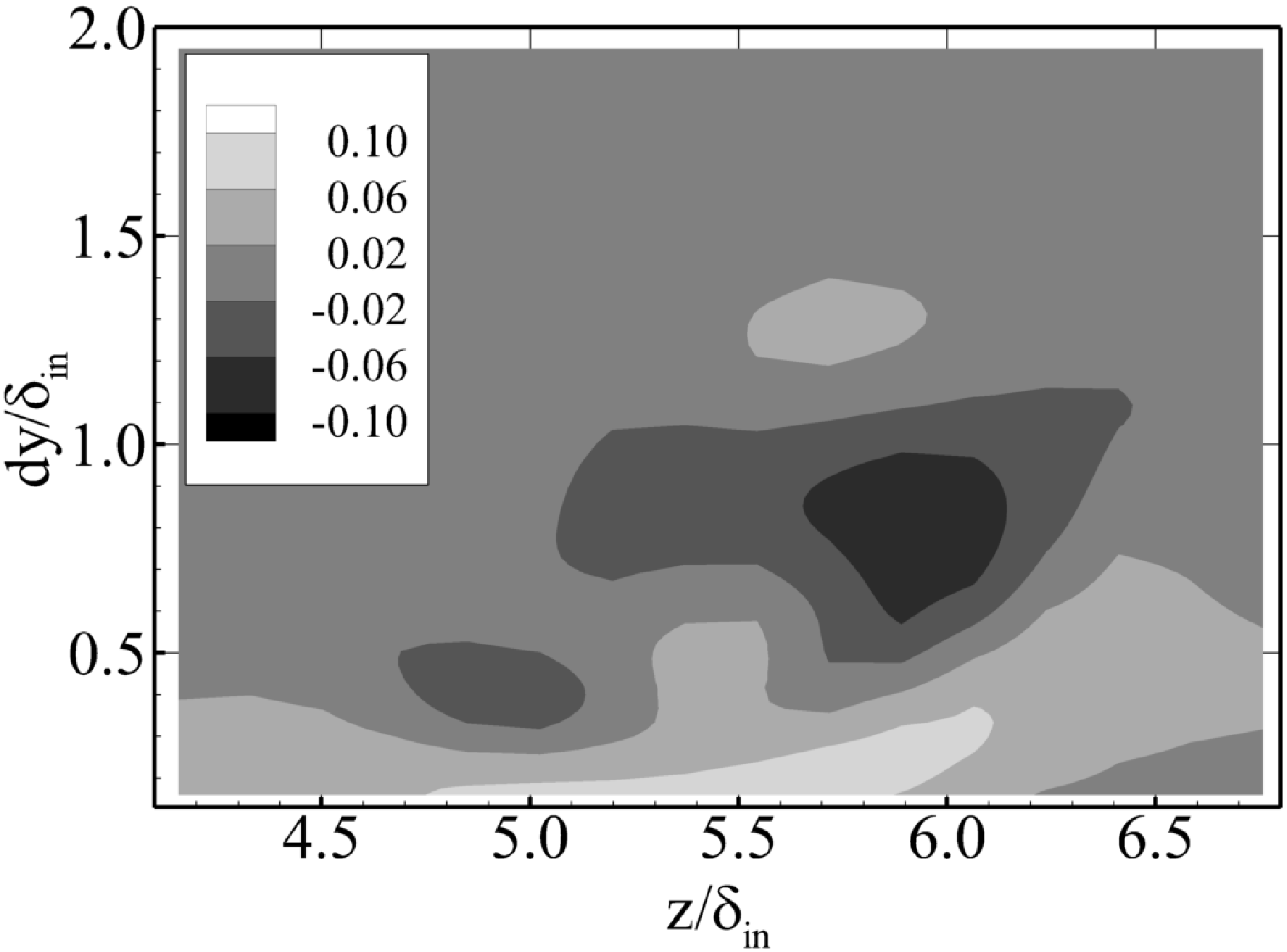}
\label{fig:53}
\end{minipage}
\begin{minipage}{0.24\hsize}
\centering
\includegraphics[width=43mm]{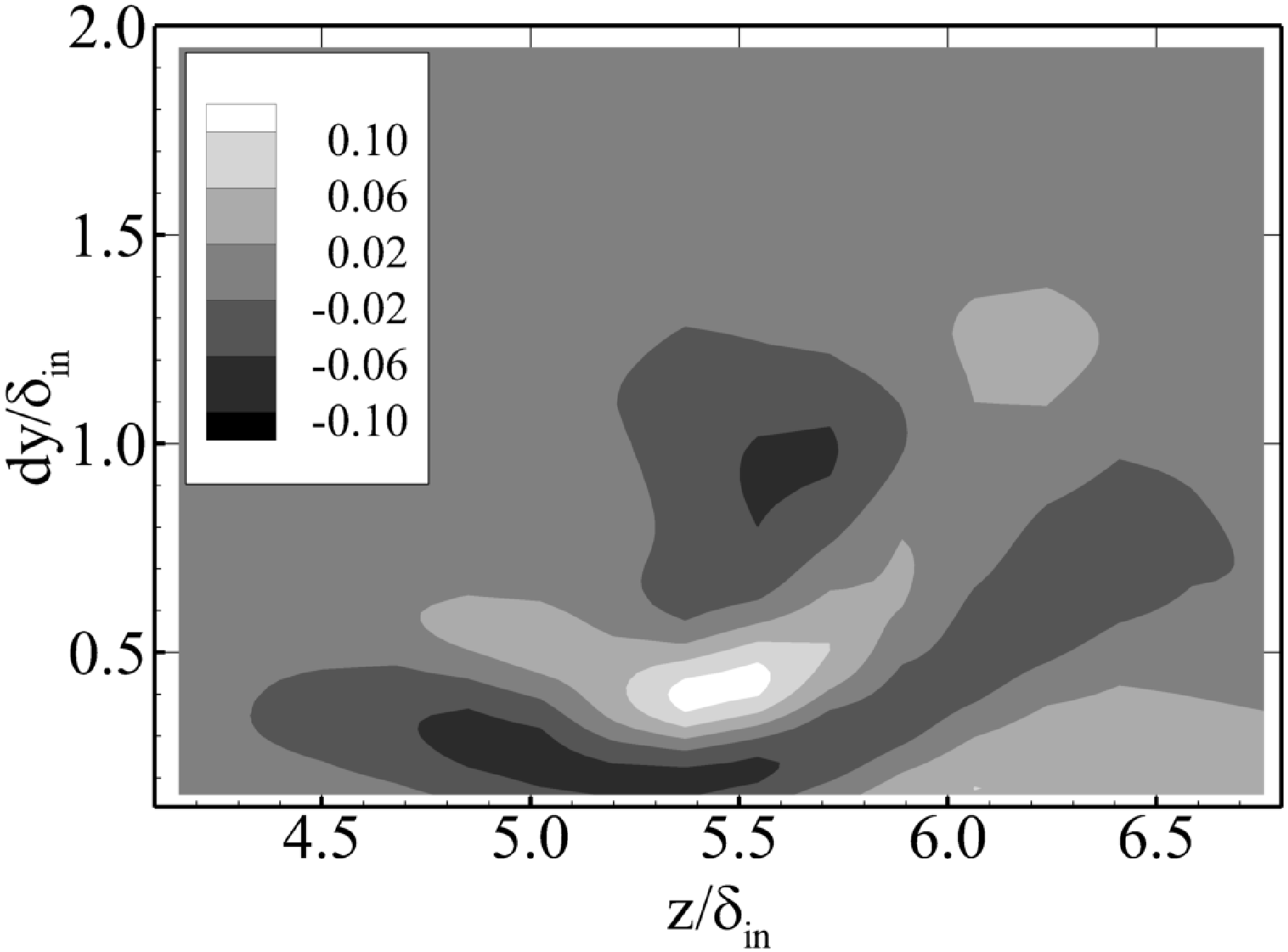}
\label{fig:54}
\end{minipage}
\begin{minipage}{0.24\hsize}
\centering
\includegraphics[width=43mm]{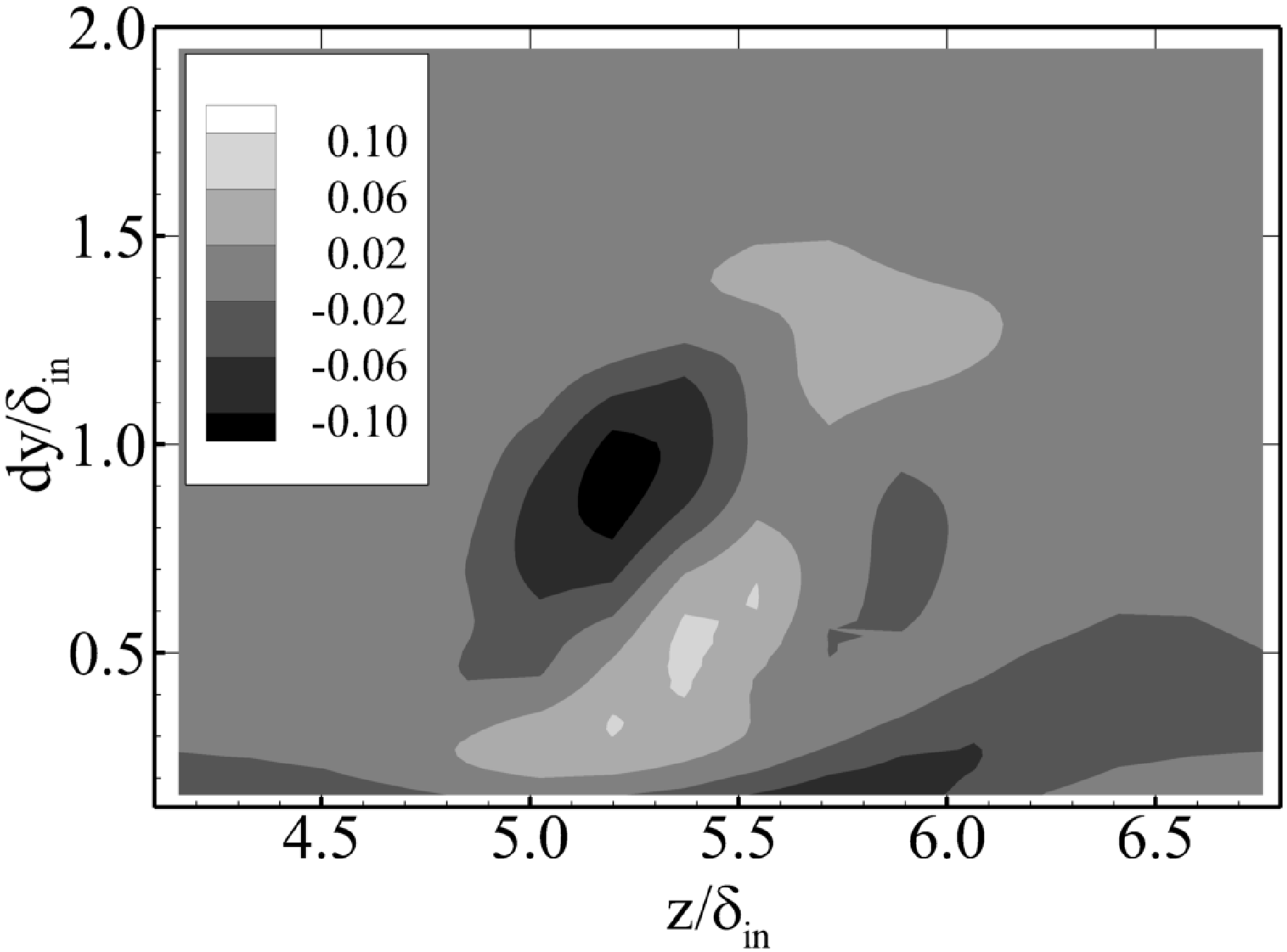}
\label{fig:55}
\end{minipage}
%
\\
\centering
(b) $t^\ast=10.6$\\
%
%
\begin{minipage}{0.24\hsize}
\centering
\includegraphics[width=43mm]{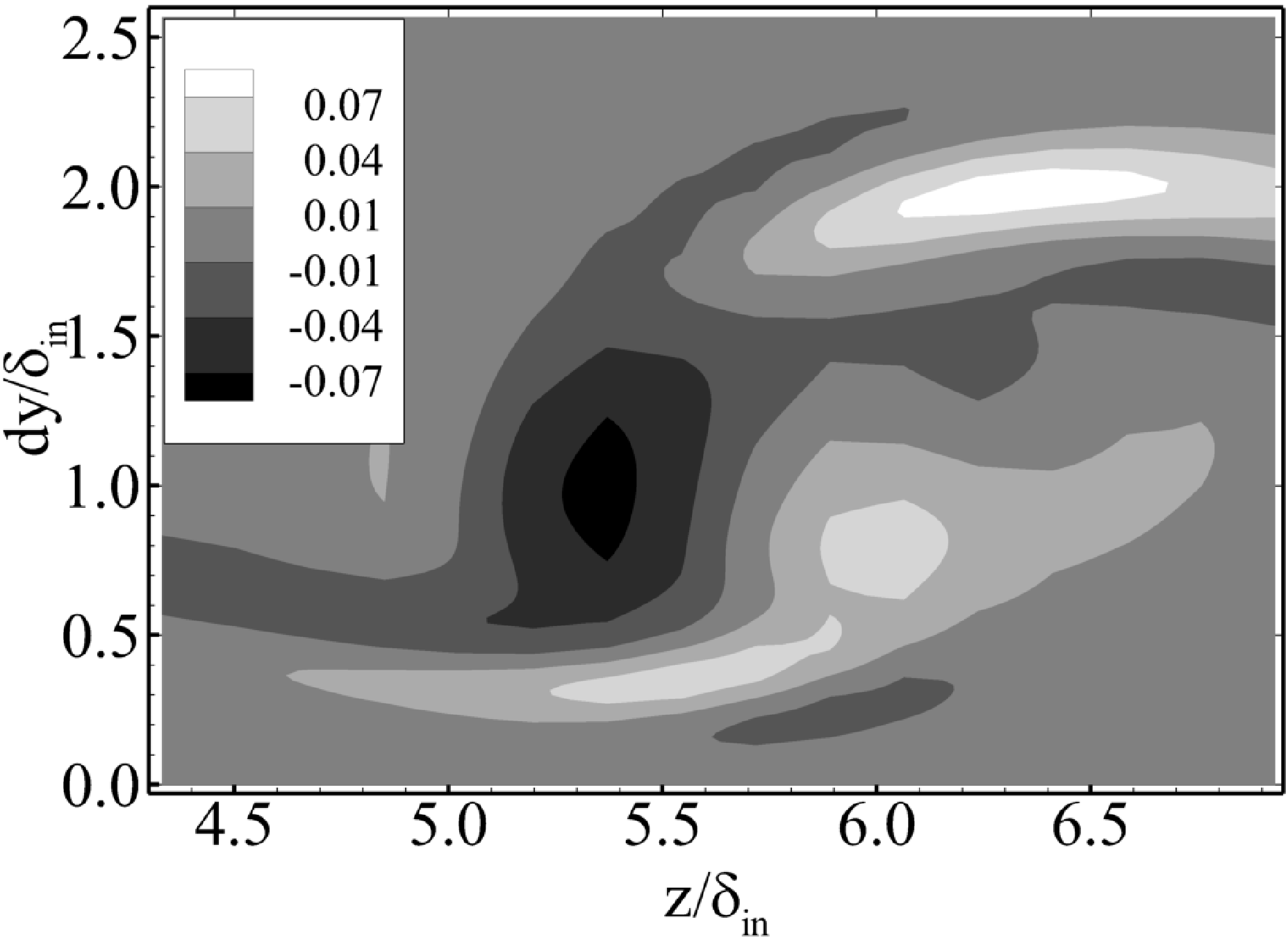}
\label{fig:56} 
\end{minipage}
\begin{minipage}{0.24\hsize}
\centering
\includegraphics[width=43mm]{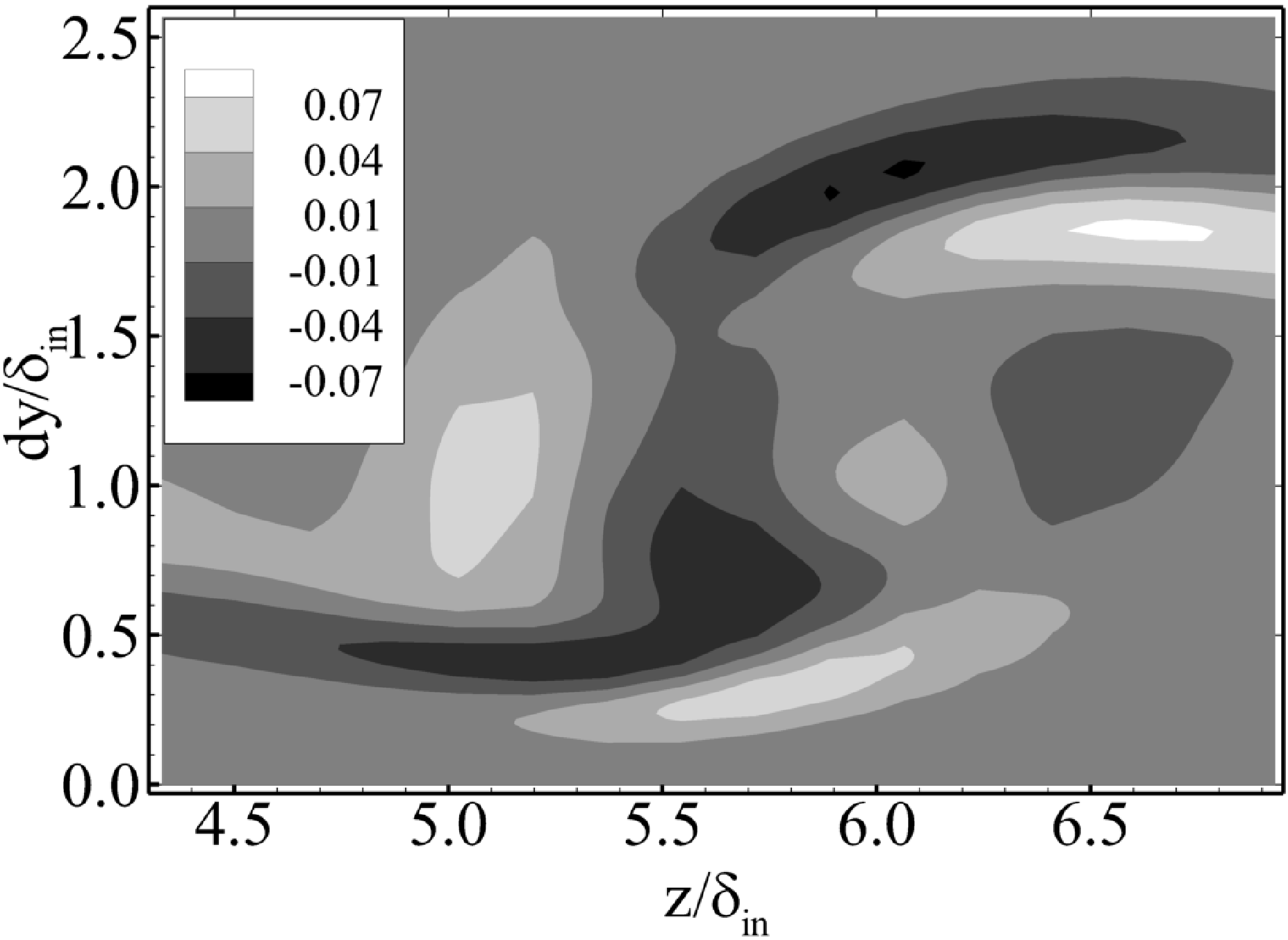}
\label{fig:57}
\end{minipage}
\begin{minipage}{0.24\hsize}
\centering
\includegraphics[width=43mm]{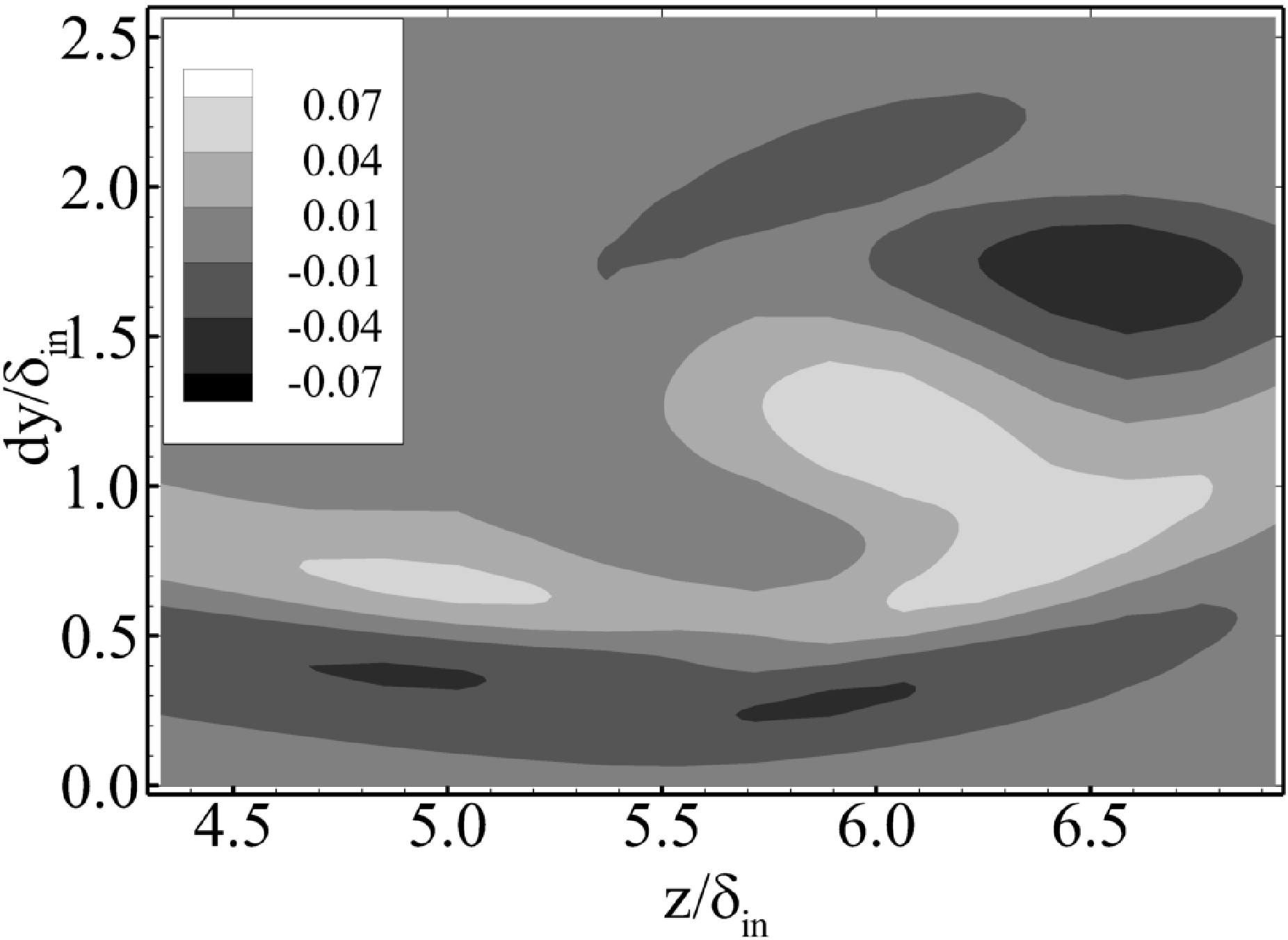}
\label{fig:58}
\end{minipage}
\begin{minipage}{0.24\hsize}
\centering
\includegraphics[width=43mm]{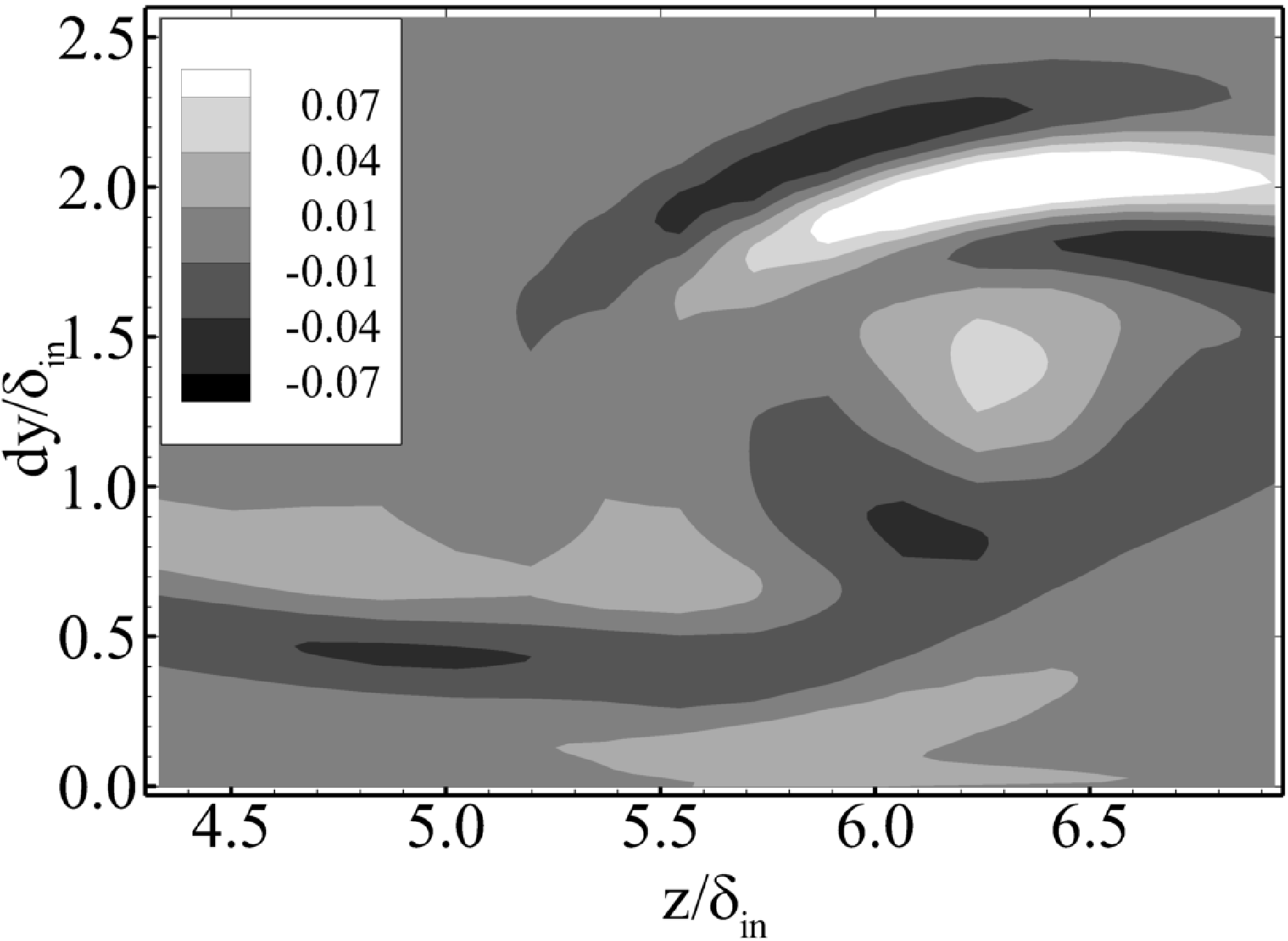}
\label{fig:59}
\end{minipage}
%
\\
\centering
(c) $t^\ast=63.8$\\
%
%
\caption{Top four eigenfunctions for $t^\ast$=2.13, 10.6 and 63.8. From the most left column to the most right column: 1st, 2nd, 3rd and 4th eigenfunctions, $dy=y-y_s$}
\end{figure}
%
%
\begin{figure}
%
\begin{center}
\includegraphics[width=90mm]{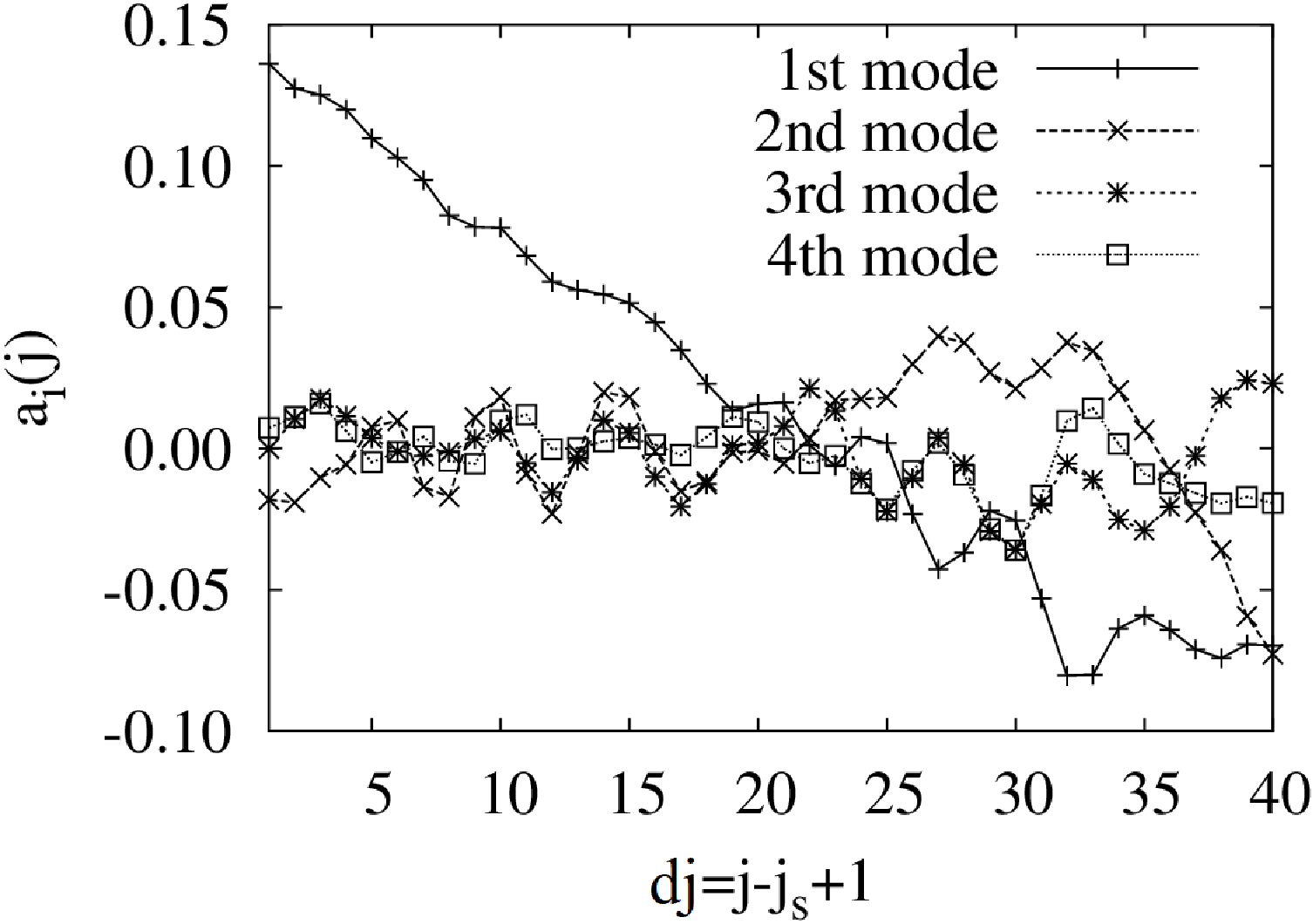}\\
(a)~$t^\ast=2.13$
\label{fig:60} 
\end{center}
%
\begin{center}
\includegraphics[width=90mm]{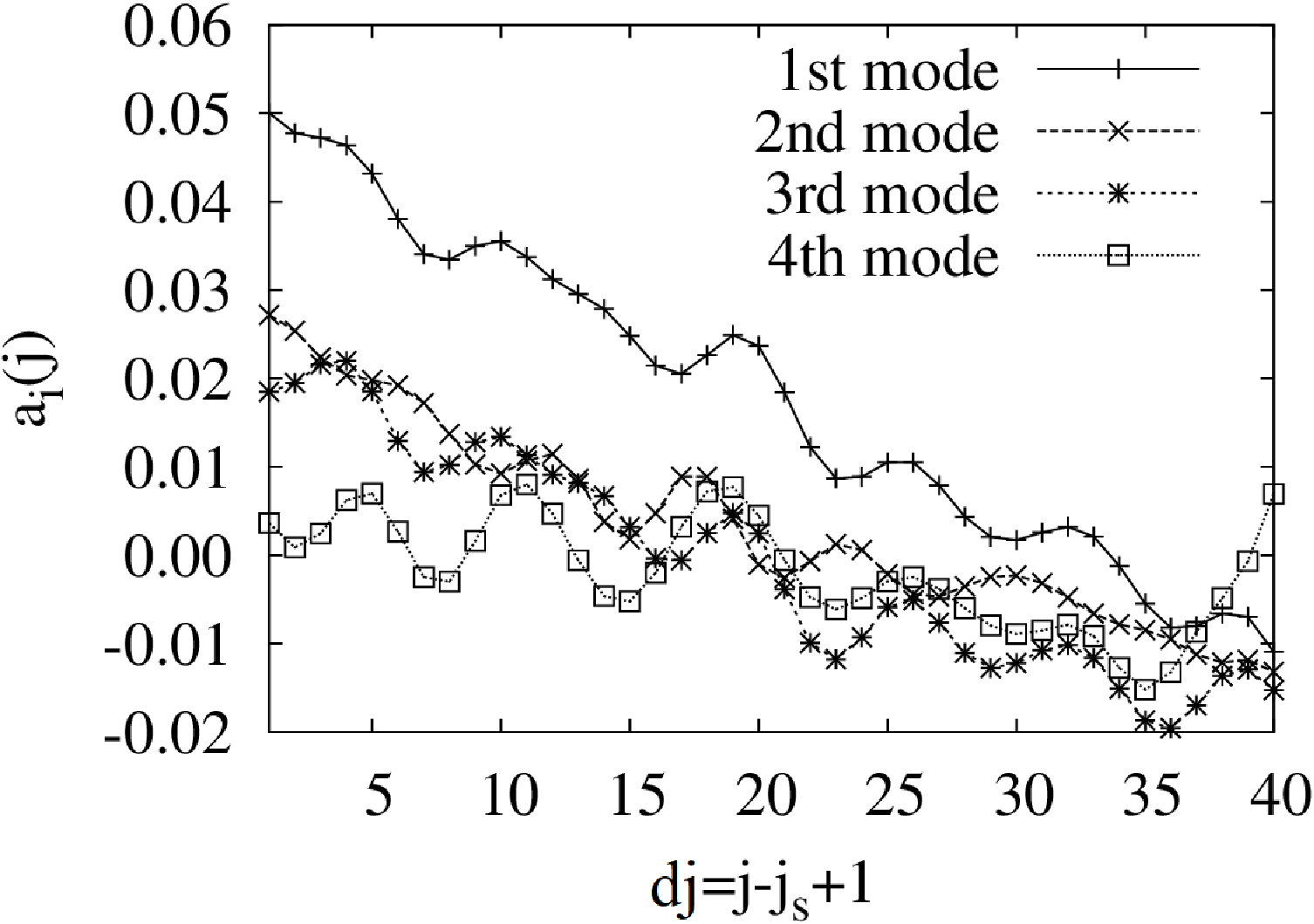}\\
(b)~$t^\ast=10.6$
\label{fig:61}
\end{center}
%
\begin{center}
\includegraphics[width=90mm]{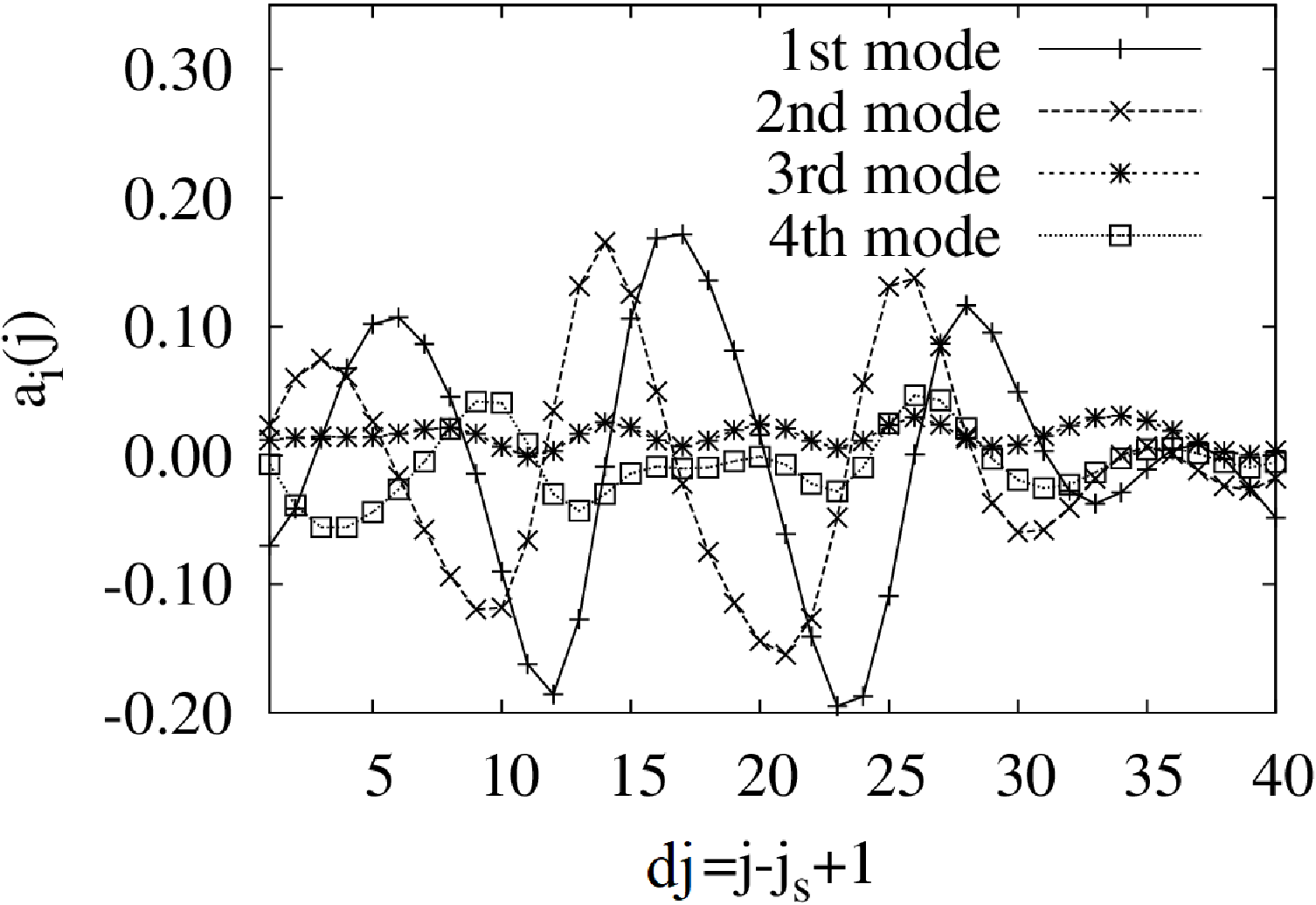}\\
(c)~$t^\ast=63.8$
\label{fig:62}
\end{center}
%
\begin{center}
\caption{Variation of the coefficients of the top four POD modes along $j$ for $t^\ast$=2.13, 10.6 and 63.8. $dj=j-j_s+1$.}
\end{center}
\end{figure}
Figure 17 shows the fractional energies of the first 15 dominant POD modes for these times. 
When $t^\ast$=2.13 and 10.6, the fractional energies of the first POD modes are more than four times larger than those of the second POD modes, 
which means that the spatial variation of $S_x$ along the vortex tube can be represented mainly by the 1st mode. 
In contrast, when $t^\ast$=63.8, the fractional energy, i.e., dominance, of the second mode increases, and becomes just 1.4 times smaller than that of the first mode. 

Figures 18 and 19 show respectively the eigenfunctions of the top four POD modes and the variation of the coefficients of the modes along $dj$ for
$t^\ast$=2.13, 10.6, 63.8. Here, $dj=j-j_s+1$.
When $t^\ast=2.13$, both regions of high and low $\pmb{\phi}_i(\cdot)$ are localized around the spanwise center,
i.e., $z/\delta_{in}=5.5$, in all the four eigenfunctions. 
The coefficient of the first mode changes significantly from positive to negative values along $j$. 
In contrast, the coefficients for other modes are relatively small and fluctuate around zero.
This distribution associated with the disturbance agrees with the fact that corkscrew-like structures that have spanwise variation are not observed in Fig. 11. 

At $t^\ast=10.6$, the corkscrew-like disturbance begins to form (Fig. 11(b)).
The regions of low and high fluctuation of $\pmb{\phi}_i(\cdot)$ corresponding to the upper and lower shear layers of the vortex tube, 
denoted as `A' and `B', respectively in Fig. 18(b), widen along with the vortex advection.
These high and low regions tend to be distributed near the periphery of the cross section of the vortex tube.
In the first and second modes' eigenfunction, the minimum and maximum peaks of $\pmb{\phi}_i(\cdot)$ are mainly
located in the right half of the cross section, i.e., $L_z/2 \le z \le L_z$.
The peaks in the third and fourth mode are almost at the center and in the left half, respectively.
Therefore, the combination of these four modes reproduces spanwise variation.
The variation in the coefficient of the second mode is almost opposite to that of the first mode for $dj \le 30$.
Although the coefficients of the third and fourth modes nearly follow the variation of the coefficient of the first mode,
slight differences in peak positions are also seen between the first and fourth modes. 

At $t^\ast=63.8$, the high and low regions intermingle, and become thin and distributed in the shear layer around the periphery of the cross section.
Differences in the $j$ locations between the coefficients of the first and second modes become almost constant (Fig. 19(c)).
The coefficients fluctuate around zero in contrast to those at earlier times $t^\ast=2.13$ and 10.6.
Thus, the structure of disturbance shifts from the variation localized near the spanwise center to the variation in the peripheral region.
This trend agrees with the development of the corkscrew-shaped periodic structure connecting the disturbances in the upper and lower shear layers through the circulation noted in Fig. 11.

From the analyses mentioned above, the development of the disturbances leading to hairpin vortex formation and its structure are clarified. 
Depending on the magnitude of $\Gamma$, the amplification or attenuation of the disturbances in the upper and lower shear layers, i.e., off-wall and near-wall modes, is determined. 
Based on the comparisons of results from DNS, LSA and POD, a corkscrew-like structure involving both near-wall and off-wall regions develops if disturbances in both shear layers are unstable. 

\section{Conclusions}
The evolution of a nearly straight vortical structure in the shear of a laminar boundary-layer flow and 
its response of the near-wall flows by varying the magnitude of circulation and the angle-to-wall were investigated by the DNS. 
Freestream Mach number is set to 0.5.
The circulation and the angle-to-wall of the vortex tube were varied over $\Gamma$=1.25-25.0 and fixed at $\phi=4^\circ$ and $10^\circ$.

When the circulation of the vortex tube is small, the vortex is either attenuated or elongated. 
With increasing circulation, multiple hairpin vortices align in the streamwise direction and evolve over the vortex tube. 
With a further increase, streamwise vortices of small cross-section appear and twist about each other. 
A hairpin arch is generated over the twisted vortices. 
In regard to shape, satellite structures, and streamwise vortices bundled under the arch, generated hairpin vortices exhibit dependencies on both the magnitude of circulation and the angle-to-wall. 

To understand the initial disturbances triggering the train of the hairpin vortices and the wavelengths of the disturbances, the LSA was performed. 
The validity of the present LSA in predicting the instability mechanism related to the wall-normal profile
was examined by the comparison with the DNS with regard to the disturbance growth in the off-wall and near-wall regions, its growth rates, 
and the distance between hairpin vortices when $\Gamma=12.5$ and $\phi=4^\circ$.
The contribution of the linear and nonlinear terms in the dynamics of the disturbance evolution was further clarified by the FNDA.
It is found that the linear stability mechanism of the wall-normal base profile has a great influence for the streamwise wavelengths of the sinuous waves,
and improvements in the agreement between the LSA and DNS were observed as time proceeds.
When the circulation is small, there is only one mode. 
When the circulation is sufficiently large, there are two unstable modes, i.e., `off-wall' and `near-wall'. 
The presence of these two modes leads to the generation of corkscrew-like disturbances around the vortex tube, and then becomes the precursor for the generation of hairpin vortices. 

The dominant structure of the disturbances appearing around the vortex tube as time proceeds was extracted by the POD. 
In the beginning stage of superposing the vortex tube on the background boundary layer profile, fluctuations are relatively localized below and above the vortex tube.
In the next stage, spanwise variation and the peripheral distribution of fluctuation appear.
Then, the corkscrew-shaped periodic structure connecting the disturbances in
the upper and lower shear layers through the circulation develops.

The number of hairpin vortices generated around the vortex tube depends on circumstances in which the vortex tube is situated inside the boundary layer such as circulation, 
angle-to-wall and the fraction of the vortex tube inside the boundary layer.  
When $\Gamma$ is increased, the curves of the eigenvalues with constant $\Gamma$ shift in the direction of larger $\omega_r$ and larger $\omega_i$. 
Thus, the distance between hairpin vortices decreased as $\Gamma$ is increased. 


Finally, it is found that the present DNS of the Hon \& Walker's model allowing the continuous variation of the characteristics of the vortex tube clarifies the new aspect of the nonlinear boundary-layer stability associated with the generation of hairpin vortices. 

\section{Acknowledgements}
This study was supported by the Sumitomo Foundation through a Grant for Basic Science Research Projects, the Ministry of Education, Culture, Sports, Science and Technology (MEXT) through a Grant-in-Aid for Scientific Research (C), 15K04759, 2015 and the Institute of Statistical Mathematics (ISM) Cooperative Research Program 2015 ISM-CRP2021 and 2016 ISM-CRP2019.  
Computational resources are provided by ISM and Japan Aerospace Exploration Agency (JAXA).

\appendix
\section{}\label{appA}

To explain the finite difference scheme used for the present LSA, the solution procedure is explained. 
In outline, the procedure consists of a rough computation of all eigenvalues by the global method, and a refinement of eigenvalues and eigenfunctions of unstable modes by the local method with Newton iteration \citep{Malik90}. 
Here, the focus was on only the stage of the global method. 
The stage of the local method is same as \citet{Malik90}.

The mesh used for the LSA is different from that used for the DNS. The LSA mesh
$\{y_j;0 \le y_j \le y_{max}, j \in \{0,\cdots,N\}\}$ is
generated by the following algebraic mapping of the equally divided computational domain
$\{\eta_j;0 \le \eta_j \le 1, j \in \{0,\cdots,N\}\}$.
\begin{equation}
y_j=\displaystyle \frac{a \eta_j}{b-\eta_j}
\end{equation}
where $b=1+a/y_{max}$ and $a=y_{max}y_i/(y_{max}-2y_i)$.
$y_i$ and $y_{max}$ are 3.0 and 8.0. With the meshed computational domain $\eta$, the linearized disturbance equation of
Navier-Stokes equation is discretized in second-order as
\begin{eqnarray}
&&f_1 A_j \displaystyle [\frac{\pmb{\phi}_{j+1}-2\pmb{\phi}_{j}+\pmb{\phi}_{j-1}}{\Delta \eta^2}]+d_1 \displaystyle [(f_2 A_j+f_3 B_j)(\frac{\pmb{\phi}_{j+1}-\pmb{\phi}_{j-1}}{2 \Delta \eta})+C_j \pmb{\phi}_j]+ \nonumber\\
&&d_2 \displaystyle [ f_3 B_j (\frac{\pmb{\phi}_{j+1/2}-\pmb{\phi}_{j-1/2}}{\Delta \eta})+C_j (\frac{\pmb{\phi}_{j+1/2}+\pmb{\phi}_{j-1/2}}{2})]=0~(j=1,\cdots,N-1).
\end{eqnarray}
Here, $\pmb{\phi}$ is a five-element vector defined by $(\hat{u}(y),\hat{v}(y),\hat{p}(y),\hat{T}(y),\hat{w}(y))^T$
where $\hat{u},\hat{v},\hat{w},\hat{p},\hat{T}$ are eigenfunctions of velocity fluctuations in $x,~y,~z$ directions,
pressure and temperature fluctuations, respectively. 
The variables are arranged on the mesh in a staggered manner. 
$\pmb{\phi}_j$ is the value of $\pmb{\phi}$ at $\eta_j=j/N,~j\in\{0,\cdots,N\}$ and has components $\phi_{kj}(k=1,\cdots,5)$.
$\Delta \eta=1/N$. 
$A_j$, $B_j$ and $C_j$ are 5$\times$5 matrices defined at $\eta_j$ whose elements are given in \citet{Malik90}.  
Also, $d_1=1,d_2=0$, except $d_1=0,d_2=1$ for the $\hat{p}$ component of $\pmb{\phi}$, and
\begin{equation}
f_1=\displaystyle \frac{(b-\eta_j)^4}{b^2 a^2},~f_2=-\displaystyle \frac{2(b-\eta_j)^3}{b^2 a^2},~f_3=\displaystyle \frac{(b-\eta_j)^2}{b a}.
\end{equation}
The first-order discretized continuity equation is
\begin{equation}
f_3 B_{j+1/2} \frac{\pmb{\phi}_{j+1}-\pmb{\phi}_{j}}{\Delta \eta}+C_{j+1/2}\pmb{\phi}_{j+1/2}=0~(j=0,N-1).
\end{equation}
The boundary conditions used are
\begin{equation}
y=0; \phi_{1,0}=\phi_{2,0}=\phi_{4,0}=\phi_{5,0}=0,
\end{equation}
\begin{equation}
y=y_{max};~\phi_{1,N}=\omega \epsilon \phi_{1,N},~\phi_{2,N}=\omega \epsilon \phi_{2,N},~\phi_{4,N}=\omega \epsilon \phi_{4,N},~\phi_{5,N}=\omega \epsilon \phi_{5,N},~\epsilon=1.
\end{equation}
Eqs. (A2) and (A4) along the eight boundary conditions (A5) and (A6) represent $5N+4$ equations for $5N+4$ unknowns. 
As the velocity and temperature disturbances are assumed to be identically zero at the solid boundary ($\eta_0=0$), 
the system reduces to $5N$ equations for $5N$ unknowns when these boundary conditions are imposed. 
The discretized compressible stability equations (A2) and (A4), along with the boundary conditions (A5) and (A6) are formulated as a matrix eigenvalue problem,
\begin{equation}
\bar{A}\pmb{\phi}=\omega \bar{B} \pmb{\phi}
\end{equation}
where $\omega$ is the eigenvalue and $\pmb{\phi}$ is the discrete representation of the eigenfunction. 
The $y$-directional derivatives of mean quantities such as
\begin{equation}
\displaystyle \frac{dU}{dy}, \frac{d^2U}{dy^2}, \frac{dT}{dy}, \frac{d^2T}{dy^2}, \frac{dW}{dy}, \frac{d^2W}{dy^2}
\end{equation}
appearing in $\bar{A}$ and $\bar{B}$ are evaluated by the following chain rule.
\begin{equation}
\displaystyle \frac{df}{dy}=(\frac{df}{d\eta})(\frac{d\eta}{dy})=(\frac{df}{d\eta})/(\frac{dy}{d\eta}).
\end{equation}
The second derivative is obtained by a differentiation of $\displaystyle (\frac{df}{dy})$.
$(\displaystyle \frac{df}{d\eta})$ and $(\displaystyle \frac{dy}{d\eta})$ are evaluated by the following 7th/8th-order finite difference schemes.
With prime (') denoting a differentiation by $\eta$, then for $j=0-3$,
\begin{eqnarray}
&&f'_0=-\displaystyle \frac{363}{140}f_0+7f_1-\frac{21}{2}f_2+\frac{35}{3}f_3-\frac{35}{4}f_4+\frac{21}{5}f_5-\frac{7}{6}f_6+\frac{1}{7}f_7, \nonumber\\
&&f'_1=-\displaystyle \frac{1}{8}f_0-\frac{223}{140}f_1+\frac{7}{2}f_2-\frac{7}{2}f_3+\frac{35}{12}f_4-\frac{7}{4}f_5+\frac{7}{10}f_6-\frac{1}{6}f_7+\frac{1}{56}f_8, \nonumber\\
&&f'_2=\displaystyle \frac{1}{56}f_0-\frac{2}{7}f_1-\frac{19}{20}f_2+2f_3-\frac{5}{4}f_4+\frac{2}{3}f_5-\frac{1}{4}f_6+\frac{2}{35}f_7-\frac{1}{168}f_8, \nonumber\\
&&f'_3=-\displaystyle \frac{1}{168}f_0+\frac{1}{14}f_1-\frac{1}{2}f_2-\frac{9}{20}f_3+\frac{5}{4}f_4-\frac{1}{2}f_5+\frac{1}{6}f_6-\frac{1}{28}f_7+\frac{1}{280}f_8.
\end{eqnarray}
For $j=4,N-4$
\begin{equation}
f'_j=\displaystyle \frac{1}{280}(f_{j-4}-f_{j+4})-\frac{4}{105}(f_{j-3}-f_{j+3})+\frac{1}{5}(f_{j-2}-f_{j+2})-\frac{4}{5}(f_{j-1}-f_{j+1}).
\end{equation}

For $j=N-3,N$,
\begin{eqnarray}
f'_{N-3}=&&\displaystyle \frac{1}{168}f_N-\frac{1}{14}f_{N-1}+\frac{1}{2}f_{N-2}+\nonumber\\
&&\frac{9}{20}f_{N-3}-\frac{5}{4}f_{N-4}+\frac{1}{2}f_{N-5}-\frac{1}{6}f_{N-6}+\frac{1}{28}f_{N-7}-\frac{1}{280}f_{N-8},\nonumber\\
f'_{N-2}=&&-\displaystyle \frac{1}{56}f_N+\frac{2}{7}f_{N-1}+\frac{19}{20}f_{N-2}-\nonumber\\
&&2f_{N-3}+\frac{5}{4}f_{N-4}-\frac{2}{3}f_{N-5}+\frac{1}{4}f_{N-6}-\frac{2}{35}f_{N-7}+\frac{1}{168}f_{N-8},\nonumber\\
f'_{N-1}=&&\displaystyle \frac{1}{8}f_N+\frac{223}{140}f_{N-1}-\frac{7}{2}f_{N-2}+\nonumber\\
&&\frac{7}{2}f_{N-3}-\frac{35}{12}f_{N-4}+\frac{7}{4}f_{N-5}-\frac{7}{10}f_{N-6}+\frac{1}{6}f_{N-7}-\frac{1}{56}f_{N-8},\nonumber\\
f'_{N}=&&\displaystyle \frac{363}{140}f_N-7f_{N-1}+\frac{21}{2}f_{N-2}-\nonumber\\
&&\frac{35}{3}f_{N-3}+\frac{35}{4}f_{N-4}-\frac{21}{5}f_{N-5}+\frac{7}{6}f_{N-6}-\frac{1}{7}f_{N-7}.
\end{eqnarray}

\section{}\label{appB}

In order to quantify the contribution of linear and nonlinear terms in the dynamics of the disturbance evolution, a FNDA is conducted concurrently with the DNS.
First, a general system of fully-nonlinear disturbance equations is derived.

The original non-dimensionalized compressible Navier-Stokes equations are:
\begin{eqnarray}
\displaystyle \frac{\partial \rho}{\partial t}+\displaystyle \frac{\partial (\rho u)}{\partial x}+\displaystyle \frac{\partial (\rho v)}{\partial y}+\displaystyle \frac{\partial (\rho w)}{\partial z}=0,
\end{eqnarray}
\begin{eqnarray}
\rho (\displaystyle \frac{\partial u}{\partial t}+u \displaystyle \frac{\partial u}{\partial x}+v \displaystyle \frac{\partial u}{\partial y}+w \displaystyle \frac{\partial u}{\partial z})=\nonumber\\
-\displaystyle \frac{\partial p}{\partial x}+
\displaystyle \frac{1}{Re} \displaystyle \frac{\partial}{\partial x}[2\mu \{-\frac{1}{3} (\nabla \cdot \pmb{u})+\frac{\partial u}{\partial x}\}]+\nonumber\\
\displaystyle \frac{1}{Re} \frac{\partial}{\partial y} \{\mu(\frac{\partial v}{\partial x}+\frac{\partial u}{\partial y})\}+
\displaystyle \frac{1}{Re} \frac{\partial}{\partial z} \{\mu(\frac{\partial u}{\partial z}+\frac{\partial w}{\partial x})\},
\end{eqnarray}
\begin{eqnarray}
\rho (\displaystyle \frac{\partial v}{\partial t}+u \displaystyle \frac{\partial v}{\partial x}+v \displaystyle \frac{\partial v}{\partial y}+w \displaystyle \frac{\partial v}{\partial z})=\nonumber\\
-\displaystyle \frac{\partial p}{\partial y}+\displaystyle \frac{1}{Re} \displaystyle \frac{\partial}{\partial x} \{\mu (\frac{\partial u}{\partial y}+\frac{\partial v}{\partial x})\}+\nonumber\\
\displaystyle \frac{1}{Re} \frac{\partial}{\partial y} [2\mu \{-\frac{1}{3} (\nabla \cdot \pmb{u})+\frac{\partial v}{\partial y}\}]+
\displaystyle \frac{1}{Re} \frac{\partial}{\partial z} \{\mu(\frac{\partial v}{\partial z}+\frac{\partial w}{\partial y})\},
\end{eqnarray}
\begin{eqnarray}
\rho (\displaystyle \frac{\partial w}{\partial t}+u \displaystyle \frac{\partial w}{\partial x}+v \displaystyle \frac{\partial w}{\partial y}+w \displaystyle \frac{\partial w}{\partial z})=\nonumber\\
-\displaystyle \frac{\partial p}{\partial z}+\displaystyle \frac{1}{Re} \displaystyle \frac{\partial}{\partial x} \{\mu (\frac{\partial w}{\partial x}+\frac{\partial u}{\partial z})\}+\nonumber\\
\displaystyle \frac{1}{Re} \frac{\partial}{\partial y} \{\mu(\frac{\partial v}{\partial z}+\frac{\partial w}{\partial y})\}+
\displaystyle \frac{1}{Re} \frac{\partial}{\partial z} [2\mu \{-\frac{1}{3} (\nabla \cdot \pmb{u})+\frac{\partial w}{\partial z}\}],
\end{eqnarray}
\begin{eqnarray}
\rho (\displaystyle \frac{\partial T}{\partial t}+u \displaystyle \frac{\partial T}{\partial x}+v \displaystyle \frac{\partial T}{\partial y}+w \displaystyle \frac{\partial T}{\partial z})=\nonumber\\
\displaystyle \frac{1}{RePr} \left[ 
\displaystyle \frac{\partial}{\partial x}(\mu \displaystyle \frac{\partial T}{\partial x})+
\displaystyle \frac{\partial}{\partial y}(\mu \displaystyle \frac{\partial T}{\partial y})+
\displaystyle \frac{\partial}{\partial z}(\mu \displaystyle \frac{\partial T}{\partial z})
\right]+\nonumber\\
(\gamma-1)M^2 \left[
\displaystyle \frac{\partial p}{\partial t}+
u \displaystyle \frac{\partial p}{\partial x}+
v \displaystyle \frac{\partial p}{\partial y}+
w \displaystyle \frac{\partial p}{\partial z}
\right]+\nonumber\\
\frac{(\gamma-1)M^2}{Re} \mu [
2 \{
(\frac{\partial u}{\partial x})^2+
(\frac{\partial v}{\partial y})^2+
(\frac{\partial w}{\partial z})^2\}+\nonumber\\
(\frac{\partial v}{\partial x}+\frac{\partial u}{\partial y} )^2+
(\frac{\partial w}{\partial y}+\frac{\partial v}{\partial z} )^2+
(\frac{\partial u}{\partial z}+\frac{\partial w}{\partial x} )^2
-\frac{2}{3} (\nabla \cdot \pmb{u})^2 ],
\end{eqnarray}
\begin{eqnarray}
%
\gamma M^2 p=\rho T.
\end{eqnarray}
Here, all lengths are scaled by $\delta_{in}$, velocities by $u_{\infty}$, density by $\rho_{\infty}$, pressure by $\rho_{\infty}u_{\infty}^2$,
temperature by $T_{\infty}$, time by $\delta_{in}/u_{\infty}$ and other variables by their corresponding boundary edge values.
The Mach number $M$ and Reynolds number $Re$ are defined as $M=u_{\infty}/\sqrt{\gamma R T_{\infty}}$ and $Re=\rho_{\infty} u_{\infty} \delta_{in}/\mu_{\infty}$, respectively.
The instantaneous values of velocities, $u,v,w$, pressure $p$, temperature $T$, density $\rho$, may be represented as the sum of 
a base and a fluctuation quantity, i.e.,
\begin{eqnarray}
\rho=\bar{\rho}(x,y,z)+\tilde{\rho},~~u=\bar{u}(x,y,z)+\tilde{u},~~v=\bar{v}(x,y,z)+\tilde{v},~~w=\bar{w}(x,y,z)+\tilde{w},\nonumber\\
T=\bar{T}(x,y,z)+\tilde{T},~~p=\bar{p}(x,y,z)+\tilde{p}=\frac{\bar{\rho}\bar{T}}{\gamma M^2}+\tilde{p},~~\mu=\bar{\mu}(x,y,z)+\tilde{\mu}.
\end{eqnarray}
After inserting the above decomposition to the original system of Navier-Stokes equations, the following general system of equations is derived for the set of the independent variables $(\tilde{\rho}, \tilde{u}, \tilde{v}, \tilde{w}, \tilde{T})$.
\begin{eqnarray}
\frac{\partial \tilde{\rho}}{\partial t}=\sum_{k=1}^{15} r_k,\nonumber\\
%
%
%
r_{1}=-\frac{\partial \bar{u}}{\partial x}\tilde{\rho},~
r_{2}=-\bar{u}\frac{\partial \tilde{\rho}}{\partial x},~
r_{3}=-\frac{\partial \bar{\rho}}{\partial x}\tilde{u},~
r_{4}=-\bar{\rho}\frac{\partial \tilde{u}}{\partial x},\nonumber\\
r_{5}=-\frac{\partial \bar{v}}{\partial y}\tilde{\rho},~
r_{6}=-\bar{v}\frac{\partial \tilde{\rho}}{\partial y},~
r_{7}=-\frac{\partial \bar{\rho}}{\partial y}\tilde{v},~
r_{8}=-\bar{\rho}\frac{\partial \tilde{v}}{\partial y},\nonumber\\
r_{9}=-\frac{\partial \bar{w}}{\partial z}\tilde{\rho},~
r_{10}=-\bar{w}\frac{\partial \tilde{\rho}}{\partial z},~
r_{11}=-\frac{\partial \bar{\rho}}{\partial z}\tilde{w},~
r_{12}=-\bar{\rho}\frac{\partial \tilde{w}}{\partial z},\nonumber\\
r_{13}=-\frac{\partial}{\partial x}(\tilde{\rho}\tilde{u}),~
r_{14}=-\frac{\partial}{\partial y}(\tilde{\rho}\tilde{v}),~
r_{15}=-\frac{\partial}{\partial z}(\tilde{\rho}\tilde{w}),
\end{eqnarray}
\begin{eqnarray}
\frac{\partial \tilde{u}}{\partial t}=\sum_{k=1}^{37} u_k,\nonumber\\
u_{1}=-\bar{u}\frac{\partial \tilde{u}}{\partial x},~
u_{2}=-\tilde{u}\frac{\partial \bar{u}}{\partial x},~
u_{3}=-\bar{v}\frac{\partial \tilde{u}}{\partial y},~\nonumber\\
u_{4}=-\tilde{v}\frac{\partial \bar{u}}{\partial y},~
u_{5}=-\bar{w}\frac{\partial \tilde{u}}{\partial z},~
u_{6}=-\tilde{w}\frac{\partial \bar{u}}{\partial z},~\nonumber\\
u_{7}=-\frac{\tilde{\rho}}{\bar{\rho}}\bar{u}\frac{\partial \bar{u}}{\partial x},~
u_{8}=-\frac{\tilde{\rho}}{\bar{\rho}}\bar{v}\frac{\partial \bar{u}}{\partial y},~
u_{9}=-\frac{\tilde{\rho}}{\bar{\rho}}\bar{w}\frac{\partial \bar{u}}{\partial z},~
u_{10}=-\frac{1}{\bar{\rho}}(\frac{\partial \tilde{p}}{\partial x})_L,\nonumber\\
u_{11}=\frac{1}{\bar{\rho}Re}\frac{\partial}{\partial x}[2\bar{\mu}\{-\frac{1}{3}(\nabla \cdot\ \tilde{\pmb{u}})+\frac{\partial \tilde{u}}{\partial x}\}],~\nonumber\\
u_{12}=\frac{1}{\bar{\rho}Re}\frac{\partial}{\partial y}\{\bar{\mu}(\frac{\partial \tilde{v}}{\partial x}+\frac{\partial \tilde{u}}{\partial y})\},~
u_{13}=\frac{1}{\bar{\rho}Re}\frac{\partial}{\partial z}\{\bar{\mu}(\frac{\partial \tilde{u}}{\partial z}+\frac{\partial \tilde{w}}{\partial x})\},~\nonumber\\
u_{14}=-\bar{u}\frac{\partial \bar{u}}{\partial x},~
u_{15}=-\bar{v}\frac{\partial \bar{u}}{\partial y},~
u_{16}=-\bar{w}\frac{\partial \bar{u}}{\partial z},~
u_{17}=-\frac{1}{\bar{\rho}}\frac{\partial \bar{p}}{\partial x},\nonumber\\
u_{18}=\frac{1}{\bar{\rho} Re}\frac{\partial}{\partial x}[2\bar{\mu}\{-\frac{1}{3}(\nabla \cdot\ \bar{\pmb{u}})+\frac{\partial \bar{u}}{\partial x}\}],~\nonumber\\
u_{19}=\frac{1}{\bar{\rho} Re}\frac{\partial}{\partial y}\{\bar{\mu}(\frac{\partial \bar{v}}{\partial x}+\frac{\partial \bar{u}}{\partial y})\},~
u_{20}=\frac{1}{\bar{\rho} Re}\frac{\partial}{\partial z}\{\bar{\mu}(\frac{\partial \bar{u}}{\partial z}+\frac{\partial \bar{w}}{\partial x})\},\nonumber\\
u_{21}=-\frac{\tilde{\rho}}{\bar{\rho}}\frac{\partial \tilde{u}}{\partial t},~
u_{22}=-\tilde{u}\frac{\partial \tilde{u}}{\partial x},~
u_{23}=-\tilde{v}\frac{\partial \tilde{u}}{\partial y},~
u_{24}=-\tilde{w}\frac{\partial \tilde{u}}{\partial z},\nonumber\\
u_{25}=-\frac{\tilde{\rho}}{\bar{\rho}}\bar{u}\frac{\partial \tilde{u}}{\partial x},~
u_{26}=-\frac{\tilde{\rho}}{\bar{\rho}}\tilde{u}\frac{\partial \bar{u}}{\partial x},~
u_{27}=-\frac{\tilde{\rho}}{\bar{\rho}}\tilde{u}\frac{\partial \tilde{u}}{\partial x},\nonumber\\
u_{28}=-\frac{\tilde{\rho}}{\bar{\rho}}\bar{v}\frac{\partial \tilde{u}}{\partial y},~
u_{29}=-\frac{\tilde{\rho}}{\bar{\rho}}\tilde{v}\frac{\partial \bar{u}}{\partial y},~
u_{30}=-\frac{\tilde{\rho}}{\bar{\rho}}\tilde{v}\frac{\partial \tilde{u}}{\partial y},\nonumber\\
%
u_{31}=-\frac{\tilde{\rho}}{\bar{\rho}}\bar{w}\frac{\partial \tilde{u}}{\partial z},~
u_{32}=-\frac{\tilde{\rho}}{\bar{\rho}}\tilde{w}\frac{\partial \bar{u}}{\partial z},~
u_{33}=-\frac{\tilde{\rho}}{\bar{\rho}}\tilde{w}\frac{\partial \tilde{u}}{\partial z},\nonumber\\
u_{34}=-\frac{1}{\bar{\rho} \gamma M^2}\frac{\partial}{\partial x}(\tilde{\rho}\tilde{T}),\nonumber\\
u_{35}=\frac{1}{\bar{\rho} Re}\frac{\partial}{\partial x}[2\tilde{\mu}\{
-\frac{1}{3}(\nabla \cdot\ \bar{\pmb{u}})
-\frac{1}{3}(\nabla \cdot\ \tilde{\pmb{u}})
+\frac{\partial \bar{u}}{\partial x}
+\frac{\partial \tilde{u}}{\partial x}
\}],\nonumber\\
u_{36}=\frac{1}{\bar{\rho} Re}\frac{\partial}{\partial y}\{
\tilde{\mu}(\frac{\partial \bar{v}}{\partial x}+\frac{\partial \bar{u}}{\partial y}+\frac{\partial \tilde{v}}{\partial x}+\frac{\partial \tilde{u}}{\partial y})\},\nonumber\\
u_{37}=\frac{1}{\bar{\rho} Re}\frac{\partial}{\partial z}\{
\tilde{\mu}(\frac{\partial \bar{u}}{\partial z}+\frac{\partial \bar{w}}{\partial x}+\frac{\partial \tilde{u}}{\partial z}+\frac{\partial \tilde{w}}{\partial x})\},
%
%
\end{eqnarray}
\begin{eqnarray}
\frac{\partial \tilde{v}}{\partial t}=\sum_{k=1}^{37} v_k,\nonumber\\
v_{1}=-\bar{u}\frac{\partial \tilde{v}}{\partial x},~
v_{2}=-\tilde{u}\frac{\partial \bar{v}}{\partial x},~
v_{3}=-\bar{v}\frac{\partial \tilde{v}}{\partial y},\nonumber\\
v_{4}=-\tilde{v}\frac{\partial \bar{v}}{\partial y},~
v_{5}=-\bar{w}\frac{\partial \tilde{v}}{\partial z},~
v_{6}=-\tilde{w}\frac{\partial \bar{v}}{\partial z},~\nonumber\\
v_{7}=-\frac{\tilde{\rho}}{\bar{\rho}}\bar{u}\frac{\partial \bar{v}}{\partial x},~
v_{8}=-\frac{\tilde{\rho}}{\bar{\rho}}\bar{v}\frac{\partial \bar{v}}{\partial y},~
v_{9}=-\frac{\tilde{\rho}}{\bar{\rho}}\bar{w}\frac{\partial \bar{v}}{\partial z},~
%
%
%
v_{10}=-\frac{1}{\bar{\rho}}(\frac{\partial \tilde{p}}{\partial y})_L,~\nonumber\\
v_{11}=\frac{1}{\bar{\rho} Re}\frac{\partial}{\partial x}\{\bar{\mu}(\frac{\partial \tilde{u}}{\partial y}+\frac{\partial \tilde{v}}{\partial x})\},\nonumber\\
v_{12}=\frac{1}{\bar{\rho} Re}\frac{\partial}{\partial y}[2\bar{\mu}\{-\frac{1}{3}(\nabla \cdot\ \tilde{\pmb{u}})+\frac{\partial \tilde{v}}{\partial y}\}],\nonumber\\
v_{13}=\frac{1}{\bar{\rho} Re}\frac{\partial}{\partial z}\{\bar{\mu}(\frac{\partial \tilde{v}}{\partial z}+\frac{\partial \tilde{w}}{\partial y})\},\nonumber\\
%
v_{14}=-\bar{u}\frac{\partial \bar{v}}{\partial x},~
v_{15}=-\bar{v}\frac{\partial \bar{v}}{\partial y},~
v_{16}=-\bar{w}\frac{\partial \bar{w}}{\partial z},~
v_{17}=-\frac{1}{\bar{\rho}}\frac{\partial \bar{p}}{\partial y},\nonumber\\
%
v_{18}=\frac{1}{\bar{\rho} Re}\frac{\partial}{\partial x}\{\bar{\mu}(\frac{\partial \bar{u}}{\partial y}+\frac{\partial \bar{v}}{\partial x})\},~\nonumber\\
v_{19}=\frac{1}{\bar{\rho} Re}\frac{\partial}{\partial y}[2\bar{\mu}\{-\frac{1}{3}(\nabla \cdot\ \bar{\pmb{u}})+\frac{\partial \bar{v}}{\partial y}\}],~\nonumber\\
v_{20}=\frac{1}{\bar{\rho} Re}\frac{\partial}{\partial z}\{\bar{\mu}(\frac{\partial \bar{v}}{\partial z}+\frac{\partial \bar{w}}{\partial y})\},~\nonumber\\
%
v_{21}=-\frac{\tilde{\rho}}{\bar{\rho}}\frac{\partial \tilde{v}}{\partial t},~
v_{22}=-\tilde{u}\frac{\partial \tilde{v}}{\partial x},~
v_{23}=-\tilde{v}\frac{\partial \tilde{v}}{\partial y},~
v_{24}=-\tilde{w}\frac{\partial \tilde{v}}{\partial z},~~\nonumber\\
%
%
v_{25}=-\frac{\tilde{\rho}}{\bar{\rho}}\bar{u}\frac{\partial \tilde{v}}{\partial x},~
v_{26}=-\frac{\tilde{\rho}}{\bar{\rho}}\tilde{u}\frac{\partial \bar{v}}{\partial x},~
v_{27}=-\frac{\tilde{\rho}}{\bar{\rho}}\tilde{u}\frac{\partial \tilde{v}}{\partial x},~\nonumber\\
%
v_{28}=-\frac{\tilde{\rho}}{\bar{\rho}}\bar{v}\frac{\partial \tilde{v}}{\partial y},~
v_{29}=-\frac{\tilde{\rho}}{\bar{\rho}}\tilde{v}\frac{\partial \bar{v}}{\partial y},~
v_{30}=-\frac{\tilde{\rho}}{\bar{\rho}}\tilde{v}\frac{\partial \tilde{v}}{\partial y},~\nonumber\\
%
v_{31}=-\frac{\tilde{\rho}}{\bar{\rho}}\bar{w}\frac{\partial \tilde{v}}{\partial z},~
v_{32}=-\frac{\tilde{\rho}}{\bar{\rho}}\tilde{w}\frac{\partial \bar{v}}{\partial z},~
v_{33}=-\frac{\tilde{\rho}}{\bar{\rho}}\tilde{w}\frac{\partial \tilde{v}}{\partial z},\nonumber\\
%
v_{34}=-\frac{1}{\bar{\rho} \gamma M^2}\frac{\partial}{\partial y}(\tilde{\rho}\tilde{T}),\nonumber\\
v_{35}=\frac{1}{\bar{\rho} Re}\frac{\partial}{\partial x}\{\tilde{\mu}(
\frac{\partial \bar{u}}{\partial y}+\frac{\partial \bar{v}}{\partial x}+
\frac{\partial \tilde{u}}{\partial y}+\frac{\partial \tilde{v}}{\partial x})\},\nonumber\\
v_{36}=\frac{1}{\bar{\rho} Re}\frac{\partial}{\partial y}[2\tilde{\mu}\{
-\frac{1}{3}(\nabla \cdot\ \bar{\pmb{u}})
-\frac{1}{3}(\nabla \cdot\ \tilde{\pmb{u}})
+\frac{\partial \bar{v}}{\partial y}
+\frac{\partial \tilde{v}}{\partial y}\}],\nonumber\\
v_{37}=\frac{1}{\bar{\rho} Re}\frac{\partial}{\partial z}\{\tilde{\mu}(
\frac{\partial \bar{v}}{\partial z}+\frac{\partial \bar{w}}{\partial y}+
\frac{\partial \tilde{v}}{\partial z}+\frac{\partial \tilde{w}}{\partial y})\},
%
%
%
\end{eqnarray}
\begin{eqnarray}
\frac{\partial \tilde{w}}{\partial t}=\sum_{k=1}^{37} w_k,\nonumber\\
w_{1}=-\bar{u}\frac{\partial \tilde{w}}{\partial x},~
w_{2}=-\tilde{u}\frac{\partial \bar{w}}{\partial x},~
w_{3}=-\bar{v}\frac{\partial \tilde{w}}{\partial y},\nonumber\\
w_{4}=-\tilde{v}\frac{\partial \bar{w}}{\partial y},~
w_{5}=-\bar{w}\frac{\partial \tilde{w}}{\partial z},~
w_{6}=-\tilde{w}\frac{\partial \bar{w}}{\partial z},\nonumber\\
%
w_{7}=-\frac{\tilde{\rho}}{\bar{\rho}}\bar{u}\frac{\partial \bar{w}}{\partial x},~
w_{8}=-\frac{\tilde{\rho}}{\bar{\rho}}\bar{v}\frac{\partial \bar{w}}{\partial y},~
w_{9}=-\frac{\tilde{\rho}}{\bar{\rho}}\bar{w}\frac{\partial \bar{w}}{\partial z},~
w_{10}=-\frac{1}{\bar{\rho}}(\frac{\partial \tilde{p}}{\partial z})_L,\nonumber\\
%
w_{11}=\frac{1}{\bar{\rho} Re}\frac{\partial}{\partial x}\{\bar{\mu}(\frac{\partial \tilde{u}}{\partial z}+\frac{\partial \tilde{w}}{\partial x})\},~
w_{12}=\frac{1}{\bar{\rho} Re}\frac{\partial}{\partial y}\{\bar{\mu}(\frac{\partial \tilde{v}}{\partial z}+\frac{\partial \tilde{w}}{\partial y})\},\nonumber\\
w_{13}=\frac{1}{\bar{\rho} Re}\frac{\partial}{\partial z}[2\bar{\mu}\{-\frac{1}{3}(\nabla \cdot\ \tilde{\pmb{u}})+\frac{\partial \tilde{w}}{\partial z}\}],\nonumber\\
%
w_{14}=-\bar{u}\frac{\partial \bar{w}}{\partial x},~
w_{15}=-\bar{v}\frac{\partial \bar{w}}{\partial y},~
w_{16}=-\bar{w}\frac{\partial \bar{w}}{\partial z},~
w_{17}=-\frac{1}{\bar{\rho}}\frac{\partial \bar{p}}{\partial z},\nonumber\\
%
%
w_{18}=\frac{1}{\bar{\rho} Re}\frac{\partial}{\partial x}\{\bar{\mu}(\frac{\partial \bar{w}}{\partial x}+\frac{\partial \bar{u}}{\partial z})\},~
w_{19}=\frac{1}{\bar{\rho} Re}\frac{\partial}{\partial y}\{\bar{\mu}(\frac{\partial \bar{v}}{\partial z}+\frac{\partial \bar{w}}{\partial y})\},\nonumber\\
w_{20}=\frac{1}{\bar{\rho} Re}\frac{\partial}{\partial z}[2\bar{\mu}\{-\frac{1}{3}(\nabla \cdot\ \bar{\pmb{u}})+\frac{\partial \bar{w}}{\partial z}\}],\nonumber\\
%
w_{21}=-\frac{\tilde{\rho}}{\bar{\rho}}\frac{\partial \tilde{w}}{\partial t},~
%
w_{22}=-\tilde{u}\frac{\partial \tilde{w}}{\partial x},~
w_{23}=-\tilde{v}\frac{\partial \tilde{w}}{\partial y},~
w_{24}=-\tilde{w}\frac{\partial \tilde{w}}{\partial z},\nonumber\\
w_{25}=-\frac{\tilde{\rho}}{\bar{\rho}}\bar{u}\frac{\partial \tilde{w}}{\partial x},~
w_{26}=-\frac{\tilde{\rho}}{\bar{\rho}}\tilde{u}\frac{\partial \bar{w}}{\partial x},~
w_{27}=-\frac{\tilde{\rho}}{\bar{\rho}}\tilde{u}\frac{\partial \tilde{w}}{\partial x},\nonumber\\
%
w_{28}=-\frac{\tilde{\rho}}{\bar{\rho}}\bar{v}\frac{\partial \tilde{w}}{\partial y},~
w_{29}=-\frac{\tilde{\rho}}{\bar{\rho}}\tilde{v}\frac{\partial \bar{w}}{\partial y},~
w_{30}=-\frac{\tilde{\rho}}{\bar{\rho}}\tilde{v}\frac{\partial \tilde{w}}{\partial y},\nonumber\\
%
w_{31}=-\frac{\tilde{\rho}}{\bar{\rho}}\bar{w}\frac{\partial \tilde{w}}{\partial z},~
w_{32}=-\frac{\tilde{\rho}}{\bar{\rho}}\tilde{w}\frac{\partial \bar{w}}{\partial z},~
w_{33}=-\frac{\tilde{\rho}}{\bar{\rho}}\tilde{w}\frac{\partial \tilde{w}}{\partial z},\nonumber\\
%
w_{34}=-\frac{1}{\bar{\rho} \gamma M^2}\frac{\partial}{\partial z}(\tilde{\rho}\tilde{T}),\nonumber\\
%
w_{35}=\frac{1}{\bar{\rho} Re}\frac{\partial}{\partial x}\{\tilde{\mu}(
\frac{\partial \bar{w}}{\partial x}+\frac{\partial \bar{u}}{\partial z}+
\frac{\partial \tilde{w}}{\partial x}+\frac{\partial \tilde{u}}{\partial z})\},\nonumber\\
%
w_{36}=\frac{1}{\bar{\rho} Re}\frac{\partial}{\partial y}\{\tilde{\mu}(
\frac{\partial \bar{v}}{\partial z}+\frac{\partial \bar{w}}{\partial y}+
\frac{\partial \tilde{v}}{\partial z}+\frac{\partial \tilde{w}}{\partial y})\},\nonumber\\
%
w_{37}=\frac{1}{\bar{\rho} Re}\frac{\partial}{\partial z}[2\tilde{\mu}\{
-\frac{1}{3}(\nabla \cdot\ \bar{\pmb{u}})
-\frac{1}{3}(\nabla \cdot\ \tilde{\pmb{u}})
+\frac{\partial \bar{w}}{\partial z}
+\frac{\partial \tilde{w}}{\partial z}\}],\nonumber\\
\end{eqnarray}
\begin{eqnarray*}
\frac{\partial \tilde{T}}{\partial t}=\sum_{k=1}^{78} T_k,\nonumber\\
%
T_{1}=-\bar{u}\frac{\partial \tilde{T}}{\partial x},~
T_{2}=-\tilde{u}\frac{\partial \bar{T}}{\partial x},~
T_{3}=-\bar{v}\frac{\partial \tilde{T}}{\partial y},\nonumber\\
T_{4}=-\tilde{v}\frac{\partial \bar{T}}{\partial y},~
T_{5}=-\bar{w}\frac{\partial \tilde{T}}{\partial z},~
T_{6}=-\tilde{w}\frac{\partial \bar{T}}{\partial z},\nonumber\\
T_{7}=-\frac{\tilde{\rho}}{\bar{\rho}}\bar{u}\frac{\partial \bar{T}}{\partial x},~
T_{8}=-\frac{\tilde{\rho}}{\bar{\rho}}\bar{v}\frac{\partial \bar{T}}{\partial y},~
T_{9}=-\frac{\tilde{\rho}}{\bar{\rho}}\bar{w}\frac{\partial \bar{T}}{\partial z},\nonumber\\
%
T_{10}=\frac{1}{\bar{\rho}RePr}\frac{\partial}{\partial x}(\bar{\mu}\frac{\partial \tilde{T}}{\partial x}),~
T_{11}=\frac{1}{\bar{\rho}RePr}\frac{\partial}{\partial y}(\bar{\mu}\frac{\partial \tilde{T}}{\partial y}),~
T_{12}=\frac{1}{\bar{\rho}RePr}\frac{\partial}{\partial z}(\bar{\mu}\frac{\partial \tilde{T}}{\partial z}),\nonumber\\
%
T_{13}=\frac{(\gamma-1)M^2}{\bar{\rho}}\frac{\partial \tilde{p}}{\partial t},~\nonumber\\
T_{14}=\frac{(\gamma-1)M^2}{\bar{\rho}}\bar{u}(\frac{\partial \tilde{p}}{\partial x})_L,~
T_{15}=\frac{(\gamma-1)M^2}{\bar{\rho}}\tilde{u}\frac{\partial \bar{p}}{\partial x},~\nonumber\\
T_{16}=\frac{(\gamma-1)M^2}{\bar{\rho}}\bar{v}(\frac{\partial \tilde{p}}{\partial y})_L,~
T_{17}=\frac{(\gamma-1)M^2}{\bar{\rho}}\tilde{v}\frac{\partial \bar{p}}{\partial y},~\nonumber\\
T_{18}=\frac{(\gamma-1)M^2}{\bar{\rho}}\bar{w}(\frac{\partial \tilde{p}}{\partial z})_L,~
T_{19}=\frac{(\gamma-1)M^2}{\bar{\rho}}\tilde{w}\frac{\partial \bar{p}}{\partial z},~\nonumber\\
%
%
%
%
T_{20}=\frac{(\gamma-1)M^2}{\bar{\rho}Re}4\bar{\mu}\frac{\partial \bar{u}}{\partial x}\frac{\partial \tilde{u}}{\partial x},~
T_{21}=\frac{(\gamma-1)M^2}{\bar{\rho}Re}4\bar{\mu}\frac{\partial \bar{v}}{\partial y}\frac{\partial \tilde{v}}{\partial y},~
T_{22}=\frac{(\gamma-1)M^2}{\bar{\rho}Re}4\bar{\mu}\frac{\partial \bar{w}}{\partial z}\frac{\partial \tilde{w}}{\partial z},~\nonumber\\
T_{23}=\frac{(\gamma-1)M^2}{\bar{\rho}Re}2\bar{\mu}(\frac{\partial \bar{v}}{\partial x}+\frac{\partial \bar{u}}{\partial y})(\frac{\partial \tilde{v}}{\partial x}+\frac{\partial \tilde{u}}{\partial y}),~\nonumber\\
T_{24}=\frac{(\gamma-1)M^2}{\bar{\rho}Re}2\bar{\mu}(\frac{\partial \bar{w}}{\partial y}+\frac{\partial \bar{v}}{\partial z})(\frac{\partial \tilde{w}}{\partial y}+\frac{\partial \tilde{v}}{\partial z}),~\nonumber\\
T_{25}=\frac{(\gamma-1)M^2}{\bar{\rho}Re}2\bar{\mu}(\frac{\partial \bar{u}}{\partial z}+\frac{\partial \bar{w}}{\partial x})(\frac{\partial \tilde{u}}{\partial z}+\frac{\partial \tilde{w}}{\partial x}),~\nonumber\\
T_{26}=\frac{(\gamma-1)M^2}{\bar{\rho}Re}4\bar{\mu}\{-\frac{1}{3}(\nabla \cdot\ \bar{\pmb{u}})(\nabla \cdot\ \tilde{\pmb{u}})\},~\nonumber\\
T_{27}=-\bar{u}\frac{\partial \bar{T}}{\partial x},~
T_{28}=-\bar{v}\frac{\partial \bar{T}}{\partial y},~
T_{29}=-\bar{w}\frac{\partial \bar{T}}{\partial z},~\nonumber\\
T_{30}=\frac{1}{\bar{\rho}RePr}\frac{\partial}{\partial x}(\bar{\mu}\frac{\partial \bar{T}}{\partial x}),~
T_{31}=\frac{1}{\bar{\rho}RePr}\frac{\partial}{\partial y}(\bar{\mu}\frac{\partial \bar{T}}{\partial y}),~
T_{32}=\frac{1}{\bar{\rho}RePr}\frac{\partial}{\partial z}(\bar{\mu}\frac{\partial \bar{T}}{\partial z}),~\nonumber\\
%
T_{33}=\frac{(\gamma-1)M^2}{\bar{\rho}}\bar{u}\frac{\partial \bar{p}}{\partial x},~
T_{34}=\frac{(\gamma-1)M^2}{\bar{\rho}}\bar{v}\frac{\partial \bar{p}}{\partial y},~
T_{35}=\frac{(\gamma-1)M^2}{\bar{\rho}}\bar{w}\frac{\partial \bar{p}}{\partial z},~\nonumber\\
%
T_{36}=\frac{(\gamma-1)M^2}{\bar{\rho}Re}2\bar{\mu}(\frac{\partial \bar{u}}{\partial x})^2,~
T_{37}=\frac{(\gamma-1)M^2}{\bar{\rho}Re}2\bar{\mu}(\frac{\partial \bar{v}}{\partial y})^2,~\nonumber\\
T_{38}=\frac{(\gamma-1)M^2}{\bar{\rho}Re}2\bar{\mu}(\frac{\partial \bar{w}}{\partial z})^2,~\nonumber\\
T_{39}=\frac{(\gamma-1)M^2}{\bar{\rho}Re}\bar{\mu}(\frac{\partial \bar{v}}{\partial x}+\frac{\partial \bar{u}}{\partial y})^2,~
T_{40}=\frac{(\gamma-1)M^2}{\bar{\rho}Re}\bar{\mu}(\frac{\partial \bar{w}}{\partial y}+\frac{\partial \bar{v}}{\partial z})^2,~\nonumber\\
T_{41}=\frac{(\gamma-1)M^2}{\bar{\rho}Re}\bar{\mu}(\frac{\partial \bar{u}}{\partial z}+\frac{\partial \bar{w}}{\partial x})^2,~\nonumber\\
T_{42}=\frac{(\gamma-1)M^2}{\bar{\rho}Re}2\bar{\mu}\{-\frac{1}{3}(\nabla \cdot\ \bar{\pmb{u}})^2\}, \nonumber\\
\end{eqnarray*}
\begin{eqnarray}
%
T_{43}=-\frac{\tilde{\rho}}{\bar{\rho}}\frac{\partial \tilde{T}}{\partial t},~
T_{44}=-\tilde{u}\frac{\partial \tilde{T}}{\partial x},~
T_{45}=-\tilde{v}\frac{\partial \tilde{T}}{\partial y},~
T_{46}=-\tilde{w}\frac{\partial \tilde{T}}{\partial z},~\nonumber\\
%
%
T_{47}=-\frac{\tilde{\rho}}{\bar{\rho}}\bar{u}\frac{\partial \tilde{T}}{\partial x},~
T_{48}=-\frac{\tilde{\rho}}{\bar{\rho}}\tilde{u}\frac{\partial \bar{T}}{\partial x},~
T_{49}=-\frac{\tilde{\rho}}{\bar{\rho}}\tilde{u}\frac{\partial \tilde{T}}{\partial x},\nonumber\\
%
T_{50}=-\frac{\tilde{\rho}}{\bar{\rho}}\bar{v}\frac{\partial \tilde{T}}{\partial y},~
T_{51}=-\frac{\tilde{\rho}}{\bar{\rho}}\tilde{v}\frac{\partial \bar{T}}{\partial y},~
T_{52}=-\frac{\tilde{\rho}}{\bar{\rho}}\tilde{v}\frac{\partial \tilde{T}}{\partial y},\nonumber\\
%
T_{53}=-\frac{\tilde{\rho}}{\bar{\rho}}\bar{w}\frac{\partial \tilde{T}}{\partial z},~
T_{54}=-\frac{\tilde{\rho}}{\bar{\rho}}\tilde{w}\frac{\partial \bar{T}}{\partial z},~
T_{55}=-\frac{\tilde{\rho}}{\bar{\rho}}\tilde{w}\frac{\partial \tilde{T}}{\partial z},\nonumber\\
%
T_{56}=\frac{1}{\bar{\rho} Re Pr}\frac{\partial}{\partial x}\{\tilde{\mu}(\frac{\partial \bar{T}}{\partial x}+\frac{\partial \tilde{T}}{\partial x})\},~
T_{57}=\frac{1}{\bar{\rho} Re Pr}\frac{\partial}{\partial y}\{\tilde{\mu}(\frac{\partial \bar{T}}{\partial y}+\frac{\partial \tilde{T}}{\partial y})\},~\nonumber\\
T_{58}=\frac{1}{\bar{\rho} Re Pr}\frac{\partial}{\partial z}\{\tilde{\mu}(\frac{\partial \bar{T}}{\partial z}+\frac{\partial \tilde{T}}{\partial z})\},\nonumber\\
%
T_{59}=\frac{\gamma-1}{\bar{\rho}\gamma}\bar{u}\frac{\partial}{\partial x}(\tilde{\rho}\tilde{T}),~
T_{60}=\frac{\gamma-1}{\bar{\rho}\gamma}\bar{v}\frac{\partial}{\partial y}(\tilde{\rho}\tilde{T}),~
T_{61}=\frac{\gamma-1}{\bar{\rho}\gamma}\bar{w}\frac{\partial}{\partial z}(\tilde{\rho}\tilde{T}),\nonumber\\
%
T_{62}=\frac{(\gamma-1)M^2}{\bar{\rho}}\tilde{u}\frac{\partial \tilde{p}}{\partial x},~
T_{63}=\frac{(\gamma-1)M^2}{\bar{\rho}}\tilde{v}\frac{\partial \tilde{p}}{\partial y},~
T_{64}=\frac{(\gamma-1)M^2}{\bar{\rho}}\tilde{w}\frac{\partial \tilde{p}}{\partial z},\nonumber\\
%
%
T_{65}=\frac{(\gamma-1)M^2}{\bar{\rho}Re}2\bar{\mu}(\frac{\partial \tilde{u}}{\partial x})^2,~
T_{66}=\frac{(\gamma-1)M^2}{\bar{\rho}Re}2\bar{\mu}(\frac{\partial \tilde{v}}{\partial y})^2,~\nonumber\\
T_{67}=\frac{(\gamma-1)M^2}{\bar{\rho}Re}2\bar{\mu}(\frac{\partial \tilde{w}}{\partial z})^2,\nonumber\\
%
T_{68}=\frac{(\gamma-1)M^2}{\bar{\rho}Re}2\tilde{\mu}(\frac{\partial \bar{u}}{\partial x}+\frac{\partial \tilde{u}}{\partial x})^2,~
T_{69}=\frac{(\gamma-1)M^2}{\bar{\rho}Re}2\tilde{\mu}(\frac{\partial \bar{v}}{\partial y}+\frac{\partial \tilde{v}}{\partial y})^2,~\nonumber\\
T_{70}=\frac{(\gamma-1)M^2}{\bar{\rho}Re}2\tilde{\mu}(\frac{\partial \bar{w}}{\partial z}+\frac{\partial \tilde{w}}{\partial z})^2,~\nonumber\\
%
T_{71}=\frac{(\gamma-1)M^2}{\bar{\rho}Re}\bar{\mu}(\frac{\partial \tilde{v}}{\partial x}+\frac{\partial \tilde{u}}{\partial y})^2,~
T_{72}=\frac{(\gamma-1)M^2}{\bar{\rho}Re}\bar{\mu}(\frac{\partial \tilde{u}}{\partial z}+\frac{\partial \tilde{w}}{\partial x})^2,~\nonumber\\
T_{73}=\frac{(\gamma-1)M^2}{\bar{\rho}Re}\bar{\mu}(\frac{\partial \tilde{w}}{\partial y}+\frac{\partial \tilde{v}}{\partial z})^2,~\nonumber\\
%
T_{74}=\frac{(\gamma-1)M^2}{\bar{\rho}Re}\tilde{\mu}(\frac{\partial \bar{v}}{\partial x}+\frac{\partial \bar{u}}{\partial y}+\frac{\partial \tilde{v}}{\partial x}+\frac{\partial \tilde{u}}{\partial y})^2,~\nonumber\\
T_{75}=\frac{(\gamma-1)M^2}{\bar{\rho}Re}\tilde{\mu}(\frac{\partial \bar{u}}{\partial z}+\frac{\partial \bar{w}}{\partial x}+\frac{\partial \tilde{u}}{\partial z}+\frac{\partial \tilde{w}}{\partial x})^2,~\nonumber\\
T_{76}=\frac{(\gamma-1)M^2}{\bar{\rho}Re}\tilde{\mu}(\frac{\partial \bar{w}}{\partial y}+\frac{\partial \bar{v}}{\partial z}+\frac{\partial \tilde{w}}{\partial y}+\frac{\partial \tilde{v}}{\partial z})^2,~\nonumber\\
%
T_{77}=\frac{(\gamma-1)M^2}{\bar{\rho}Re}2\bar{\mu}\{-\frac{1}{3}(\nabla \cdot\ \tilde{\pmb{u}})^2\},~
T_{78}=\frac{(\gamma-1)M^2}{\bar{\rho}Re}2\tilde{\mu}\{-\frac{1}{3}(\nabla \cdot\ \bar{\pmb{u}}+\nabla \cdot\ \tilde{\pmb{u}})^2\}.\nonumber\\
\end{eqnarray}
Here, $\bar{p}$ and $\tilde{p}$ are constructed from $\bar{\rho},\tilde{\rho},\bar{T}$ and $\tilde{T}$.
Because viscosity is evaluated by Sutherland's formula as mentioned in Subsection 2.1, $\mu, \bar{\mu}$ and $\tilde{\mu}$ are evaluated as 
\begin{eqnarray}
\mu(T)=T^{1.5}\frac{1+\frac{S}{T_{\infty}}}{T+\frac{S}{T_{\infty}}},~\bar{\mu}(\bar{T})=\bar{T}^{1.5}\frac{1+\frac{S}{T_{\infty}}}{\bar{T}+\frac{S}{T_{\infty}}},~\tilde{\mu}=\mu(T)-\bar{\mu}(\bar{T}),
~S=111.0.
\end{eqnarray} 
$(\frac{\partial \tilde{p}}{\partial x})_L,(\frac{\partial \tilde{p}}{\partial y})_L,(\frac{\partial \tilde{p}}{\partial z})_L$ are defined as 
the linear parts of $(\frac{\partial \tilde{p}}{\partial x}), (\frac{\partial \tilde{p}}{\partial y}), (\frac{\partial \tilde{p}}{\partial z})$, respectively. For example, 
\begin{eqnarray}
(\frac{\partial \tilde{p}}{\partial x})_L=(\frac{\partial \tilde{p}}{\partial x})-\frac{1}{\gamma M^2}\frac{\partial}{\partial x}(\tilde{\rho}\tilde{T})
=\frac{1}{\gamma M^2}(
\frac{\partial \bar{\rho}}{\partial x}\tilde{T}+
\bar{\rho}\frac{\partial \tilde{T}}{\partial x}+
\frac{\partial \tilde{\rho}}{\partial x}\bar{T}+
\tilde{\rho}\frac{\partial \bar{T}}{\partial x}).
\end{eqnarray}

In this study, the base quantities $\{\bar{\rho},\bar{u},\bar{v},\bar{w},\bar{T},\bar{p},\bar{\mu}\}$ are functions of only $y$ as a special case, and therefore the $x$ and $z$ derivatives of the base quantities vanish.     
This system is just a splitting of the original system of the compressible Navier-Stokes equations, and no modelling is introduced in its derivation.
\end{document}